\documentclass[11pt,a4paper,twoside]{book}

\usepackage{comment}

\usepackage[
  pdftitle={Chiral-Transport-Induced Collective Modes in Strong Magnetic Fields and Their Implications for Neutron Star Phenomenology},
  pdfauthor={Sota Hanai},
  bookmarks=false,
  hidelinks
]{hyperref}

\usepackage{graphicx,color,hyperref}
\usepackage{tikz}
\usepackage[compat=1.1.0]{tikz-feynhand}
\usepackage{multirow}
\usepackage{subfigure}
\usepackage{wrapfig}
\usepackage{here}

\usepackage{amsmath,amssymb}
\usepackage{bm}
\usepackage{braket}
\usepackage{slashed}

\usepackage{ascmac}
\usepackage{makeidx}
\usepackage[noadjust]{cite}
\usepackage[top=25truemm,bottom=20truemm,left=20truemm,right=20truemm]{geometry}
\makeatletter
\renewcommand{\chaptermark}[1]{%
  \markboth{\@chapapp\ \thechapter.\ #1}{}%
}

\makeatother

\usepackage{xcolor}
\hypersetup{
    colorlinks=true,
    citecolor=teal,
    linkcolor=blue,
    urlcolor=blue,
    }

\newcommand{\im}{\mathrm{i}}
\newcommand{\e}{\mathrm{e}}
\newcommand{\dif}{\mathrm{d}}
\newcommand{\del}{\partial} 
\newcommand{\tr}{\mathrm{tr}}

\newcommand{\vev}[1]{\left\langle #1 \right\rangle}

\newcommand{\bs}{\boldsymbol}
\newcommand{\fr}{\frac}
\newcommand{\lb}{\left}
\newcommand{\rb}{\right}

\newcommand{\mcl}{\mathcal}

\newcommand{\nom}{\nonumber \\}

\makeindex

\begin{document}
\begin{titlepage}
\centering 
\vspace*{5cm}
\begin{huge}
Chiral-Transport-Induced Collective Modes in\\[0.1em]
Strong Magnetic Fields and Their Implications for \\[0.25em]
Neutron Star Phenomenology    
\end{huge}
\\[10cm]
{\Large February 2026}
\\[2cm]
{\huge  Sota Hanai}

\end{titlepage}
\thispagestyle{empty}

\begin{titlepage}
\centering 
\vspace*{2cm}
{\Large A Thesis for the Degree of Ph.D. in Science} \\[3cm]
\begin{huge}
Chiral-Transport-Induced Collective Modes in\\[0.1em]
Strong Magnetic Fields and Their Implications for \\[0.25em]
Neutron Star Phenomenology    
\end{huge}
\\[7cm]
{\Large February 2026}
\\[2cm]
{\Large Science and Technology}
\\[1em]
{\Large Keio University}
\\[1cm]
{\huge  Sota Hanai}

\end{titlepage}

\thispagestyle{empty}

\frontmatter
\begin{flushleft}
    {\Huge{\textbf{Abstract}}}
\end{flushleft}
\vspace{4em}
Neutron stars are compact objects with both a high-density environment and strong magnetic fields.
Elucidating their internal structure remains an important problem spanning particle physics, nuclear physics, and astrophysics.
In recent years, motivated by detections of the gravitational waves from binary neutron star mergers, asteroseismology has attracted attention.
In asteroseismology, seismic oscillations and gravitational waves are classified according to their physical origins.
Meanwhile, in the context of particle and nuclear physics, it is known that the chirality of relativistic fermions leads to chiral transport phenomena.
One of the prototypical examples is the chiral magnetic effect, in which electric currents are generated along a magnetic field.
This further induces a density wave propagating along the field called the chiral magnetic wave.

In this thesis, noting that quark matter inside strongly magnetized neutron stars and neutrino matter inside rotating supernova cores are many-body systems of relativistic fermions, we first perform a hydrodynamic analysis incorporating the chiral transport.
In both cases, we theoretically predict the emergence of new types of seismic oscillations, resulting in gravitational wave emission.
In particular, the frequency of the gravitational wave from the quark matter reflects the information of the internal magnetic fields of the neutron star, providing an observational probe.
Also, finite quark masses cause the chirality flipping, which suppresses the chiral transport. Using kinetic theory, we derive the damping rate of the seismic oscillations and evaluate the observationally possible range of the frequency.

Second, we study the response of a magnetized relativistic collisionless plasma to dynamical electromagnetic fields. To properly incorporate the chiral anomaly, we derive the dispersion relation of the collective mode using the chiral kinetic theory in the weak magnetic field limit and employing the lowest Landau level approximation in the strong magnetic field limit. As a result, we find that a dynamical screening distinct from the conventional Landau damping appears. Furthermore, we show that the relaxation time parametrically varies with the magnetic field strength and that dynamical magnetic fields are partially screened. The former modifies the temperature and magnetic field dependence of the shear viscosity. We then suggest that in the intermediate magnetic field region, the r-mode instability driven by gravitational wave emission from neutron star seismic oscillations may be more likely to occur.

\tableofcontents

\mainmatter

\chapter{Introduction}
Revealing the internal structure of neutron stars, in which dense matter exceeding the nuclear density $\rho_0\sim10^{14}~{\rm g/cm^3}$ exists, is an important problem spanning particle physics, nuclear physics, and astrophysics.
Neutrons were confirmed by Chadwick in 1932~\cite{Chadwick:1932wcf}, and only two years later, in 1934, Baade and Zwicky theoretically predicted the existence of neutron stars~\cite{Baade:1934wuu}.
A neutron star was first observed about thirty years later.
In 1967, Bell, then a graduate student, detected regular pulsed signals from space~\cite{Hewish:1968bj}.
Because the signals were remarkably regular, they were initially suspected to originate from extraterrestrial intelligent life, and the radio source was nicknamed “Little Green Men.”
It was soon turned out, however, that the source was a star emitting periodic pulsed electromagnetic radiation called the pulsar.
Subsequently, the pulsar was identified as a rapidly rotating neutron star emitting beams of electromagnetic radiation~\cite{Gold:1968zf}.
Just as the light from a lighthouse appears to blink when viewed from a distance, a rotating neutron star emitting beams of radiation appears to pulse.
Such rapid rotation is possible because neutron stars are extremely dense objects.

The core of a neutron star is believed to reach such an extreme density that even nucleons can no longer maintain their individual identities, and quarks are expected to be deconfined from within the nucleons.
The dynamics of these quarks, together with the gluons that mediate the strong force, are described by quantum chromodynamics (QCD), which is the fundamental theory of the strong interaction.
A remarkable property of QCD is asymptotic freedom, namely, the fact that the coupling constant becomes progressively weaker as the characteristic energy scale increases~\cite{Gross:1973id,Politzer:1973fx}.
The intrinsic energy scale of QCD is set by $\Lambda_{\rm QCD}\simeq200,{\rm MeV}$, which characterizes the onset of nonperturbative phenomena such as confinement.
At baryon densities sufficiently higher than this scale, weak-coupling analyses become applicable, and they indicate the emergence of nontrivial phases of matter, including various forms of color superconductivity~\cite{Barrois:1977xd,Bailin:1979nh,Alford:1997zt,Rapp:1997zu,Alford:1998mk}.
Although such phases have not been directly observed, they have important implications for the properties of neutron stars.
At lower densities, however, the strong-coupling nature of QCD severely limits our ability to perform controlled analytical calculations directly from the fundamental theory.
In addition, first-principles numerical approaches based on lattice QCD suffer from the notorious sign problem at finite baryon density, which makes direct simulations in this regime extremely challenging.
As a result, the properties of strongly interacting matter in this intermediate-density region remain largely unexplored, despite their potential relevance to neutron star physics.
Understanding this regime therefore continues to be a central and outstanding problem in nuclear and particle physics.

However, to understand the fundamental nature of such dense matter, it is essential to investigate not only the equilibrium properties but also the non-equilibrium transport phenomena.
Since these responses are closely related to observable quantities such as the neutron star cooling rate, studying them can provide observational ways to verify the theoretical predictions.
In particular, a neutron star is a many-body system of relativistic fermions including electrons, quarks, and neutrinos.
The chirality of such relativistic fermions plays a crucial role in transport phenomena and has attracted considerable attention.
A prototypical example induced by a magnetic field is the chiral magnetic effect (CME)~\cite{Vilenkin:1980fu,Nielsen:1983rb,Alekseev:1998ds,Fukushima:2008xe}, in which an electric current is generated along the magnetic field in the presence of a chirality imbalance.
Similarly, fluid vorticity can induce a number current, known as the chiral vortical effect (CVE)~\cite{Vilenkin:1979ui,Son:2009tf,Landsteiner:2011cp}.
These effects are related to the chiral anomaly~\cite{Adler:1969gk,Bell:1969ts,Fujikawa:1980eg} and represent macroscopic manifestations of quantum phenomena.
Such chiral transport can lead to various collective excitations as well.
In the presence of a magnetic field, a density wave known as the chiral magnetic wave (CMW) propagates along the field direction.
Even without a net chirality imbalance, the CMW can arise due to the chiral separation effect (CSE)~\cite{Son:2004tq,Metlitski:2005pr}, whereby fluctuations of the number density and the chiral charge density couple each other.
The CVE also gives rise to the chiral vortical wave (CVW)~\cite{Jiang:2015cva}.%
\footnote{Other examples of the collective excitations include the chiral Alfvén wave~\cite{Yamamoto:2015ria}, chiral heat wave~\cite{Chernodub:2015gxa}, and chiral shock wave~\cite{Sen:2016jzl}.}
In plasmas with a chirality imbalance, the collective modes of dynamical electromagnetic fields can exhibit an instability known as the chiral plasma instability (CPI)~\cite{Akamatsu:2013pjd,Akamatsu:2014yza}.
The collective behavior of magnetized relativistic plasmas, including dynamical electromagnetic effects, has been investigated within hydrodynamics~\cite{Rybalka:2018uzh,Shovkovy:2018tks}.

Observations of neutron stars are indispensable for testing theoretical predictions of dense matter.
Historically, such observations began with radio astronomy through the detection of pulsars, which established neutron stars as astrophysical objects.
With the advent of X-ray astronomy in the 1970s, neutron stars also became accessible through their high-energy electromagnetic emissions, providing important information on their surface properties and accretion phenomena.
In addition to electromagnetic observations, neutrinos offer another crucial probe of neutron star physics.
The detection of neutrinos from the supernova explosion SN~1987A~\cite{Kamiokande-II:1987idp} marked the beginning of neutrino astronomy and demonstrated the potential of neutrinos as messengers of dense and hot astrophysical environments.
Since neutron stars emit a large amount of neutrinos during their cooling processes, neutrino observations are also expected to play an important role in probing their internal dynamics and thermal evolution.
More recently, gravitational-wave observations have opened a new window on neutron stars.
In 2015, gravitational waves from a binary black hole merger were detected for the first time by the Laser Interferometer Gravitational-Wave Observatory (LIGO)~\cite{LIGOScientific:2016aoc}.
This was followed in 2017 by the first detection of gravitational waves from a binary neutron star merger~\cite{LIGOScientific:2017vwq}, firmly establishing gravitational-wave astronomy as a realistic and powerful observational tool for neutron stars.
Until recently, studies of neutron stars primarily relied on electromagnetic waves such as X-rays and gamma rays.
However, with the rapid development of gravitational-wave astronomy, together with advances in neutrino observations, we have entered an era in which neutron stars can be explored using a variety of observational probes.
In this context, so-called multi-messenger astronomy has become a central framework for neutron star studies, combining information from electromagnetic waves, neutrinos, and gravitational waves.
Within this framework, asteroseismology, studying stellar oscillations and their associated gravitational or electromagnetic emissions, is regarded as a particularly promising method to probe the internal structure and dynamics of compact stars.

Motivated by these developments, this thesis investigates collective excitations induced by chiral transport phenomena in strongly magnetized relativistic plasmas and explores their implications for neutron star phenomenology.
Through this study, we aim to bridge the microscopic quantum aspects of chirality with the macroscopic dynamics of dense matter, such as seismic oscillations.
In neutron stars, relativistic quarks under strong magnetic fields can induce the CMW as a new type of seismic oscillation.
In rotating supernova cores, neutrino matter can give rise to the CVW.
We thus explore a novel framework of asteroseismology incorporating chiral effects, which we call ``chiral asteroseismology.''

Collective excitations in relativistic plasmas can occur not only in the hydrodynamic regime but also in the collisionless regime~\cite{Gorsky:2012gi,Stephanov:2014dma}, as described by the chiral kinetic theory~\cite{Son:2012wh,Stephanov:2012ki,Son:2012zy}.
Previous studies have investigated the zero sound of chiral fermions in static background magnetic fields and chiral chemical potentials.
In this thesis, we extend these analyses by including the dynamical electromagnetic fields and fluctuations of the chiral charge density in magnetized collisionless plasmas.
In particular, the presence of a dynamical electric field affects the chiral charge through the chiral anomaly in a magnetic background.

This thesis is organized as follows.
In Chap.~\ref{chap:QCD}, we briefly review QCD, starting from its construction and discussing color superconductivity.
In Chap.~\ref{chap:Chi_Ph}, we overview the chiral transport phenomena, introducing the chiral anomaly and reproducing the derivations of the CME and CVE.
We also review the related collective modes and the chiral kinetic theory.
Chap.~\ref{chap:NS_Astero} provides an introduction to neutron stars and asteroseismology, including their fundamental properties such as mass, radius, and magnetic field, followed by the basic concepts of stellar oscillations.
In Chap.~\ref{chap:Chi_Astero}, we develop a framework for chiral asteroseismology, focusing on quark matter in magnetized neutron stars and neutrino matter in rotating supernovae.
We demonstrate that new types of seismic oscillations and gravitational waves can arise in these systems.
We also discuss chirality flipping induced by the quark mass, as well as the angular dependence of gravitational-wave emission.
In Chap.~\ref{chap:Anom_Dyn_Res}, we investigate the dynamical electromagnetic response of relativistic magnetized plasmas in the weak and strong magnetic field limits.
Employing chiral kinetic theory in the weak field regime and the lowest Landau level (LLL) approximation in the strong field regime, we show that collective modes propagating perpendicular to the background magnetic field receive corrections from the back reaction induced by the chiral anomaly.
We demonstrate that the resulting collective modes exhibit a distinctive form of dynamical screening.
We further discuss the implications of this anomalous dynamical screening for neutron star phenomenology.
Finally, our conclusions and outlook are presented in Chap.~\ref{chap:Summary}.

Throughout this thesis, we adopt natural units with $\hbar=c=k_{\rm B}=\varepsilon_0=\mu_0=1$.

\chapter{Quantum Chromodynamics}
\label{chap:QCD}
In this chapter, we review the fundamental properties of QCD, with a particular focus on dense QCD.
In Sec.~\ref{sec:QCD:Lagangian}, we introduce the QCD Lagrangian and discuss its symmetries.
In Sec.~\ref{sec:QCD:CSC}, we provide a brief overview of color superconductivity.

\section{Lagrangian}
\label{sec:QCD:Lagangian}
The Lagrangian is the most fundamental object that determines the dynamics of the system.
Here, we construct the QCD Lagrangian based on the gauge principle.

\subsection{Gauge principle}
\label{subsec:Gauge_principle}
For humans to describe nature, a gauge such as a coordinate system needs to be introduced.
The freedom in choosing the reference should be understood as a redundant degree of freedom, distinct from the actual physical degrees of freedom.
The laws of nature should be expressed in a form that is independent of these gauge choices.
This very natural requirement is known as the gauge principle.
A theory based on the gauge principle is called a gauge theory.
By imposing the gauge principle, the interaction between gauge fields and matter fields is uniquely determined at low energy, which is why a gauge theory can be regarded as a theory of fundamental forces.

Here, let us note the terminology.
It is sometimes said that ``gauge symmetry" refers to the invariance under gauge transformation.
However, this usage can be misleading.
The concept of gauge symmetry must be clearly distinguished from that of symmetry in the usual sense.
While the usual symmetry can be broken, a physical theory and its observables must always be invariant under gauge transformations~\cite{Elitzur:1975im}.

\subsection{Construction of Lagrangian}
QCD is a gauge theory describing the strong interactions of quarks.
The quarks have various quantum numbers.
One is the spin because the quarks are fermions.
Another is the color charge: red, green, and blue.
The strong force acts on particles with the color charge.
The other is the flavor: up, down, strange, charm, bottom, and top.
The mass of each quark increases from the up quark to the top quark.
While the strange quark mass is about $100~{\rm MeV}$, the charm quark mass is beyond $1~{\rm GeV}$.
The charm, bottom, and top quarks are then called heavy quarks.
In this thesis, we do not treat the heavy quarks and limit the flavor to the up, down, and strange, implying that we set the heavy quark masses to be infinite.
From the above, we can write the quark field as $q^{a}_{\alpha i}(x)$ with the spinor index $\alpha=1,\cdots,4$, the color index $a=1,\cdots, N_{\rm c}$, and the flavor index $i=1,\cdots, N_{\rm f}$.
In the following, we omit these indices otherwise needed.

We now construct the QCD Lagrangian based on the SU(3) gauge invariance.
The SU(3) gauge transformation acts in color space.
Using the SU(3) generators $t^{a}~(a=1,2,\cdots,8)$,%
\footnote{Quarks are in the fundamental representation of SU(3), while the generators are in the adjoint representation. However, we use the same index $a$ to denote their indices.}
we can write the SU(3) gauge transformation as
\begin{align}
    q(x)\to U(x)q(x),
    \qquad
    U(x)\equiv\e^{-\im \theta^{a}(x)t^{a}}\,,
\end{align}
where $\theta^a(x)$ is a local parameter.
The generators satisfy the algebraic relation with the structure constant $f^{abc}$: $[t^{a},t^{b}]=\im f^{abc}t^{c}$.
We impose the normalization $\tr(t^at^b)=\delta^{ab}/2$.
Since the gauge transformation is defined locally, we have to introduce a connection to compare the fields at different points, which is the gauge field (or gluon field) $A_{\mu}^a(x)$.
Corresponding to the number of the generators, there are eight gluons.
To maintain the gauge invariance, we have to replace the derivative with the covariant derivative,
\begin{align}
    D_{\mu}
    =
    \del_{\mu}+\im gA_{\mu}^{a}t^{a}
    \equiv \del_{\mu}+\im gA_{\mu}\,.
\end{align}
where $g$ is the strong coupling constant.
The SU(3) transformation of the gauge field is
\begin{align}
    A_{\mu}
    &\to
    UA_{\mu}U^{\dagger}-\fr{\im}{g}U(\del_{\mu}U^{\dagger})\,.
\end{align}
The covariant derivative transforms as $D_{\mu}\to UD_{\mu}U^{\dagger}$.
We can write down the gauge invariant Lagrangian,
\begin{align}
    \mcl{L}
    =
    \bar{q}(\im\slashed{D}-M)q\,,
\end{align}
where $\bar{q}\equiv q^{\dagger}\gamma^{0}, \slashed{D}\equiv\gamma^{\mu}D_{\mu}$, and $\gamma^{\mu}$ is the Dirac gamma matrix.
In the case of $N_{\rm f}=3$, the mass matrix $M$ is
\begin{align}
M=\lb(
    \begin{array}{ccc}
    m_{\rm u} & 0 & 0 \\
    0 & m_{\rm d} & 0 \\
    0 & 0 & m_{\rm s}
    \end{array}
    \rb)\,,
\end{align}
with  $m_{\rm u}, m_{\rm d}, m_{\rm s}$ being up, down, and strange quark masses respectively.

We next consider the Lagrangian describing the dynamics of the gluons.
We define the field strength as
\begin{align}
    G_{\mu\nu}
    \equiv
    -\frac{\im}{g}[D_{\mu},D_{\nu}]
    =
    \del_{\mu}A_{\nu}-\del_{\nu}A_{\mu}-gf^{abc}A_{\mu}^{a}A_{\nu}^{b}t^{c}\,.
\end{align}
The field strength represents the curvature of the gauge field, analogous to the electromagnetic field in quantum electrodynamics (QED).
Unlike the Abelian U(1) gauge field, the non-Abelian nature of SU(3) gives rise to interaction terms between the gluons.
Physically, this is because gluons carry color charge, whereas photons do not carry electric charge.
The field strength transforms as $G_{\mu\nu}\to UG_{\mu\nu}U^{\dagger}$ and is not gauge invariant.
To make it gauge invariant, we have to take the trace in color space.
Considering the Lorentz symmetry as well, the Lagrangian describing the dynamics of the gauge field is given by
\begin{align}
    \mcl{L}
    =
    -\frac{1}{2}\tr(G_{\mu\nu}G^{\mu\nu})
    =
    -\frac{1}{4}G_{\mu\nu}^{a}G^{a\mu\nu}\,.
\end{align}
From the above, we finally obtain the QCD Lagrangian,
\begin{align}
\label{eq:QCD:QCD_Lagrangian}
    \mcl{L}_{\rm QCD}
    =
    -\frac{1}{2}\tr(G_{\mu\nu}G^{\mu\nu})
    +\bar{q}(\im\gamma^{\mu}D_{\mu}-M)q\,.
\end{align}

\subsection{Symmetry of QCD}
\label{subsec:QCD:QCD_symmetry}
We overview the symmetry of QCD.
The QCD Lagrangian (\ref{eq:QCD:QCD_Lagrangian}) is invariant under the U(1)$_{\rm B}$ transformation,
\begin{align}
    q
    \to
    \e^{-\im\theta_{\rm B}}q\,,
\end{align}
where $\theta_{\rm B}$ is a constant parameter.
This symmetry corresponds to the baryon number conservation.
The baryon number is a quantity in which quarks carry a value of $+1/3$, antiquarks carry $-1/3$, and represents the number of nucleons, such as protons and neutrons.

We introduce the chirality operator $\gamma_{5}\equiv \im\gamma^0\gamma^1\gamma^2\gamma^3$.
By the property of the gamma matrix, we have $(\gamma_{5})^2=1$, implying that the eigenvalue is $\lambda=\pm1$.
According to each eigenvalue, we can define the right-handed fermion $(\lambda=+1~\text{or R})$ and left-handed fermion $(\lambda=-1~\text{or L})$.%
\footnote{We use $\lambda=\pm 1$ for mathematical operations, while writing $\lambda=\text{R, L}$ for labeling the chirality.}
Using the projection operator,
\begin{align}
\label{eq:QCD:Chirality_projection_op}
    \mcl{P}_{\lambda}
    \equiv
    \fr{1+\lambda\gamma_5}{2}\,,
\end{align}
the right- and left-handed fermions are expressed as $q_{\lambda}=\mcl{P}_{\lambda}q$.
In terms of the right- and left-handed fermions, the matter part of the QCD Lagrangian~(\ref{eq:QCD:QCD_Lagrangian}) can be written as
\begin{align}
\label{eq:QCD:QCD_Lagrangian_matter_RL}
     \mcl{L}
    =
    \bar{q}_{\rm R}\im\slashed{D}q_{\rm R}
    +\bar{q}_{\rm L}\im\slashed{D}q_{\rm L}
    -\bar{q}_{\rm R}Mq_{\rm L}
    -\bar{q}_{\rm L}Mq_{\rm R}\,.
\end{align}
As we can see from this expression, the mass term mixes the right- and left-handed fermions.
Consequently, the QCD Lagrangian is not invariant under the U(1)$_{\rm A}$ transformation,
\begin{align}
    q\to \e^{-\im\theta_{\rm A}\gamma_{5}}q\,,
\end{align}
and the ${\rm SU}_{\rm R}(3)\times{\rm SU}_{\rm L}(3)$ transformation,
\begin{align}
    q_{\lambda}\to V_{\lambda}q_{\lambda}\,,
    \qquad
    V_{\lambda}\equiv\e^{-\im\theta_{\lambda}^it^i}\,,
\end{align}
where $\theta_{\rm A},\theta_{\lambda}^i$ are constant parameters.
Note that $t^i$ is the ${\rm SU}(3)$ generator in flavor space.

In the massless limit ($M=0$), QCD possesses the U(1)$_{\rm A}$ classically and the chiral symmetries.
Although the real quarks have a finite mass $(m_{\rm u}\simeq3~{\rm MeV}, m_{\rm d}\simeq5~{\rm MeV}, m_{\rm s}\simeq100~{\rm MeV})$, these symmetries are approximately valid if the typical energy scale of systems is much larger than them.
However, even in the massless limit, the U(1)$_{\rm A}$ symmetry is known to be explicitly broken due to quantum effects.
One is the QCD anomaly~\cite{Belavin:1975fg,tHooft:1976rip}.
In (3+1)-dimensional spacetime, gluons can form topologically non-trivial configurations called instantons, and these give rise to the breaking of ${\rm U}(1)_{\rm A}$ symmetry.
When background electromagnetic fields exist, the U(1)$_{\rm A}$ symmetry is subject to the chiral anomaly~\cite{Adler:1969gk,Bell:1969ts,Fujikawa:1980eg}, as will be discussed in Chap.~\ref{chap:Chi_Ph}.
Although we have mainly considered the case of $N_{\rm f}=3$, the case of $N_{\rm f}=2$ can be discussed in a similar manner.
Therefore, the symmetry of QCD in the massless limit is summarized as
\begin{align}
     \mcl{G}
     =
     {\rm SU}(3)_{\rm c}\times{\rm SU}(N_{\rm f})_{\rm R}\times{\rm SU}(N_{\rm f})_{\rm L}\times{\rm U}(1)_{\rm B}\,.
\end{align}

\section{Color superconductivity}
\label{sec:QCD:CSC}
Superconductivity is one of the prototypical phenomena in physics near the Fermi surface.
Due to the Cooper instability~\cite{Cooper:1956zz}, if an attractive interaction exists between fermions near the Fermi surface, the Cooper pairs are formed and condense.
In the case of metals, electrons give rise to superconductivity.
The Cooper instability is not specific to metals but universal and independent of the species of fermions.
Therefore, in dense matter, quarks can also undergo Cooper pairing, resulting in color superconductivity~\cite{Barrois:1977xd,Bailin:1979nh,Alford:1997zt,Rapp:1997zu,Alford:1998mk}, which is one of the main subjects of this thesis.

\subsection{QCD phase diagram}
\begin{figure}
\centering
\begin{tikzpicture}[scale=1.2, >=stealth]

\draw[->, thick] (0,0) -- (8,0) node[anchor=north] {$\mu_{\rm B}$};
\draw[->, thick] (0,0) -- (0,4) node[anchor=east] {$T$};

\node at (1.5,1.2) {Hadron};
\node at (4.5,3) {Quark-Gluon Plasma};
\node at (6.2,0.8) {Color Superconductivity};
\node at (6.2,-0.5) {Nuclear Superfluidity};
\draw[->, thick] (5.5,-0.3)--(3.5,0.5);

\draw[thick] (1,3) to [out=-10,in=100] (4,0.7);
\filldraw[black] (1,3) circle (2pt);
\filldraw[black] (4,0.7) circle (2pt);

\draw[thick] (3.85,1.2) to [out=20,in=180] (8,2);

\draw[thick] (2.5,0) to [out=80,in=210] (3.7,1.5);

\node at (-1,3) {$\sim 10^{12}~{\rm K}$};
\node at (3.5,-0.5) {$\sim 10^{14}~{\rm g/cm^3}$};

\end{tikzpicture}
\caption{Schematic QCD phase diagram.}
\label{fig:QCD:QCD_phase_diagram}
\end{figure}
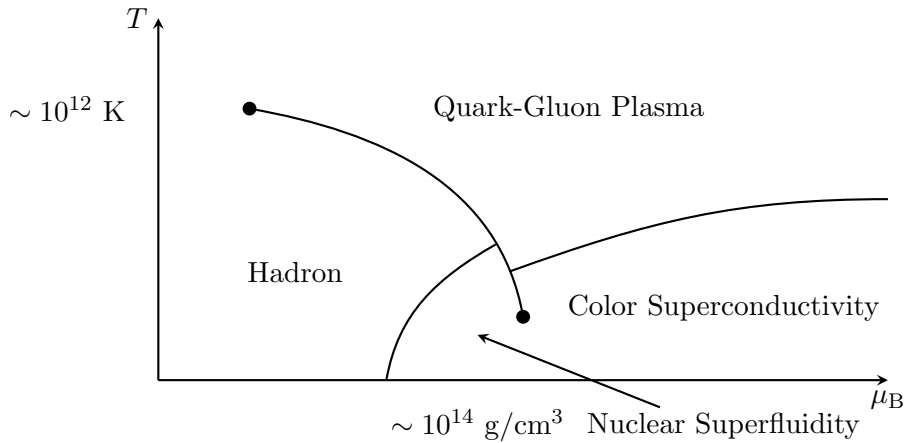

Before discussing the color superconductivity, we describe the overall structure of the QCD phase diagram (see Ref.~\cite{Fukushima:2010bq} for a review).
The QCD phase diagram shows the various forms of matter in extreme environments with the typical energy scale of QCD $(\Lambda_{\rm QCD}\simeq200~{\rm MeV})$.
This energy scale corresponds to the temperature of $T\sim10^{12}~{\rm K}$ and the mass density of $\rho\sim10^{14}~{\rm g/cm^3}$.
Although the phase diagram has been extensively studied, its complete picture remains elusive.
We show a schematic QCD phase diagram in Fig.~\ref{fig:QCD:QCD_phase_diagram}.
The horizontal line denotes the baryon chemical potential, and the vertical line represents the temperature.
There are mainly three regions: hadronic phase, quark-gluon plasma (QGP) phase, and color superconductivity phase.

The hadronic phase is located in the low-temperature and low-density region, in which quarks can exist only in color-singlet combinations, namely baryons and mesons, due to confinement.
The order parameter of the hadronic phase is the chiral condensate,
\begin{align}
    \sigma=\bar{q}q=\bar{q}_{\rm R}q_{\rm L}+\bar{q}_{\rm L}q_{\rm R}\,.
\end{align}
This represents the pair of the right- and left-handed quarks.
The expectation value of the chiral condensate is finite in the hadronic phase $(\vev{\sigma}\neq0)$, implying that the chiral symmetry is then spontaneously broken in the massless limit:
\begin{align}
    \mcl{G}\to{\rm SU}(3)_{\rm c}\times{\rm SU}(3)_{\rm V}\times{\rm U}(1)_{\rm B}\,,
\end{align}
where ${\rm SU}(3)_{\rm V}$ corresponds to the group that acts identically on right- and left-handed components.
The chiral condensate is the main origin of the mass of nucleons~\cite{Nambu:1961tp}.

As temperature increases, quarks and gluons bound in hadrons become deconfined, and the system enters the QGP phase. 
Results by the lattice QCD indicate that the transition is a crossover~\cite{Aoki:2006br}. 
In addition, the chiral condensate decreases, and chiral symmetry is restored. 
In the massless limit, the order of the chiral transition remains unsettled~(see, e.g., \cite{Pisarski:1983ms,Cuteri:2021ikv,Fejos:2024bgl}), while it becomes a crossover for finite quark masses~\cite{Brown:1990ev}.

At the low temperature region, as the baryon density increases, nucleons form Cooper pairs, leading to a nuclear superfluid phase~\cite{Migdal:1962,Larkin:1963}, where the superfluid gap serves as the order parameter; this transition is generally considered to be of second order (see Ref.~\cite{Sedrakian:2018ydt} for a review).
At even higher densities, quarks form a color-superconducting phase, with diquark condensates as the order parameter (see the details below).
Furthermore, the transition from hadronic superfluid to color-superconducting matter can occur smoothly, which is known as the quark-hadron continuity~\cite{Schafer:1998ef,Alford:1999pa}.
The critical point may appear due to the QCD anomaly~\cite{Hatsuda:2006ps,Yamamoto:2007ah}.

\subsection{Diquark condensate}
In the case of metallic superconductivity, there is a repulsive interaction between electrons, but phonons can mediate an effective attractive force.
This effective attraction arises as a consequence of complex effects such as the band structure.
On the other hand, in QCD, even without such complicated circumstances, gluons themselves provide attractive channels.

\begin{figure}[tb]
\centering
\begin{tikzpicture}
\begin{feynhand}
    \vertex (a) at (0,0) node [left]{$t^{A}_{aa'}$}; 
    \vertex[particle] (b) at (-1,-1.5);
    \vertex[particle] (c) at (-1,1.5);
    \propag[fer] (b) to (a);
    \propag[fer] (a) to (c);
    \vertex (d) at (1.5,0) node [right] at (1.5,0) {$t^{A}_{bb'}$};
    \propag[gluon] (a) to (d);
    \vertex[particle] (e) at (2.5,1.5);
    \vertex[particle] (f) at (2.5,-1.5);
    \propag[fer] (d) to (e);
    \propag[fer] (f) to (d);
\end{feynhand}
\end{tikzpicture}
\caption{Feynman diagram of the quark scattering mediated by a single gluon. The solid lines and the spiral line denote quarks and a gluon, and $t^{A}_{ab}$ is the SU(3) generator. Here, we use $A$ to denote the index of the adjoint representation, to distinguish it from that of the fundamental representation.}
\label{fig:QCD:One-gluon_exchange}
\end{figure}
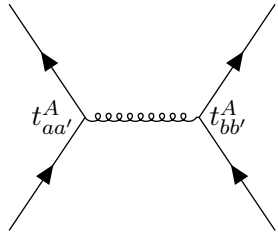

In the sufficiently dense region, perturbation theory is applicable because of asymptotic freedom~\cite{Gross:1973id,Politzer:1973fx}.
There, quark scattering mediated by a single gluon (Fig.~\ref{fig:QCD:One-gluon_exchange}) dominates.
The scattering amplitude is proportional to
\begin{align}
    t^{A}_{aa'}t^{A}_{bb'}
    =
    -\fr{1}{3}\lb(\delta_{aa'}\delta_{bb'}-\delta_{ab'}\delta_{a'b}\rb)
    +\fr{1}{6}\lb(\delta_{aa'}\delta_{bb'}+\delta_{ab'}\delta_{a'b}\rb)\,.
\end{align}
The negative sign in the first term implies the attractive channel.
By exchanging the color indices $a$ and $b$ of the incoming quarks, one can see that the sign of the first term is reversed.
This indicates that an attractive interaction arises when the colors of the quarks are antisymmetric, resulting in the diquark condensate.

We now consider the possible combinations of quark quantum numbers.
For the attractive interaction, the color indices are antisymmetric.
In addition, we assume that the diquark condensate is isotropic, indicating that the spinor labels are antisymmetric.
Due to the Pauli principle, the total state of the quarks must be antisymmetric under the exchange of all quantum numbers, and the remaining flavor indices have to be antisymmetric.

We express the diquark condensate so that the above conditions for the quantum numbers are satisfied.
We will consider specific cases of color and flavor later, so for now, we focus on the spinor.
Using an operator in spinor space $O_{\alpha\beta}$, the diquark condensate is written as $q_{\alpha}O_{\alpha\beta}q_{\beta}$.
The Lorentz covariant operator can be expressed as a combination of the gamma matrices.
From the antisymmetry of the spinors, $O=-O^{\rm T}$ must hold.
The charge conjugation operator satisfies this condition:
\begin{align}
    C=\im\gamma^{2}\gamma^{0}\,.
\end{align}
Under the parity transformation, the gamma matrices transforms $\gamma^{0}\to\gamma^{0},~\gamma^{i}\to-\gamma^{i}$, and the operator $C$ is the pseudoscalar.
Since $\gamma_5$ is a pseudoscalar, we can express the scalar diquark condensate using $O=C\gamma_5$.
For simplicity, we consider the massless case.
By decomposing the quarks into the right- and left-handed ones, we can express the diquark condensate,
\begin{align}
    \Delta^{ab}_{ij}
    \equiv
    \vev{q_{{\rm L}i}^{a}(\bs{p})Cq_{{\rm L}j}^{b}(-\bs{p})}
    =
    -\vev{q_{{\rm R}i}^{a}(\bs{p})Cq_{{\rm R}j}^{b}(-\bs{p})}\,.
\end{align}
According to the combinations of the color and flavor indices, there are various kinds of color superconductivity.
In the following, we discuss two prototypical cases. 

\subsection{Color-flavor locked phase}
We first discuss the maximally symmetric color superconductivity, namely the color-flavor locked (CFL) phase~\cite{Alford:1998mk}.
We consider ideal QCD with $N_{\rm f}=3$ and $m_{\rm u}=m_{\rm d}=m_{\rm s}=0$.
In this case, up, down, and strange quarks can be treated equivalently, except for their electric charges.
Such a situation can be realized when the typical chemical potential is much larger than the strange quark mass.
As mentioned before, the color and flavor indices have to be antisymmetric.
Such indices can be constructed only with the antisymmetric tensor.
For the case of $N_{\rm c}=3$ and $N_{\rm f}=3$, since both have three components, we use $\epsilon^{abc}$ and $\epsilon^{ijk}$.
The diquark condensate is then given by
\begin{align}
    \Delta^{ab}_{ij}
    =
    \Delta\epsilon^{abc}\epsilon_{ijk}\varphi^{c}_{k}\,,
\end{align}
where $\Delta$ is the gap size and $\varphi^{c}_{k}$ is a tensor characterizing the paring of quarks.
Since we are interested in the maximally symmetric pattern for both color and flavor $(\varphi^{c}_{k}=\delta^{c}_{k})$, the diquark condensate reduces to
\begin{align}
    \Delta^{ab}_{ij}
    =
    \Delta_{\rm CFL}\epsilon^{abA}\epsilon_{ijA}
    =\Delta_{\rm CFL}(\delta^a_i\delta^b_j-\delta^a_j\delta^b_i)\,.
\end{align}
Via the Kronecker delta, the color and flavor indices become locked, which is the origin of the name CFL.

We examine the symmetry-breaking pattern of the CFL phase.
In the present case, QCD possesses the symmetry
\begin{align}
    \mcl{G}={\rm SU}(3)_{\rm c}\times{\rm SU}(3)_{\rm R}\times{\rm SU}(3)_{\rm L}\times{\rm U}(1)_{\rm B}\,.
\end{align}
Under this transformation, the diquark condensate of right-handed quarks transforms as
\begin{align}
    \Delta^{ab}_{ij}
    &\to
    \e^{-2\im\theta_{\rm B}}\Delta_{\rm CFL}\lb[(UV_{\rm R})^a_{i}(UV_{\rm R})^b_{j}-(UV_{\rm R})^a_{j}(UV_{\rm R})^b_{i}\rb]\,.
\end{align}
This expression is invariant when $U^{\dagger}=V_{\rm R}$ and $\theta_{\rm B}=0,\pi$.
Similarly, for the left-handed condensate, the invariance holds under the transformation $U^{\dagger}=V_{\rm L}$ with $\theta_{\rm B}=0,\pi$.
Thus, the CFL phase is characterized by the symmetry breaking pattern
\begin{align}
    \mcl{G}
    \to
    {\rm SU}(3)_{\rm c+R+L}\times Z_2\,,
\end{align}
where ${\rm SU}(3)_{\rm c+R+L}$ denotes the diagonal subgroup generated by the transformation $U^{\dagger}=V_{\rm R}=V_{\rm L}$ and $Z_2$ represents a cyclic group of order 2.

From the symmetry breaking pattern, we can capture some physical implications.
First, the broken $\rm{U(1)_{\rm B}}$ symmetry leads to the superfluidity, and a single Nambu--Goldstone (NG) mode appears.
In addition, the breaking of the chiral symmetry $\rm{SU(3)_{L}\times SU(3)_{R}}$ induces eight NG modes.
The ``spontaneous breaking'' of the $\rm{SU(3)_{\rm c}}$ gauge symmetry indicates that eight gluons all become massive due to the Higgs mechanism.%
\footnote{Strictly speaking, the gauge symmetry is not broken as mentioned in Subsec.~\ref{subsec:Gauge_principle}. It appears that the gauge symmetry is broken because we have fixed the gauge. Here, for convenience of explanation, we refer to this as ``gauge symmetry breaking.''}

\subsection{Two-flavor color superconductivity phase}
We now discuss another prototypical color superconductivity called the two-flavor color superconductivity (2SC)~\cite{Barrois:1977xd,Bailin:1979nh,Alford:1997zt,Rapp:1997zu}.
We focus on the case of $N_{\rm f}=2$ and $m_{\rm u}=m_{\rm d}=0$.
The masses of the up and down quarks in reality are only a few MeV, whereas the strange quark mass is of order 100 MeV.
The present assumption corresponds to focusing on a density scale where the strange quark mass can be regarded as effectively infinite.
Unlike the three-flavor case, we have to use $\epsilon_{ij}$ instead of $\epsilon_{ijk}$ to make the flavor indices antisymmetric.
Thus, the diquark condensate can be written as
\begin{align}
    \Delta^{ab}_{ij}
    =
    \Delta_{\rm 2SC}\epsilon^{ab3}\epsilon_{ij}\,,
\end{align}
where we fixed $A=3$, implying that only red and green quarks take part in the Cooper pairing.

The symmetry of QCD for the two-flavor case is
\begin{align}
    \mcl{G}={\rm SU}(3)_{\rm c}\times{\rm SU}(2)_{\rm R}\times{\rm SU}(2)_{\rm L}\times{\rm U}(1)_{\rm B}\,.
\end{align}
Under this transformation, the diquark condensate of right-handed quarks transforms as
\begin{align}
\label{eq:QCD:2SCdiquark_transformation}
    \Delta^{ab}_{ij}
    &\to
    \e^{-2\im\theta_{\rm B}}\Delta_{\rm 2SC}U^{ac}U^{bd}\epsilon^{cd3}\epsilon_{ij}\,,
\end{align}
where we used $(V_{\rm R})_{ik}(V_{\rm R})_{jl}\epsilon_{kl}=\det(V_{\rm R})\epsilon_{ij}=\epsilon_{ij}$.
This is also the case for left-handed quarks, and the chiral symmetry is unbroken.
The ${\rm SU}(3)_{\rm c}$ gauge symmetry is partially broken.
For the ${\rm SU}(2)_{\rm c}$ part of the ${\rm SU}(3)_{\rm c}$ denoted by $\tilde{U}^{ab}$, we have $\tilde{U}^{ac}\tilde{U}^{bd}\epsilon^{cd3}=\epsilon^{ab3}$, and the expression of the condensate is invariant, implying that the ${\rm SU}(2)_{\rm c}$ gauge symmetry remains.
Equation~(\ref{eq:QCD:2SCdiquark_transformation}) also appears to break the $\rm{U(1)_{\rm B}}$ symmetry.
However, there exists a transformation that keeps it invariant in combination with the ${\rm SU}(3)_{\rm c}$ transformation.
The generator of the the ${\rm SU}(3)_{\rm c}$ is $t^a=\lambda^a/2$ where $\lambda^a$ is the Gell-Mann matrix.
One of the Gell-Mann matrix is diagonal: $\lambda^8={\rm diag}(1,1,-2)/\sqrt{3}$.
Using this component, we can find that the ${\rm SU}(3)_{\rm c}\times{\rm U}(1)_{\rm B}$ transformation contains a factor in form of
\begin{align}
    \e^{-2\im\theta_{\rm B}}\lb[\e^{-\im\theta^8t^8}\rb]^{ac}\lb[\e^{-\im\theta^8t^8}\rb]^{bd}\epsilon^{cd3}
    &=
    \exp\lb[-\im\lb(2\theta_{\rm B}+\fr{1}{\sqrt{3}}\theta^8\rb)\rb]\epsilon^{ab3}\,,
\end{align}
leading to the modified U(1) symmetry with the phase $\theta^8=-2\sqrt{3}\theta_{\rm B}$.
We denote this representation as $\rm{U(1)_{\tilde{\rm B}}}$.
This is analogous to the symmetry breaking in electroweak theory $(\rm SU(2)_L\times U(1)_Y\to U(1)_{em})$.
In this case, the generators of $\rm SU(2)_L$ and $\rm U(1)_{Y}$ are mixed into the generator of $\rm U(1)_{em}$.
Consequently, in the 2SC phase, the symmetry is broken as
\begin{align}
    \mcl{G}\to{\rm SU}(2)_{\rm c}\times{\rm SU}(2)_{\rm R}\times{\rm SU}(2)_{\rm L}\times{\rm U}(1)_{\tilde{\rm B}}\,.
\end{align}

Unlike the CFL, not all quarks gain the gap.
In our fixing (A=3), red and green quarks form the Copper pairs, while blue quarks do not. 
Thus, the blue quark remains massless in the 2SC phase.
This is important in the discussion of the main part.

\chapter{Chiral Transport Phenomena}
\label{chap:Chi_Ph}
In this chapter, we explain the chiral transport phenomena.
In Sec.~\ref{sec:Chi_Ph:Chi_Anom}, we introduce the chiral anomaly, which is one of the most typical chiral effects.
In Secs.~\ref{sec:Chi_Ph:CME} and \ref{sec:Chi_Ph:CMW}, we review the magnetic-induced chiral phenomenon, namely the CME, and the resulting collective mode called the CMW.
In Sec.~\ref{sec:Chi_Ph:CVE_CVW}, we also explain the vorticity-driven transport, CVE, and the CVW.
We finally construct the chiral kinetic theory in Sec.~\ref{sec:Chi_Ph:CKT}.

\section{Chiral anomaly}
\label{sec:Chi_Ph:Chi_Anom}
The breaking of a classical symmetry due to quantum effects is called an anomaly.
Among various types of anomalies, we focus on the chiral anomaly, which is the quantum violation of the $\rm U(1)_{A}$ symmetry induced by background electromagnetic fields.
The chiral anomaly plays a crucial role in understanding chiral transport phenomena, which will be addressed in subsequent sections.

In the following, we generally consider a massless Dirac fermion $\psi(x)$ with electric charge $e$.
Since the chiral anomaly is caused by the electromagnetic field, we here include only the U(1) gauge field $A^{\mu}(x)$.
The Lagrangian is given by
\begin{align}
\label{eq:Chi_Ph:Lagrangian_elemag}
    \mcl{L}
    =
    \bar{\psi}\im\slashed{D}\psi\,,
\end{align}
where the covariant derivative is defined as $D_{\mu}\equiv\del_{\mu}+\im eA_{\mu}$.
Equation~(\ref{eq:Chi_Ph:Lagrangian_elemag}) is invariant under the global U(1)$_{\rm V}$ and U(1)$_{\rm A}$ transformations:
\begin{align}
    \psi(x) \to \e^{-\im\theta_{\rm V}}\psi(x)\,,
    \qquad
    \psi(x) \to \e^{-\im\theta_{\rm A}\gamma_{5}}\psi(x)\,,
\end{align}
where $\theta_{\rm V,A}$ are constant parameters.
By Noether's theorem, two continuity equations are satisfied:
\begin{align}
    \del_{\mu}j^\mu
    =
    0\,,
    \qquad
    \del_{\mu}j_{5}^\mu
    =
    0\,,
\end{align}
where $j^\mu=\bar{\psi}\gamma^{\mu}\psi$ and $j_{5}^\mu=\bar{\psi}\gamma^{\mu}\gamma_{5}\psi$ are the Noether currents.
The first continuity equation means the particle number conservation, and the second one represents the conservation of the chiral charge (difference between the right- and left-handed fermion numbers).
However, as will be shown below, U(1)$_{\rm A}$ symmetry is explicitly broken due to the quantum effect $(\del_{\mu}j_{5}^{\mu}\neq0)$.

\subsection{Spectral flow}
\label{sec:Chi_Ph:Intuitive_discussion}
Historically, the chiral anomaly was found in the perturbative calculations~\cite{Adler:1969gk,Bell:1969ts} and the path integral formalism~\cite{Fujikawa:1980eg} in the framework of quantum field theory.
The details of the derivation in quantum field theory are deferred to App.~\ref{app:Chi_Anom_Der}, and we here examine the appearance of the chiral anomaly by following a physically transparent argument of the spectral flow~\cite{Nielsen:1983rb}.

\subsubsection{(1+1) dimensions}
\begin{figure}[htb]
\begin{tabular}{p{0.46\textwidth}p{0.03\textwidth}p{0.46\textwidth}}
\centering
\begin{tikzpicture}
    \draw[->,>=stealth,thick](-2,0)--(2,0) node [right]{$p_x$};
    \draw[->,>=stealth,thick](0,-2)--(0,2) node [right]{$\varepsilon_{\lambda}$};
    \draw[thick,black](-2,-2)--(2,2) node [right] {R};
    \draw[thick,black](2,-2)--(-2,2) node [left] {L};
    \filldraw [fill=black](0,0) circle [radius=0.15];
    \filldraw [fill=black](-0.5,-0.5) circle [radius=0.15];
    \filldraw [fill=black](-1,-1) circle [radius=0.15]; 
    \filldraw [fill=black](-1.5,-1.5) circle [radius=0.15];
    \filldraw [fill=black](0.5,-0.5) circle [radius=0.15];
    \filldraw [fill=black](1,-1) circle [radius=0.15]; 
    \filldraw [fill=black](1.5,-1.5) circle [radius=0.15];
\end{tikzpicture}
\caption{The energy level of the massless fermion. R and L denote the right- and left-handed fermions, respectively. The negative energy levels are occupied due to the Pauli principle (Dirac sea).}
\label{fig:Chi_Ph:fermion_dispersion}
&&
\centering
\begin{tikzpicture}
    \draw[->,>=stealth,thick](-2,0)--(2,0) node [right]{$p_x$};
    \draw[->,>=stealth,thick](0,-2)--(0,2) node [right]{$\varepsilon_{\lambda}$};
    \draw[thick,black](-2,-2)--(2,2) node [right] {R};
    \draw[thick,black](2,-2)--(-2,2) node [left] {L};
    \filldraw [fill=black](0,0) circle [radius=0.15];
    \filldraw [fill=black](-0.5,-0.5) circle [radius=0.15];
    \filldraw [fill=black](-1,-1) circle [radius=0.15]; 
    \filldraw [fill=black](-1.5,-1.5) circle [radius=0.15];
    \filldraw [fill=white](0.5,-0.5) circle [radius=0.15];
    \filldraw [fill=white](1,-1) circle [radius=0.15]; 
    \filldraw [fill=black](1.5,-1.5) circle [radius=0.15];
    \filldraw [fill=black](0.5,0.5) circle [radius=0.15];
    \filldraw [fill=black](1,1) circle [radius=0.15];
\end{tikzpicture}
\caption{The energy level under the electric field. The right-handed fermions are created while the left-handed fermions are annihilated.}
\label{fig:Chi_Ph:fermion_dispersion_electric_field}
\end{tabular}
\end{figure}
As the simplest case, we first consider massless Dirac fermions in (1+1) dimensions.
We note that we can define only the electric field in the (1+1) dimensions.
The gamma matrices in (1+1) dimensions can be expressed as%
\footnote{The chirality operator $\gamma_5$ can be defined only in even dimensions.}
\begin{align}
    \gamma^{0}
    =
    \sigma^{1},
    \qquad
    \gamma^{1}
    =
    \im\sigma^{2},
    \qquad
    \gamma_{5}
    =
    \gamma^{0}\gamma^{1}
    =
    -\sigma^{3}\,.
\end{align}
We define the right- and left-handed fermions in the same way as the case of (3+1) dimensions using the projection operator (\ref{eq:QCD:Chirality_projection_op}).
The Lagrangian reduces to
\begin{align}
    \mcl{L}
    &=
    \psi_{\rm R}^{\dagger}\im\lb(D_{t}+D_{x}\rb)\psi_{\rm R}
    +\psi_{\rm L}^{\dagger}\im\lb(D_{t}-D_{x}\rb)\psi_{\rm L}\,.
\end{align}
The dispersion relations of the right- and left-handed fermions are given by
\begin{align}
    \varepsilon_{\lambda}=\lambda p_x\,,
\end{align}
where $\varepsilon_{\lambda}$ is the energy of each fermion.

In the ground state, the negative energy levels are occupied due to the Pauli principle, which is called the Dirac sea (see Fig.~\ref{fig:Chi_Ph:fermion_dispersion}).
As will be discussed in detail later, the fact that the Dirac sea is filled up to a semi-infinite is essential for the chiral anomaly.

Next, we apply an electric field $E_x$ along the $x$-axis for a time interval $\varDelta t$.
During this process, the right-handed fermions acquire momentum in the positive $x$-direction, while the left-handed fermions acquire momentum in the opposite direction.
This implies that right-handed fermions are effectively created and left-handed fermions are annihilated (see Fig.~\ref{fig:Chi_Ph:fermion_dispersion_electric_field}).
The change of fermion momentum in this interval is
\begin{align}
    \varDelta p_x
    =
    eE_x\varDelta t\,.
\end{align}
We impose the periodic boundary conditions with the period $L$, and the change in the number of right- and left-handed fermions, $\varDelta N_{\lambda}$, is given by
\begin{align}
    \varDelta N_{\lambda}
    &=
    \lambda\fr{\varDelta p_x}{2\pi/L}
    =
    \lambda\fr{e}{2\pi}E_xL\varDelta t\,.
\end{align}
Therefore, the number of fermions is conserved:
\begin{align}
    \varDelta N
    &\equiv
    \varDelta N_{\rm R}+\varDelta N_{\rm L}
    =
    0\,.
\end{align}
However, the change of the chiral charge is finite:
\begin{align}
    \varDelta N_{5}
    &\equiv
    \varDelta N_{\rm R}-\varDelta N_{\rm L}
    =
    \frac{e}{\pi}E_xL\varDelta t\,.
\end{align}
Assuming that the electric field is constant, the time derivative becomes
\begin{align}
\label{eq:Cho_Ph:Anomaly_int_1-1}
    \fr{\dif}{\dif t}N_{5}
    =
    \int \dif x \frac{e}{\pi}E_x\,.
\end{align}
Thus, it turns out that the chiral charge is not conserved.
We can express the above equation locally as
\begin{align}
\label{eq:Chi_Ph:Anomaly_eq_1+1}
    \del_{\mu}j_{5}^{\mu}
    &=
    \fr{e}{\pi}E_x
    =
    \fr{e}{2\pi}\epsilon^{\mu\nu}F_{\mu\nu}\,,
\end{align}
where $\epsilon^{\mu\nu}$ is the anti-symmetric tensor in (1+1) dimensions.
From this equation, we can find that the $\rm U(1)_{A}$ symmetry is broken by the quantum effect.
This is the chiral anomaly in (1+1) dimensions.

The chiral anomaly arises due to the quantum effect of the negative energy levels being filled up infinitely.
If the occupation is finite, the electric field just shifts the occupation.
The numbers of the right- and left-handed fermions do not change $(\varDelta N_{\lambda}=0)$, and the chiral charge is conserved.

We here give two points to be noted.
At first, it is essential in this intuitive argument that the fermions are semi-infinitely occupied.
A related example is Hilbert's infinite hotel.
In this thought experiment, the hotel has an infinite number of rooms, but all of the rooms are occupied.
Then, a new guest has come.
To make an empty room for the guest, the person in the first room must move to the second room, the person in the second room must move to the third, and so on, with everyone shifting to the next room in sequence, thereby freeing up the first room.
However, we can make the empty room since the first room exists as the edge one.
If there is no edge, we cannot make it available.

The second point is that we naively regard ``infinity minus infinity'' as zero in this discussion.
The change in the chiral charge under the applied electric field is calculated as the difference between the contributions from right- and left-handed chirality.
This implicitly assumes that the infinite number of right- and left-handed particles perfectly cancel each other out.
In fact, as discussed in App.~\ref{app:Chi_Anom_Der}, it is necessary to consider regularization seriously.

\subsubsection{(3+1) dimensions}
We here discuss massless Dirac fermions in (3+1) dimensions.
In this discussion, we use the gamma matrices in the chiral representation as
\begin{align}
    \gamma^{\mu}
    =
    \lb(
    \begin{array}{cc}
        0 & \sigma^{\mu} \\
         \bar{\sigma}^{\mu} & 0
    \end{array}\rb)\,,
    \qquad
    \gamma_5
    =
    \lb(
    \begin{array}{cc}
        -1 & 0 \\
         0 & 1
    \end{array}\rb)\,.
\end{align}
According to the eigenvalue of the chirality operator, we can write the Dirac fermion field as
\begin{align}
    \psi
    =
    \lb(
    \begin{array}{c}
    \chi_{\rm L} \\
    \chi_{\rm R}
    \end{array}
    \rb)\,,
\end{align}
where $\chi_{\lambda}$ are the two-component spinors of the right- and left-handed chiral fermion fields, respectively.
The Lagrangian is given by
\begin{align}
    \mcl{L}
    &=
    \chi_{\rm L}^{\dagger}\im\lb(D_{0}+\sigma^{i}D_{i}\rb)\chi_{\rm L}
    +\chi_{\rm R}^{\dagger}\im\lb(D_{0}-\sigma^{i}D_{i}\rb)\chi_{\rm R}\,.
\end{align}

\begin{figure}[tb]
\centering
\begin{tikzpicture}
    \draw[->,>=stealth,thick](-3,0)--(3,0) node [right]{$p_z$};
    \draw[->,>=stealth,thick](0,-3)--(0,3) node [right]{$\varepsilon_{n,s}$};
    \draw[thick,red](-3,-3)--(3,3) node [right,black] {R};
    \draw[thick,red](3,-3)--(-3,3) node [left,black] {L};
    \draw [thick,domain=-2.8:2.8,samples=100]plot(\x,{sqrt(pow(\x,2)+1)});
    \draw [thick,domain=-2.6:2.6,samples=100]plot(\x,{sqrt(pow(\x,2)+2)});
    \draw [thick,domain=-2.4:2.4,samples=100]plot(\x,{sqrt(pow(\x,2)+3)});
    \draw [thick,domain=-2.17:2.17,samples=100]plot(\x,{sqrt(pow(\x,2)+4)});
    \draw [thick,domain=-1.93:1.93,samples=100]plot(\x,{sqrt(pow(\x,2)+5)});
    \draw [thick,domain=-2.8:2.8,samples=100]plot(\x,-{sqrt(pow(\x,2)+1)});
    \draw [thick,domain=-2.6:2.6,samples=100]plot(\x,-{sqrt(pow(\x,2)+2)});
    \draw [thick,domain=-2.4:2.4,samples=100]plot(\x,-{sqrt(pow(\x,2)+3)});
    \draw [thick,domain=-2.17:2.17,samples=100]plot(\x,-{sqrt(pow(\x,2)+4)});
    \draw [thick,domain=-1.93:1.93,samples=100]plot(\x,-{sqrt(pow(\x,2)+5)});
\end{tikzpicture}
    \caption{The Landau level in (3+1) dimensions. Only the lowest Landau level (red line) has the linear dispersion relation. The states with the positive and negative slopes correspond to the right- and left-handed fermions, respectively. Higher excited levels, on the other hand, do not exhibit a clear separation between right- and left-handed fermions.}
\label{fig:Chi_Ph:Landau_dispersion}
\end{figure}
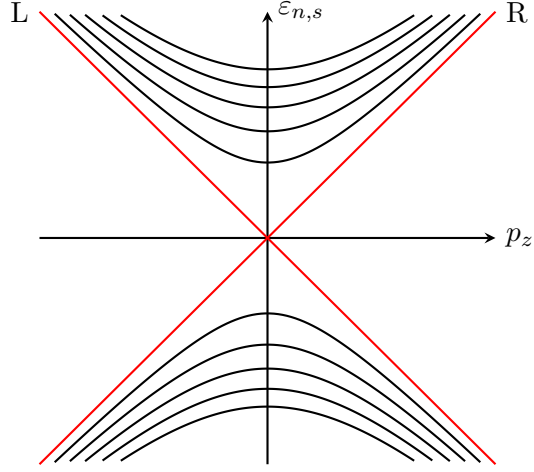

Let us consider the presence of an external magnetic field.
Without loss of generality, we set the magnetic field to be along the $z$-axis $(\bs{B}=B\bs{e}_{z})$.
We can take the temporal gauge $(A_{0}=0)$, and the equation of motion is
\begin{align}
    \im\del_{t}\chi_{\lambda}
    =
    \lambda\im\bs{\sigma}\cdot\bs{D}\chi_{\lambda}\,.
\end{align}
We define Hamiltonian operator of the right- and left-handed fermions as $\hat{H}_{\lambda}\equiv\lambda\im\bs{\sigma}\cdot\bs{D}$.
Squaring the Hamiltonian, we have
\begin{align}
\label{eq:Chi_Ph:Square_Hamiltonian_op}
    \hat{H}_{\lambda}^2
    &=
    -\bs{D}^2
    +e\epsilon^{ijk}\sigma^kF_{ij}
    \nom
    &=
    (\hat{\bs{p}}-e\bs{A})^2-2e\bs{B}\cdot\bs{s}\,,
\end{align}
where the field strength is defined as $F_{ij}\equiv-\fr{\im}{e}[D_i,D_j]$ and the spin is $s^i=\fr{1}{2}\sigma^i$.
We denote the momentum operator as $\hat{p}^i=\im\del^i$.
Taking the Landau gauge, $\bs{A}=(0,Bx,0)$, we can further rewrite the Hamiltonian (\ref{eq:Chi_Ph:Square_Hamiltonian_op}) as
\begin{align}
\label{eq:Chi_Ph:H^2_Landau_energy}
    \hat{H}_{\lambda}^2
    &=
    \hat{p}_{x}^2+(eB)^2\lb(x-\frac{\hat{p}_{y}}{eB}\rb)^2+\hat{p}_{z}^2-2eBs\,.
\end{align}
Since the system has the translational symmetry along the $y$- and $z$-axes, we have the commutation relations,
\begin{align}
    [\hat{H}_{\lambda}^2,\hat{p}_{y}]
    =
    [\hat{H}_{\lambda}^2,\hat{p}_{z}]
    =
    0\,.
\end{align}
These indicate that the momenta $\hat{p}_{y},\hat{p}_{z}$ are conserved quantities respectively, and we can replace them with the eigenvalues.
In addition, the first and second terms in Eq.~(\ref{eq:Chi_Ph:H^2_Landau_energy}) are mathematically equivalent to the Hamiltonian of the harmonic oscillator.
Therefore, we obtain the Landau levels labeled by a quantum number $n~(n\geq 0)$ and the spin $s$:
\begin{align}
\label{eq:Chi_Ph:Landau_energy}
    {\varepsilon}^2_{n,s}
    =
    (2n+1)eB+p_{z}^2-2eBs\,.
\end{align}
Explicitly writing down the energies in order from the lowest, we have
\begin{align}
    \varepsilon^2_{0,\uparrow}
    =
    p_{z}^2,
    \qquad
    \varepsilon^2_{0,\downarrow}
    =
    \varepsilon^2_{1,\uparrow}
    =
    p_{z}^2+eB,
    \qquad
    \varepsilon^2_{1,\downarrow}
    =
    \varepsilon^2_{2,\uparrow}
    =
    p_{z}^2+2eB,
    \quad\cdots\,.
\end{align}
Here, we labeled the spin using $\uparrow~(s=+\fr{1}{2})$ and $\downarrow~(s=-\fr{1}{2})$.
The energy spectrum is schematically shown in Fig.~\ref{fig:Chi_Ph:Landau_dispersion}.
The dispersion relation of the lowest Landau level (LLL) is linear, similar to the case in (1+1) dimensions:
\begin{align}
    \varepsilon_{{\lambda},0,\uparrow}
    =\lambda p_z\,.
\end{align}
Thus, applying electric fields along the $z$-axis adiabatically, the creation and annihilation of the fermion arise, and the chiral charge is no longer conserved.
On the other hand, in the higher Landau levels (HLLs), the right- and left-handed fermions are degenerate,%
\footnote{Degeneracy generally implies the presence of a symmetry behind. In this case, we have the symmetry for interchanging the right- and left-handed fermions except the LLL.}
and only the LLL contributes to the chiral anomaly.

Unlike the (1+1)-dimensional case, we also have the degrees of freedom in the $xy$-plane, which must be taken into account.
Let $L_{x}$ and $L_{y}$ denote the sizes of the system in the $x$- and $y$-directions, respectively.
Imposing the periodic boundary condition with respect to the $y$-direction, the momentum is quantized as
\begin{align}
    p_{y}
    =
    \frac{2\pi}{L_{y}}m\,,
\end{align}
where $m$ is an integer.
In the $x$-direction, the center of oscillation lies within the region $0 \leq x \leq L_{x}$:
\begin{align}
    0 \leq \frac{p_{y}}{eB}
    =
    \frac{2\pi}{eB L_{y}} m \leq L_{x}\,.
\end{align}
This imposes the following condition on the quantum number $m$:
\begin{align}
    0 \leq m \leq \frac{eB}{2\pi} L_{x} L_{y}.
\end{align}
The degeneracy per unit area is given by $\frac{eB}{2\pi}$, which is referred to as the Landau degeneracy.
Therefore, multiplying the (1+1) dimensional anomaly relation (\ref{eq:Cho_Ph:Anomaly_int_1-1}) by the Landau degeneracy, we obtain the chiral anomaly in (3+1) dimensions:
\begin{align}
\label{eq:Chi_Ph:Anomaly_Int}
    \fr{\dif}{\dif t}N_{5}
    =
    \int\dif^3\bs{x}\fr{e^2}{2\pi^2}EB\,.
\end{align}
The local expression of the chiral anomaly is
\begin{align}
    \label{eq:Chi_Ph:anomaly_4d}
    \del_{\mu}j_{5}^{\mu}
    =
    \fr{e^2}{2\pi^2}\bs{E}\cdot\bs{B}
    =
    -\fr{e^2}{8\pi^2}F_{\mu\nu}\tilde{F}^{\mu\nu}\,,
\end{align}
where $\tilde{F}^{\mu\nu}\equiv\fr{1}{2}\epsilon^{\mu\nu\alpha\beta}F_{\alpha\beta}$.
From this expression, we can find that the chiral charge is not conserved in the presence of non-orthogonal electric and magnetic fields.

\subsection{Topological property}
One of the important properties of the chiral anomaly is that the coefficient is topologically quantized.
We verify this fact in (3+1) dimensions~\cite{Vazifeh:2010pq}.

\begin{figure}[tbh]
\centering
\begin{tikzpicture}[yscale=cos(70),samples=400]
    \draw[double distance=40] (0:3) arc (0:180:3);
    \draw[double distance=40] (180:3) arc (180:360:3);
    \draw[->,>=stealth,ultra thick, blue] (0,-1)--(0,8);
    \draw[ultra thick, blue] (0,-8)--(0,-5);
    \draw (0,7) node [right]{$\bs{B}(t)$};
    \draw[->,>=stealth,ultra thick,red] (-3.7,0) to [out=-70,in=180](0.1,-3);
    \draw[ultra thick,red] (0,-3) to [out=0,in=-110](3.7,0);
    \draw (0.3,-3.2) node [below]{$\bs{E}$};
\end{tikzpicture}
\caption{The figure representing the relation between the two-dimensional torus and electromagnetic field. The red and blue lines denote the electric field and the magnetic field, respectively. Due to the magnetic field penetrating the hole of the torus, a constant electric field is generated along the torus.}
\label{fig:Chi_Ph:torus_field}
\end{figure}
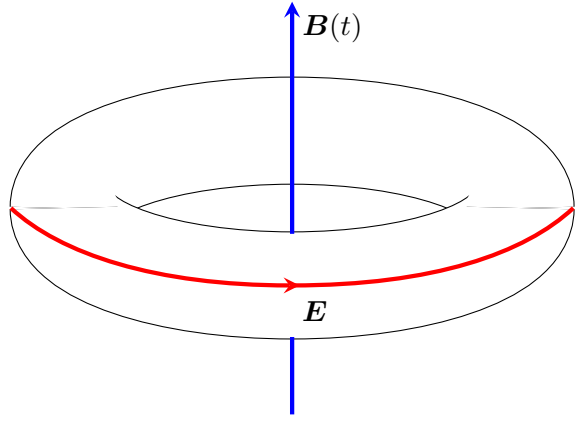

The change in the number of right-handed particles is given by
\begin{align}
\label{eq:Chi_Ph:anomaly_4d_R}
    \varDelta N_{\rm R}
    =
    \int_x\del_{\mu}j_{\rm R}^{\mu}
    =
    \fr{e^2}{4\pi^2}\int\bs{E}\cdot\bs{B}\,.
\end{align}
First, the right-hand side of Eq.~(\ref{eq:Chi_Ph:anomaly_4d_R}) can be written in the form of a total derivative:
\begin{align}
    \bs{E}\cdot\bs{B}
    &=
    -\fr{1}{2}\epsilon^{\mu\nu\alpha\beta}\del_{\mu}(A_{\nu}\del_{\alpha}A_{\beta})\,.
\end{align}
This indicates that the integral in four-dimensional spacetime is determined by the boundary configuration, independent of the bulk.
The integral can yield a finite value because of the non-periodicity of the gauge field that generates a periodic electromagnetic field.
For instance, when a magnetic field penetrating the hole of the two-dimensional torus $\rm T^2$ increases linearly in time, a uniform electric field is induced on $\rm T^2$ (Fig.~\ref{fig:Chi_Ph:torus_field}).
In this case, the line integral of the gauge field along the two-dimensional torus satisfies
\begin{align}
    \int_{\rm T^2}\dif\bs{l}\cdot\bs{A}
    =
    \int_{\rm T^2}\dif\bs{S}\cdot\bs{B} \neq 0\,,
\end{align}
which implies that the gauge field is not periodic.

To show that the integration is topologically quantized, we consider a periodic spacetime, namely, a four-dimensional torus $\rm T^{4}$.
We further assume that constant electric and magnetic fields point in the $z$-direction, i.e., $\bs{E}=(0,0,E)$ and $\bs{B}=(0,0,B)$.
In this case, the electric field can be expressed in terms of the temporal and $z$ components of the gauge field, while the magnetic field can be written using the $x$ and $y$ components.
For convenience of notation, we decompose the gauge field $A_{\mu}$ into two parts defined on two-dimensional tori as
\begin{align}
    \bs{A}_{\rm E}
    \equiv
    (A_{t},A_{z}),
    \qquad
    \bs{A}_{\rm M}
    \equiv
    (A_{x},A_{y})\,.
\end{align}
Using these definitions, the electric and magnetic fields under consideration can be written as 
\begin{align}
    &E
    =
    -(\del_{t}A_{z}-\del_{z}A_{t})
    \equiv
    -\bs{\nabla}\times\bs{A}_{\rm E}\,,
    \\
    &B
    =
    \del_{x}A_{y}-\del_{y}A_{x}
    \equiv
    \bs{\nabla}\times\bs{A}_{\rm M}\,,
\end{align}
where $\bs{\nabla}$ denotes the differential operator defined on each of the respective two-dimensional tori.
The integral in the right-hand side of Eq.~(\ref{eq:Chi_Ph:anomaly_4d_R}) can be decomposed into the product of integrals over the two tori:
\begin{align}
    \int_{\rm T^4}\dif^{4}x
    \bs{E}\cdot\bs{B}
    &=
    \int_{\rm T^2}\dif t\dif zE
    \int_{\rm T^2}\dif x\dif yB
    \nom
    &=
    -\int_{\rm T^2}\dif t\dif z\bs{\nabla}\times\bs{A}_{\rm E}
    \int_{\rm T^2}\dif x\dif y\bs{\nabla}\times\bs{A}_{\rm M}\,.
\end{align}

We now examine that each integral is topologically quantized.
At first, we focus on the electric part.
We consider two gauge fields connected by the gauge transformation:
\begin{align}
    \bs{A}_{\rm E}-\bs{A}'_{\rm E}
    =
    \fr{1}{e}\bs{\nabla}\theta(t,z)\,,
\end{align}
where $\theta$ is a gauge parameter defined on the two-dimensional torus.
Both gauge fields give the physically equivalent electric fields.
We divide the two-dimensional torus into two regions $\rm S_{1}$ and $\rm S_{2}$ by a contractible closed path $\rm C$, such that $\rm{T^2 = S_{1} + S_{2}}$. 
We suppose that $\bs{A}_{\rm E}$ is in the region $\rm S_1$, while $\bs{A}'_{\rm E}$ is in the region $\rm S_2$.
Making use of the Stokes theorem, we have%
\footnote{We note that positive directions along the path $\rm C$ for $\rm S_{1}$ and $\rm S_{2}$ are opposite to each other.}
\begin{align}
    \int_{\rm T^2}\dif t\dif z\bs{\nabla}\times\bs{A}_{\rm E}
    &=
    \int_{\rm S_1}\dif t\dif z\bs{\nabla}\times\bs{A}_{\rm E}
    +\int_{\rm S_2}\dif t\dif z\bs{\nabla}\times\bs{A}'_{\rm E}
    \nom
    &=
    \oint_{\rm C}\dif\bs{l}\cdot\bs{A}_{\rm E}
    -\oint_{\rm C}\dif\bs{l}\cdot\bs{A}'_{\rm E}
    \nom
    &=
    \fr{1}{e}\oint_{\rm C}\dif\bs{l}\cdot\bs{\nabla}\theta
    \nom
    &=
    \fr{2\pi}{e}m\,,
\end{align}
where $m$ is an integer.

The same argument applies to the integration of the magnetic part, and we get
\begin{align}
    \int_{\rm T^2}\dif x\dif y\bs{\nabla}\times\bs{A}_{\rm M}
    =
    \fr{2\pi}{e}n\,,
\end{align}
where $n$ is an integer.
Therefore, we finally obtain
\begin{align}
    \varDelta N_{\rm R}
    &=
    -\fr{e^2}{4\pi^2}\lb(\frac{2\pi}{e}\rb)^{2}mn
    \equiv
    N\,.
\end{align}
An important consequence of the topological nature is that the above result is exact and it receives no higher-order corrections in perturbation theory, which is also known as the Adler--Bardeen theorem~\cite{Adler:1969er}.
This is because topologically quantized quantities generally remain invariant under small perturbations or continuous deformations.

\section{Chiral magnetic effect}
\label{sec:Chi_Ph:CME}
In this section, we give two derivations of the CME: the energy balance argument and the Landau level description.

\subsection{Symmetry of transport}
Before we derive the CME, let us take a look at the symmetry of transports.
One of the most famous transport phenomena is Ohm's law,
\begin{align}
    e\bs{j}
    =
    \sigma\bs{E}\,,
\end{align}
where $\sigma$ is the electric conductivity.
The electric current and the electric field are vectors and flip their sign under the parity transformation, while the conductivity does not: 
\begin{align}
    e\bs{j}\to-e\bs{j},
    \qquad
    \bs{E}\to-\bs{E},
    \qquad
    \sigma\to\sigma\,.
\end{align}
This means that both sides have the same transformation property, and we find that the Ohm's law is not prohibited by symmetry.

Similarly, we next consider the electric current induced by the magnetic field.
Writing the coefficient as $\sigma_{\rm m}$, the current would be given by
\begin{align}
    e\bs{j}
    =
    \sigma_{\rm m}\bs{B}\,.
\end{align}
Under the parity transformation, these quantities transform as
\begin{align}
    e\bs{j}\to-e\bs{j},
    \qquad
    \bs{B}\to\bs{B},
    \qquad
    \sigma_{\rm m}\to\sigma_{\rm m}\,.
\end{align}
Unlike Ohm's law, the magnetic field is an axial vector and does not change its sign under parity.
Then, both sides have different transformations, which means that such a transport is not allowed by symmetry.
This is the reason why we do not find the electric current driven by the magnetic field in daily life.
However, in the above discussion, we implicitly supposed the coefficient to be invariant under the parity transformation.
If the conductivity flips its sign, the parity symmetry is not violated.
Since the coefficient has the information of the matter, we can expect that ``parity-odd matter'' can cause the electric current induced by the magnetic field.
In the case of the CME, the chiral matter where the numbers of the right- and left-handed fermions are different plays the role.

\subsection{Intuitive picture}
\begin{figure}[tb]
\centering
\begin{tikzpicture}[samples=400]
    \draw[->,>=stealth,ultra thick, blue] (0,-2)--(0,2);
    \draw (0,0) node [left]{$\bs{B}$};
    \draw (1.5,1) circle [radius=0.3];
    \draw[->,>=stealth,ultra thick, black] (1.5,0.4)--(1.5,1.6);
    \draw (1.8,1) node [right]{$\bs{s}$};
    \draw (1.5,-1) circle [radius=0.3];
    \draw[->,>=stealth,ultra thick, black] (1.5,-1.6)--(1.5,-0.4);
    \draw (1.8,-1) node [right]{$\bs{s}$};
    \draw[->,>=stealth,ultra thick, red] (3,0.4)--(3,1.6);
    \draw (3.1,1) node [right]{$\bs{p}$};
    \draw[->,>=stealth,ultra thick, red] (3,-0.4)--(3,-1.6);
    \draw (3.1,-1) node [right]{$\bs{p}$};
    \draw (3.8,1) node [right]{R};
    \draw (3.8,-1) node [right]{L};
\end{tikzpicture}
\caption{The schematic figure representing the relation between the chirality and the magnetic field. The right-handed fermion moves along the magnetic field, and the left-handed one goes in the opposite direction.}
\label{fig:Chi_Ph:CME_image}
\end{figure}
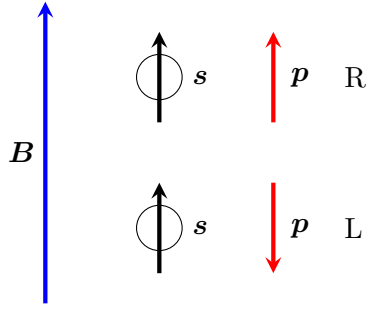

We here give an intuitive picture of the CME.
We consider right- and left-handed massless fermions in magnetic fields (see Fig.~\ref{fig:Chi_Ph:CME_image}).
Since the chirality is the same as the helicity in the massless limit,%
\footnote{For antifermion, the sign is opposite.}
the right-handed fermion has the spin and momentum in the same direction, and the left-handed one has these in the opposite direction.
The spins are aligned along the magnetic field.
Then, the right-handed fermion flows parallel to the magnetic field, and the left-handed one moves antiparallel to it.

When the system has no chirality imbalance, i.e., the numbers of right- and left-handed fermions are the same, the right- and left-handed currents are canceled, and the net current is zero.
However, if the imbalance exists, the current can be generated, which is the CME.
The conductivity is parity-odd because of the chirality imbalance.

\subsection{Energy balance}
We first derive the CME in the simplest way using the energy balance argument~\cite{Nielsen:1983rb}.
We consider the amount of energy to create the change of the chiral charge $\varDelta N_{5}$.
Introducing the chiral chemical potential $\mu_5$ as the Lagrange multiplier for the chiral charge, the energy shift is $\mu_{5}\varDelta N_{5}$.
We assume that this energy is provided by the electric current $e\bs{j}$.
The change in the energy per unit time can be written as
\begin{align}
    \int\dif^3 \bs{x}e\bs{j}\cdot\bs{E}
    &=
    \mu_{5}\fr{\dif N_{5}}{\dif t}
    \nom
    &=
    \mu_{5}\int\dif^3 \bs{x}\fr{e^2}{2\pi^2}\bs{E}\cdot\bs{B}\,,
\end{align}
where we used the chiral anomaly relation (\ref{eq:Chi_Ph:Anomaly_Int}).
Since this equality holds for arbitrary electric fields, we obtain the CME~\cite{Vilenkin:1980fu,Alekseev:1998ds,Fukushima:2008xe}:
\begin{align}
\label{eq:Chi_Ph:CME}
    \bs{j}
    =
    \fr{e\mu_{5}}{2\pi^2}\bs{B}\,.
\end{align}
The chiral chemical potential is finite only when systems have the chirality imbalance.
Thus, the CME can be seen in the chiral matter.
From the derivation, we can find that the transport coefficient is related to the chiral anomaly, which is topologically quantized.
This implies that the coefficient is independent of the details of the system and the thermal effect.
In the next discussion, we will show that the CME does not receive the thermal correction.

It is worth noting that the CME is not an equilibrium current.
In general, electric currents do not appear at equilibrium, which is known as the Bloch theorem.
Since the CME is driven by external magnetic fields, it seems to contradict the Bloch theorem.%
\footnote{If the magnetic field is dynamical, the system is unstable due to the CPI~\cite{Akamatsu:2013pjd,Akamatsu:2014yza}.
Thus, it does not contradict the Bloch theorem.}
However, as discussed in Ref.~\cite{Yamamoto:2015fxa}, the chiral chemical potential $\mu_5$ plays the role of voltage.
To keep the chirality imbalance, we need to inject energy, meaning that the CME is a non-equilibrium current.

\subsection{Transport from Landau levels}
We derive the CME based on the Landau levels.
The magnetic field is set to be along the $z$-axis.
Although the motion perpendicular to the magnetic field is quantized, we can treat the longitudinal component classically.
Since we are interested in the current along the magnetic field, we can use the kinetic approach to obtain the current.
We have to note that the treatment here is different from the chiral kinetic theory, which we will discuss in Sec.~\ref{sec:Chi_Ph:CKT}.
The CME can also be derived in the chiral kinetic theory, but it is applicable in the weak magnetic field limit, in which the distinction between the levels becomes irrelevant.

\begin{figure}[tb]
\centering
\begin{tikzpicture}
    \draw[->,>=stealth,thick](-2,0)--(2,0) node [right]{$p_z$};
    \draw[->,>=stealth,thick](0,-2)--(0,2) node [right]{$\varepsilon_{\rm R}$};
    \draw[thick,black](-2,-2)--(2,2) node [right] {R};
    \filldraw [fill=black](0,0) circle [radius=0.15];
    \filldraw [fill=black](-0.5,-0.5) circle [radius=0.15];
    \filldraw [fill=black](-1,-1) circle [radius=0.15]; 
    \filldraw [fill=black](-1.5,-1.5) circle [radius=0.15];
    \filldraw [fill=black](0.5,0.5) circle [radius=0.15];
    \filldraw [fill=black](1,1) circle [radius=0.15];
    \draw[dashed,black](0,1) node [left] {$\mu_{\rm R}$}--(1,1);
\end{tikzpicture}
\caption{The dispersion relation of the right-handed fermion at finite density.}
\label{fig:Chi_Ph:CME_dispersion_R}
\end{figure}
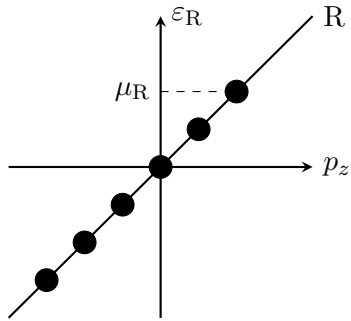

At first, we consider the right-handed fermion at zero temperature $T=0$.
We introduce the right-handed chemical potential $\mu_{\rm R}$, which corresponds to the Fermi energy of the right-handed fermion.
Applying a magnetic field along the $z$-axis, the Landau levels are formed.
We focus on the LLL with the linear dispersion shown in Fig.~\ref{fig:Chi_Ph:CME_dispersion_R}.
Similarly to Sec.~\ref{sec:Chi_Ph:Chi_Anom}, the degrees of freedom perpendicular to the magnetic field are included in the Landau degeneracy $eB/(2\pi)$.
Using $\dot{z}=\fr{\del\varepsilon_{\rm R}}{\del p_z}=1$, the $z$-component of the right-handed number current is expressed as
\begin{align}
    j_{\rm R}^{z}
    &=
    \fr{eB}{2\pi}\int_{0}^{\mu_{\rm R}}\fr{\dif p^{z}}{2\pi}\dot{z}
    =
    \fr{e\mu_{\rm R}}{4\pi^2}B\,.
\end{align}
The left-handed fermion flows in the opposite direction:
\begin{align}
    j_{\rm L}^{z}
    =
    -\fr{e\mu_{\rm L}}{4\pi^2}B\,,
\end{align}
where $\mu_{\rm L}$ is the left-handed chemical potential.
Summing these currents, we can reproduce the CME for the arbitrary direction of the magnetic field as~\cite{Vilenkin:1980fu,Alekseev:1998ds,Fukushima:2008xe}
\begin{align}
\label{eq:Chi_Ph:CME_current}
    \bs{j}
    &=
    \bs{j}_{\rm R}+\bs{j}_{\rm L}
    =
    \fr{e\mu_{5}}{2\pi^2}\bs{B}\,,
\end{align}
where the chiral chemical potential is defined as
\begin{align}
    \mu_{5}
    \equiv
    \fr{\mu_{\rm R}-\mu_{\rm L}}{2}\,.
\end{align}
We also consider the axial current, which is the difference between the right- and left-handed currents.
In this case, we obtain the CSE~\cite{Son:2004tq,Metlitski:2005pr},
\begin{align}
\label{eq:Chi_Ph:CSE_current}
    \bs{j}_{5}
    &=
    \bs{j}_{\rm R}-\bs{j}_{\rm L}
    =
    \fr{e\mu}{2\pi^2}\bs{B}\,,
\end{align}
where we defined the chemical potential as
\begin{align}
    \mu
    \equiv
    \fr{\mu_{\rm R}+\mu_{\rm L}}{2}\,.
\end{align}

So far, we assumed zero temperature and ignored the contributions from the HLLs.
We then examine if these effects change the CME.

We first consider the temperature effect.
At finite temperature, we have to include the antifermions.
The distribution functions for right-handed fermions and antifermions are, respectively, given by
\begin{align}
    f_{\rm R}(p_z)
    =
    \fr{1}{\e^{\beta(|p_z|-\mu_{\rm R})}+1}\,,
    \qquad
    \bar{f}_{\rm R}(p)
    =
    \fr{1}{\e^{\beta(|p_z|+\mu_{\rm R})}+1}\,.
\end{align}
Then, the electric current is expressed as
\begin{align}
\label{eq:Chi_Ph:R_current_CME_int}
    ej_{\rm{R}}^{z}
    &=
    \fr{e^2B}{2\pi}
    \int_{0}^{\infty}
    \fr{\dif p_z}{2\pi}
    \lb[f_{\rm R}(p_z)-\bar{f}_{\rm R}(p_z)\rb]\,.
\end{align}
To perform the integration, we use the formula \cite{Loganayagam:2012pz},
\begin{align}
\label{eq:Chi_Ph:Bernoulli_formula}
    \int_{0}^{\infty}\fr{\dif p}{2\pi}\lb(\fr{p}{2\pi}\rb)^n\lb[f(p)-(-1)^n\bar{f}(p)\rb]
    =
    \fr{(\im T)^{n+1}}{n+1}B_{n+1}\lb(\fr{1}{2}+\fr{\mu}{2\pi\im T}\rb)\,,
\end{align}
where $n=0, 1, 2, \cdots$ and $B_{n}(x)$ is the Bernoulli polynomial defined as
\begin{align}
    \fr{t\e^{xt}}{e^{t}-1}
    \equiv
    \sum_{n=0}^{\infty}\fr{t^n}{n!}B_{n}(x)\,.
\end{align}
The Bernoulli polynomials for $n=0, 1,2$ are
\begin{align}
    &B_{0}(x)=1\,,
    \\
    &B_{1}(x)=x-\fr{1}{2}\,,
    \\
    &B_{2}(x)=x^2-x+\fr{1}{6}\,.
\end{align}
From the case of $n=0$, we can find that the current has the same expression as Eq.~(\ref{eq:Chi_Ph:CME_current}).
Thus, the CME is not affected by the temperature.
This property originates from the fact that the chiral anomaly is topologically quantized.

We next investigate the contribution of the HLLs.
Using the dispersion relation (\ref{eq:Chi_Ph:Landau_energy}), we obtain the right-handed current as
\begin{align}
    j_{\rm R}^{z}
    &=
    \fr{eB}{2\pi}
    \int_{-\infty}^{\infty}
    \fr{\dif p_z}{2\pi}
    \fr{\del\varepsilon_{\rm R}}{\del p_z}
    \lb[f_{\rm R}(p_z)-\bar{f}_{\rm R}(p_z)\rb]
    =
    0\,.
\end{align}
Here, we used the fact that $\fr{\del\varepsilon_{\rm R}}{\del p_z}$ is an odd function and the Fermi distribution function is an even function with respect to the momentum $p_z$.
Thus, only the LLL contributes to the CME similarly to the chiral anomaly.

\section{Chiral magnetic wave}
\label{sec:Chi_Ph:CMW}
The chiral effects manifest in hydrodynamics as well.
In this section, we derive the CMW.

\subsection{Chiral effects in effective theories}
\begin{figure}[tb]
\centering
\begin{tikzpicture}[scale=1.2, >=stealth]

\draw[->, thick] (0,0) -- (10,0);
\node at (0,0) [above] {micro};
\node at (10,0) [above] {macro};

\node[draw, rounded corners=6pt, thick, inner sep=8pt] at (1,1.5) {Quantum Field Theory};
\node[draw, rounded corners=6pt, thick, inner sep=8pt] at (5,1.5) {Kinetic Theory};
\node[draw, rounded corners=6pt, thick, inner sep=8pt] at (9,1.5) {Hydrodynamics};
\node[draw, ellipse, thick, right of=qft, inner sep=6pt] at (4.2,3.5) {Chiral Anomaly};

\draw[->, thick] (3.5,3.25) -- (1,1.9);
\draw[->, thick] (5,3.05) -- (5,1.9);
\draw[->, thick] (6.6,3.25) -- (9,1.9);

\end{tikzpicture}
\caption{Schematic figure representing hierarchy of theoretical frameworks.}
\label{fig:Chi_Ph:Hierarchy}
\end{figure}
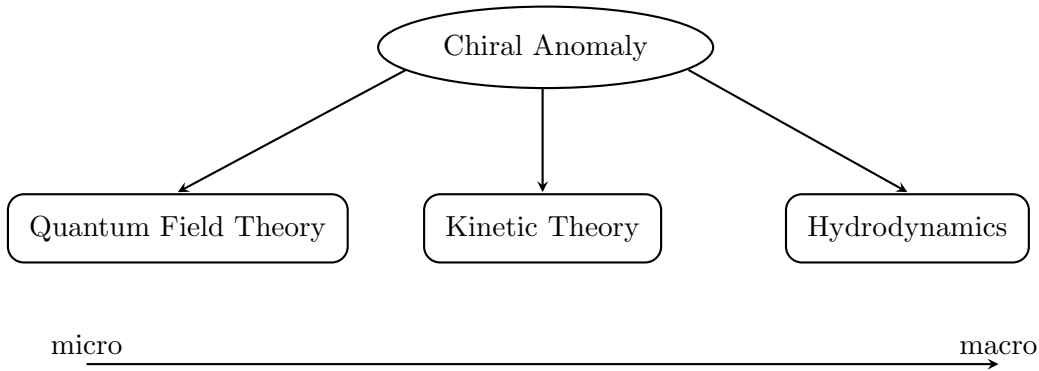

To investigate macroscopic dynamics of a system, we can use low-energy effective theories obtained by coarse-graining quantum field theory, such as kinetic theory and hydrodynamics.
As mentioned in Sec.~\ref{sec:Chi_Ph:Chi_Anom}, the chiral anomaly is exact due to the Adler--Bardeen theorem~\cite{Adler:1969er} and is generally scale independent, indicating its presence in both the UV and the IR scales equally, which is known as the anomaly matching~\cite{tHooft:1979rat}.
Therefore, the chiral anomaly should be included in kinetic theory and hydrodynamics (see Fig.~\ref{fig:Chi_Ph:Hierarchy}).
In this section, we focus on chiral effects in hydrodynamics.
We will discuss kinetic theory, including chiral effects in Sec.~{\ref{sec:Chi_Ph:CKT}}.

The chiral anomaly was first incorporated into hydrodynamics in Ref.~\cite{Son:2009tf}.
There, the electromagnetism is treated as the background, and the extension to magnetohydrodynamics (MHD) was investigated in Refs.~\cite{Giovannini:2013oga,Boyarsky:2015faa,Hattori:2017usa}.
Collective modes induced by the chiral effects are generally called chiral waves.
Among them, we here focus on the CMW~\cite{Newman:2005hd,Kharzeev:2010gd}, which plays a crucial role in the main part of this thesis.

\subsection{Derivation of CMW}
Before we derive the wave equation of the CMW, we give an intuitive explanation.
If the number density fluctuation $\delta n$ occurs due to perturbations, then the axial current fluctuation $\delta\bs{j}_5$ is induced by the CSE (\ref{eq:Chi_Ph:CSE_current}).
The fluctuation $\delta\bs{j}_5$ causes the chiral charge density fluctuation $\delta n_5$, resulting in the density fluctuation $\delta n$ again by the CME (\ref{eq:Chi_Ph:CME_current}).
Repeating this process, the density fluctuation propagates along the magnetic field, which is the CMW.
As this argument shows, the CMW can appear even without the chirality imbalance at equilibrium ($\bar{n}_{5}=0$), unlike the CME.

In general, hydrodynamic equations are formulated as follows~(see, e.g., Ref.~\cite{Landau:6}):
First, we identify the hydrodynamic variables, which are conserved quantities and NG modes.
We use the conservation laws based on the symmetry of a system.
To close the equations, we have to write down the currents in terms of the hydrodynamic variables, which are called the constitutive equations.

We now derive the wave equation of the CMW.
In our purpose to derive the CMW, it is sufficient to consider only the number density and the chiral charge density as the hydrodynamic variables in the presence of a background magnetic field.
We assume that the system possesses the ${\rm U(1)}_{\rm V}$ and ${\rm U(1)}_{\rm A}$ symmetries, leading to the conservation laws:
\begin{align}
    \del_{t}n+\bs{\nabla}\cdot\bs{j}=0\,,
    \qquad
    \del_{t}n_5+\bs{\nabla}\cdot\bs{j}_5=0\,.
\end{align}
The chiral anomaly does not appear due to the absence of electric fields.
For simplicity, we here do not consider the diffusion, and the constitutive equations are just given by the CME (\ref{eq:Chi_Ph:CME_current}) and CSE (\ref{eq:Chi_Ph:CSE_current}).
To examine the collective behavior, we decompose the hydrodynamic variables into the equilibrium value and the fluctuation: 
\begin{align}
    n(t,\bs{x})=\bar{n}+\delta n(t,\bs{x})\,,
    \qquad
    n_5(t,\bs{x})=\bar{n}_5+\delta n_5(t,\bs{x})\,.
\end{align}
Using the number susceptibilities,
\begin{align}
    \chi\equiv\fr{\del n}{\del\mu}\,,
    \qquad
    \chi_{5}\equiv\fr{\del n_5}{\del\mu_5}\,,
\end{align}
the fluctuations of the currents can be expressed as
\begin{align}
    \delta\bs{j}
    =
    \fr{e\bs{B}}{2\pi\chi_5}\delta n_{5}\,,
    \qquad
    \delta\bs{j}_5
    =
    \fr{e\bs{B}}{2\pi\chi}\delta n\,.
\end{align}
We now suppose the absence of the chirality imbalance $(\bar{n}_5=0)$.
In this case, the susceptibilities of the number density and the chiral charge density have the same form ($\chi=\chi_5$).%
\footnote{From the symmetry viewpoint, the number density and the chiral charge density are in general expressed as $n\sim\#\mu^3+\#T^2\mu$ and $n_{5}\sim\#\mu^2\mu_5+\#T^2\mu_5$, respectively. Here, we ignored the higher-order terms in $\mu_5$. We can then find that the susceptibilities are common for each density.}
Thus, the hydrodynamic equations are
\begin{align}
    \del_t\delta n+\fr{e\bs{B}}{2\pi\chi}\cdot\bs{\nabla}\delta n_{5}
    =0\,,
    \qquad
    \del_t\delta n_5+\fr{e\bs{B}}{2\pi\chi}\cdot\bs{\nabla}\delta n
    =0\,.
\end{align}
Eliminating $\delta n_5$ from these equations, we finally obtain the wave equation for the CMW:
\begin{align}
    \lb[\del_t^2-\lb(\fr{e\bs{B}}{2\pi^2\chi}\cdot\bs{\nabla}\rb)^2\rb]\delta n=0\,.
\end{align}
From the wave equation, we can derive the dispersion relation of the CMW as
\begin{align}
    \omega
    =
    \pm\fr{e\bs{B}}{2\pi^2\chi}\cdot\bs{k}\,.
\end{align}
In this derivation, we have set $\bar{n}_5=0$, and the CMW can propagate even without the chirality imbalance.
This result is consistent with the above intuitive explanation.
The velocity of the CMW includes the coefficient of the CME, or equivalently, the anomaly coefficient.

\section{Chiral vortical effect and chiral vortical wave}
\label{sec:Chi_Ph:CVE_CVW}
We have seen that, in a system with a chirality imbalance, the number current is induced by the magnetic field.
The current can also be generated by vorticity, which is known as the CVE.
In this section, we discuss the CVE and the resulting collective excitations, the CVW.

\subsection{Correspondence between magnetic fields and rotation}
We first give an explanation of the relation between the magnetic field $\bs{B}$ and the rotation $\bs{\Omega}$.
In the non-relativistic case, charged particles are subject to the Lorentz force,
\begin{align}
    \bs{F}
    =
    e\bs{v}\times\bs{B}\,,
\end{align}
where $\bs{v}$ is the velocity of a particle. 
On the other hand, in a system with the angular velocity $\bs{\Omega}$, the inertia force acts on the particle is
\begin{align}
    \bs{F}
    =
    2m\bs{v}\times\bs{\Omega}
    -m\bs{\Omega}\times(\bs{\Omega}\times\bs{x})\,.
\end{align}
The first term represents the Coriolis force, and the second term is the centrifugal force.
Up to $\mcl{O}(\bs{\Omega})$, we can ignore the latter.
From these equations, we can find the correspondence,
\begin{align}
    e\bs{B}\leftrightarrow 2m\bs{\Omega}\,.
\end{align}
For a relativistic particle, we can obtain the correspondence by replacing the mass $m$ with the energy $\varepsilon=|\bs{p}|$ as
\begin{align}
\label{eq:Chi_Ph:Mag_rot_correspondence_rel}
    e\bs{B}\leftrightarrow 2|\bs{p}|\bs{\Omega}\,.
\end{align}

\subsection{Chiral vortical effect}
We now derive the CVE at finite temperature. 
We set the angular velocity of the system to be along the $z$-axis $(\bs{\Omega}=\Omega\bs{e}_{z})$.
Applying the relation (\ref{eq:Chi_Ph:Mag_rot_correspondence_rel}) to the electric current for the right-handed fermion (\ref{eq:Chi_Ph:R_current_CME_int}), we obtain the CVE~\cite{Vilenkin:1979ui,Son:2009tf,Landsteiner:2011cp},
\begin{align}
    j_{\rm{R}}^{z}
    &=
    \fr{\Omega}{2\pi}
    \int_{0}^{\infty}
    \fr{\dif p_z}{2\pi}
    \lb[(+2p_z)f_{\rm R}(p_z)-(-2p_z)\bar{f}_{\rm R}(p_z)\rb]
    \nom
    &=
    \lb(\fr{\mu_{\rm R}^2}{4\pi^2}+\fr{T^2}{12}\rb)\Omega\,,
\end{align}
where we used the formula (\ref{eq:Chi_Ph:Bernoulli_formula}).
For the left-handed fermion, the current is given by
\begin{align}
    j_{\rm{L}}^{z}
    &=
    -\lb(\fr{\mu_{\rm L}^2}{4\pi^2}+\fr{T^2}{12}\rb)\Omega\,,
\end{align}
From these, we obtain the vector and axial currents as
\begin{align}
    \bs{j}
    =\fr{\mu\mu_5}{\pi^2}\bs{\Omega}\,,
    \qquad
    \bs{j}_{5}
    =\lb(\fr{\mu^2+\mu_5^2}{2\pi^2}+\fr{T^2}{6}\rb)\bs{\Omega}\,.
\end{align}

\subsection{Chiral vortical wave}
The CVE can induce the collective mode called the CVW~\cite{Jiang:2015cva}.
In the same way as the CMW, we derive the dispersion relation of the CVW.
Since we will treat matter composed of neutrinos in the main part, we treat chiral fermions instead of Dirac fermions.
We then consider only the number densities of the right- and left-handed fermions as the hydrodynamic variable.
Due to the absence of the chiral anomaly, these numbers are conserved:
\begin{align}
    \del_tn_{\lambda}+\bs{\nabla}\cdot\bs{j}_{\lambda}=0\,.
\end{align}
By decomposing the number densities as $n_{\lambda}(t,\bs{x})=\bar{n}_{\lambda}+\delta n_{\lambda}(t,\bs{x})$, we obtain the fluctuations of the currents,
\begin{align}
\label{eq:ChiTra:linearized_CVE_RL}
    \delta\bs{j}_{\lambda}
    =
    \pm\fr{\bar{\mu}_{\lambda}\bs{\Omega}}{2\pi^2\chi_{\lambda}}\delta n_{\lambda}\,,
    \qquad
    \chi_{\lambda}\equiv\fr{\del n_{\lambda}}{\del\mu_{\lambda}}\,.
\end{align}
From the above, the wave equations of the CVW are
\begin{align}
    \lb(\del_{t}\pm\frac{\bar{\mu}_{\lambda}\bs{\Omega}}{2\pi^2\chi_{\lambda}}\cdot\bs{\nabla}\rb)\delta n_{\lambda}=0\,.
\end{align}
The dispersion relation is
\begin{align}
    \omega_{\lambda}
    =
    \pm\fr{\bar{\mu}_{\lambda}\bs{\Omega}}{2\pi^2\chi_{\lambda}}\cdot\bs{k}\,.
\end{align}
The CVW of the right-handed fermions propagates in the direction of the rotation vector, whereas that of left-handed fermions propagates in the opposite direction.

\section{Chiral kinetic theory}
\label{sec:Chi_Ph:CKT}
To study the dynamics of the system out of equilibrium, one can use kinetic theory.
Kinetic theory is an effective theory derived by coarse-graining quantum field theory and describes physics at more microscopic scales of hydrodynamics.
As explained in Sec.~{\ref{sec:Chi_Ph:CMW}}, chiral effects should be included in effective theories, while the classical Boltzmann equation cannot reproduce the chiral anomaly and the CME.
By taking the Berry phase~\cite{Berry:1984jv} into account, we can construct the kinetic theory capturing such quantum effects, which is the chiral kinetic theory~\cite{Son:2012wh,Stephanov:2012ki,Son:2012zy}.

\subsection{Berry phase}
We first provide a brief review of the Berry phase in quantum mechanics.%
\footnote{Our discussion here follows Ref.~\cite{Sakurai:2011zz}.}
We consider a time-dependent Hamiltonian $H(t)$.
The eigenvalue equation is given by
\begin{align}
    H(t)\ket{n(t)}=\varepsilon_{n}(t)\ket{n(t)}\,,
\end{align}
where $\ket{n(t)}$ and $\varepsilon_{n}(t)$ are the eigenstate and eigenvalue labeled by $n$, respectively.
The time evolution is governed by the Schr\"{o}dinger equation,
\begin{align}
    \im\fr{\del}{\del t}\ket{\Psi(t)}
    =H(t)\ket{\Psi(t)}\,.
\end{align}
To solve the equation, we assume the solution in the form of
\begin{align}
    \ket{\Psi(t)}
    =\sum_{n}c_n(t)\e^{\im\theta_n(t)}\ket{n(t)}\,,
    \qquad
    \theta_{n}(t)
    \equiv
    -\int_{0}^t\dif t'\varepsilon_{n}(t')\,.
\end{align}
Substituting this into the Schr\"{o}dinger equation and multiplying it by $\bra{m(t)}$, the time evolution of the coefficient $c_m(t)$ is determined by
\begin{align}
    \dot{c}_{m}(t)
    &=
    -\sum_{n}c_n(t)\e^{\im[\theta_n(t)-\theta_m(t)]}\bra{m(t)}\fr{\del}{\del t}\ket{n(t)}
    \nom
    &=
    -c_m(t)\bra{m(t)}\fr{\del}{\del t}\ket{m(t)}
    -\sum_{n\neq m}c_n(t)\e^{\im[\theta_n(t)-\theta_m(t)]}\fr{\bra{m(t)}\dot{H}\ket{n(t)}}{\varepsilon_{n}(t)-\varepsilon_{m}(t)}\,,
\end{align}
where we used $\bra{m(t)}\dot{H}\ket{n(t)}=[\varepsilon_{n}(t)-\varepsilon_m(t)]\bra{m(t)}\del_t\ket{n(t)}$ for $m\neq n$.
We now take the adiabatic approximation, in which we neglect the second term due to the absence of mixing between energy levels. 
The coefficient is then written as
\begin{align}
    c_m(t)
    =\e^{\im\gamma_m(t)}c_m(0)\,,
    \qquad
    \gamma_m(t)
    \equiv
    \im\int_0^t\dif t'\bra{m(t')}\fr{\del}{\del t'}\ket{m(t')}\,.
\end{align}
We can show that the phase $\gamma_m$ is a real-valued function using the relation $\del_t\braket{m(t)|m(t)}=0$.

We consider the case where the time dependence of the Hamiltonian comes from a set of parameters $\bs{R}(t)$.
Assuming that the parameters move along a closed path C, we obtain the Berry phase:
\begin{align}
\label{eq:Chi_Ph:Berry_phase}
    \gamma_{m}({\rm C})
    =\int_{\rm C}\dif\bs{R}\cdot\bs{a}_{\bs{R}}
    =\int_{\rm S}\dif\bs{S}\cdot\bs{\Omega}_{\bs{R}}\,,
\end{align}
where S is a surface closed by the loop C in parameter space.
We defined the Berry connection and the Berry curvature as
\begin{align}
    \bs{a}_{\bs{R}}\equiv\im\bra{m(\bs{R})}\bs{\nabla}_{\bs{R}}\ket{m(\bs{R})}\,,
    \qquad
    \bs{\Omega}_{\bs{R}}\equiv\bs{\nabla_{\bs{R}}}\times\bs{a}_{\bs{R}}\,.
\end{align}
In the second equality in Eq.~(\ref{eq:Chi_Ph:Berry_phase}), we used the Stokes theorem.
The mathematical relations above are analogous to those of gauge fields, but in parameter space.

\subsection{Path integral formalism}
In kinetic theory, chiral fermions are treated semi-classically and described by the distribution function $f_{\lambda}(t,\bs{x},\bs{p})$ where $\lambda=\pm$ denotes the chirality.
The dynamics of the distribution function is governed by the Boltzmann equation,
\begin{align}
\label{eq:Chi_Ph:Boltzmann_collisionless}
    \fr{\del f_{\lambda}}{\del t}
    +\dot{\bs{x}}\cdot\fr{\del f_{\lambda}}{\del\bs{x}}
    +\dot{\bs{p}}\cdot\fr{\del f_{\lambda}}{\del\bs{p}}
    =C[f_\lambda]\,,
\end{align}
where $C[f_\lambda]$ is the collision term.
For chiral fermions, the equations of motion are modified by the Berry curvature.
To derive the chiral kinetic theory, we focus on right-handed chiral fermions $(\lambda=+)$ in a background electromagnetic field. 
The chiral kinetic theory can be constructed in various ways: the Hamiltonian formalism~\cite{Son:2012wh,Son:2012zy}, the path integral formalism~\cite{Stephanov:2012ki,Chen:2014cla}, and the Wigner function approach~\cite{Son:2012zy,Chen:2012ca,Hidaka:2016yjf,Hidaka:2017auj}.
Here, we follow the path integral approach.

The Hamiltonian of the right-handed chiral fermion is given by
\begin{align}
\label{eq:Chi_Ph:Hamiltonian_chiral_BGEM}
    H=\bs{\sigma}\cdot[\bs{\pi}-e\bs{A}(x)]+e\phi(x)\,,
\end{align}
where $\bs{\sigma}$ are the Pauli matrices and $A^{\mu}=(\phi,\bs{A})$ is the gauge field.
We denote the canonical momentum as $\bs{\pi}=\bs{p}+e\bs{A}$ with $\bs{p}$ being the kinetic momentum.
To take the quantum effect into account, we introduce the transition matrix as
\begin{align}
    K(t_{\rm f},t_{\rm i})
    &=\bra{\bs{x}_{\rm f}}\e^{-\im H(t_{\rm f}-t_{\rm i})}\ket{\bs{x}_{\rm i}}
    \nom
    &=\int\mcl{D}\bs{x}\mcl{D}\bs{\pi}P\exp\lb[\im\int_{t_{\rm i}}^{t_{\rm f}}\dif t\lb(\bs{\pi}\cdot\dot{\bs{x}}-\bs{\sigma}\cdot\bs{p}-e\phi\rb)\rb]\,,
\end{align}
where $P$ represents the path-ordering operator.

Since $(\bs{\sigma}\cdot\bs{p})^2=|\bs{p}|^2$, we can diagonalize the Hamiltonian (\ref{eq:Chi_Ph:Hamiltonian_chiral_BGEM}) using a unitary matrix $U_{\bs{p}}$ as
\begin{align}
    U_{\bs{p}}^{\dagger}HU_{\bs{p}}
    &=\lb(
    \begin{array}{cc}
        +|\bs{p}|+e\phi &  0\\
        0 & -|\bs{p}|+e\phi
    \end{array}\rb)
    \nom
    &=|\bs{p}|\sigma_z+e\phi\,.
\end{align}
For the eigenvalues, there are two helicity states: one is the fermion, and the other is the antifermion.
To focus on the trajectory of each state, we write the transition matrix in the helicity basis.
This corresponds to tracing a specific eigenstate of the time-dependent Hamiltonian under the adiabatic approximation.
To avoid the crossing of the two energy levels, the momentum must be sufficiently large (see also below).

By inserting the identity matrices, the transition matrix is diagonalized as
\begin{align}
    \int\mcl{D}\bs{x}\mcl{D}\bs{\pi}P\exp\lb(-\im\int_{t_{\rm i}}^{t_{\rm f}}\dif t\bs{\sigma}\cdot\bs{p}\rb)
    &=\lim_{N\to\infty}\prod_{j=1}^{N-1}\exp\lb(-\im\bs{\sigma}\cdot\bs{p}_{j}\varDelta t\rb)
    \nom
    &=\lim_{N\to\infty}\prod_{j=1}^{N-1}U_{\bs{p}_j}\exp\lb(-\im|\bs{p}_{j}|\sigma_z\varDelta t\rb)U^{\dagger}_{\bs{p}_j}\,,
\end{align}
where $\varDelta t\equiv (t_{\rm f}-t_{\rm i})/N$.
Between the exponential factors, we encounter the additional matrices $U^{\dagger}_{\bs{p}_j}U_{\bs{p}_{j-1}}$.
Assuming that $\varDelta\bs{p}_{j}\equiv\bs{p}_j-\bs{p}_{j-1}$ is small enough, we can expand it as
\begin{align}
    U^{\dagger}_{\bs{p}_j}U_{\bs{p}_{j-1}}
    &\simeq
    1-\varDelta\bs{p}_j\cdot U^{\dagger}_{\bs{p}_j}\bs{\nabla}_{\bs{p}_j}U_{\bs{p}_{j}}
    \nom
    &\simeq
    \exp\lb(-\im\bs{a}_{\bs{p}_j}\cdot\varDelta\bs{p}_j\rb)\,,
\end{align}
where $\bs{a}_{\bs{p}_j}\equiv-\im U^{\dagger}_{\bs{p}_j}\bs{\nabla}_{\bs{p}_j}U_{\bs{p}_{j}}$ is the Berry connection.
Therefore, the transition matrix reduces to
\begin{align}
    K(t_{\rm f},t_{\rm i})
    &=\int\mcl{D}\bs{x}\mcl{D}\bs{\pi}U_{\bs{p}_{\rm f}}P\exp\lb[\im\int_{t_{\rm i}}^{t_{\rm f}}\dif t
    \lb(\bs{p}\cdot\dot{\bs{x}}+e\bs{A}\cdot\dot{\bs{x}}-\bs{a}_{\bs{p}}\cdot\dot{\bs{p}}-\varepsilon_{\bs{p}}\sigma_z-e\phi\rb)
    \rb]U^{\dagger}_{\bs{p}_{\rm i}}\,,
\end{align}
where $\varepsilon_{\bs{p}}$ is the energy of the chiral fermion.
From the expression, we can read the analogous relation between the gauge field and the Berry connection.

We now focus on the fermion sector ((1,1)-component of the matrix).
This is justified when the diagonal component of $\bs{a}_{\bs{p}}$ is sufficiently larger than the off-diagonal component.
The semi-classical action is
\begin{align}
    S=\int\lb[(\bs{p}+e\bs{A})\cdot\dif\bs{x}-\bs{a}_{\bs{p}}\cdot\dif\bs{p}-(\varepsilon+e\phi)\dif t\rb]\,,
\end{align}
leading to the equations of motion as
\begin{align}
    \dot{\bs{x}}=\bs{v}+\dot{\bs{p}}\times\bs{\Omega}_{\bs{p}}\,,
    \qquad
    \dot{\bs{p}}=e(\bs{E}+\dot{\bs{x}}\times\bs{B})\,,
\end{align}
where $\bs{v}\equiv\bs{p}/|\bs{p}|$.
For the right-handed chiral fermions, the Berry curvature is given by
\begin{align}
    \bs{\Omega}_{\bs{p}}=\fr{\bs{p}}{2|\bs{p}|^3}\,.
\end{align}
This has the same expression as the magnetic field of a magnetic monopole.
The evolution of the coordinate $\bs{x}$ in phase space is then driven by the ``Lorentz force'' in momentum space. 
By solving these equations of motion, the Boltzmann equation (\ref{eq:Chi_Ph:Boltzmann_collisionless}) gives the chiral kinetic equation.
It is known that the Lorentz symmetry requires the correction by the magnetic moment as~\cite{Son:2012zy,Chen:2014cla}
\begin{align}
\label{eq:Chi_Ph:Dispersion_relation_chiral_magnetic_moment}
    \varepsilon_{\bs{p}}=|\bs{p}|(1-e\bs{B}\cdot\bs{\Omega}_{\bs{p}})\,.
\end{align}
Including the modification, we finally obtain the equations of motion for the chiral fermions:
\begin{align}
\label{eq:Chi_Ph:EoM_x}
    &\sqrt{G}\dot{\bs{x}}
    =
    \tilde{\bs{v}}+e\tilde{\bs{E}}\times\bs{\Omega}_{\lambda}+(\tilde{\bs{v}}\cdot\bs{\Omega}_{\lambda})e\bs{B}\,,
    \\
\label{eq:Chi_Ph:EoM_p}
    &\sqrt{G}\dot{\bs{p}}
    =
    e(\tilde{\bs{E}}+\tilde{\bs{v}}\times\bs{B})+e^2(\tilde{\bs{E}}\cdot\bs{B})\bs{\Omega}_{\lambda}\,,
\end{align}
where
\begin{align}
    \sqrt{G}\equiv1+e\bs{B}\cdot\bs{\Omega}_{\bs{p}}\,,
    \qquad
    \tilde{\bs{v}}\equiv\fr{\del\varepsilon_{\bs{p}}}{\del\bs{p}}\,,
    \qquad
    e\tilde{\bs{E}}\equiv e\bs{E}-\fr{\del\varepsilon_{\bs{p}}}{\del\bs{x}}\,,
    \qquad
    \bs{\Omega}_{\lambda}\equiv\lambda\fr{\bs{p}}{2|\bs{p}|^3}\,.
\end{align}

\subsection{Chiral anomaly revisited}
We examine whether the chiral kinetic theory can reproduce the chiral anomaly.
Since the invariant phase space is modified as $\sqrt{G}\dif^3\bs{x}\dif^3\bs{p}/(2\pi)^3$, we can define the number density and the number current as
\begin{align}
    &n_{\lambda}(t,\bs{x})
    =
    \int\fr{\dif^3\bs{p}}{(2\pi)^3}\sqrt{G}f_{\lambda}
    +(\text{antifermion})\,,
    \\
    &\bs{j}_{\lambda}(t,\bs{x})
    =
    \int\fr{\dif^3\bs{p}}{(2\pi)^3}\sqrt{G}\dot{\bs{x}}f_{\lambda}
    +(\text{antifermion})
    \nom
\label{eq:Chi_Ph:CKT_current}
    &\hspace{3.15em}
    =\int\fr{\dif^3\bs{p}}{(2\pi)^3}
    \lb[\tilde{\bs{v}}
    +(\tilde{\bs{v}}\cdot\bs{\Omega}_{\lambda})e\bs{B}
    -\varepsilon_{\bs{p}}\bs{\Omega}_{\lambda}\times\fr{\del}{\del\bs{x}}\rb]f_{\lambda}
    +e\bs{E}\times\bs{\sigma}_{\lambda}
    +(\text{antifermion})\,,
\end{align}
where we defined
\begin{align}
    \bs{\sigma}_{\lambda}
    \equiv
    \int\fr{\dif^3\bs{p}}{(2\pi)^3}\bs{\Omega}_{\lambda}f_{\lambda}\,.
\end{align}
Multiplying Eq.~(\ref{eq:Chi_Ph:Boltzmann_collisionless}) by $\sqrt{G}$ and using Eqs.~(\ref{eq:Chi_Ph:EoM_x}) and (\ref{eq:Chi_Ph:EoM_p}), we can reproduce the chiral anomaly
\begin{align}
\label{eq:Chi_Ph:Chiral_anomaly}
    \del_t n_{\lambda}
    +\bs{\nabla}\cdot\bs{j}_{\lambda}
    &=
    -e^2\int\fr{\dif^3\bs{p}}{(2\pi)^3}\lb(\bs{\Omega}_{\lambda}\cdot\fr{\del}{\del\bs{p}}[f_{\lambda}+\bar{f}_{\lambda}]\rb)\bs{E}\cdot\bs{B}
    \nom
    &=
    \lambda\fr{e^2}{4\pi^2}\bs{E}\cdot\bs{B}\,,
\end{align}
where we used $\bs{\nabla}_{p}\cdot\bs{\Omega}_{\lambda}=2\pi\lambda\delta(\bs{p})$.
At low temperatures, antifermion contributions are suppressed, and we have $f_{\lambda}(\bs{p}=0)=1$ \cite{Son:2012zy}.
Conversely, at high temperatures, antifermions need to be included to reproduce the chiral anomaly \cite{Manuel:2013zaa}.
This fact is also examined using the Wigner function \cite{Gao:2019zhk}.

\chapter{Neutron Stars and Asteroseismology}
\label{chap:NS_Astero}
In this chapter, we provide an overview of neutron stars and asteroseismology.
In Sec.~\ref{sec:NS_Astero:Int_structure_NS}, we introduce the internal structure of neutron stars.
In the following sections, we describe their fundamental properties, such as mass and radius (Sec.~\ref{sec:NS_Astero:MR}), rotation period (Sec.~\ref{sec:NS_Astero:Rotation_density}), and magnetic field (Sec.~\ref{sec:NS_Astero:Magnetic_field}).
We also discuss neutrino trapping in supernovae in Sec.~\ref{sec:NS_Astero:Neutrino_trapping}.
Finally, we review asteroseismology in the frameworks of Newtonian gravity (Sec.~\ref{sec:NS_Astero:Astero_NG}) and general relativity (Sec.~\ref{sec:NS_Astero:Astero_GR}).

\section{Internal structure of neutron stars}
\label{sec:NS_Astero:Int_structure_NS}
In this section, we give a schematic picture of a neutron star and explain why a neutron star is mainly composed of neutrons~\cite{Shapiro:1983du,Haensel:2007yy,Reisenegger:2015crq}.

\subsection{Overview}
Stars shine because nuclear fusion reactions occur in their interiors.
They maintain hydrostatic equilibrium through the balance between the pressure generated by nuclear burning and their self-gravity.
Inside stars, heavier elements such as helium, carbon, oxygen, $\cdots$ are progressively synthesized through nuclear fusion.
In stars with masses greater than about ten times the solar mass, nucleosynthesis proceeds up to iron.
Once iron is produced, nuclear fusion can no longer occur, and as no further pressure is generated, the star collapses due to its own gravity.
At this stage, a supernova explosion takes place.
The light generated in the explosion is extraordinarily bright, rivaling that of an entire galaxy.
Because such an event appears as a suddenly shining star, it is called a supernova.

In the core of the supernova, a dense object is formed by the gravitational collapse.
This remnant is a proto-neutron star.
Inside a proto-neutron star, neutrinos are produced, and as they are emitted, the star gradually cools and becomes a neutron star.
A neutron star can retain its form owing to the degeneracy pressure of neutrons and nuclear forces.
For even more massive stars, however, neither neutron degeneracy pressure nor nuclear forces can support their own gravity, and the star collapses into a spacetime singularity, namely a black hole.
Therefore, neutron stars are the densest matter in the universe.

\begin{figure}[t]
\centering
\begin{tikzpicture}[scale=1.5, font=\sffamily, >=stealth]

\definecolor{innercore}{RGB}{60, 80, 120}
\definecolor{outercore}{RGB}{90, 120, 170}
\definecolor{innercrust}{RGB}{150, 180, 220}
\definecolor{outercrust}{RGB}{210, 220, 240}

\filldraw[thick, fill=outercrust, draw=black] (0,0) circle (2);
\filldraw[thick, fill=innercrust, draw=black] (0,0) circle (1.7);
\filldraw[thick, fill=outercore, draw=black] (0,0) circle (1.2);
\filldraw[thick, fill=innercore, draw=black] (0,0) circle (0.5);

\node[draw, rounded corners=6pt, thick, inner sep=8pt, align=left] at (-3.5,2.6) 
    {{\large Outer Crust}\\
    \hspace{1em}$\cdot$~neutron-rich nuclei\\
    \hspace{1em}$\cdot$~electrons};
\node[draw, rounded corners=6pt, thick, inner sep=8pt, align=left] at (-4,0.9) 
    {{\large Inner Crust}\\
    \hspace{1em}$\cdot$~neutron-rich nuclei\\
    \hspace{1em}$\cdot$~dripped neutrons\\
    \hspace{1em}$\cdot$~electrons};
\node[draw, rounded corners=6pt, thick, inner sep=8pt, align=left] at (-4,-0.9) 
    {{\large Outer Core}\\
    \hspace{1em}$\cdot$~neutrons\\
    \hspace{1em}$\cdot$~protons\\
    \hspace{1em}$\cdot$~electrons, muons};
\node[draw, rounded corners=6pt, thick, inner sep=8pt, align=left] at (-3.5,-2.6) 
    {{\large Inner Core}\\
    \hspace{1em}$\cdot$~hyperons\\
    \hspace{1em}$\cdot$~quarks};

\node at (3,1.2) {$0.001n_0$};
\node at (3,0) {$0.5n_0$};
\node at (3,-1.2) {$2n_0$};

\draw[->, thick] (-2.1,2.6) -- (-1,1.55);
\draw[->, thick] (-2.6,0.9) -- (-1.3,0.6);
\draw[->, thick] (-2.7,-0.9) -- (-0.9,-0.4);
\draw[->, thick] (-2.64,-2.6) -- (-0.2,-0.2);

\draw[->, thick] (2.5,1.2) -- (1.5,0.8);
\draw[->, thick] (2.63,0) -- (1.2,0);
\draw[->, thick] (2.75,-1.2) -- (0.4,-0.3);

\end{tikzpicture}
\caption{Schematic structure of a neutron star interior. The boundaries represent typical densities.}
\label{fig:NS_Astero:neutron_star_structure}
\end{figure}
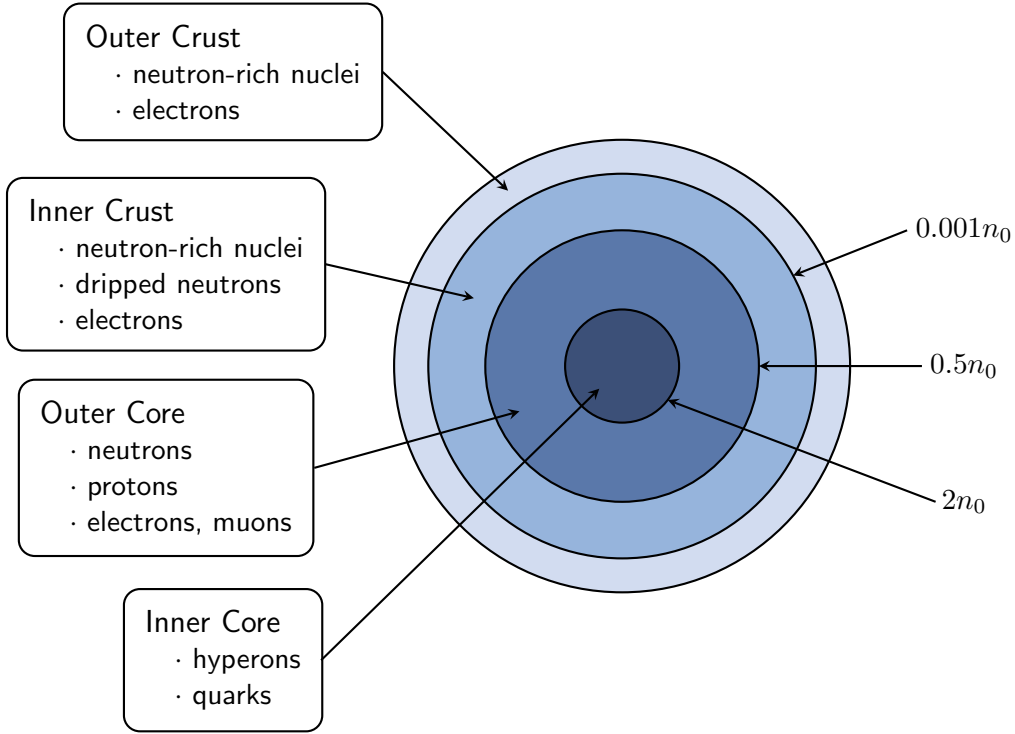

The internal structure of a neutron star consists of four layers (see Fig.~\ref{fig:NS_Astero:neutron_star_structure}): outer crust, inner crust, outer core, and inner core.
The outer crust consists of neutron-rich nuclei and electrons.
The nuclei are solidified by the Coulomb force.
The electrons form a non-degenerate gas.
When the density exceeds $0.001n_0$, where $n_0\simeq 0.17~{\rm fm^{-3}}$ is the nuclear saturation density, neutrons start to drip out of neutron-rich nuclei.
This layer is called the inner crust.
In this region, neutron-rich nuclei form so-called ``pasta'' structures.
The electrons constitute a degenerate gas.
The region from $0.5n_0$ to $2n_0$ corresponds to the outer core.
In this region, neutrons and protons behave as Fermi liquids, and the superfluid and superconducting phases emerge.
In addition to electrons, muons also appear.
Finally, in the inner core, where the density exceeds $2n_0$, hyperons, which are baryons containing strange quarks, may appear.
Furthermore, quark matter, such as color-superconducting phases, is expected to exist, although its detailed properties remain uncertain.

\subsection{Chemical composition}
\begin{figure}[tb]
\centering
\begin{tikzpicture}
    \draw[->,>=stealth,very thick](0,0)--(0,3) node [above]{$E$};
    \draw[thick](0.5,2.3)--(2.5,2.3) node [above] at  (1.5,2.3) {n};
    \draw[thick](1.5,0.5)--(4,0.5);
    \draw[<->,>=stealth,thick](2.0,0.5)--(2.0,2.3) node [left] at (2.0,1.4) {$Q$};
    \node[above] at (2.5,0.5) {p};
    \node[above] at (3.5,0.55) {$\e^{-}$};
\end{tikzpicture}
\caption{The energy relation of the $\beta$-decay.}
\label{fig:NS:Beta_decay}
\end{figure}
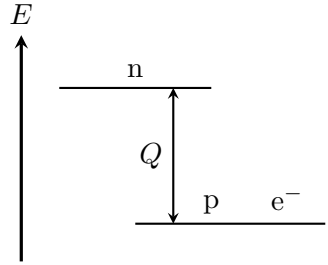

We now discuss why a neutron star is mainly composed of neutrons.
A neutron in the vacuum decays in about 15 min by the $\beta$-decay,
\begin{align}
\label{eq:NS_Astero:Beta_decay}
    {\rm n}\to{\rm p}+{\rm e^{-}}+\bar{\rm \nu}_{\rm e}\,,
\end{align}
where n, p, ${\rm e^{-}}$, and $\bar{\rm \nu}_{\rm e}$ denote the neutron, proton, electron, and anti-electron neutrino.
The mass difference between the final and initial states is%
\footnote{Since the neutrino mass is sufficiently small, we ignored it.}
\begin{align}
    Q
    &\equiv
    m_{\rm n}-(m_{\rm p}+m_{\rm e})
    \nom
    &\simeq
    939.57~{\rm MeV}-938.28~{\rm MeV}-0.51~{\rm MeV}
    \nom
    &=
    0.78~{\rm MeV}\,.
\end{align}
Since $Q$ is positive (see Fig.~\ref{fig:NS:Beta_decay}), the decay of the neutron is the more likely process.
On the other hand, if the kinetic energy of the protons and electrons exceeds the energy $Q$, the inverse process of the $\beta$-decay can occur:
\begin{align}
\label{eq:NS_Astero:Electron_capture}
    {\rm p}+{\rm e^{-}}\to{\rm n}+{\rm \nu}_{\rm e}\,.
\end{align}
This is called the electron capture.
In this process, the protons and electrons are combined into neutrons.
Thus, if the electron capture is dominant, the amount of neutrons is increased.
As shown below, at high density, the electron chemical potential is much higher than the proton chemical potential, and the electron capture is the preferred process to decrease the electrons.

For simplicity, we consider the degenerate fermions. 
The number density of the fermion is given by
\begin{align}
\label{eq:NS_Astero:Number_density_Fermi_momentum}
    n_i
    =
    2\int\fr{\dif^3\bs{p}}{(2\pi)^3} \theta(|\bs{p}_{{\rm F},i}|-|\bs{p}|)
    =
    \fr{|\bs{p}_{\rm F}|^3}{3\pi^2}\,,
\end{align}
where $i$ denotes the spices of the fermions $(i={\rm n},{\rm p},{\rm e})$ and $\bs{p}_{{\rm F},i}$ is the Fermi momentum.
The factor $2$ comes from the spin degree of freedom.
In the degenerate matter, the Fermi energy is equal to the chemical potential:
\begin{align}
\label{eq:NS_Astero:Chemical_potential}
    \mu_i
    =
    \sqrt{|\bs{p}_{{\rm F},i}|^2+{m_{i}}^2}\,.
\end{align}
As the density increases, the Fermi momentum becomes larger.
By the $\beta$-decay (\ref{eq:NS_Astero:Beta_decay}) and electron capture $(\ref{eq:NS_Astero:Electron_capture})$, the neutrinos carry away the energy, and the neutron star is cooled until it reaches equilibrium.
Consequently, with respect to each chemical potential, the following relation is satisfied:
\begin{align}
\label{eq:NS_Astero:Chemical_equiliblium}
    \mu_{\rm n}
    =
    \mu_{\rm p}+\mu_{\rm e}\,.
\end{align}

The typical Fermi momentum is $|\bs{p}_{{\rm F},i}|\sim 10^{2}~{\rm MeV}$.
While the electron mass is much smaller ($m_{\rm e}\simeq 0.5~{\rm MeV}
$), the nucleon mass is larger ($m_{\rm n}\simeq m_{\rm p}\equiv m_{\rm N}\simeq 940~{\rm MeV}
$).%
\footnote{The subscript of $m_{\rm N}$ means the nucleon.}
We then assume that there are relativistic electrons and non-relativistic protons and neutrons in the neutron star.
We can then rewrite Eq.~(\ref{eq:NS_Astero:Chemical_equiliblium}) as
\begin{align}
\label{eq:NS_Astero:Momentum_n_p_e}
    \fr{|\bs{p}_{\rm F,n}|^2}{2m_{\rm N}}
    =
    \fr{|\bs{p}_{\rm F,p}|^2}{2m_{\rm N}}+|\bs{p}_{\rm F,e}|\,.
\end{align}
Additionally, we impose charge neutrality,
\begin{align}
\label{eq:NS_Astero:Charge_neutrality}
    n_{\rm p}
    =
    n_{\rm e}\,.
\end{align}
This means that the proton Fermi momentum is equal to that of the electron $(|\bs{p}_{\rm F,p}|=|\bs{p}_{\rm F,e}|)$.
Because $|\bs{p}_{\rm F,e}|\ll m_{\rm N}$, we can ignore the first term on the right-hand side in Eq.~(\ref{eq:NS_Astero:Momentum_n_p_e}) and obtain
\begin{align}
    \fr{|\bs{p}_{\rm F,n}|^2}{2m_{\rm N}}
    \simeq
    |\bs{p}_{\rm F,e}|\,.
\end{align}
This leads to the electron fraction as
\begin{align}
    y_{\rm e}
    \equiv
    \fr{n_{\rm e}}{n_{\rm n}}
    =
    \lb(\fr{|\bs{p}_{\rm F,e}|}{|\bs{p}_{\rm F,n}|}\rb)^{3}
    \simeq
    \lb(\fr{|\bs{p}_{\rm F,e}|}{2m_{\rm N}}\rb)^{3/2}
    \ll1\,.
\end{align}
Therefore, we find that a neutron star is mainly composed of neutrons.

\section{Mass and radius}
\label{sec:NS_Astero:MR}
Neutron stars provide us with various observational clues.
Among them, the mass and radius are the most typical observable quantities.

\subsection{Simple estimates}
We estimate the mass and radius of a neutron star in the simplest setup where the neutron star is a non-interacting neutron Fermi gas.
The main idea is to find the radius $R$ minimizing the energy of the neutron star~\cite{Reisenegger:2015crq}.

The energy of the neutron star can be written as a function of the mass $M$ and radius $R$ as
\begin{align}
    E_{\rm NS}(M,R)
    =
    \fr{4\pi}{3}R^3\varepsilon_{\rm n}+U(M,R)\,,
\end{align}
where $\varepsilon_{\rm n}$ denotes the neutron energy density,
\begin{align}
\label{eq:NS_Astero:Neutron_energy_density}
    \varepsilon_{\rm n}
    =
    2\int\fr{\dif^3\bs{p}}{(2\pi)^3}\sqrt{|\bs{p}|^2+m_{\rm n}}\theta(|\bs{p}_{\rm F}|-|\bs{p}|)\,,
\end{align}
and $U$ is the self-gravitational potential.

We first consider the neutron energy density.
The Fermi momentum can be expressed using the mass and radius:
\begin{align}
    |\bs{p}_{\rm F}|
    =
    (3\pi^2 n)^{1/3}
    =
    \lb(\fr{9\pi}{4}\rb)^{1/3}\lb(\fr{M}{m_{\rm n}}\rb)^{1/3}\fr{1}{R}\,.
\end{align}
Thus, the energy density in the relativistic limit becomes
\begin{align}
\label{eq:NS_Astero:relativistic_neutron_energy}
    \varepsilon_{\rm n,R}
    &=
    2\int\fr{\dif^3\bs{p}}{(2\pi)^3}|\bs{p}|\theta(|\bs{p}_{\rm F}|-|\bs{p}|)
    \nom
    &=
    \fr{9}{16\pi}\lb(\fr{9\pi}{4}\rb)^{1/3}\lb(\fr{M}{m_{\rm n}}\rb)^{4/3}\fr{1}{R^4}\,.
\end{align}
In the non-relativistic limit, we have
\begin{align}
    \varepsilon_{\rm n,NR}
    &=
    2\int\fr{\dif^3\bs{p}}{(2\pi)^3}\fr{|\bs{p}|^2}{2m_{\rm n}}\theta(|\bs{p}_{\rm F}|-|\bs{p}|)
    \nom
    &=
    \fr{9}{40\pi m_{\rm n}}\lb(\fr{9\pi}{4}\rb)^{2/3}\lb(\fr{M}{m_{\rm n}}\rb)^{5/3}\fr{1}{R^5}\,.
\end{align}

Next, we derive the self-gravity potential.
The mass within the radius $r$ can be written as
\begin{align}
    m(r)=M\lb(\fr{r}{R}\rb)^{3}\,.
\end{align}
Here, we assumed that the density of the neutron star is constant.
Writing the mass of the spherical shell with small width $\dif r$ as $\dif m(r)$, the gravitational potential of this shell is given by
\begin{align}
    \dif U(r)
    &=
    -\fr{Gm(r)\dif m(r)}{r}
    \nom
    &=
    -\fr{3GM^2}{R^6}r^4\dif r\,.
\end{align}
By integrating this expression over $r$ from $0$ to $R$, we obtain the desired potential,
\begin{align}
    U(M,R)
    =-\fr{3GM^2}{5R}\,.
\end{align}

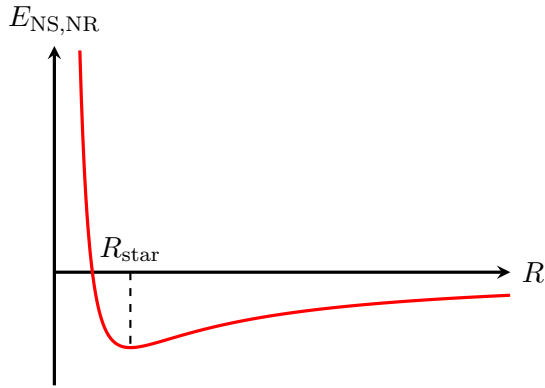
\begin{figure}[tb]
\centering
\begin{tikzpicture}[scale=1,samples=300]
    \draw[->,>=stealth,very thick](0,0)--(6,0) node [right]{$R$};
    \draw[->,>=stealth,very thick](0,-1.5)--(0,3) node [above]{$E_{\rm NS,NR}$};
    \draw[dashed,thick](1,0)--(1,-1) node [above] at (1,0){$R_{\rm star}$};
    \draw[very thick, red, domain=0.335:6] plot(\x,{1/pow(\x,2)-2/\x});
\end{tikzpicture}
\caption{The total energy of the neutron star in the non-relativistic case. $R_{\rm star}$ is the radius minimizing the energy.}
\label{fig:NS:NS_NR_energy}
\end{figure}

From the above, we can write the total energy of the neutron star for the relativistic and non-relativistic cases, respectively, as
\begin{align}
\label{eq:NS_Astero:NS_energy_rel}
    &E_{\rm NS,R}(M,R)
    =
    \lb[\fr{3}{4}\lb(\fr{9\pi}{4}\rb)^{1/3}\lb(\fr{M}{m_{\rm n}}\rb)^{4/3}-\fr{3GM^2}{5}\rb]\fr{1}{R}\,,
    \\
    &E_{\rm NS,NR}(M,R)
    =
    \lb[\fr{3}{10m_{\rm n}}\lb(\fr{9\pi}{4}\rb)^{2/3}\lb(\fr{M}{m_{\rm n}}\rb)^{5/3}\fr{1}{R^2}-\fr{3GM^2}{5}\fr{1}{R}\rb]\,.
\end{align}
In the relativistic limit, the energy is proportional to $R^{-1}$.
To avoid the collapse, the coefficient in Eq.~(\ref{eq:NS_Astero:NS_energy_rel}) is required to be positive.
This gives the maximum mass of the neutron star:
\begin{align}
    M<\fr{15\sqrt{5\pi}}{16m_{\rm n}^2G^{3/2}}\simeq7M_{\odot}\,,
\end{align}
where $M_{\odot}$ is the solar mass.
For a fixed mass smaller than the maximum value, the radius becomes larger to lower the energy.
Therefore, the density is decreased, and the picture of the relativistic neutrons is no longer valid.
In the non-relativistic case, the energy has a stable point.
The schematic figure is shown in Fig.~\ref{fig:NS:NS_NR_energy}.

We can then derive the radius as
\begin{align}
    R_{\rm star}
    =
    \lb(\fr{9\pi}{4}\rb)^{2/3}\fr{1}{Gm_{\rm n}^{8/3}M^{1/3}}
    \simeq
    12\lb(\fr{M_{\odot}}{M}\rb)~{\rm km}\,.
\end{align}
The typical mass and radius of neutron stars are $M\sim M_{\odot}$ and $R\sim 10~{\rm km}$.
Although this is a simple estimate, it agrees with the actual value well.

\subsection{M--R relation}
In general, a star keeps its shape by balancing the pressure and gravitational force.
To obtain a more precise value of mass and radius, we have to consider the hydrostatic equilibrium equation.
In the case of neutron stars, it is so compact that we have to include the general relativistic correction.
As an index to know whether the general relativistic effect is relevant or not, one can introduce the compactness defined as
\begin{align}
    C\equiv
    \fr{GM}{R}
    \sim
    10^{-1}\lb(\fr{M}{M_{\odot}}\rb)\lb(\fr{R}{10~{\rm km}}\rb)^{-1}\,.
\end{align}
The general relativistic effect is dominant when the compactness parameter is large.
The order of the compactness parameter for typical neutron stars is about $C\sim10^{-1}$, and the general relativistic correction can affect determining the mass and radius of a neutron star.

The hydrostatic equilibrium equation with the general relativistic correction is known as the Tolman--Oppenheimer--Volkoff (TOV) equation~\cite{Tolman:1939jz,Oppenheimer:1939ne}:
\begin{align}
\label{eq:NS_Astero:TOV_eq}
    &\fr{\dif P(r)}{\dif r}
    =
    -\fr{G\varepsilon(r)m(r)}{r^2}
    \lb[1-\fr{2Gm(r)}{r}\rb]^{-1}
    \lb[1+\fr{P(r)}{\varepsilon(r)}\rb]
    \lb[1+\fr{4\pi r^3P(r)}{m(r)}\rb]\,,
\end{align}
where $m(r)$ is a mass within the distance $r$, $\varepsilon(r)$ and $P(r)$ are the energy density and the pressure at the distance $r$.
The last three factors represent the general relativistic corrections.
The equation for the mass is given by
\begin{align}
\label{eq:NS_Astero:Mass_eq}
    \fr{\dif m(r)}{\dif r}
    =
    4\pi\varepsilon(r) r^2\,.
\end{align}
The mass of a star is defined as $M=m(R)$.
There are three unknown functions, $P(r)$, $\varepsilon(r)$, and $m(r)$.
Thus, we need another equation called the equation of state,
\begin{align}
\label{eq:NS_Astero:EoS_general}
    P=P(\varepsilon)\,.
\end{align}

The microscopic details of matter inside a neutron star are reflected in the equation of state.
One of the most important problems in studying neutron stars is determining the equation of state.
Once we choose an equation of state, we can obtain the relation between the mass and radius called the mass-radius relation (M--R relation). 
In other words, there is a one-to-one correspondence between the equation of state and the M--R relation.
Thus, by comparing the obtained M--R relation with the observational data, we can constrain the equation of state.

\section{Rotation period and density}
\label{sec:NS_Astero:Rotation_density}
The period of the first-discovered neutron star is about one second \cite{Hewish:1968bj}.
A more rapidly rotating neutron star called the millisecond pulsar was also found~\cite{Backer:1982owp}.
The reason why a neutron star can rotate so rapidly is related to its density.
In this section, we show that the minimum period of stars is determined by the density~\cite{Reisenegger:2015crq}.

We consider a spherically symmetric star of mass $M$ and radius $R$ rotating with the constant angular velocity $\Omega$.
In order for the matter on the surface at the equator not to be ejected into space, the gravitational force acting on a small mass element there must exceed the centrifugal force:
\begin{align}
R\Omega^2\leq\fr{GM}{R^2}.
\end{align}
Therefore, the maximum angular velocity can be written as
\begin{align}
    \Omega
    \leq
    \lb( \fr{GM}{R^3} \rb)^{\fr{1}{2}}
    =
    \lb(\fr{4\pi}{3}G\rho\rb)^{\fr{1}{2}}
    \equiv\Omega_{\rm max}\,.
\end{align}
The rotation period is defined as $P\equiv 2\pi/\Omega$, and the minimum rotation period $P_{\rm min}$ is given by
\begin{align}
    P_{\rm min}
    &=
    \fr{2\pi}{\Omega_{\rm max}}
    =
    \lb(\fr{3\pi}{G\rho}\rb)^{\fr{1}{2}}.
\end{align}

This relation indicates that the minimum rotation period of a star can be estimated from its mean density.
Several examples are listed in Tab.~\ref{tab:NS:min_period}.
From this minimum period, the observation of pulses with periods on the order of 1~s suggests that the source is likely a neutron star.
\renewcommand{\arraystretch}{1.3}
\begin{table}[tb]
\centering
\caption{Examples of mean densities and minimum rotation periods of various celestial objects.}
\label{tab:NS:min_period}
\vspace{1em}

\begin{tabular}{|c|c|c|}
    \hline
    Object & Mean density & Minimum period \\ \hline
    Earth & 5.5~g/cm$^3$ & 1.4~h \\ \hline
    Sun & 1.4~g/cm$^3$ & 2.8~h \\ \hline
    White dwarf & $\sim 10^6$~g/cm$^3$ & $\sim 10$~s \\ \hline
    Neutron star & $\sim 10^{15}$~g/cm$^3$ & $\sim 10^{-4}$~s \\ \hline
\end{tabular}
\end{table}

\section{Magnetic fields}
\label{sec:NS_Astero:Magnetic_field}
A neutron star is not only dense but also has a strong magnetic field.
This fact makes the neutron star a fascinating object to examine physics in extreme environments.
In this section, we give a brief review of the magnetic fields of neutron stars~\cite{Shapiro:1983du,Reisenegger:2015crq}.

\subsection{Theoretical maximum intensity}
In the interior of a neutron star, the theoretical upper limit of the magnetic field strength is estimated to be $10^{18}~{\rm G}$~\cite{Shapiro:1991,Cardall:2000bs,Ferrer:2010wz}.
Here, we demonstrate this estimate through a simple argument~\cite{Reisenegger:2015crq}.
For simplicity, we assume that the magnetic field is homogeneous inside the neutron star.
The energy of the magnetic field is then
\begin{align}
    E_{B}
    =
    \fr{2\pi}{3}B^2 R^3\,,
\end{align}
where $R$ is the radius of the neutron star.
If the magnetic energy exceeds the gravitational potential, the neutron star cannot keep its shape.
Thus, for the neutron star to be stable, the inequality,
\begin{align}
    E_{B}-|U(R)|
    \simeq
    \fr{2\pi}{3}B^2 R^3-\fr{3}{5}\fr{GM^2}{R}
    <
    0\,,
\end{align}
should be satisfied.
Therefore, we obtain the maximum intensity of the magnetic field as
\begin{align}
    B
    \lesssim
    \lb(\fr{9G}{10\pi}\rb)^{1/2}\fr{M}{R^2}
    &\sim
    10^{18}~{\rm G}
    \lb(\fr{M}{M_{\odot}}\rb)
    \lb(\fr{R}{10~{\rm km}}\rb)^{-2}\,.
\end{align}
From this expression, we find that the more compact a neutron star is, the larger the maximum magnetic field becomes. 
In the case of the typical neutron star, the maximum value is about $10^{18}~{\rm G}$.
In natural units, we can convert $1~{\rm G}\sim (10^{-7}~{\rm MeV})^{2}$.
Thus, the maximum magnetic field corresponds to about the square of the QCD scale $(10^{18}~{\rm G}\sim \Lambda_{\rm QCD}^2)$.

\subsection{Surface magnetic field}
We can estimate the magnetic fields on the surface of a neutron star by observing the rotation period and its time variation.
Here, we discuss the relation between the magnetic field and the rotation period~\cite{Shapiro:1983du}.

The rotational energy of a neutron star, $E_{\rm rot}$, is expressed in terms of the moment of inertia $I$ and the angular velocity $\Omega$ as
\begin{align}
    E_{\rm rot}
    =
    \fr{1}{2}I\Omega^2\,.
\end{align}
Taking the time derivative of this expression yields
\begin{align}
    \dot{E}_{\rm rot}
    =
    I\Omega\dot{\Omega}\,.
\end{align}

Now, let us assume that the neutron star is a rotating magnetic dipole.
The vector potential $\bs{A}(\bs{x})$ generated by an electric current density $e\bs{j}$ is given by
\begin{align}
    \bs{A}(\bs{x})
    =
    \fr{1}{4\pi}\int\dif^{3}\bs{x}'\fr{e\bs{j}(\bs{x}')}{|\bs{x}-\bs{x}'|}\,.
\end{align}
Hereafter, we denote the radial distance as $r\equiv|\bs{x}|$.
In the limit $r\gg r'$, one can perform the multipole expansion of the integrand, leading to
\begin{align}
    \bs{A}(\bs{x})
    &=
    \sum_{l=0}^{\infty}\bs{A}_l(\bs{x})
    \nom
    &=
    \fr{1}{r}\sum_{l=0}^{\infty}\int\dif^{3}\bs{x}'e\bs{j}(\bs{x}')
    \lb(\fr{r'}{r}\rb)^{l}P_{l}(\cos{\theta})\,,
\end{align}
where $P_{l}$ denotes the Legendre polynomial, and we used $\bs{x}\cdot\bs{x}'=rr'\cos{\theta}$ with $\theta$ being the angle between $\bs{x}$ and $\bs{x}'$.
Defining the magnetic dipole moment as
\begin{align}
    \bs{m}
    \equiv
    \int \dif^{3}\bs{x}' \bs{x}'\times e\bs{j}\,,
\end{align}
the dipole ($l=1$) component of the vector potential becomes
\begin{align}
    \bs{A}_1(\bs{x})
    =
    \fr{1}{4\pi}\fr{\bs{m}\times \bs{x}}{r^3}\,.
\end{align}
From this vector potential, the magnetic field of the dipole is expressed as
\begin{align}
    \bs{B}(\bs{x})
    &=
    \bs{\nabla}\times\bs{A}_1(\bs{x})
    \nom
    &=
    \fr{1}{4\pi}\fr{3(\bs{m}\cdot\hat{\bs{x}})\hat{\bs{x}}-\bs{m}}{r^3}\,,
\end{align}
where $\hat{\bs{x}}$ is the unit vector in the direction of $\bs{x}$.
This expression shows that the magnetic field reaches its maximum when the magnetic dipole moment is aligned with the position vector, giving
\begin{align}
    B_{\rm max}(r)
    =
    \fr{m}{2\pi r^3}\,.
\end{align}
Hence, the maximum magnetic field strength at the surface of the neutron star is
\begin{align}
    B_{\rm max}(R)
    =
    \fr{m}{2\pi R^3}
    \equiv B_{\rm s}\,.
\end{align}

The energy emitted per unit time via magnetic dipole radiation is given by
\begin{align}
    \dot{E}_{\rm MD}
    =
    \fr{1}{6\pi}|\ddot{\bs{m}}|^2\,.
\end{align}
Assuming that the magnetic dipole moment of the neutron star makes an angle $\alpha$ with the rotation axis, it can be written as
\begin{align}
    \bs{m}
    =
    m
    \lb(\sin{\alpha}\cos{(\Omega t)}\bs{e}_{x}
    +\sin{\alpha}\sin{(\Omega t)}\bs{e}_{y}
    +\cos{\alpha}\bs{e}_{z}\rb)\,.
\end{align}
Therefore, using the surface magnetic field, the magnetic dipole radiation power becomes
\begin{align}
    \dot{E}_{\rm MD}
    &=
    \fr{m^{2}\Omega^{4}\sin^{2}{\alpha}}{6\pi}
    \nom
    &=
    \fr{2\pi}{3}
    B_{\rm s}^2\Omega^{4} R^6 \sin^{2}{\alpha}\,.
\end{align}

Assuming that the rotational energy is entirely converted into magnetic dipole radiation, i.e., $\dot{E}_{\rm rot}=-\dot{E}_{\rm MD}$, and defining the effective magnetic field as $B_{\rm eff}\equiv B_{\rm s}\sin{\alpha}$, one obtains
\begin{align}
    B_{\rm eff}
    &=
    \lb(\fr{3}{8\pi^3}\rb)^{1/2}\fr{I^{1/2}}{R^3}\sqrt{P\dot{P}}
    \nom
    &\sim
    10^{12}~{\rm G}
    \lb(\fr{I}{M_{\odot}\times(10~{\rm km})^2}\rb)^{1/2}
    \lb(\fr{R}{10~{\rm km}}\rb)^{-3}
    \sqrt{\lb(\fr{P}{1~{\rm s}}\rb)\lb(\fr{\dot{P}}{10^{-13}}\rb)}\,.
\end{align}
Thus, the surface magnetic field strength of a neutron star can be inferred from its rotation period and its time derivative.
For instance, the Crab pulsar is known to have $P\sim 10^{-1}{\rm s}$ and $\dot{P}\sim 10^{-13}$, which yields a surface magnetic field strength of
\begin{align}
    B_{\rm Crab}\sim 10^{12}~{\rm G}\,.
\end{align}

\begin{figure}[tb]
 \begin{center}
 \includegraphics[width=8cm]{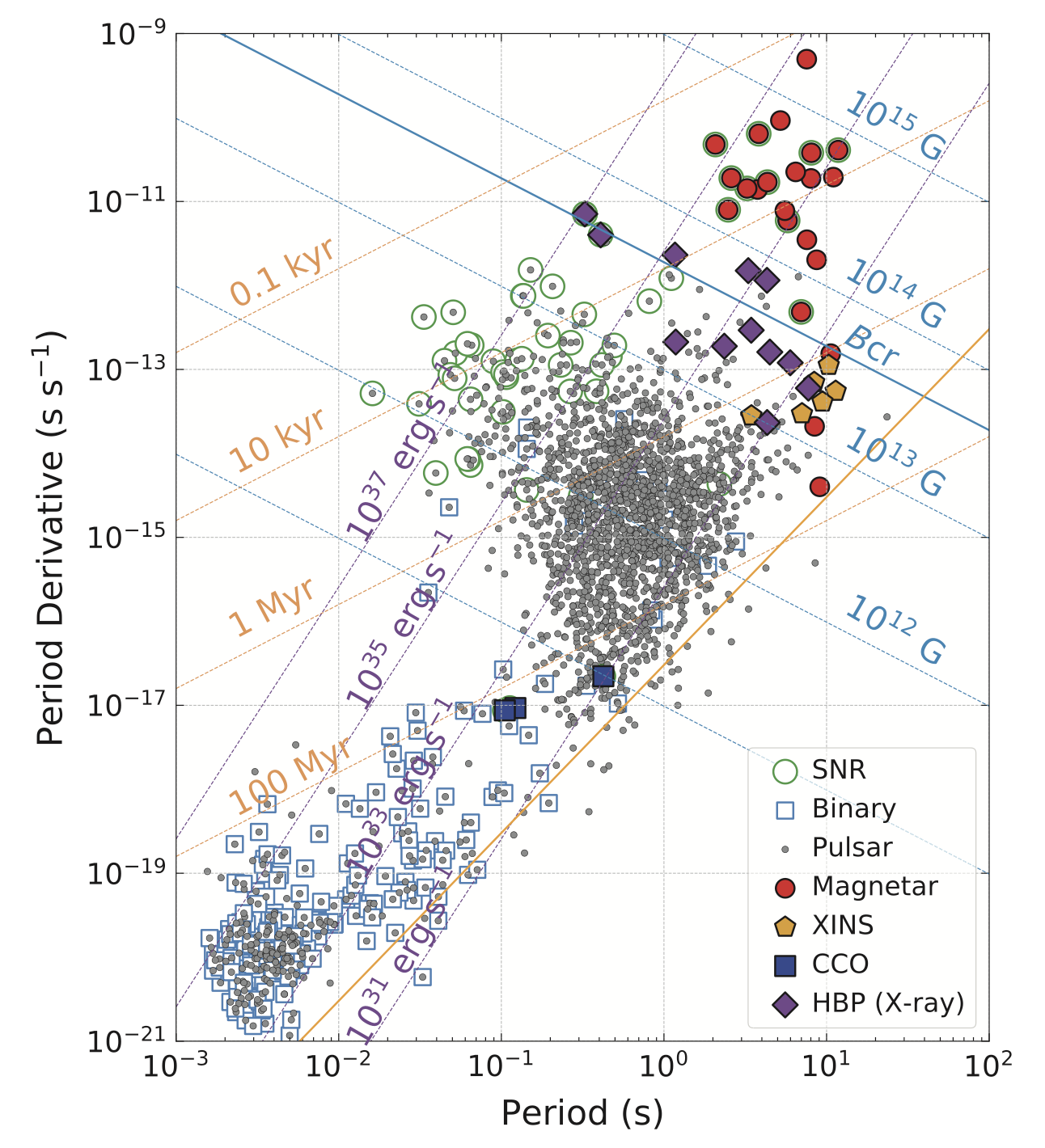}
 \end{center}
\caption{An example of the $P\text{--}\dot{P}$ diagram cited from Ref.~\cite{Enoto:2019vcg}. In this diagram, pulsars (gray dots), magnetars (red filled circles), x-ray isolated neutron stars (XINSs; orange pentagons), compact central objects (CCOs; blue filled squares), and high-B pulsars with x-ray emission reported (HBP; purple diamonds) are shown.
Additional markers are added for sources associated with supernova remnants (SNRs; green circles) or found in binary systems (blue squares).
Blue and orange lines indicate the surface magnetic field strengths and characteristic ages, respectively.}
\label{fig:NS:P-Pdot}
\end{figure}

Since the period and its time derivative give us various information about neutron stars, the observational data are often plotted in Fig.~\ref{fig:NS:P-Pdot} called the $P\text{--}\dot{P}$ diagram~\cite{Enoto:2019vcg}.
The upper-right region corresponds to neutron stars with stronger magnetic fields, and the red points are magnetars.
The surface magnetic fields of the magnetars reach $10^{15}~\rm G$.
Although the magnetic field inside the magnetar has not been observed directly yet, it would be stronger.
Such a strong magnetic field is crucial in the discussion in Chap.~\ref{chap:Chi_Astero} and Chap.~\ref{chap:Anom_Dyn_Res}.

\section{Neutrino trapping in supernovae}
\label{sec:NS_Astero:Neutrino_trapping}
In the core of a supernova, a proto-neutron star is formed.
Since the density of a proto-neutron star is high, even neutrinos interacting only weakly with matter are trapped and cannot freely diffuse~\cite{Sato:1975,Sato:1975vu}.
In this section, we compute the timescale of the neutrino diffusion and gravitational collapse and show that the neutrinos are trapped in the proto-neutron star.

The mean free path of the neutrinos scattered by the nucleons in a proto-neutron star is given by
\begin{align}
    l_{\nu}
    \simeq
    10^{7}~{\rm cm}
    \lb(\fr{\rho}{3\times10^{10}~{\rm g/cm^3}}\rb)^{-5/3}
    \lb(\fr{A}{56}\rb)^{-1}
    \lb(\fr{y_{\rm e}}{26/56}\rb)^{-2/3}\,,
\end{align}
where $\rho$ is the mass density of the core, $A$ denotes the mass number, and $y_{\rm e}$ is the electron fraction~\cite{Sato:1975,Sato:1975vu,Kotake:2005zn}.
The number 26/56 comes from the number of protons and neutrons of iron ($^{56}_{26}$Fe).
As we can find from this expression, when the density of the core exceeds $\rho\sim10^{11}~{\rm g/cm^3}$, the mean free path becomes $l_{\nu}\lesssim10~{\rm km}$, which is smaller than the typical size of a proto-neutron star.
This means that the neutrinos in the proto-neutron star are scattered by the nucleons and diffuse.

We first estimate the diffusion timescale of the neutrino using the random walk model.
In $N$ steps, the neutrino propagates a distance 
\begin{align}
    r
    \simeq
    l_{\nu}\sqrt{N}\,.
\end{align}
This reaches the radius of the proto-neutron star, $R$, when the number of steps is
\begin{align}
    N
    \simeq
    \lb(\fr{R}{l_{\nu}}\rb)^2\,.
\end{align}
Thus, the length of the path is
\begin{align}
    d
    \simeq
    l_{\nu}N
    \simeq
    \fr{R^2}{l_{\nu}}\,.
\end{align}
The radius of a proto-neutron star with a solar mass is 
\begin{align}
    R
    \simeq
    \lb(\fr{3M_{\odot}}{4\pi\rho}\rb)^{1/3}
    \sim
    10^{7}~{\rm cm}\lb(\fr{\rho}{10^{10}~{\rm g/cm^3}}\rb)^{-1/3}\,.
\end{align}
Since the speed of the neutrino is almost equal to the speed of light, the diffusion timescale of the neutrino is written as
\begin{align}
    t_{\rm dif}
    \sim
    d
    \simeq
    \fr{R^2}{l_{\nu}}
    \sim
    1~{\rm s}
    \lb(\fr{\rho}{10^{12}~{\rm g/cm^3}}\rb)
    \lb(\fr{A}{56}\rb)
    \lb(\fr{y_{\rm e}}{26/56}\rb)^{2/3}\,.
\end{align}
Meanwhile, the timescale of the gravitational collapse is called the dynamical timescale and corresponds to the free-fall timescale as
\begin{align}
    t_{\rm dyn}
    \simeq
    \sqrt{\fr{R}{g}}
    =
    \fr{1}{\sqrt{G\rho}}
    \sim
    10^{-2}~{\rm s}\lb(\fr{\rho}{10^{12}~{\rm g/cm^3}}\rb)^{-1/2}\,,
\end{align}
where we used the gravitational acceleration
\begin{align}
    g
    =
    \fr{GM}{R^2}
    =
    \fr{4\pi G\rho}{3}R\,.
\end{align}
From these, if the mass density exceeds $\rho\sim10^{11}~{\rm g/cm^3}$, the dynamical timescale is much longer than the diffusion timescale.
Thus, the neutrinos are trapped in the proto-neutron star.

Due to the neutrino trapping, we can treat them as a fluid in the core of a proto-neutron star.
The unique property of neutrino matter is that it is composed only of left-handed neutrinos.
All the neutrinos inside a proto-neutron star created by the electron capture are left-handed:
\begin{align}
    {\rm p}+{\rm e^{-}_{\rm L}}\to{\rm n}+{\rm \nu}_{\rm e,L}\,.
\end{align}
Thus, the neutrino matter has a chirality imbalance, namely a chiral matter~\cite{Yamamoto:2015gzz}.
This property is crucial when we discuss the chiral transport in the core of a supernova in Chap.~\ref{chap:Chi_Astero}.

\section{Asteroseismology in Newtonian gravity}
\label{sec:NS_Astero:Astero_NG}
In general, it is difficult to observe the interior of stars because electromagnetic waves, such as radio waves and X-rays, are emitted only from their surfaces.
In seismology, we infer the interior structure of the Earth by analyzing earthquakes.
This idea can be extended to the seismic oscillations of stars, known as asteroseismology.
In this section, we introduce the basic concepts of asteroseismology within the framework of Newtonian gravity (see, e.g., Ref.~\cite{Aerts:2010}).
The counterpart of general relativity will be discussed in the next section.

\subsection{Hydrodynamic equations}
We consider an ideal fluid in a gravitational field.
Hydrodynamic equations can be formulated from conservation laws.
In the non-relativistic case, the mass is conserved as
\begin{align}
\label{eq:NS_Astero:Mass_conservation}
    \fr{\del\rho}{\del t}+\bs{\nabla}\cdot(\rho\bs{v})=0\,,
\end{align}
where $\bs{v}$ is the fluid velocity.
The momentum conservation law leads to the Euler equation,
\begin{align}
\label{eq:NS_Astero:Euler_equation}
    \fr{\del \bs{v}}{\del t}+(\bs{v}\cdot\bs{\nabla})\bs{v}
    =
    -\fr{1}{\rho}\bs{\nabla}P-\bs{\nabla}\Phi\,,
\end{align}
where $\Phi$ is the gravitational potential, which satisfies the Poisson equation,
\begin{align}
\label{eq:NS_Astero:Poisson_eqiaton}
    \bs{\nabla}^2\Phi=4\pi G\rho\,.
\end{align}

To close the equations, we need the equation of state connecting $P$ and $\rho$ as well.
Since the equation of state depends on the details of matter, we then introduce the adiabatic exponents as
\begin{align}
    \Gamma_1\equiv\lb(\fr{\del\ln P}{\del\ln\rho}\rb)_{s}\,,
    \qquad
    \fr{\Gamma_2-1}{\Gamma_2}\equiv\lb(\fr{\del\ln T}{\del \ln P}\rb)_{s}\,,
    \qquad
    \Gamma_3-1\equiv\lb(\fr{\del \ln T}{\del\ln\rho}\rb)_{s}\,,
\end{align}
where $s$ is the entropy per unit mass.
Expressing the pressure as $P=P(\rho,s)$, its total derivative is
\begin{align}
\label{eq:NS_Astero:Pressure_total_derivative}
    \dif P
    &=
    \lb(\fr{\del P}{\del \rho}\rb)_{s}\dif\rho
    +\lb(\fr{\del P}{\del s}\rb)_{\rho}\dif s
    \nom
    &=
    \fr{\Gamma_1P}{\rho}\dif\rho+T\rho(\Gamma_3-1)\dif s\,.
\end{align}
In the second line, we used the first law of thermodynamics,
\begin{align}
    \dif e=\fr{P}{\rho^2}\dif\rho+T\dif s\,,
\end{align}
where $e$ is the energy per unit mass.
Using the heat per unit mass, $\dif q=T\dif s$, we rewrite Eq.~(\ref{eq:NS_Astero:Pressure_total_derivative}) as
\begin{align}
    \dif q
    =
    \fr{1}{\rho(\Gamma_3-1)}\lb(\dif P-\fr{\Gamma_1P}{\rho}\dif\rho\rb)\,.
\end{align}
In the following, we assume that the fluid motion is adiabatic $(\dif q=0)$.
Therefore, the equation of state reduces to
\begin{align}
\label{eq:NS_Astero:EoS_adiabatic}
    \dif P
    =
    \fr{\Gamma_1P}{\rho}\dif\rho\,.
\end{align}
The hydrodynamics variables $\rho$, $\bs{v}$, $P$, and $\Phi$ are completely determined by Eqs.~(\ref{eq:NS_Astero:Mass_conservation}), (\ref{eq:NS_Astero:Euler_equation}), (\ref{eq:NS_Astero:Poisson_eqiaton}), and (\ref{eq:NS_Astero:EoS_adiabatic}).

\subsection{Linear analysis}
To investigate fluid oscillations, we linearize the hydrodynamic equations.
We introduce two types of perturbations.
The first one is the Eulerian perturbation describing the change of quantity $Q$ at a point:
\begin{align}
    \delta Q(t,\bs{x})\equiv Q(t,\bs{x})-\bar{Q}(\bs{x})\,,
\end{align}
where $\bar{Q}$ denotes the equilibrium value.
The other is the Lagrangian perturbation following the fluid motion:
\begin{align}
    \varDelta Q
    &\equiv Q(\bs{x}+\varDelta\bs{x})-\bar{Q}(\bs{x})
    \nom
    &=\delta Q+(\bs{\xi}\cdot\bs{\nabla})Q\,,
\end{align}
where $\bs{\xi}\equiv\varDelta\bs{x}$ is the Lagrangian displacement vector.

In linear order in the perturbations, the governing equations become
\begin{align}
\label{eq:NS_Astero:Mass_conservation_linear}
    &\fr{\del\delta\rho}{\del t}+\bs{\nabla}\cdot(\bar{\rho}\delta\bs{v})=0\,,
    \\
\label{eq:NS_Astero:Euler_equation_linear}
    &\fr{\del^2\bs{\xi}}{\del t^2}
    =
    -\fr{1}{\bar{\rho}}\bs{\nabla}\delta P
    -\fr{\delta\rho}{\bar{\rho}}\bs{\nabla}\bar{\Phi}
    -\bs{\nabla}\delta\Phi\,,
    \\
\label{eq:NS_Astero:Poisson_eqiaton_linear}
    &\bs{\nabla}^2\delta\Phi=4\pi G\delta\rho\,,
    \\
\label{eq:NS_Astero:EoS_adiabatic_linear}
    &\varDelta P
    =
    \fr{\Gamma_1\bar{P}}{\bar{\rho}}\varDelta\rho\,.
\end{align}
In the following discussion, we assume that the equilibrium state is spherically symmetric, and it is convenient to use the spherical coordinates $(r,\theta,\phi)$.
We separate the displacement into the radial and the horizontal components as
\begin{align}
    \bs{\xi}=\xi^r\bs{e}_{r}+\bs{\xi}_{\rm h}\,,
\end{align}
where $\bs{\xi}_{\rm h}=(\xi^\theta,\xi^\phi)$\,.
The radial component of Eq.~(\ref{eq:NS_Astero:Euler_equation_linear}) is
\begin{align}
    \fr{\del^2\xi^r}{\del t^2}
    =
    -\fr{1}{\bar{\rho}}\fr{\del\delta P}{\del r}
    -\fr{\delta\rho}{\bar{\rho}}\fr{\del \bar{\Phi}}{\del r}
    -\fr{\del\delta\Phi}{\del r}\,.
\end{align}
Since $\bs{\nabla}_{\rm h}\bar{\rho}=0$, multiplying the horizontal component of Eq.~(\ref{eq:NS_Astero:Euler_equation_linear}) by $\bs{\nabla}_{\rm h}$, we have
\begin{align}
\label{eq:NS_Astero:Euler_equation_linear_horizontal_div}
    \fr{\del^2}{\del t^2}(\bs{\nabla}_{\rm h}\cdot\bs{\xi}_{\rm h})
    =
    -\fr{1}{\bar{\rho}}\bs{\nabla}_{\rm h}^2\delta P
    -\bs{\nabla}_{\rm h}^2\delta\Phi\,.
\end{align}
The continuity equation (\ref{eq:NS_Astero:Mass_conservation_linear}) can be written as
\begin{align}
    \fr{\del\delta\rho}{\del t}
    &=
    -\fr{\del}{\del t}\lb[\bs{\nabla}\cdot(\bar{\rho}\bs{\xi})\rb]
    \nom
    &=
    -\fr{\del}{\del t}\lb[\fr{1}{r^2}\fr{\del}{\del r}\lb(r^2\bar{\rho}\xi^r\rb)+\bar{\rho}\bs{\nabla}_{\rm h}\cdot\bs{\xi}_{\rm h}\rb]\,.
\end{align}
Using this expression, we can eliminate $\bs{\nabla}_{\rm h}\cdot\bs{\xi}_{\rm h}$ from Eq.~(\ref{eq:NS_Astero:Euler_equation_linear_horizontal_div}) as
\begin{align}
\label{eq:NS_Astero:Euler_equation_linear_horizontal_div_}
    \fr{\del^2}{\del t^2}\lb[\delta\rho+\fr{1}{r^2}\fr{\del}{\del r}\lb(r^2\bar{\rho}\xi^r\rb)\rb]
    =
    \fr{1}{\bar{\rho}}\bs{\nabla}_{\rm h}^2\delta P
    +\bs{\nabla}_{\rm h}^2\delta\Phi\,.
\end{align}
The Poisson equation (\ref{eq:NS_Astero:Poisson_eqiaton_linear}) is given by
\begin{align}
    \fr{1}{r^2}\fr{\del}{\del r}\lb(r^2\fr{\del\delta\Phi}{\del r}\rb)
    +\bs{\nabla}_{\rm h}^2\delta\Phi=4\pi G\delta\rho\,.
\end{align}

To proceed with the analysis, we expand the perturbations in the spherical harmonics $Y_{l}^{m}(\theta,\phi)$ satisfying
\begin{align}
    \bs{\nabla}_{\rm h}^2Y_l^m=
    -\fr{l(l+1)}{r^2}Y_{l}^{m}(\theta,\phi)\,.
\end{align}
One can then express the perturbations in the form of
\begin{align}
    &\xi^r=\xi^r(r)\e^{-\im\omega t}Y_{l}^{m}(\theta,\phi)\,,
    \\
    &\delta\rho=\delta\rho(r)\e^{-\im\omega t}Y_{l}^{m}(\theta,\phi)\,,
    \\
    &\delta P=\delta P(r)\e^{-\im\omega t}Y_{l}^{m}(\theta,\phi)\,,
    \\
    &\delta\Phi=\delta\Phi(r)\e^{-\im\omega t}Y_{l}^{m}(\theta,\phi)\,.
\end{align}
In addition, Eq.~(\ref{eq:NS_Astero:EoS_adiabatic_linear}) is rewritten as
\begin{align}
    \delta\rho
    =
    \fr{\rho}{\Gamma_1P}\delta P+\bar{\rho}\xi^r\lb(\fr{1}{\Gamma_1P}\fr{\dif \bar{P}}{\dif r}-\fr{1}{\rho}\fr{\dif \bar{\rho}}{\dif r}\rb)\,,
\end{align}
and we can eliminate $\delta\rho$ from the above equations.
Therefore, the fundamental equations describing the linear stellar oscillations are summarized as
\begin{align}
    &\fr{\dif\xi^r}{\dif r}
    =-\lb(\fr{2}{r}+\fr{1}{\Gamma_1\bar{P}}\fr{\dif\bar{P}}{\dif r}\rb)\xi^r
    +\fr{1}{\bar{\rho}c_{\rm s}^2}\lb(\fr{S_l^2}{\omega^2}-1\rb)\delta P
    +\fr{l(l+1)}{\omega^2r^2}\delta\Phi\,,
    \\
    &\fr{\dif\delta P}{\dif r}
    =\bar{\rho}(\omega^2-N^2)\xi^r
    +\fr{1}{\Gamma_1\bar{P}}\fr{\dif\bar{P}}{\dif r}\delta P-\bar{\rho}\fr{\dif\delta\Phi}{\dif r}\,,
    \\
    &\fr{1}{r^2}\fr{\dif}{\dif r}\lb(r^2\fr{\dif\delta\Phi}{\dif r}\rb)
    =4\pi G\lb(\fr{\delta P}{c_{\rm s}^2}+\fr{\bar{\rho}\xi^r}{g}N^2\rb)+\fr{l(l+1)}{r^2}\delta\rho\,,
\end{align}
where we defined the adiabatic sound speed $c_{\rm s}$, the gravitational acceleration $g$, the Lamb frequency $S_l$, and the Brunt--V\"{a}is\"{a}l\"{a} frequency $N$, respectively, as 
\begin{align}
    c_{\rm s}^2\equiv\fr{\Gamma_1\bar{P}}{\bar{\rho}}\,,
    \qquad
    g\equiv\lb|\fr{\dif\bar{\Phi}}{\dif r}\rb|\,,
    \qquad
    S_l^2\equiv\fr{l(l+1)c_{\rm s}^2}{r^2}\,,
    \qquad
    N^2\equiv g\lb(\fr{1}{\Gamma_1\bar{P}}\fr{\dif \bar{P}}{\dif r}-\fr{1}{\bar{\rho}}\fr{\dif \bar{\rho}}{\dif r}\rb)\,.
\end{align}
The Lamb frequency represents the horizontal component of the sound wave, while the Brunt--V\"{a}is\"{a}l\"{a} frequency characterizes the oscillation driven by buoyancy.

\subsection{Cowling approximation and stellar oscillation modes}
Although the full set of hydrodynamic equations is available, an approximate analytical treatment can provide valuable insight into the behavior of stellar oscillations.
We here neglect the recoil of the star due to the gravitational potential by setting $\delta \Phi=0$, which is called the Cowling approximation~\cite{Cowling:1941nqk}.
In this approximation, the hydrodynamic equations reduce to
\begin{align}
\label{eq:Hydro_eq_Cowling_xi}
    &\fr{\dif\xi^r}{\dif r}
    =-\lb(\fr{2}{r}-\fr{1}{\Gamma_1\bar{P}}\fr{\dif\bar{P}}{\dif r}\rb)\xi^r
    +\fr{1}{\bar{\rho}c_{\rm s}^2}\lb(\fr{S_l^2}{\omega^2}-1\rb)\delta P\,,
    \\
\label{eq:Hydro_eq_Cowling_P}
    &\fr{\dif\delta P}{\dif r}
    =\bar{\rho}(\omega^2-N^2)\xi^r
    +\fr{1}{\Gamma_1\bar{P}}\fr{\dif\bar{P}}{\dif r}\delta P\,.
\end{align}
To capture the overall properties of stellar oscillations, we further employ a rough approximation. 
Assuming that the spatial scale of the perturbations is much shorter than the gradient of the equilibrium quantities due to gravity, we neglect the derivatives other than those acting on the perturbations.
Then, Eq.~(\ref{eq:Hydro_eq_Cowling_xi}) and Eq.~(\ref{eq:Hydro_eq_Cowling_P}) are combined into a single second-order differential equation:
\begin{align}
    \fr{\dif^2\xi^r}{\dif r^2}
    =
    -K(r)\xi^r\,,
    \qquad
    K(r)\equiv\fr{\omega^2}{c_{\rm s}^2}\lb(1-\fr{S_l^2}{\omega^2}\rb)\lb(1-\fr{N^2}{\omega^2}\rb)\,.
\end{align}
From this expression, we can find that the oscillation modes are characterized by two types of frequencies, $S_l$ and $N$.
The local behavior of the displacement $\xi^r$ depends on the sign of $K(r)$.
While the solution oscillates for the positive sign, the displacement exhibits exponential behavior for the negative sign.
We focus on the oscillation modes, and there are two cases satisfying $K>0$:
\begin{align}
\label{eq:NS_Astero:High_freq_condition}
    &({\rm a})~\omega>S_l\quad{\text{and}}\quad\omega>|N|\,,
    \\
\label{eq:NS_Astero:Low_freq_condition}
    &({\rm b})~\omega<S_l\quad{\text{and}}\quad\omega<|N|\,.
\end{align}
The high-frequency mode (a) is called the p-mode, and the low-frequency mode (b) is called the g-mode.
The f-mode, which has no nodes, exists in the intermediate frequency region.

We first consider the p-mode.
Since the Lamb frequency is generally larger than the Brunt--V\"{a}is\"{a}l\"{a} frequency $(S_l>|N|)$, we assume $\omega^2\gg N^2$.
Under this condition, the coefficient $K(r)$ can be written as
\begin{align}
    K(r)\simeq\fr{1}{c_{\rm s}^2}\lb(\omega^2-S_l^2\rb)\,.
\end{align}
The p-mode comes from the $S_l$, and the driving force is the pressure.
The dispersion relation of the p-mode is
\begin{align}
    \omega^2=c_{\rm s}^2\lb(k_r^2+k_{\rm h}^2\rb)\,,
    \qquad
    k_{\rm h}^2\equiv\fr{l(l+1)}{r^2}\,.
\end{align}
This coincides exactly with the dispersion relation of sound waves.
For the g-mode, we assume $\omega^2\ll S_l^2$ and use
\begin{align}
    K(r)\simeq\fr{1}{\omega^2}\lb(N^2-\omega^2\rb)S_l^2\,.
\end{align}
In this case, the oscillation is driven by buoyancy.
The dispersion relation of the g-mode is expressed as
\begin{align}
    \omega^2
    =
    \fr{N^2}{1+(k_r^2/k_{\rm h}^2)}\,.
\end{align}

In the above discussion, we considered simple cases in order to extract the typical behavior of the modes. 
In realistic situations, however, the distinction between the p- and g-modes is not always clear, and mixed modes exhibiting characteristics of both can exist.

\subsection{Rotational effect}
We now consider a rotating star.
When a star rotates, the Coriolis force can give rise to an oscillation called the r-mode~\cite{Papaloizou:1978zz,Saio:1982}.
The rotation breaks the spherical symmetry, and the hydrodynamic equations for a spherical star are not applicable.
In this subsection, instead of deriving the complete set of the hydrodynamic equations in a rotating system, we derive the r-mode in a simple discussion~\cite{Saio:1982}.

Assuming a uniformly rotating star, we set the angular velocity $\bs{\Omega}$ to be along the $z$-axis.
The velocity in the inertial frame, $\bs{v}_{\rm i}$, and the velocity in the rotating frame, $\bs{v}_{\rm r}$, are related as follows:
\begin{align}
    \bs{v}_{\rm i}
    =
    \bs{v}_{\rm r}+\bs{\Omega}\times \bs{x}\,.
\end{align}
From this relation, for a vector $\bs{V}$, the time derivative in the inertial frame, $\fr{\dif}{\dif t_{\rm i}}$, and the time derivative in the rotating frame, $\fr{\dif}{\dif t_{\rm r}}$, satisfy
\begin{align}
    \fr{\dif}{\dif t_{\rm i}}\bs{V}
    =
    \fr{\dif}{\dif t_{\rm r}}\bs{V}
    +\bs{\Omega}\times\bs{V}\,.
\end{align}
Thus, the time derivative of the velocity is
\begin{align}
    \fr{\dif}{\dif t_{\rm i}}\bs{v}_{\rm i}
    &=
    \fr{\dif}{\dif t_{\rm i}}\bs{v}_{\rm r}
    +\bs{\Omega}\times \fr{\dif}{\dif t_{\rm i}}\bs{x}
    \nom
    &=
    \fr{\dif}{\dif t_{\rm r}}\bs{v}_{\rm r}
    +2\bs{\Omega}\times\bs{v}_{\rm r}
    +\mathcal{O}(\Omega^2)\,.
\end{align}
The second term corresponds to the Coriolis force.
Up to first order in $\bs{\Omega}$, the lineralized Euler equation becomes
\begin{align}
\label{eq:NS_Astero:Linearized_Euler_rotation}
    \del_{t}\delta \bs{v}
    +2(\bs{\Omega}\times\delta\bs{v})
    =
    -\fr{\delta\rho}{\bar{\rho}}\bs{\nabla}\bar{\Phi}
    -\fr{1}{\bar{\rho}}\bs{\nabla}\delta P\,,
\end{align}
where we omitted the subscript ``r'' for notational convenience.
In the following, we describe the dynamics in the rotating frame.
We also assume that the horizontal components of the velocity $\delta\bs{v}_{\rm h}$ are sufficiently larger than the radial component of the velocity $\delta v^{r}$.
Taking the rotation of Eq.~(\ref{eq:NS_Astero:Linearized_Euler_rotation}), the radial component is
\begin{align}
\label{eq:NS_Astero:Rot_hydro_eq_radial}
    \del_{t}\varpi^{r}
    +2[\bs{\nabla}\times(\bs{\Omega}\times\delta \bs{v})]^{r}
    =0\,,
\end{align}
where $\bs{\varpi}\equiv\bs{\nabla}\times\delta\bs{v}$ is the vorticity.
The second term can be rewritten as
\begin{align}
    2[\bs{\nabla}\times(\bs{\Omega}\times\delta \bs{v})]^{r}
    &=
    2\lb[
    (\bs{\nabla}\cdot\delta\bs{v})\bs{\Omega}
    +(\delta\bs{v}\cdot\bs{\nabla})\bs{\Omega}
    -(\bs{\nabla}\cdot\bs{\Omega})\delta\bs{v}
    -(\bs{\Omega}\cdot\bs{\nabla})\delta\bs{v}
    \rb]^r
    \nom
    &=
    2(\delta \bs{v}_{\rm h}\cdot\bs{\nabla}_{\rm h})\Omega^r\,.
\end{align}
Equation~(\ref{eq:NS_Astero:Rot_hydro_eq_radial}) can then be written in the form of the conservation of the vorticity:
\begin{align}
\label{eq:NS_Astero:Conservation_voticity_radial}
    \fr{\dif}{\dif t}(\varpi^r+2\Omega^r)
    =0\,.
\end{align}

\begin{figure}[tb]
\centering
\begin{tikzpicture}[scale=1, samples=300]

\draw[->, very thick] (0,0) -- (8,0) node[right]{Rotation};

\draw[thick, domain=0.2:7.8] plot(\x,{sin(\x r)});
\draw[red, very thick,domain=2.2:9.8] plot({\x-2},{sin(\x r)});

\foreach \x in {180,360}{
  \fill[gray] ({pi*\x/180},{sin(\x)}) circle (5pt);
}

\foreach \x in {45,90,135,405}{
  \fill[green] ({pi*\x/180},{sin(\x)}) circle (5pt);
}

\foreach \x in {225,270,315}{
  \fill[blue] ({pi*\x/180},{sin(\x)}) circle (5pt);
}

\draw[->, red, ultra thick, arrows = {-Stealth[length=10pt]}] (4.5,0.6)--(3.3,0.6);

\fill[thick, fill=green] (1,-2) circle (5pt);
\draw[<-, thick] (1.15,-2.26)  arc (-70:70:8pt);
\draw[<-, thick] (0.85,-1.74)  arc (110:250:8pt);
\node[right] at (1.5,-2) {Clockwise};

\fill[thick, fill=blue] (5,-2) circle (5pt);
\draw[->, thick] (5.15,-2.26)  arc (-70:70:8pt);
\draw[->, thick] (4.85,-1.74)  arc (110:250:8pt);
\node[right] at (5.5,-2) {Counterclockwise};

\node[left] at (-0.3,0.7) {North};
\node[left] at (-0.1,0) {Equator};
\node[left] at (-0.3,-0.7) {South};
\node at (1.55,1.5) {A};
\node at (0.75,1.15) {B};
\node at (2.35,1.15) {C};

\end{tikzpicture}
\caption{Schematic illustration of the r-mode. The filled circles denote the local fluid elements at each position. The fluid elements colored green rotate clockwise, and the blue one rotates anticlockwise. The red line represents the r-mode.}
\label{fig:NS_Astero:r-mode}
\end{figure}
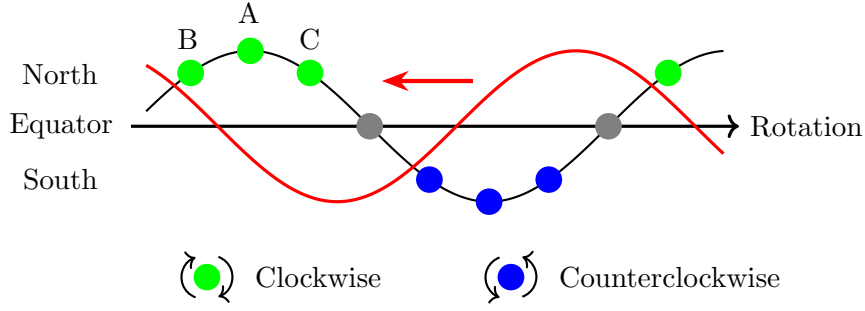

The conservation law (\ref{eq:NS_Astero:Conservation_voticity_radial}) provides us with a qualitative understanding of the mechanism inducing the r-mode.
For the axisymmetry, we can set $\phi=0$ without loss of generality.
Also, we focus on the fluid behavior at the stellar surface $(r=R)$.
We impose the boundary condition at the equator $(\theta=\pi/2)$ as
\begin{align}
    \varpi^r+2\Omega^r=0\,.
\end{align}
Since $\Omega^r = \Omega \cos\theta$, the radial component of the vorticity is negative in the northern hemisphere ($0 < \theta < \pi/2$) and positive in the southern hemisphere ($\pi/2 < \theta < \pi$).
In other words, fluid elements rotate clockwise in the northern region and counterclockwise in the southern region.
A schematic illustration is shown in Fig.~\ref{fig:NS_Astero:r-mode}.


Let us focus on a fluid element labeled by A.
Due to its clockwise motion, the western side of the element, labeled by B, moves northward, thereby increasing its vorticity.
Conversely, the eastern side of A, labeled by C, moves southward, which leads to a decrease in vorticity.
As a result, the total force acting on A tends to restore it to its original position, and a wave propagating westward is induced.
This is the r-mode.

As can be seen from the above discussion, the $r$-mode is not driven by perturbations of pressure or density, and therefore its radial displacement vanishes ($\xi^r = 0$).
In this sense, the r-mode differs from the p- and g-modes.
The class including the p- and g-modes is called the spheroidal modes, whereas the r-mode belongs to the second class, namely, the toroidal modes.
The toroidal mode exhibits a twisting motion without involving any deformation of the star.

We now derive the frequency of the r-mode.
As a preparation, let us first discuss the parity properties of the spherical harmonics~\cite{Regge:1957td,Zerilli:1970wzz}.
Under the parity transformation, a spherical harmonic transforms as
\begin{align}
    Y_{l}^{m}(\theta,\phi)
    \to
    Y_{l}^{m}(\pi-\theta,\pi+\phi)
    =
    (-1)^{l}Y_{l}^{m}(\theta,\phi)\,.
\end{align}
We introduce the polar and axial vector spherical harmonics as
\begin{align}
    &\bs{Y}_l^{m({\rm pol})}\equiv\bs{\nabla}_{\rm h}Y_l^m(\theta,\phi)
    =\fr{1}{r}\fr{\del Y_l^m}{\del\theta}\bs{e}_{\theta}+\fr{1}{r\sin\theta}\fr{\del Y_l^m}{\del\phi}\bs{e}_{\phi}\,,
    \\
    &\bs{Y}_l^{m({\rm axi})}\equiv\bs{\nabla}_{\rm h}\times Y_l^m(\theta,\phi)
    =\fr{1}{r\sin\theta}\fr{\del Y_l^m}{\del\phi}\bs{e}_{\theta}
    -\fr{1}{r}\fr{\del Y_l^m}{\del\theta}\bs{e}_{\phi}\,,
\end{align}
where the scalar spherical harmonics are treated as the radial component in performing the rotation.
Under the parity transformation, the former transforms as
$\bs{Y}_l^{m({\rm pol})}\to(-1)^l\bs{Y}_l^{m({\rm pol})}$,
while the latter transforms as
$\bs{Y}_l^{m({\rm axi})}\to(-1)^{l+1}\bs{Y}_l^{m({\rm axi})}$,
which explains the naming convention.%
\footnote{The parity transformations of the basis vectors are $\bs{e}_{\theta}\to-\bs{e}_{\theta}$ and $\bs{e}_{\phi}\to\bs{e}_{\phi}$.}

The toroidal modes satisfy $\bs{\nabla}_{\rm h}\cdot\bs{\xi}_{\rm h}=0$.
Accordingly, we assume a solution in the form of 
\begin{align}
    \bs{\xi}_{\rm h}=\xi_{\rm h}(r)\e^{-\im\omega t}\bs{Y}_l^{m({\rm axi})}(\theta,\phi)\,.
\end{align}
Using this expression, the conservation of the vorticity (\ref{eq:NS_Astero:Conservation_voticity_radial}) reduces to
\begin{align}
    0
    &=
    \del_{t}^2(\bs{\nabla}_{\rm h}\times\bs{\xi}_{\rm h})^{r}
    +2\del_t(\bs{\xi}_{\rm h}\cdot\bs{\nabla}_{\rm h})\Omega^{r}
    \nom
    &=
    \omega\lb[l(l+1)\omega-2m\Omega\rb]\xi_{\rm h}(r)\e^{-\im\omega t}Y_l^m\,.
\end{align}
Thus, we obtain the dispersion relation of the r-mode:
\begin{align}
\label{eq:NS_Astero:r-mode_frequency_rot_frame}
    \omega_{\rm r}
    &=
    \fr{2m\Omega}{l(l+1)}\,.
\end{align}
The subscript ``r'' denotes the frequency in the rotating frame.
In the inertial frame, the azimuthal angle shifts as $\phi\to\phi+\Omega t$.
For the spherical harmonics, this implies that the phase factor changes as $\e^{-\im(\omega_{\rm r}t+m\phi)}\to\e^{-\im[(\omega_{\rm i}+m\Omega)t+m\phi]}$, where $\omega_{\rm i}$ is the frequency in the inertial frame.
The relation, $\omega_{\rm i}=\omega_{\rm r}-m\Omega$, leads to the r-mode frequency in the inertial frame:
\begin{align}
\label{eq:NS_Astero:r-mode_frequency_inertial_frame}
    \omega_{\rm i}
    =-\fr{(l-1)(l+2)m\Omega}{l(l+1)}\,.
\end{align}

\section{Asteroseismology in general relativity}
\label{sec:NS_Astero:Astero_GR}
In the previous section, we discussed the typical seismic oscillations in the framework of Newtonian gravity.
For neutron stars, however, the high compactness requires the inclusion of general relativistic effects, such as gravitational wave emissions.

\subsection{Cowling approximation in general relativity}
Here, we give an overview of deriving the frequencies of seismic oscillations and gravitational waves in the framework of general relativity.
In general relativity, matter and spacetime interact with each other, and one must consider the spacetime geometry instead of the Newtonian gravitational potential.
The spacetime metric is determined by the Einstein equation,
\begin{align}
\label{eq:NS_Astero:Einstein_eq}
    R_{\mu\nu}-\frac{1}{2}Rg_{\mu\nu}
    =
    8\pi GT_{\mu\nu}\,,
\end{align}
where $R_{\mu\nu}$ is the Ricci tensor, $R$ is the scalar curvature, and $T_{\mu\nu}$ is the energy-momentum tensor.
Equation~(\ref{eq:NS_Astero:Einstein_eq}) is not independent of the energy-momentum conservation law.
Using the Bianchi identity,
\begin{align}
    \nabla_{\mu}R^{\alpha}_{~\beta\nu\rho}
    +\nabla_{\nu}R^{\alpha}_{~\beta\rho\mu}
    +\nabla_{\rho}R^{\alpha}_{~\beta\mu\nu}
    =0\,,
\end{align}
where $\nabla_{\mu}$ denotes the covariant derivative in general relativity, one finds
\begin{align}
    0=
    \nabla_{\mu}\lb(R^{\mu\nu}-\frac{1}{2}Rg^{\mu\nu}\rb)
    =
    8\pi G\nabla_{\mu}T^{\mu\nu}\,.
\end{align}

Before analyzing perturbations, we have to determine the background metric.
In the following discussion, we assume a static and spherically symmetric star without dissipation.
Outside the star $(T^{\mu\nu}=0)$, the metric is given by the Schwarzschild solution to the Einstein equation (\ref{eq:NS_Astero:Einstein_eq}):
\begin{align}
    &\bar{g}_{\mu\nu}^{({\rm out})}\dif x^{\mu}\dif x^{\nu}
    =
    A(r)\dif t^2
    -A^{-1}(r)\dif r^2
    -r^{2}\dif \theta^2
    -r^{2}\sin^2\theta\dif \phi^2\,,
\end{align}
where we defined 
\begin{align}
    A(r)
    \equiv
    1-\fr{2GM}{r}\,,
\end{align}
with $M$ being the stellar mass.
Inside the star $(T^{\mu\nu}\neq0)$, the Einstein equation leads to the TOV equation (\ref{eq:NS_Astero:TOV_eq}) and the mass equation (\ref{eq:NS_Astero:Mass_eq}).
As mentioned in Sec.~\ref{sec:NS_Astero:MR}, combining these with the equation of state (\ref{eq:NS_Astero:EoS_general}) allows one to determine the stellar mass and radius.

We now consider perturbations.
In addition to the perturbations of the hydrodynamic variables, such as pressure and energy density, the metric itself is perturbed as
\begin{align}
    g_{\mu\nu}=\bar{g}_{\mu\nu}+h_{\mu\nu}\,,
\end{align}
where the perturbation $h_{\mu\nu}$ corresponds to gravitational waves.
The linear stellar oscillations in general relativity are governed by 
\begin{align}
\label{eq:NS_Astero:Hydro_eq_GR}
    \delta(\nabla_{\mu}T^{\mu\nu})=0\,.
\end{align}
Note that the covariant derivative itself also contains the perturbation due to the perturbed metric.

By solving Eq.~(\ref{eq:NS_Astero:Hydro_eq_GR}) under appropriate boundary conditions, we can obtain the frequencies of the seismic oscillations and gravitational waves.
However, it is difficult to solve these equations analytically.
Similar to the case of Newtonian gravity, we here employ the general relativistic Cowling approximation.
In Newtonian gravity, we ignore the perturbation of the gravitational potential.
This corresponds to neglecting the perturbation of the metric in general relativity.
Although we set $h_{\mu\nu}=0$, this does not mean that gravitational waves are not emitted.
We neglect the recoil of gravitational waves on the star.
Practically, we derive the dispersion relation of the seismic oscillation, focusing only on the matter part, and assume that the frequency of the gravitational wave is equal to the seismic frequency.

The frequency under the Cowling approximation, $\omega_{\rm Cowling}$, is different from the frequency obtained by solving the full hydrodynamic equations, $\omega_{\rm mode}$.
To examine the validity of the general relativistic Cowling approximation in practice, it is necessary to compare its results with those obtained without the approximation.
For neutron stars and supernovae, calculations of the f-, p-, and g-modes have been performed both with and without the general relativistic Cowling approximation, showing that the deviation in the oscillation frequencies is at most about 20\%~\cite{Yoshida:1997bf,Sotani:2020mwc}.
Therefore, when performing an order-of-magnitude estimate of the oscillation frequency of a mode, the following relation is expected to provide a reasonable result:
\begin{align}
    \omega_{\rm Cowling}
    \sim
    \omega_{\rm mode}\,.
\end{align}

Gravitational waves are also classified according to their corresponding seismic oscillations.
The typical modes and their frequencies are summarized in Tab.~\ref{tab:NS_Astero:Modes}.
\begin{table}[tb]
\begin{center}
\caption{Characteristic frequencies of the oscillation modes for typical neutron stars~\cite{Kokkotas:1999bd}.}
\label{tab:NS_Astero:Modes}
\medskip
\begin{tabular}{|c|c|c|c|}
\hline
     & Mode & Physical origin & Typical frequency 
    \\
    \hline
    \multirow{4}{*}{Fluid mode} & f-mode & Pressure & $\sim10^{3}$~Hz 
    \\
    \cline{2-4}
     & p-mode & Pressure & $\sim10^{3}$~Hz 
    \\
    \cline{2-4}
     & g-mode & Buoyancy & $\sim10$~Hz 
    \\
    \cline{2-4}
     & r-mode & Coriolis force & $\sim1$~Hz 
    \\
    \hline
    Spacetime mode & w-mode & General relativity & $\sim10^{4}$~Hz 
    \\
    \hline
\end{tabular}
\end{center}
\end{table}
As already discussed, f-, p-, g-, and r-modes come from the hydrodynamic oscillations.
Thus, they are also called ``fluid modes.''
In addition, there exists a mode specific to general relativity called the w-mode~\cite{Kokkotas:1992xak,Andersson:1996ua}.
This is also referred to as a ``spacetime mode,'' in contrast to fluid modes.
The w-mode is strongly damped with the timescale $\sim10^{-5}~$sec~\cite{Kokkotas:1999bd} and the frequency is independent of the structure inside stars.
The frequencies of the modes above are, in general, complex numbers because these oscillations are damped by the emission of gravitational waves, and they are called quasi-normal modes.

\subsection{r-mode instability}
\label{subsec:NS_Astero:r-mode_instability}
\begin{figure}[tb]
\centering
\begin{tikzpicture}[scale=1, decoration={snake, amplitude=4pt, segment length=15pt}]

\draw[draw=black,very thick] (0,0) circle [radius=1.2];
\fill[black] (0,0) circle [radius=2pt];
\draw[draw=black,very thick] (7,0) circle [radius=1.2];
\fill[black] (7,0) circle [radius=2pt];

\draw[draw=black,very thick] (0,-4) circle [radius=1.2];
\fill[black] (0,-4) circle [radius=2pt];
\draw[draw=black,very thick] (7,-4) circle [radius=1.2];
\fill[black] (7,-4) circle [radius=2pt];

\draw[decorate, ultra thick, ->, >=stealth] (1.1,0.55) -- (3,1.5);

\draw[decorate, ultra thick, ->, >=stealth] (1.1,-3.45) -- (3,-2.5);

\draw[<-,>=stealth,blue,line width=1.8pt] ([shift={(0,0)}]-10:0.7) arc [radius=0.7, start angle = -10, end angle=50];
\draw[<-,>=stealth,blue,line width=1.8pt] ([shift={(7,0)}]-10:0.7) arc [radius=0.7, start angle = -10, end angle=50];
\draw[->,>=stealth,blue, line width=1.8pt] ([shift={(7,0)}]-10:0.9) arc [radius=0.9, start angle = -10, end angle=50];
\draw[->,>=stealth,line width=10pt] (2.5,0)--(4.5,0);

\draw[<-,>=stealth,blue,line width=1.8pt] ([shift={(0,-4)}]-10:0.7) arc [radius=0.7, start angle = -10, end angle=50];
\draw[->,>=stealth,very thick] ([shift={(0,-4)}]100:1.5) arc [radius=1.5, start angle = 100, end angle=170];
\draw[->,>=stealth,very thick] ([shift={(0,-4)}]280:1.5) arc [radius=1.5, start angle = 280, end angle=350];
\draw[<-,>=stealth,blue, line width=1.8pt] ([shift={(7,-4)}]-10:0.7) arc [radius=0.7, start angle = -10, end angle=50];
\draw[<-,>=stealth,blue, line width=1.8pt] ([shift={(7,-4)}]-10:0.9) arc [radius=0.9, start angle = -10, end angle=50];
\draw[->,>=stealth,line width=10pt] (2.5,-4)--(4.5,-4);

\node at (1.5,1.5) {$\delta J<0$};
\node at (1.5,-2.5) {$\delta J>0$};

\end{tikzpicture}
\caption{Schematic illustration of the CFS instability. The stars are viewed from the north. The blue arrows indicate the directions of the fluid perturbations, while the wavy lines represent the emitted gravitational waves. The upper panel shows the non-rotating case, and the lower one shows the rotating case.}
\label{fig:NS_Astero:CFS}
\end{figure}
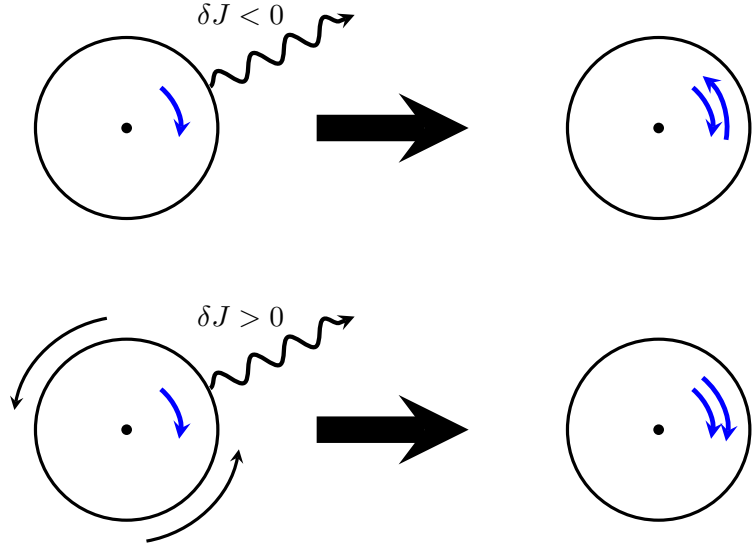

In general, a star with an angular velocity above a certain threshold becomes unstable due to the emission of gravitational waves.
This phenomenon is known as the Chandrasekhar--Friedman--Schutz (CFS) instability~\cite{Chandrasekhar:1970pjp,Friedman:1978hf}.
Here, we provide an intuitive explanation of the CFS instability (see Fig.~\ref{fig:NS_Astero:CFS}).

When a star emits gravitational waves, they carry away angular momentum, denoted by $\delta J$ in the figure.
The sign of $\delta J$ is determined by the angular momentum of the fluid perturbation.
In the upper panel, a gravitational wave carrying negative angular momentum is emitted.
As a result of the reaction to the gravitational wave emission, the fluid perturbation acquires positive angular momentum, leading to its suppression.

In contrast, when the star is rotating rapidly, a gravitational wave can carry away positive angular momentum (see the lower panel).
Since the direction of the reaction coincides with that of the original perturbation, the amplitude increases, resulting in the CFS instability.
This occurs when the mode frequency satisfies the following condition:
\begin{align}
\label{eq:NS_Astero:CFS_condition}
    \omega_{\rm r}\omega_{\rm i}
    =
    \omega_{\rm r}(\omega_{\rm r}-m\Omega)<0\,.
\end{align}

If the condition (\ref{eq:NS_Astero:CFS_condition}) is satisfied, the CFS instability can occur regardless of the type of mode.
Gravitational waves are expected to be emitted more efficiently by pressure perturbations, such as the f- and g-modes.
However, as shown in Tab.~\ref{tab:NS_Astero:Modes}, the typical frequency is of order $\sim10^3~{\rm Hz}$, and rapid rotation with a period shorter than a millisecond is required to satisfy the condition.

On the other hand, for the r-mode frequencies given by Eqs.~(\ref{eq:NS_Astero:r-mode_frequency_rot_frame}) and (\ref{eq:NS_Astero:r-mode_frequency_inertial_frame}), the condition reduces to
\begin{align}
    \omega_{\rm r}\omega_{\rm i}
    =-\fr{2(l-1)(l+2)m^2\Omega^2}{l^2(l+1)^2}<0\,.
\end{align}
If $l\geq2$, this condition is satisfied for any angular momentum.
Thus, the r-mode is more likely to be unstable than other modes, which is known as the r-mode instability~\cite{Andersson:1997xt}.
We have not considered shear viscosity suppressing the r-mode instability in realistic systems.
Whether the instability can occur depends on the competition between the timescales of gravitational wave emission and viscous damping.

\subsection{Amplitude of gravitational waves}
To discuss the observability of gravitational waves, it is necessary to examine not only their frequency but also their amplitude.
The frequencies of the oscillation modes are determined by the restoring forces, whereas the amplitude of the gravitational waves depends on the explosive event that excites the seismic oscillations.
This situation is analogous to a drum, where the pitch of the sound is determined by its shape and material, while the loudness depends on how strongly it is struck.
More precisely, the amplitude of the gravitational waves depends on the energy emitted as gravitational radiation during the explosive event, $E_{\rm GW}$.
We here derive a formula for the effective amplitude of gravitational waves~\cite{Andersson:1996pn,Andersson:2002ch}.

At a distance $d$ from the source, the energy flux of the gravitational waves is given by
\begin{align}
    \fr{1}{16\pi G}|\dot{h}|^2
    =
    \fr{1}{4\pi d^2}|\dot{E}_{\rm GW}|\,,
\end{align}
where $h$ denotes the gravitational wave amplitude.
In the following, we assume that the frequency of the gravitational wave, $f=2\pi\omega$, remains constant.
The time derivative of the amplitude can then be approximated as $\dot{h}\simeq2\pi f h$,
and the time derivative of the energy as $\dot{E}_{\rm GW}\simeq\ E_{\rm GW}/\tau$, where $\tau$ is the damping timescale associated with the gravitational wave emission.
Substituting these relations into the above equation, we obtain
\begin{align}
\label{eq:NS_Astero:amplitude_GW_formula}
    h
    \simeq
    \fr{1}{\pi d f}\sqrt{\fr{GE_{\rm GW}}{\tau}}\,.
\end{align}
To extract gravitational signals from observational data, the amplitude after matched filtering is of greater practical importance.
For an observed cycle $n\simeq f\tau$, the effective amplitude becomes
\begin{align}
\label{eq:NS_Astero:Eff_amp_GW_formula}
    h_{\rm eff}
    =
    h\sqrt{n}
    \simeq
    \fr{1}{\pi d}\sqrt{\fr{GE_{\rm GW}}{f}}\,.
\end{align}
This formula is useful to make an order-of-magnitude estimate.

\subsection{Giant flare}
\begin{figure}[tb]
\vspace{-1cm}
 \begin{center}
 \includegraphics[width=13cm]{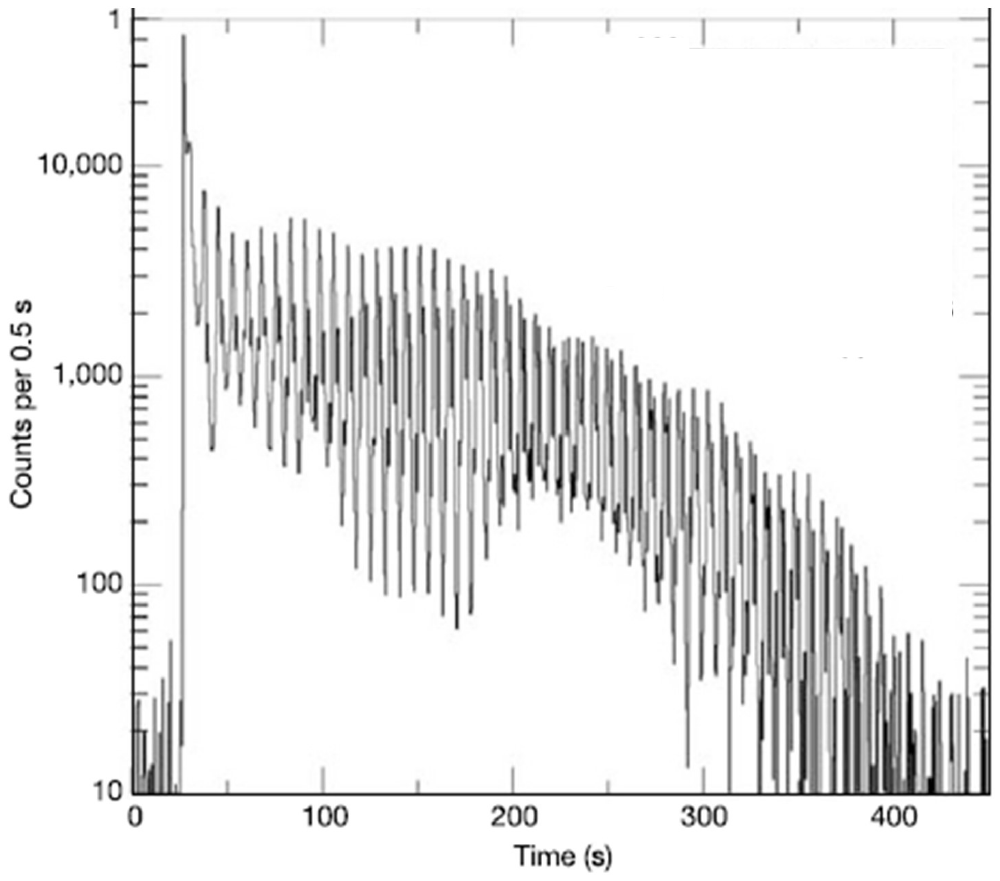}
 \end{center}
\vspace{-1.2cm}
\caption{Light curve of the 2004 giant flare from SGR~1806$-$20. 
The horizontal axis denotes time, and the vertical axis shows the number of detected photons per 0.5~s on a logarithmic scale~\cite{Hurley:2005zs}.}
\label{fig:NS_Astero:Giant-flare}
\end{figure}

As an example of an explosive event that can excite gravitational waves, we here discuss a giant flare.
A neutron star that repeatedly emits gamma-ray bursts is called a soft gamma-ray repeater (SGR).%
\footnote{The term ``soft'' indicates that the energy of the gamma rays is lower than that of ordinary gamma-ray bursts.}
An SGR occasionally undergoes an intense gamma-ray outburst known as a giant flare.
The first giant flare was observed in 1979~\cite{Mazets:1979uu}, followed by subsequent events in 1998~\cite{Feroci:2000ga} and 2004~\cite{Hurley:2005zs}.
The light curve of the giant flare observed in 2004 is shown in Fig.~\ref{fig:NS_Astero:Giant-flare}.
The initial burst of gamma rays lasted for about $0.2~{\rm s}$.
The oscillation period of $7.56~{\rm s}$ observed in the tail corresponds to the rotation period of the neutron star.
The total released energy is estimated to be $\sim10^{46}~{\rm erg}$, and the giant flare is the second most energetic phenomenon in the Universe after supernovae and gamma-ray bursts.%
\footnote{The energies of the three giant flares observed so far range from $10^{44}$ to $10^{46}$~erg.}
Consequently, seismic oscillations can be excited, leading to the emission of gravitational waves.
Indeed, quasi-periodic X-ray oscillations at $92.5~{\rm Hz}$ were detected~\cite{Israel:2005av}, indicating the excitation of seismic oscillations.

In most cases, a neutron star loses its rotational energy through the emission of electromagnetic radiation.
This type of neutron star is called a rotation-powered pulsar.
However, the energy of a giant flare cannot be supplied by rotation alone.
Its origin is instead believed to be the magnetic energy of a magnetar, a class of neutron stars with extremely strong magnetic fields.
One possible mechanism is magnetic reconnection, which releases a large amount of the magnetar’s magnetic energy~\cite{Thompson:1995gw}.
Below, we estimate the energy of the gravitational waves released by a giant flare~\cite{Kashiyama:2011fs}.

First, we derive the energy change due to a small deformation of a star with constant density $\rho$.
We assume that a spherical star is deformed into a spheroidal one with semi-major and semi-minor axes $a$ and $b$, respectively.
The volume is assumed to be conserved, giving $R^3 = a^2 b$.
Then, the moment of inertia of the deformed star, $I$, can be written as
\begin{align}
    I
    =
    \fr{2}{5}Ma^2
    =
    \bar{I}(1+\epsilon)\,,
    \qquad
    \bar{I}\equiv\fr{2}{5}MR^2\,,
    \qquad
    \epsilon\equiv(1-e^2)^{\fr{1}{3}}-1\,,
\end{align}
where $\bar{I}$ is the moment of inertia of the spherical star, and $e$ is the eccentricity defined as $e^2 \equiv 1 - (b^2/a^2)$. 
If the deformation is small $(\epsilon \ll 1)$, the gravitational energy of the spheroidal star is given by~\cite{Shapiro:1983du}
\begin{align}
    E_{\rm G}
    &=
    -\fr{3}{5}\fr{GM^2}{a}\fr{\sin^{-1}e}{e}
    \nom
    &=
    -\fr{3}{5}\left(\fr{4\pi}{3}\right)^{1/3}
    GM^{5/3}\rho^{1/3}
    \fr{\sin^{-1}e}{e}(1-e^2)^{1/6}
    \nom
    &\simeq
    \bar{E}_{\rm G}\lb(1-\fr{1}{5}\epsilon^2\rb)\,,
\end{align}
where the gravitational energy of the spherical star is
\begin{align}
    \bar{E}_{\rm G}
    =
    -\fr{3}{5}\fr{GM^2}{R}\,.
\end{align}
We assume that all the energy associated with the deformation is converted into gravitational wave energy:
\begin{align}
    E_{\rm GW}
    =
    -\fr{1}{5}\bar{E}_{\rm G}\epsilon^2\,.
\end{align}

We now consider the case in which the deformation is caused solely by the magnetic energy of a magnetar.
The parameter $\epsilon$ is determined by the competition between the gravitational energy $\bar{E}_{\rm G}$ and the magnetic energy $E_{\rm M}$~\cite{Ioka:2000hs}.
The upper limit of this parameter is estimated as
\begin{align}
    \epsilon
    \sim
    \fr{E_{\rm M}}{|\bar{E}_{\rm G}|}
    \leq
    \fr{(B^2/2)\times(4\pi R^3/3)}{3GM^2/5R}
    =
    \fr{10\pi R^4}{9GM^2}B^2\,.
\end{align}
Thus, the energy of gravitational waves released by a giant flare is given by
\begin{align}
    E_{\rm GW}
    \lesssim
    \fr{4\pi^2 R^7}{27GM^2}B^4
    \sim
    10^{43}~{\rm erg}
    \lb(\fr{R}{10~{\rm km}}\rb)^{7}
    \lb(\fr{M}{M_{\odot}}\rb)^{-2}
    \lb(\fr{B}{10^{16}~{\rm G}}\rb)^{4}\,.
\end{align}

\chapter{Chiral Asteroseismlogy}
\label{chap:Chi_Astero}
In this chapter, we investigate a novel form of asteroseismology that incorporates chiral transport phenomena, namely, ``chiral asteroseismology.''
In Sec.~\ref{sec:Chi_Astero:CM-mode_q}, we examine quark matter inside a magnetized neutron star and demonstrate the existence of a new type of seismic oscillation induced by the CME.
In Sec.~\ref{sec:Chi_Astero:CV-mode}, we discuss another type of seismic oscillation that occurs in neutrino matter inside a supernova, which is driven by the CVE.
Finally, in Sec.~\ref{sec:Chi_Astero:Angular}, we discuss the angular dependence of the gravitational wave emission associated with these modes.
This chapter is based on our original work~\cite{Hanai:2022yfh}.

\section{CM-mode in magnetized quark matter}
\label{sec:Chi_Astero:CM-mode_q}
We consider quark matter in a uniform magnetic field~$\bm{B}$.
This assumption is reasonable because, as discussed below, the typical wavelength of the seismic oscillations is much shorter than the spatial variation scale of the magnetic field inside the neutron star.
Without loss of generality, we set the magnetic field along the $z$-axis $(\bm{B}=B\bm{e}_{z})$.
The presence of the magnetic field breaks the spherical symmetry.
We therefore adopt cylindrical coordinates~$(t, r, \phi, z)$, whose central axis is aligned with the magnetic field.
The background metric is given by
\begin{align}
\label{eq:Chi_Astero:CM-metric}
    \dif s^2
    =
    \bar{g}_{\mu\nu}\dif x^{\mu}\dif x^{\nu}
    =
    \e^{2\lambda}\dif t^2
    -\e^{2\nu}\dif r^2
    -r^2\dif \phi^{2}
    -\e^{2\rho}\dif z^{2}\,,
\end{align}
where $\lambda$, $\nu$, and $\rho$ are the metric functions.
In general, these functions depend on $r$ and $z$ for stationary and axisymmetric systems.
However, as shown later, the typical length scale of collective excitations $(\sim10^{-1}~{\rm km})$ is much smaller than that of the spacetime variation inside neutron stars $(\sim1~{\rm km})$.
We therefore assume that the metric functions are constant.
Throughout this chapter, we use the notation $\bs{e}_{\mu}=e^{\hat{\alpha}}_{~\mu}\bs{e}_{\hat{\alpha}}$ where $\bs{e}_{\hat{\alpha}}$ denotes the orthonormal basis vectors and $e^{\hat{\alpha}}_{~\mu}={\rm diag}(\sqrt{|\bar{g}_{tt}|},\sqrt{|\bar{g}_{rr}|},\sqrt{|\bar{g}_{\phi\phi}|},\sqrt{|\bar{g}_{zz}|})$ is the vierbein satisfying $\bar{g}_{\mu\nu}=\eta_{\hat{\alpha}\hat{\beta}}e^{\hat{\alpha}}_{~\mu}e^{\hat{\beta}}_{~\nu}$.

\subsection{Wave equation and dispersion relation}
We focus on fluctuations in the number density, chiral charge density, and energy density, which are written as
\begin{align}
    n=\bar{n}+\delta n\,,
    \qquad
    n_5=\delta n_5\,,
    \qquad
    \varepsilon=\bar{\varepsilon}+\delta\varepsilon\,.
\end{align}
For the $N_{\rm f}$-flavor quark field~$\psi$, the quark number density and chiral charge density are expressed as $n=\bar{\psi}\gamma^{0}V\psi$ and $n_5 = \bar{\psi}\gamma^{0}\gamma_{5}A\psi$, respectively, where $V=A=\bm{1}_{N{\rm f}}$, with $\bm{1}_{N{\rm f}}$ being the $N_{\rm f}\times N_{\rm f}$ identity matrix.
We also denote the corresponding chemical potentials by $\mu$ and $\mu_5$.
Here we assume that the system is chirally symmetric $(\bar{n}_5=0)$.
Using the relativistic Cowling approximation (see Sec.~\ref{sec:NS_Astero:Astero_GR}), we neglect metric perturbations $(g_{\mu\nu}=\bar{g}_{\mu\nu})$.

Let us now derive the hydrodynamic equations.
As in Sec.~\ref{sec:Chi_Ph:CMW}, we consider the conservation laws.
Although the chiral anomaly does not appear in the absence of electric fields,%
\footnote{Strictly speaking, the QCD anomaly (or instanton effect) also induces chirality flipping.
This effect, however, is suppressed by a high power of $\Lambda_{\rm QCD}/\mu$ at sufficiently large densities where a weak-coupling analysis is applicable~\cite{Schafer:2002ty}.
Its relevance in the density range of interest, $\mu\simeq 500~{\rm MeV}$, is uncertain (see below).
Here we assume that, by extrapolating the above formula to intermediate densities, the instanton effect is negligibly small compared with the quark mass effect.}
we include the quark mass, which explicitly breaks the $\mathrm{U_{A}(1)}$ symmetry and induces chirality flipping.
The continuity equations are then given by
\begin{align}
\label{eq:Chi_Astero:Continuity_eqs_n_n5}
    \nabla_{\mu}j^{\mu}=0\,,
    \qquad
    \nabla_{\mu}j_5^{\mu}=-\Gamma_{\rm m}j_5^t\,,
\end{align}
where $\nabla_{\mu}$ is the covariant derivative in general relativity and $\Gamma_{\rm m}$ denotes the chirality-flipping rate.
Note that $j^\mu$ is the number current, not the electric current.

As discussed in Sec.~\ref{sec:Chi_Ph:CME}, (nearly) gapless fermions induce the CME~\cite{Vilenkin:1980fu,Nielsen:1983rb,Fukushima:2008xe} and the CSE~\cite{Son:2004tq,Metlitski:2005pr} in a background magnetic field.
Including diffusion, the constitutive relations in normalized bases can be written as
\begin{align}
\label{eq:Chi_Astero:CME}
    &\bm{j}
    =
    N_{\rm c}{\rm tr}(VAQ)\fr{e\mu_{\rm 5}}{2\pi^{2}}\bm{B}
    -D\bs{\nabla}n\,,
    \\
\label{eq:Chi_Astero:CSE}
    &\bm{j}_{\rm 5}
    =
    N_{\rm c}{\rm tr}(AVQ)\fr{e\mu}{2\pi^{2}}\bm{B}
    -D\bs{\nabla}n_5\,,
\end{align}
where $N_{\rm c}$ is the number of colors and $Q$ is the quark electric charge matrix, given by $Q={\rm diag}(2/3,-1/3)$ for $N_{\rm f}=2$ and $Q={\rm diag}(2/3,-1/3,-1/3)$ for $N_{\rm f}=3$.
The diffusion coefficient $D$ is common to both $\bs{j}$ and $\bs{j}_5$ in the chirally symmetric phase~\cite{Kharzeev:2010gd}.
The coefficients of the CME and CSE are related to the chiral anomaly~\cite{Adler:1969gk,Bell:1969ts,Nielsen:1983rb} and are topologically quantized~\cite{Son:2009tf,Son:2012wh,Son:2012zy} (see also Sec.~\ref{sec:Chi_Ph:Chi_Anom}).%
\footnote{Although the CSE generally has a fermion mass correction \cite{Metlitski:2005pr,Gorbar:2013upa,Guo:2016dnm}, the contribution is subleading. We then do not include the mass in the coefficient of the CSE.}
Therefore, these coefficients remain unaffected even in curved spacetime, which is characterized by differential geometry.

In the three flavor case $(N_{\rm f}=3)$, one finds ${\rm tr}(VAQ)=0$~\cite{Kharzeev:2010gr}.
Furthermore, since the strange quark mass $(m_{\rm s}\simeq 100~{\rm MeV})$ is comparable to the typical chemical potential in neutron stars, we focus on two-flavor quark matter in $\beta$ equilibrium, for which, e.g., $\bar{\mu}_{\rm u}\simeq 0.80\bar{\mu}_{\rm d}$ at $T=0$.
Here, $T$ denotes the temperature and $\bar{\mu}_{\rm q}$ (${\rm q}={\rm u,d}$) are the equilibrium chemical potentials of the up and down quarks, respectively.
As discussed in Sec.~\ref{sec:QCD:CSC}, such a situation can occur in the two-flavor color superconducting (2SC) phase~\cite{Barrois:1977xd,Bailin:1983bm,Alford:1997zt,Rapp:1997zu}, in which one of the three colors does not participate in Cooper pairing, leaving nearly gapless fermions.

The fluctuations of the currents are then expressed as
\begin{align}
\label{eq:Chi_Astero:Fluctuation_CME}
    &\delta \bm{j}
    =
    \fr{N_{\rm c} e\bm{B}}{6\pi^{2}\chi}\delta n_{\rm 5}
    -D\bm{\nabla}\delta n\,,
    \\
\label{eq:Chi_Astero:Fluctuation_CSE}
    &\delta \bm{j}_{\rm 5}
    =
    \fr{N_{\rm c} e\bm{B}}{6\pi^{2}\chi}\delta n
    -D\bm{\nabla}\delta n_{5}\,.
\end{align}
As with the diffusion coefficient, $n$ and $n_5$ share a common number susceptibility~$\chi$ in chirally symmetric matter~\cite{Kharzeev:2010gd}.
In the ideal-gas approximation, this can be written as
\begin{align}
\label{eq:Chi_Astero:Susceptibility_q}
    \chi\equiv\fr{\del n}{\del \mu}
    =\fr{\del n_5}{\del\mu_5}
    =2N_{\rm c}\lb(\fr{\bar{\mu}^2}{\pi^2}+\fr{T^2}{3}\rb)\,.
\end{align}
Since we are interested in fluctuations along the magnetic field, we express the densities as
\begin{align}
    &\delta n(t,z)=\delta n(\omega,k_z)\e^{-\im(\omega t-k_zz)}\,,
    \\
    &\delta n_5(t,z)=\delta n_5(\omega,k_z)\e^{-\im(\omega t-k_zz)}\,.
\end{align}
Using the relations $j^{\mu}=e^{~\mu}_{\hat{\alpha}}j^{\hat{\alpha}}$ and $j_5^{\mu}=e^{~\mu}_{\hat{\alpha}}j_5^{\hat{\alpha}}$ with $e^{~\mu}_{\hat{\alpha}}$ being the inverse of $e_{~\mu}^{\hat{\alpha}}$, the linearized continuity equations in (\ref{eq:Chi_Astero:Continuity_eqs_n_n5}) become
\begin{align}
    &\lb(
    \begin{array}{cc}
       \omega+\im\e^{\lambda-\rho}D |k_z|^2  &\displaystyle -\e^{\lambda-\rho}\fr{N_{\rm c} eB}{6\pi^2 \chi}k_{z} \\
       \displaystyle-\e^{\lambda-\rho}\fr{N_{\rm c} eB}{6\pi^2 \chi}k_{z}  & \omega+\im\e^{\lambda-\rho}D |k_z|^2+\im\Gamma_{\rm m}
    \end{array}\rb)\lb(
    \begin{array}{c}
        \delta n \\
        \delta n_5
    \end{array}\rb)
    =
    0\,.
\end{align}
This wave equation gives rise to a new type of seismic oscillation mode, which we call the chiral magnetic mode (CM-mode):
\begin{align}
\label{eq:Chi_Astero:Dispersion_CM}
    \omega_{\rm CM}
    &=
    \pm\sqrt{\lb(V_{\rm CM}k_{z}\rb)^2
    -\lb(\fr{\Gamma_{\rm m}}{2}\rb)^2}
    -\im\fr{\Gamma_{\rm m}}{2}
    -\im\e^{\lambda-\rho}D|k_z|^2
    \nom
    &\simeq
    \pm V_{\rm CM}|k_{z}|
    -\im\fr{\Gamma_{\rm m}}{2}
    -\im\e^{\lambda-\rho}D|k_z|^2\,,
\end{align}
where we have defined the speed of the CM-mode as
\begin{align}
    V_{\rm CM}
    \equiv
    \e^{\lambda-\rho}\fr{N_{\rm c} eB}{6\pi^2 \chi}\,.
\end{align}
In addition, for the CM-mode to propagate, the condition $V_{\rm CM}|k_{z}|\gg\Gamma_{\rm m}/2$ has to be satisfied.
The sign in front of the speed represents the propagating direction, and we focus only on the positive sign for simplicity.
Similarly to the chiral magnetic wave (CMW)~\cite{Newman:2005hd,Kharzeev:2010gd}, the CM-mode can appear even when the system does not have the chirality imbalance at equilibrium.
In the Cowling approximation, Eq.~(\ref{eq:Chi_Astero:Dispersion_CM}) also corresponds to the dispersion relation of the gravitational wave.
Indeed, the energy density fluctuation $\delta\varepsilon$ is proportional to $\delta n$ for $\mu \gg T$, resulting in the oscillations of the energy density and generates gravitational waves; see also Sec.~\ref{sec:Chi_Astero:Angular}.

Let us estimate the possible range of the CM-mode frequency~$f_{\rm CM}=\omega_{\rm CM}/(2\pi)$.
In the case of neutron stars or supernovae, the magnitude of the metric functions is characterized by the compactness parameter,
\begin{align}
    \lambda
    \sim
    \rho
    \sim
    \fr{GM}{R}
    \sim
    10^{-1}\,,
\end{align}
where $M$ and $R$ are the stellar mass and radius, respectively.
Thus, for an order-of-magnitude estimate, we may set the metric factor as $\e^{\lambda-\rho} \sim 1$.
The CM mode can propagate when $|{\rm Re}(\omega_{\rm CM})|\gg|{\rm Im}(\omega_{\rm CM})|$.
This condition yields the following possible range for the frequency:
\begin{align}
\label{eq:Chi_Astero:CM-range_quark}
    \fr{\Gamma_{\rm m}}{4\pi}
    \ll
    f_{\rm CM}
    \ll
    \fr{V_{\rm CM}^2}{2\pi D}
    =
    \fr{3V_{\rm CM}^2}{2\pi\tau}\,,
\end{align}
where we used $D=\tau/3$ with $\tau$ being the relaxation time~\cite{Heiselberg:1993cr}.
This inequality is more restrictive than the hydrodynamic limit $|k_{z}|\ll 2\pi/l_{\rm mfp}$,
where $l_{\rm mfp}$ denotes the mean free path in quark matter.
To obtain explicit numerical estimates for these quantities, one needs to evaluate the chirality-flipping rate~$\Gamma_{\rm m}$ and the relaxation time~$\tau$.

\subsection{Chirality flipping rate}
\label{subsec:Chi_Astero:chirality_flipping}

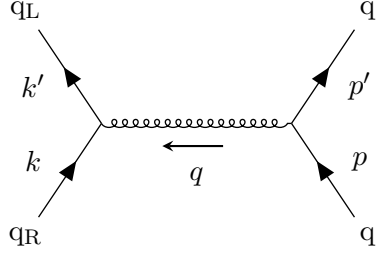
\begin{figure}[tb]
\centering
\begin{tikzpicture}
\begin{feynhand}
    \vertex (a) at (0,0); 
    \vertex[particle] (b) at (-1,-1.5) {${\rm q_R}$};
    \vertex[particle] (c) at (-1,1.5){${\rm q_L}$};
    \propag[fer] (b) to (a);
    \propag[fer] (a) to (c);
    \vertex (d) at (2.5,0);
    \propag[gluon] (a) to (d);
    \vertex[particle] (e) at (3.5,1.5) {q};
    \vertex[particle] (f) at (3.5,-1.5) {q};
    \propag[fer] (d) to (e);
    \propag[fer] (f) to (d);
\end{feynhand}

\node at (-0.9,-0.5) {$k$};
\node at (-0.9,0.5) {$k'$};
\node at (3.4,-0.5) {$p$};
\node at (3.4,0.5) {$p'$};
\node at (1.25,-0.7) {$q$};
\draw[->,>=stealth,thick] (1.6,-0.3)--(0.8,-0.3);
\end{tikzpicture}

\caption{The Feynman diagram for chirality flipping via quark-quark scattering.
In this process, a right-handed quark with momentum $k$ flips to a left-handed one with momentum $k'$, while the quark with momentum~$p$ retains its chirality.
Here, $q=k'-k$ denotes the four-momentum transfer.}

\label{fig:Chi_Astero:q-q_flip_diagram}
\end{figure}

Let us compute the chirality flipping rate of quark matter in neutron stars and supernovae. Our derivation here partly follows Ref.~\cite{Heiselberg:1993cr}.
Although we will eventually consider the 2SC phase as a concrete example, we here derive the expression of the chirality flipping rate for (nearly) gapless quarks with the generic number of colors, $N_{\rm c}$.
The leading-order contribution on the chirality flipping is the quark-quark scattering in Fig.~\ref{fig:Chi_Astero:q-q_flip_diagram}. 

The chirality flipping rate of the quark chiral charge is given by
\begin{align}
\label{eq:Chi_Astero:Gamma_flip_def_B}
    \Gamma_{\rm m}
    \equiv
    -\fr{\dot{n}_{5}}{n_{5}} \simeq
    -\fr{\dot{n}_{5,\rm q}}{n_{5,\rm q}}
    \,,
\end{align}
where $n_{5,\rm q}$ is the quark chiral charge density for ${\rm q}={\rm u,d}$ and we have assumed that $n_{\rm 5,u} \simeq n_{\rm 5,d}$.
Below, we omit the index $\rm q$ for $n_{5,{\rm q}}$, $\bar \mu_{\rm q}$, etc.~for notational simplicity unless otherwise stated.
To derive the chirality flipping rate $\Gamma_{\rm m}$, we have to know the dynamics of the chiral charge.
Introducing the distribution functions of right- and left-handed quarks as
\begin{align}
    &f_{\rm{R}}(k,t)
    =
    \fr{1}{\exp[\beta(|\bs{k}|-\bar{\mu}-\mu_{5})]+1}\,,
    \\
    &f_{\rm{L}}(k,t)
    =
    \fr{1}{\exp[\beta(|\bs{k}|-\bar{\mu}+\mu_{5})]+1}\,,
\end{align}
with $\beta=1/T$, we can write the chiral charge density as
\begin{align}
\label{eq:Chi_Astero:n5}
    n_{5}(t)
    &=
    N_{\rm c}\int \fr{\dif^{3}\bm{k}}{(2\pi)^{3}}\lb[f_{\rm R}(k,t)-f_{\rm L}(k,t)\rb]
    \nom
    &\simeq
    2N_{\rm c}\int \fr{\dif^{3}\bm{k}}{(2\pi)^{3}}\fr{\del f(k)}{\del \bar{\mu}}\mu_{5}(t)
    \nom
    &\simeq
    \fr{N_{\rm c} \bar{\mu}^{2}}{\pi^{2}}\mu_{5}(t)\,,
\end{align}
where $f$ is the equilibrium quark distribution function,
\begin{equation}
    f(k)
    =
    \fr{1}{\exp[\beta(|\bs{k}|-\bar{\mu})]+1}\,.
\end{equation}
In Eq.~(\ref{eq:Chi_Astero:n5}), we assumed that only the chiral chemical potential depends on time as $\mu_{5}(t)$.
We also used the fact that $\del f/\del \bar{\mu}$ is peaked at $|\bs{k}|=\bar{\mu}$ for degenerate quarks.

The dynamics of the distribution functions, $f_{\rm{R}}$ and $f_{\rm{L}}$, is described by the Boltzmann equations,
\begin{align}
\label{eq:Chi_Astero:Boltzmann-fR}
    &\dot{f}_{\rm R}(k,t)
    =
    -C(k,t)\,,
    \\
\label{eq:Chi_Astero:Boltzmann-fL}
    &\dot{f}_{\rm L}(k,t)
    =
    +C(k,t)\,,
\end{align}
where $C(k,t)$ is the collision integral.
The signs of the collision integrals in Eqs.~(\ref{eq:Chi_Astero:Boltzmann-fR}) and (\ref{eq:Chi_Astero:Boltzmann-fL}) are opposite for ${f}_{\rm R}$ and ${f}_{\rm L}$ such that $\dot{f}_{\rm R} + \dot{f}_{\rm L} = 0$.
We can then express $\dot{n}_{\rm5}$ as
\begin{align}
    \dot{n}_{5}(t)
    =
    -2N_{\rm c}\int \fr{\dif^{3}\bm{k}}{(2\pi)^{3}}C(k,t)\,.
\end{align}

In our case, the collision integral is given by
\begin{align}
    C(k,t)
    &=
    \int\fr{\dif^3 \bm{k}'}{(2\pi)^3 2|\bs{k}'|}\fr{\dif^3 \bm{p}}{(2\pi)^3 2|\bs{p}|}\fr{\dif^3 \bm{p}'}{(2\pi)^3 2|\bs{p}'|}
    \fr{|\mcl{M}_{\rm m}|^2}{2|\bs{k}|}
    (2\pi)^4 \delta^{(4)}(k^{\mu}+p^{\mu}-k'^{\mu}-p'^{\mu})
    \nom
    &\times
    \lb(
    f_{\rm R}(k)f(p)[1-f_{\rm L}(k')][1-f(p')]
    -f_{\rm L}(k')f(p')[1-f_{\rm R}(k)][1-f(p)]
    \rb)\,,
\end{align}
where $\mcl{M}_{\rm m}$ is the amplitude of the chirality flipping due to the mass effect averaged over initial colors.
Expanding $f_{\rm{R}}$ and $f_{\rm{L}}$ to first order in $\beta \mu_{5}$,
\begin{align}
    f_{\rm{R/L}}(k,t)
    \simeq
    f(k)\pm f(k)[1-f(k)]\beta\mu_{5}(t)\,,
\end{align}
the chirality flipping rate (\ref{eq:Chi_Astero:Gamma_flip_def_B}) reduces to
\begin{align}
\label{eq:Chi_Astero:Gamma_flip_simeq}
    \Gamma_{\rm m}
    \simeq
    \fr{4\pi^{2}\beta}{\bar{\mu}^{2}}
    &\int\fr{\dif^3 \bm{k}}{(2\pi)^3 2|\bs{k}|}\fr{\dif^3 \bm{k}'}{(2\pi)^3 2|\bs{k}|'}\fr{\dif^3 \bm{p}}{(2\pi)^3 2|\bs{p}|}\fr{\dif^3 \bm{p}'}{(2\pi)^3 2|\bs{p}'|}
    |\mcl{M}_{\rm m}|^2
    (2\pi)^4 \delta^{(4)}(k^{\mu}+p^{\mu}-k'^{\mu}-p'^{\mu})
    \nom
    &\times f(k)f(p)[1-f(k')][1-f(p')]\,.
\end{align}
To obtain this expression, we used
\begin{align}
     &
     f_{\rm R}(k)f(p)[1-f_{\rm L}(k')][1-f(p')]
     -
     f_{\rm L}(k')f(p')[1-f_{\rm R}(k)][1-f(p)]
     \nom
     &\simeq
     f(k)f(p)[1-f(k')][1-f(p')]\beta\mu_{5}
     +f(k')f(p')[1-f(k)][1-f(p)]\beta\mu_{5}
     \nom
     &\to
     2f(k)f(p)[1-f(k')][1-f(p')]\beta\mu_{5}\,.
 \end{align}
In the third line, we utilized the fact that the collision integral remains unchanged under the replacement of momenta $k\leftrightarrow k'$ and $p\leftrightarrow p'$ for the equilibrium distribution functions in the second term.

The scattering amplitude in Eq.~(\ref{eq:Chi_Astero:Gamma_flip_simeq}) is given by
\begin{align}
    |\mcl{M}_{\rm m}|^{2}
    =
    F_{\rm c}\sum_{h}|(\im g)^2 j^{\mu}_{-+} D_{\mu\nu} j ^{\nu}_{h}|^{2}\,,
    \qquad
    F_{\rm c}\equiv
    \fr{1}{N_{\rm c}^2}{\rm tr}(t^a t^b){\rm tr}(t^a t^b) = \fr{N_{\rm c}^2-1}{4N_{\rm c}^2}\,,
\end{align}
where $j^{\mu}_{-+} = \bar{u}_{-}(k')\gamma^{\mu}u_{+}(k)$ and $j ^{\mu}_{h} = \bar{u}_{h}(p')\gamma^{\mu}u_{h}(p)$ are the quark currents that do and do not involve the chirality flipping, respectively, with $u_{h}$ being the Dirac spinors with helicity $h$ (see App.~\ref{app:Helicity_Spinor} for the explicit expressions of the spinors).
The in-medium gluon propagator $D_{\mu\nu}$ is given by
\begin{align}
    D_{\mu\nu}
    &=
    \fr{P_{{\rm L},\mu\nu}}{q^{2}+\Pi_{\rm L}}
    +\fr{P_{{\rm T},\mu\nu}}{q^{2}+\Pi_{\rm T}}\,,
\end{align}
where $q^{\mu}=k'^{\mu}-k^{\mu}$ is the energy-momentum transfer, and $\Pi_{\rm L}$ and $\Pi_{\rm T}$ are the longitudinal and transverse self-energies.
The longitudinal and transverse projection operators $P_{{\rm L},\mu\nu},P_{{\rm T},\mu\nu}$ are defined as
\begin{align}
    &P_{{\rm T},00}=P_{{\rm T},0i}=P_{{\rm T},i0}=0\,,
    \\
    &P_{{\rm T},ij}=\delta_{ij}-\fr{q_{i}q_{j}}{|\bs{q}|^{2}}\,,
    \\
    &P_{{\rm L},\mu\nu}+P_{{\rm T},\mu\nu}=-\eta_{\mu\nu}+\fr{q_{\mu}q_{\nu}}{q^2}\,.
\end{align}

We decompose the square of the amplitude of the chirality flipping by the quark-quark scattering,  $|\mcl{M}_{\rm m}|^{2}$, into the following form:
\begin{align}
    |\mcl{M}_{\rm m}|^{2}
    =
    |\mcl{M}_{\rm m,L}|^2
    +|\mcl{M}_{\rm m,T}|^2
    +2|\mcl{M}_{\rm m,cross}|^2,
\end{align}
where 
\begin{align}
    &|\mcl{M}_{\rm m,L}|^2
    =
    F_{\rm c}\sum_{h}\lb|\fr{j_{-+}^{0}j_{h}^{0}}{|\bs{q}|^{2}-\Pi_{\rm L}}\rb|^2,
    \qquad
    |\mcl{M}_{\rm m,T}|^2
    =
    F_{\rm c}\sum_{h}\lb|\fr{\bm{j}_{-+,\rm T}\cdot \bm{j}_{h,\rm T}}{(q^{0})^2-|\bs{q}|^{2}-\Pi_{\rm T}}\rb|^2, \nom
    &|\mcl{M}_{\rm m,cross}|^2 = F_{\rm c}\sum_{h}\lb|\fr{j_{-+}^{0}j_{h}^{0}}{|\bs{q}|^2-\Pi_{\rm L}} \cdot \fr{\bm{j}_{-+,\rm T}\cdot \bm{j}_{h,\rm T}}{(q^{0})^2-|\bs{q}|^2-\Pi_{\rm T}}\rb|\,,
\end{align}
with $\bm{j}_{\rm T}$ being the transverse component of the current with respect to $\bm{q}$. 
The cross part $|\mcl{M}_{\rm m,cross}|^2$ is proportional to $\cos{\varphi}$, where $\varphi$ is the angle between $\bm{j}_{-+,\rm T}$ and $\bm{j}_{h,\rm T}$.
Then, the integration over $\varphi$ vanishes.
In the following, we simply focus on the contribution of $|\mcl{M}_{\rm m,L}|^2$ and $|\mcl{M}_{\rm m,T}|^2$. 

We set $\bm{k}$ to be along the $z$-axis.
For simplicity of calculation, we focus on the special case $\theta_{pk} = 0$, where $\theta_{pk}$ is the angle between $\bm{p}$ and $\bm{k}$, since the resulting parameter dependence of the chirality flipping rate does not depend on the particular choice of $\theta_{pk}$.
As the integration over $\varphi$ gives a constant and does not change the physical parameter dependence of the chirality flipping, we also assume that the scattering occurs in the same plane, namely,  $\varphi=0$.
In this case, we can express the currents with and without the chirality flipping as
\begin{align}
\label{eq:Chi_Astero:general_currents_flip}
    &j_{-+}^{\mu}
    \simeq
    -\fr{gm}{\sqrt{|\bs{k}||\bs{k}'|}}
    \lb(
    (|\bs{k}|+|\bs{k}'|)\sin{\fr{\theta_{kk'}}{2}}, \
    q^{0}\cos{\fr{\theta_{kk'}}{2}}, \
    \im q^{0}\cos{\fr{\theta_{kk'}}{2}},\
    -q^{0}\sin{\fr{\theta_{kk'}}{2}}
    \rb),
    \\
\label{eq:Chi_Astero:general_currents_no_flip}
    &j_{h}^{\mu}
    \simeq
    2g\sqrt{|\bs{p}||\bs{p}'|}
    \lb(
    \cos{\fr{\theta_{pp'}}{2}}, \ 
    \sin{\fr{\theta_{pp'}}{2}}, \ 
    \im h\sin{\fr{\theta_{pp'}}{2}}, \ 
    \cos{\fr{\theta_{pp'}}{2}}
    \rb).
\end{align}
Here, we expanded $j_{-+}^{\mu}$ and $j_{h}^{\mu}$ up to the leading first order and zeroth order of the quark mass $m$, respectively. 

We first consider the longitudinal contribution.
Using Eqs.~(\ref{eq:Chi_Astero:general_currents_flip}) and (\ref{eq:Chi_Astero:general_currents_no_flip}), the longitudinal amplitude can be written as
\begin{align}
    |\mcl{M}_{\rm m,L}|^{2}
    &=
    2F_{\rm c}\lb|\fr{j_{-+}^{0}j_{+}^0}{|\bs{q}|^2-\Pi_{\rm L}}\rb|^2
    \nom
    &=
    32\pi^{2}F_{\rm c}\alpha_{\rm s}^2 m^{2}\fr{|\bs{p}||\bs{p}|'}{|\bs{k}||\bs{k}'|}\fr{(|\bs{k}|+|\bs{k}'|)^2}{||\bs{q}|^2-\Pi_{\rm L}|^2}
    (1-\cos{\theta_{kk'}})(1+\cos{\theta_{pp'}})\,.
\end{align}
Here, the factor 2 comes from the summation over $h$ with $j_{+}^0=j_{-}^{0}$, and $\alpha_{\rm s}\equiv g^{2}/(4\pi)$.
We assumed that the up and down quark mass have the same mass as $m_{\rm u}\simeq m_{\rm d}\equiv m$.
Note that the factor $(1-\cos{\theta_{kk'}})$ is characteristic of the chirality flipping~\cite{Schafer:2001za} in the so-called high density effective theory near the Fermi surface~\cite{Hong:1998tn,Hong:1999ru} (see App.~\ref{app:HDET_Mass}).
We rewrite the delta function as
\begin{align}
    \delta^{(4)}(k+p-k'-p')
    =
    \int\dif q^{0}\dif^{3}\bm{q}
    \delta(|\bs{k}'|-|\bs{k}|-q^{0})\delta(|\bs{p}'|-|\bs{p}|+q^{0})
    \delta^{(3)}(\bm{k}'-\bm{k}-\bm{q})\delta^{(3)}(\bm{p}'-\bm{p}+\bm{q}),
\end{align}
and introduce the functions $S_{1,\rm L}$ and $S_{2,\rm L}$:
\begin{align}
    S_{1,\rm L}(q^{0},\bm{q})
    &\equiv
    \int \fr{\dif^3 \bm{k}}{(2\pi)^3}\fr{\dif^3 \bm{k}'}{(2\pi)^3}
    \fr{(k+k')^{2}}{|\bs{k}|^2|\bs{k}'|^2}(1-\cos{\theta_{kk'}})
    \nom
    &\quad\times(2\pi)^4 \delta(|\bs{k}'|-|\bs{k}|-q^{0})\delta^{(3)}(\bm{k}'-\bm{k}-\bm{q})
    f(k)[1-f(k')]\,,
    \\
    S_{2,\rm L}(q^{0},\bm{q})
    &\equiv
    \int \fr{\dif^3 \bm{p}}{(2\pi)^3}\fr{\dif^3 \bm{p}'}{(2\pi)^3}
    (1+\cos{\theta_{pp'}})
    \nom
    &\quad\times(2\pi)^4 \delta(|\bs{p}'|-|\bs{p}|+q^{0})\delta^{(3)}(\bm{p}'-\bm{p}+\bm{q})
    f(p)[1-f(p')]\,.
\end{align}
Using these functions, we can express the longitudinal contribution to the chirality flipping rate as
\begin{align}
\label{eq:Chi_Astero:CFR_qq_l}
    \Gamma_{\rm m,L}
    &\simeq
    F_{\rm c}\fr{\alpha_{\rm s}^2 m^2 \beta}{2\bar{\mu}^2}
    \int \dif q^{0}\dif^3 \bm{q} 
    \fr{1}{||\bs{q}|^2-\Pi_{\rm L}|^2}
    S_{1,\rm L}S_{2,\rm L}\,.
\end{align}

We assume that $q^{0}~(\sim T)\ll |\bs{q}|~(\sim q_{\rm IR})\ll |\bs{k}|~(\sim \bar{\mu})$, where $q_{\rm IR}$ is the IR cutoff of the momentum.%
\footnote{Since the momentum $|\bs{q}|$ around the IR cutoff has the dominant contribution to the integral over $\bm{q}$, we can assume that $|\bs{q}|\sim q_{\rm IR}$.}
In this case, the gluon self-energies are expressed as
\begin{align}
    \Pi_{\rm L}\simeq-m_{\rm D,s}^2\,,
    \qquad
    \Pi_{\rm T}\simeq \im\fr{\pi m_{\rm D,s}^2}{4}\fr{q^{0}}{|\bs{q}|}\,,
\end{align}
where $m_{\rm D,s}$ is the Debye screening mass of the gluon defined by $m_{\rm D,s}^2\equiv N_{\rm f}g^2\bar{\mu}^2/\pi^2$.
We can integrate over $\bm{k}$ and $\bm{k}'$ in $S_{1,\rm L}$ as
\begin{align}
    S_{1,\rm L}(q^{0},\bm{q})
    &=
    \int \fr{\dif^3 \bm{k}}{(2\pi)^3}
    \fr{(k+k')^{2}}{k^2 k'^2}(1-\cos{\theta_{kk'}})
    \nom
    &\quad\times\lb.(2\pi)\delta(|\bs{k}'|-|\bs{k}|-q^{0})
    f(k)[1-f(k')]\rb|_{\bm{k}'=\bm{k}+\bm{q}}
    \nom
    &\simeq
    \fr{1}{2\pi q}
    \int_{|\bs{k}|>q^{0}}\dif|\bs{k}| \dif\cos{\theta_{kq}}\fr{(k+k')^2}{|\bs{k}'|^2}
    (1-\cos{\theta_{kk'}})
    \nom
    &\quad\times\lb.\delta\lb(\cos{\theta_{kq}-\fr{q^0}{|\bs{q}|}+\fr{|\bs{q}|}{2|\bs{k}|}}\rb)
    f(k)[1-f(k')]
    \rb|_{\bm{k}'=\bm{k}+\bm{q}}
    \nom
    &\simeq
    \fr{1}{4\pi |\bs{q}|}\int_{|\bs{k}|>q^{0}}\dif|\bs{k}|\lb(\fr{2|\bs{k}|+q^{0}}{|\bs{k}|+q^{0}}\rb)^{2}\lb(\fr{|\bs{q}|}{|\bs{k}|}\rb)^{2}
    \delta(|\bs{k}|-\bar{\mu})\lb(T+\fr{q^{0}}{2}\rb)
    \nom
    &\simeq
    \fr{|\bs{q}|}{\pi \bar{\mu}^2}\lb(T+\fr{q^{0}}{2}\rb).
\end{align}
Here, we used the following relations:
\begin{align}
    &\delta(|\bm{k}+\bm{q}|-|\bs{k}|-q^{0})
    \simeq
    \fr{1}{|\bs{q}|}\delta\lb(\cos{\theta_{kq}}-\fr{q^0}{|\bs{q}|}+\fr{|\bs{q}|}{2|\bs{k}|}\rb)
    \theta(|\bs{k}|+q^{0})\,,
    \\
    &\cos{\theta_{kk'}}
    =
    \fr{|\bs{k}|+q\cos{\theta_{kq}}}{\sqrt{|\bs{k}|^2 +q^2 +2|\bs{k}|q\cos{\theta_{kq}}}}\,,
    \\
    &f(k)[1-f(k+q^{0})]
    \simeq
    \delta(|\bs{k}|-\bar{\mu})\lb(T+\fr{q^{0}}{2}\rb)\,.
\end{align}
In the same way, we can calculate $S_{2,\rm L}$ as
\begin{align}
    S_{2,\rm L}(q^{0},\bm{q})
    &\simeq
    \fr{\bar{\mu}^2}{\pi |\bs{q}|}\lb(T-\fr{q^{0}}{2}\rb).
\end{align}

Substituting $S_{1,\rm L}$ and $S_{2,\rm L}$ into Eq.~(\ref{eq:Chi_Astero:CFR_qq_l}), the longitudinal contribution to the chirality flipping rate can be obtained as
\begin{align}
\label{eq:Chi_Astero:CFR_l_parameter}
    \Gamma_{\rm m,L}
    &\simeq
    F_{\rm c}\fr{2\alpha_{\rm s}^2 m^2 \beta}{\pi\bar{\mu}^2}
    \int \dif q^{0}\dif|\bs{q}|
    \fr{|\bs{q}|^2}{||\bs{q}|^2-\Pi_{\rm L}|^2}
    \lb[T^2-\lb(\fr{q^{0}}{2}\rb)^2\rb]
    \nom
    &\simeq
    F_{\rm c}\fr{2\alpha_{\rm s}^2 m^2 \beta}{\pi\bar{\mu}^2 m_{\rm D,s}}
    \int_{0}^{T}\dif q^{0}\lb[T^2-\lb(\fr{q^{0}}{2}\rb)^2\rb]
    \int_{0}^{1}\fr{\dif \zeta}{(1+\zeta^2)^2}
    \nom
    &\simeq
    F_{\rm c}\fr{11(\pi+2)}{48\pi}\fr{\alpha_{\rm s}^2 m^2}{\bar{\mu}^2 m_{\rm D,s}}T^2\,.
\end{align}
In the second line, we defined $\zeta\equiv m_{\rm D,s}/q$.
The temperature dependence of the chirality flipping rate is consistent with the behavior of the Fermi liquid and is different from that of the Rutherford scattering in Refs.~\cite{Grabowska:2014efa,Dvornikov:2015iua}.
This difference comes from their assumption that the proton is so heavy that its recoil is negligible, which leads to the isoenergetic (or $q^0=0$) scattering.
In that case, since the contribution of the energy width $T$ of the proton is neglected, the temperature dependence of $\Gamma_{\rm m,L}$ changes from $\sim T^2$ to $\sim T$.

We next evaluate the transverse contribution to the chirality flipping rate. The amplitude is given by
\begin{align}
    |\mcl{M}_{\rm m,T}|^{2}
    =
    F_{\rm c}\lb|\fr{\bm{j}_{-+,{\rm T}}\cdot\bm{j}_{+,{\rm T}}}{q^2-\Pi_{\rm T}}\rb|^2
    +F_{\rm c}\lb|\fr{\bm{j}_{-+,{\rm T}}\cdot\bm{j}_{-,{\rm T}}}{q^2-\Pi_{\rm T}}\rb|^2.
\end{align}
Though the expressions of the two terms are different, their parameter dependence is the same, and we focus only on the first term.
Using  Eqs.~(\ref{eq:Chi_Astero:general_currents_flip}) and (\ref{eq:Chi_Astero:general_currents_no_flip}), the transverse amplitude can be written as
\begin{align}
\label{eq:Chi_Astero:amplitude_qq_t}
    |\mcl{M}_{\rm m,T}|^{2}
    &\simeq
    F_{\rm c}\lb|\fr{j^{i}_{-+}\lb[\delta_{ij}-(q_{i}q_{i}/|\bs{q}|^2)\rb]j_{+}^{j}}{q^2-\Pi_{\rm T}}\rb|^2
    \nom
    &=
    F_{\rm c}\lb|\fr{j^{i}_{-+}j_{+}^{i}-\lb(q^{0}/|\bs{q}|\rb)^2 j^{0}_{-+} j_{+}^{0}}{q^2-\Pi_{\rm T}}\rb|^2
    \nom
    &=
    16\pi^2 F_{\rm c}\alpha_{\rm s}^2 m^2\fr{|\bs{p}||\bs{p}'|}{|\bs{k}||\bs{k}'|}
    \fr{\lb[q^{0}-(|\bs{k}|+|\bs{k}'|)\lb(q^{0}/|\bs{q}|\rb)^2\rb]^2}{|q^2-\Pi_{\rm T}|^2}
    (1-\cos{\theta_{kk'}})(1+\cos{\theta_{pp'}})\,.
\end{align}
In the second line, we have used the current conservation, $\bm{q}\cdot\bm{j}=q^{0}j^{0}$.

We decompose the transverse contribution to the chirality flipping rate into three parts:
\begin{align}
    \Gamma_{\rm m,T}
    =
    \Gamma_{\rm m,T}^{(1)}
    +
    \Gamma_{\rm m,T}^{(2)}
    +
    \Gamma_{\rm m,T}^{(3)}\,,
\end{align}
where
\begin{align}
    \Gamma_{\rm m,T}^{(1)}
    &\simeq
    F_{\rm c}\fr{\alpha_{\rm s}^2 m^2 \beta}{4\bar{\mu}^2}
    \int \dif q^{0}\dif^3 \bm{q} 
    \fr{(q^{0})^2}{|q^2-\Pi_{\rm T}|^2}
    S_{1,\rm T}^{(1)}S_{2,\rm T}\,,
    \nom
    \Gamma_{\rm m,T}^{(2)}
    &\simeq
    F_{\rm c}\fr{\alpha_{\rm s}^2 m^2 \beta}{4\bar{\mu}^2}
    \int \dif q^{0}\dif^3 \bm{q} 
    \fr{-2q^{0}\lb(q^{0}/|\bs{q}|\rb)^2}{|q^2-\Pi_{\rm T}|^2}
    S_{1,\rm T}^{(2)}S_{2,\rm T}\,,
    \nom
    \Gamma_{\rm m,T}^{(3)}&\simeq
    F_{\rm c}\fr{\alpha_{\rm s}^2 m^2 \beta}{4\bar{\mu}^2}
    \int \dif q^{0}\dif^3 \bm{q} 
    \fr{\lb(q^{0}/|\bs{q}|\rb)^4}{|q^2-\Pi_{\rm T}|^2}
    S_{1,\rm T}^{(3)}S_{2,\rm T}\,.
\end{align}
The functions $S_{1,\rm T}^{(i)}$ ($i=1,2,3$) and $S_{2,\rm T}$ are defined and computed as
\begin{align}
    S_{1,\rm T}^{(i)}(q^{0},\bm{q})
    &\equiv
    \int \fr{\dif^3 \bm{k}}{(2\pi)^3}\fr{\dif^3 \bm{k}'}{(2\pi)^3}
    \fr{(|\bs{k}|+|\bs{k}'|)^{i-1}}{|\bs{k}|^2|\bs{k}'|^2}(1-\cos{\theta_{kk'}})
    \nom
    &\quad\times(2\pi)^4 \delta(|\bs{k}'|-|\bs{k}|-q^{0})\delta^{(3)}(\bm{k}'-\bm{k}-\bm{q})
    f(k)[1-f(k')]
    \nom
    &\simeq
    \fr{q}{4\pi\bar{\mu}^{5-i}}\lb(T+\fr{q^{0}}{2}\rb)\,,
    \\
    S_{2,\rm T}(q^{0},\bm{q})
    &\equiv
    S_{2,\rm L}(q^{0},\bm{q})\,.
\end{align}

In the same way as the longitudinal case, we can obtain the parameter dependence as
\begin{align}
    \Gamma_{\rm m,T}^{(1)}
    \sim
    F_{\rm c}\fr{\alpha_{\rm s}^2 m^2}{\bar{\mu}^4 m_{\rm D,s}^{2/3}}T^{\fr{11}{3}}\,, \quad 
    \Gamma_{\rm m,T}^{(2)}
    \sim
    -F_{\rm c}\fr{\alpha_{\rm s}^2 m^2}{\bar{\mu}^3 m_{\rm D,s}^{2}}T^{4}\,, \quad
    \Gamma_{\rm m,T}^{(3)}
    \sim
    F_{\rm c}\fr{\alpha_{\rm s}^2 m^2}{\bar{\mu}^2 m_{\rm D,s}^{10/3}}T^{\fr{13}{3}}\,.
\end{align}
From the ratios between the longitudinal and transverse contributions, such as
\begin{align}
    \fr{\Gamma_{\rm m,T}^{(1)}}{\Gamma_{\rm m,L}}
    \sim
    10^{-11}\lb(\fr{T}{10^{6}~{\rm K}}\rb)^{\fr{5}{3}},
    \qquad
    \fr{\Gamma_{\rm m,T}^{(3)}}{\Gamma_{\rm m,L}}
    \sim
    10^{-14}\lb(\fr{T}{10^{6}~{\rm K}}\rb)^{\fr{7}{3}},
\end{align}
we find that the transverse contribution is suppressed when $T\lesssim 10^{11}~{\rm K}$ and the longitudinal contribution is dominant in our setup ($T\sim 10^{6}\text{--}10^{11}~{\rm K}$).
This fact can be understood from the high density effective theory based on the systematic expansion of $T/\mu$~\cite{Hong:1998tn,Hong:1999ru}, where the chirality flipping is caused by the temporal component of the gauge field $A^{0}$~\cite{Schafer:2001za}, and only the longitudinal propagator is relevant (see App.~\ref{app:HDET_Mass}). 

We now consider the case of the 2SC phase and focus on the nearly gapless unpaired quark whose energy measured from the Fermi energy is much smaller than the color superconducting gap $\Delta(\sim 10^{2}~{\rm MeV})$~\cite{Alford:1997zt}.
Since the unpaired quarks have only one color out of three, the quark-quark scattering does not contribute to the chirality flipping at tree level.
The other possible candidates for the chirality flipping due to the quark-quark scattering are the QCD process at second order in $\alpha_{\rm s}$ and the QED process at first order in $\alpha_{\rm e}(\equiv e^2/(4\pi))$.
The chirality flipping rate in the former QCD process may be parametrically estimated by multiplying Eq.~(\ref{eq:Chi_Astero:CFR_l_parameter}) by $\alpha_{\rm s}^2$ as
\begin{align}
    \Gamma_{\rm m,QCD}^{(\rm 2SC)}
     &\sim
     \fr{\alpha_{\rm s}^4 m^2}{\bar{\mu}^2 m_{\rm D,s}}T^2\,,
\end{align}
while that in the latter QED process is obtained with the replacement, $F_{\rm c}\to 1$ and $g\to e~(\alpha_{\rm s}\to\alpha_{\rm e})$ in the above discussion:
\begin{align}
\label{eq:Chi_Astero:CFR_2SC}
    \Gamma_{\rm m,QED}^{(\rm 2SC)}\sim
    \fr{\alpha_{\rm e}^2 m^2}{\bar{\mu}^2 m_{\rm D,e}}T^2\,,
\end{align}
where $m_{\rm D,e}$ is the Debye screening mass of the photon given by $m_{\rm D,e}^2 \equiv 5e^2\bar{\mu}^2/(9\pi^2)$.
Comparing these two processes, e.g., at $\bar{\mu}\simeq 
500~{\rm MeV}$ and $\alpha_{\rm s}\simeq 0.7$, we find that the QCD process is dominant in our setup:
\begin{align}
    \fr{\Gamma_{\rm m,QCD}^{(\rm 2SC)}}{\Gamma_{\rm m,QED}^{(\rm 2SC)}}
    \sim
    \fr{\alpha_{\rm s}^4 m_{\rm D,e}}{\alpha_{\rm e}^2 m_{\rm D,s}}
    \sim
    10^{2}\,.
\end{align}

\subsection{Relaxation time of quark matter}

The relaxation time in quark matter was estimated in ref.~\cite{Heiselberg:1993cr} and has two parts originating from the longitudinal and transverse gluon self-energies:
\begin{align}
\label{eq:Chi_Astero:tau_1st_order}
    \begin{array}{ll}
    \displaystyle
    \fr{1}{\tau_{\rm L}}\sim F_{\rm c}\alpha_{\rm s}^2 \fr{T^2}{m_{\rm D,s}} & (\text{longitudinal contribution})\,,
    \vspace{1em}
    \\
    \displaystyle
    \fr{1}{\tau_{\rm T}}\sim F_{\rm c}\alpha_{\rm s}^2 \fr{T^{5/3}}{m_{\rm D,s}^{2/3}} & (\text{transverse contribution})\,.
    \end{array}
\end{align}
While the longitudinal contribution is consistent with the typical Fermi-liquid behavior $\sim T^2$, the transverse one has the different temperature dependence as the Landau damping results in the IR cutoff like $q_{\rm IR}\sim (m_{\rm D,s}^2 T)^{1/3}$ instead of $q_{\rm IR}\sim m_{\rm D,s}$.
In our setup ($T\sim 10^{6}\text{--}10^{11}~\rm K$), the transverse contribution is dominant.
Similarly to the chirality flipping rate above, the relaxation time in the 2SC phase is dominated by the QCD process and is given by multiplying the transverse contribution in Eq. (\ref{eq:Chi_Astero:tau_1st_order}) by $\alpha_{\rm s}^2$ as
\begin{align}
\label{eq:Chi_Astero:tau_2SC}
    \fr{1}{\tau^{(\rm 2SC)}_{\rm QCD}}\sim\alpha_{\rm s}^4\fr{T^{5/3}}{m_{\rm D,s}^{2/3}}\,.
\end{align}

\subsection{Estimate of the frequency}
Now we estimate the order of magnitude of the frequency of the CM-mode in the 2SC phase by substituting Eqs.~(\ref{eq:Chi_Astero:CFR_2SC}) and (\ref{eq:Chi_Astero:tau_2SC}) into Eq.~(\ref{eq:Chi_Astero:CM-range_quark}).
Here, the quark chemical potential coupled to the quark number density $n$ will be denoted by $\bar{\mu}$ again. 
Using $m_{\rm q}\simeq 3\text{--}5~{\rm MeV},~\bar{\mu}\simeq 
500~{\rm MeV},~\alpha_{\rm s}\simeq 0.7$, we obtain
\begin{align}
\label{eq:Chi_Astero:CM_freq_quark}
    10^{4}\,{\rm Hz}
    \lb(\fr{T}{10^{6}\,{\rm K}}\rb)^{2}
    \ll
    f_{\rm CM}
    \ll
    10^{6}\,{\rm Hz}
    \lb(\fr{B}{10^{18}\,{\rm G}}\rb)^{2}
    \lb(\fr{T}{10^{6}\,{\rm K}}\rb)^{5/3}.
\end{align}
Since the power of the temperature of the upper limit is smaller than that of the lower limit, the CM-mode would not exist in high temperature ($T\gtrsim 10^{7}~\rm K$) environments such as supernovae.

From the dispersion relation~(\ref{eq:Chi_Astero:Dispersion_CM}), the order of magnitude of the frequency of the CM-mode is expressed as
\begin{align}
\label{eq:Chi_Astero:frequency-CM-parameter}
    f_{\rm CM}
    \sim
    10^{5}\,{\rm Hz}
    \lb(\fr{B}{10^{18}~\rm G}\rb)
    \lb(\fr{\bar{\mu}}{500\,{\rm MeV}}\rb)^{-2}
    \lb(\fr{k}{10^{-4}\,{\rm /cm}}\rb)\,,
\end{align}
where we ignored the second term in Eq.~(\ref{eq:Chi_Astero:Susceptibility_q}) for $T/\mu\ll1$.
Since the CMW cannot exist in electron matter (see App.~\ref{app:CM-mode_Electron}) and the frequency of the CM-mode can be distinguished from the conventional modes, the CM-mode provides a possible new probe for quark matter in neutron stars.
Also, as Eq.~(\ref{eq:Chi_Astero:frequency-CM-parameter}) depends on the magnetic field, possible detection of this mode would provide information on the magnetic field inside neutron stars.

\subsection{Estimate of the amplitude}
We estimate the amplitude of the gravitational wave of the CM-mode.
Generally, the characteristic amplitude is given in terms of the distance to a source $d$, the released energy as gravitational waves $E_{\rm GW}$, and the frequency $f$ regardless of the details of the production mechanism.
Using the formula (\ref{eq:NS_Astero:Eff_amp_GW_formula}), the amplitude of the gravitational wave with the typical frequency of the CM-mode is given by
\begin{align}
\label{eq:Chi_Astero:CM_amp_quark}
        h_{\rm CM}
    \sim
    10^{-22}
    \lb(\fr{E_{\rm GW}}{10^{44}\,{\rm erg}}\rb)^{1/2}
    \lb(\fr{f_{\rm CM}}{10^{5}\,{\rm Hz}}\rb)^{-1/2}
    \lb(\fr{d}{10\,{\rm kpc}}\rb)^{-1}.
\end{align}
Here, considering possible sources of the CM mode in neutron stars and supernovae, we take $E_{\rm GW}$ and $d$ to correspond to approximately $1\%$ of the energy released by a giant flare in a neutron star ($\sim10^{46}~{\rm erg}$) and to the typical radius of our galaxy ($\sim10~{\rm kpc}$), respectively.
The typical frequency of the CM-mode is higher than that of conventional modes.
Thus, for the same amount of energy emitted as gravitational waves, the amplitude of the CM-mode is smaller.
For comparison, the typical frequencies and amplitudes are summarized in Tab.~\ref{tab:Chi_Astero:Modes_freq_amp}.

\begin{table}[htb]
\begin{center}
\caption{The typical frequencies of the oscillation modes and the corresponding ratios of their gravitational-wave amplitudes, with the amplitudes normalized to the r-mode.}
\label{tab:Chi_Astero:Modes_freq_amp}
\medskip
\begin{tabular}{|c|c|c|}
\hline
     Mode & Typical frequency & Amplitude ratio
    \\
    \hline
    f-mode & $\sim10^{3}$~Hz & $\sim10^{-2}$
    \\
    \hline
    p-mode & $\sim10^{3}$~Hz & $\sim10^{-2}$ 
    \\
    \hline
    g-mode & $\sim10$~Hz & $\sim10^{-1}$
    \\
    \hline
    r-mode & $\sim1$~Hz & $1$
    \\
    \hline
    CM-mode & $\sim10^{5}$~Hz & $\sim10^{-3}$
    \\
    \hline
\end{tabular}
\end{center}
\end{table}

\section{CV-mode in rotating neutrino matter}
\label{sec:Chi_Astero:CV-mode}
As shown in Sec.~\ref{sec:NS_Astero:Neutrino_trapping}, the neutrinos are trapped in the core of supernovae~\cite{Sato:1975,Kotake:2005zn}. Due to the left-handedness of neutrinos, the neutrino matter there is the ``chiral matter'' with chirality imbalance~\cite{Yamamoto:2015gzz}.
We consider the neutrino matter rotating with the angular velocity $\bm{\Omega}$.
We use the cylindrical coordinates $\{t,r,\phi,z\}$ and orient the rotational axis along the $z$-axis, $\bm{\Omega}=\Omega\bm{e}_{z}$.
The relation of the angular coordinate between the inertial and the corotating frame is $\psi=\phi-\Omega t$, where $\psi$ is the angular coordinate of the corotating frame.
In the corotating frame, the small line element is given by
\begin{align}
    \dif s^2
    &=
    \e^{2\lambda}\dif t^2
    -\e^{2\nu}\dif r^2
    -r^2\dif \psi^{2}
    -\e^{2\rho}\dif z^{2}
    \nom
    &\simeq
    \e^{2\lambda}\dif t^2
    +2r^{2}\Omega \dif t\dif \phi
    -\e^{2\nu}\dif r^2
    -r^2\dif \phi^{2}
    -\e^{2\rho}\dif z^{2}.
\end{align}
In the second line, we ignored the terms of order $\mathcal{O}(\Omega^{2})$.
For the same reason mentioned in Sec.~\ref{sec:Chi_Astero:CM-mode_q}, we assume that the metric functions are constant.
We also ignore the fluctuation of the metric in the Cowling approximation.

\subsection{Wave equation and dispersion relation}
We consider the fluctuations of the neutrino density and the energy density:
\begin{align}
    n_{\nu}=\bar{n}_{\nu}+\delta n_{\nu}\,,
    \qquad
    \varepsilon=\bar{\varepsilon}+\delta\varepsilon\,.
\end{align}
The continuity equation is
\begin{align}
\label{eq:Chi_Astero:Continuity_eqs_n_nu}
    &\nabla_{\mu}j_{\nu}^{\mu}=0\,.
\end{align}
Since the neutrino matter can induce the CVE~\cite{Vilenkin:1979ui,Son:2009tf,Landsteiner:2011cp}, the constitutive equation is given by
\begin{align}
\label{eq:Chi_Astero:Constitutive_eq_nu}
    \bs{j}_{\nu}
    =
    -\lb(\fr{\mu_{\nu}^{2}}{4\pi^{2}}+\fr{T^{2}}{12}\rb)\bm{\Omega}-D_{\nu}\bs{\nabla}n_{\nu}\,,
\end{align}
where $\mu_{\nu}$ is the neutrino chemical potential and $D_{\nu}$ is the diffusion coefficient of neutrinos.
The $\mu_{\nu}$-dependent term of the CVE is fixed by the chiral anomaly~\cite{Son:2009tf} and it does not receive corrections in curved space.%
\footnote{Equation~(\ref{eq:Chi_Astero:Constitutive_eq_nu}) has also corrections proportional to the fermion mass and scalar curvature for massive fermions in curved space~\cite{Flachi:2017vlp}. However, as we are interested in nearly massless neutrinos, these corrections are negligibly small.}

Since we are interested in the behavior along the rotation axis, the solution can be written as
\begin{align}
    \delta n_{\nu}(t,z)=\delta n_{\nu}(\omega,k_z)\e^{-\im(\omega t-k_zz)}\,,
\end{align}
In the Cowling approximation, Eq.~(\ref{eq:Chi_Astero:Continuity_eqs_n_nu}) is linearized as
\begin{align}
    \lb(\omega
    +\e^{\lambda-\rho}\fr{\bar{\mu}_{\nu}\Omega}{2\pi^2\chi_{\nu}}k_{z}
    +\im \e^{\lambda-\rho}D_{\nu}|k_z|^2\rb)\delta n_{\nu}
    =
    0\,.
\end{align}
Therefore, we arrive at another new seismic oscillation caused by the chiral vortical wave (CVW)~\cite{Jiang:2015cva}.
We call it the CV-mode, and the dispersion relation is given by
\begin{align}
\label{eq:Chi_Astero:Dispersion_CV}
    \omega = -V_{\rm CV}k_{z}-\im\e^{\lambda-\rho}D_{\nu}|k_z|^2\,,
\end{align}
where 
\begin{equation}
    V_{\rm CV}\equiv 
    \e^{\lambda-\rho}\fr{\bar{\mu}_{\nu}\Omega}{2\pi^2\chi_{\nu}}
\end{equation}
is the speed of the CV-mode. Note that the propagation of the CV-mode requires the presence of nonzero left-handed neutrino chemical potential (or chirality imbalance) at equilibrium unlike the CM-mode.

\subsection{Possible existence of the CV-mode in the core of supernovae}
We estimate the frequency of the CV-mode, $f_{\rm CV}=\omega_{\rm CV}/(2\pi)$, at the core of supernovae.
In the same way as Sec.~\ref{sec:Chi_Astero:CM-mode_q}, we set $\e^{\lambda-\rho}\sim 1$.
The wavelength has to be shorter than the radius $R$, and the condition $|{\rm Re}(\omega_{\rm CV})|\gg |{\rm Im}(\omega_{\rm CV})|$ is necessary for the propagation of the CV-mode.
Thus, the possible range of the frequency is
\begin{align}
    \fr{V_{\rm CV}}{R}\ll f_{\rm CV} \ll \fr{V_{\rm CV}^2}{2\pi D_{\nu}}
    =\fr{3V_{\rm CV}^2}{2\pi\tau_{\nu}}\,,
\end{align}
where $\tau_{\nu}$ is the relaxation time of the neutrino.
Assuming that neutrino matter is an ideal neutrino gas, the susceptibility is given as
\begin{align}
    \chi_{\nu}
    \equiv
    \fr{\del n_{\nu}}{\del\mu_{\nu}}=
    \fr{\bar{\mu}_{\nu}^2}{2\pi^2}
    +\fr{T^2}{6}\,.
\end{align}

Taking the radius of the core of supernovae, $R\sim10\,\mathrm{km}$, neutrino chemical potential $\mu_{\nu}\sim10^2\,{\rm MeV}$, and  temperature $T\sim10\,{\rm MeV}$, the lower limit is
\begin{align}
    \fr{V_{\rm CV}}{R}
    \sim
    10^{-40}~{\rm MeV}\lb(\fr{\Omega/2\pi}{1~{\rm Hz}}\rb)\,.
\end{align}
Assuming that the relaxation timescale is comparable to the mean free path $(D_{\nu} \sim l_{\nu}/3)$, the upper limit is estimated as
\begin{align}
    \fr{3V_{\rm CV}^2}{2\pi l_{\nu}}
    \sim 10^{-57}~{\rm MeV}\lb(\fr{\Omega/2\pi}{1~{\rm Hz}}\rb)^2,
\end{align}
where we have substituted $l_{\nu}\sim1\,{\rm cm}$~\cite{Kotake:2005zn,Yamamoto:2015gzz}.
Since the upper limit is much smaller than the lower limit, the CV-modes cannot propagate in the core of supernovae.

One main reason why the CV-mode cannot appear in supernovae while the CM-mode can is that the energy scale of the rotation ($\sim10^{-21}\,{\rm MeV}$) is much smaller than the possible scale of magnetic fields in neutron stars and supernovae ($\sim10^2 \,{\rm MeV}$).

\section{Angular dependence of the gravitational radiation}
\label{sec:Chi_Astero:Angular}
Since the CM- and CV-modes propagate along specific directions, such as the magnetic-field or rotational axis, the resulting gravitational waves are expected to be emitted anisotropically.
In this section, we examine the angular dependence of these gravitational wave emissions.

For completeness, we first briefly summarize the quadrupole formula.
We introduce $\tilde{h}_{\mu\nu} \equiv h_{\mu\nu}-\fr{1}{2}h\eta_{\mu\nu}$, where $h \equiv h^{\mu}_{~\mu}$ and $\eta_{\mu\nu}={\rm diag}(+1,-1,-1,-1)$.
In Minkowski space,, the equation of motion for $\tilde{h}_{\mu\nu}$ is
\begin{align}
    \Box \tilde{h}_{\mu\nu}
    =
    -16\pi GT_{\mu\nu},
\end{align}
where $\Box$ is the d'Alembertian and $T_{\mu\nu}$ is the energy-momentum tensor.
The retarded solution is expressed as
\begin{align}
    \tilde{h}_{\mu\nu}(t,\bm{x})
    =
    4G\int \dif^{3}\bm{x}' \fr{T_{\mu\nu}(t-|\bm{x}-\bm{x}'|,\bm{x}')}{|\bm{x}-\bm{x}'|}\,.
\end{align}
We assume that the observation point is located at a distance $r=|\bm{x}-\bm{x}'|$ much larger than the size of the source, i.e., $r\gg R$, where $R$ is the characteristic radius of the source.
In this case, the amplitude of the gravitational wave can be approximated as
\begin{align}
\label{eq:Chi_Astero:Amplitude-quadrupole}
    \tilde{h}_{ij}
    =
    \fr{2G}{r}\fr{\del^{2}}{\del t^{2}}I_{ij}(t-r)\,,
    \qquad
    I_{ij}
    \equiv
    \int \dif^{3}\bm{x}' T^{00}(t,\bm{x}')x'_{i}x'_{j}\,,
\end{align}
where $I_{ij}$ is the quadrupole moment.

Let us consider a source propagates in a certain direction.
Without loss of generality, we take the propagation direction to be along the $z$-axis.
Assuming that the CM- and CV-modes constant propagation speeds $V_{\chi}$ $(\chi={\rm CM, CV})$, the source position can be written as
\begin{align}
    X^{i}(t)=(0,0,V_{\chi}t)\,.
\end{align}
For a small volume element $v$, the fluctuation of the energy-momentum tensor can be expressed as
\begin{align}
    \delta T^{00}(t,\bm{x})
    =
    \delta\varepsilon v\delta^{(3)}(\bm{x}-\bm{X}(t))\,.
\end{align}
Substituting this into the quadrupole moment in Eq.~(\ref{eq:Chi_Astero:Amplitude-quadrupole}), we have
\begin{align}
    I_{ij}
    &=
    \int \dif^{3}\bm{x} \delta\varepsilon v\delta^{(3)}(\bm{x}-\bm{X}(t))x_{i}x_{j}
    \nom
    &=
    \delta\varepsilon vX_{i}X_{j}.
\end{align}
Since $X_{i}$ has only the $z$-component, all the components of the quadrupole moment except for $I_{zz}$ vanish, giving
\begin{align}
    I_{zz}=\delta\varepsilon v V_{\chi}^{2}t^{2}.
\end{align}
Then, the amplitude of the gravitational wave reduces to
\begin{align}
    \tilde{h}_{zz}
    =
    \fr{4G \delta\varepsilon v}{r}V_{\chi}^{2}\,.
\end{align}
Using the correspondence between the Cartesian and spherical coordinates,
$x = r\sin{\theta}\cos{\phi}$, $y = r\sin{\theta}\sin{\phi}$, $z = r\cos{\theta}$, we can write the amplitude as
\begin{align}
    \tilde{h}_{\theta\theta}
    =
    4Gvr \delta\varepsilon V_{\chi}^{2}\sin^{2}{\theta}.
\end{align}
Therefore, the radiation exhibits a clear angular dependence: the gravitational waves from the CM and CV modes are most strongly emitted in the equatorial direction ($\theta=\pi/2$).

\chapter{Anomalous Dynamical Screening of Magnetized Relativistic Plasma}
\label{chap:Anom_Dyn_Res}
In the previous chapter, we studied chiral charge density fluctuations through gravitational waves.
In this chapter, we investigate the dynamical electromagnetic response of a relativistic magnetized plasma, taking into account fluctuations of the chiral charge density.
In Sec.~\ref{sec:Anom_Dyn_Res:EM_mag}, we review the electromagnetism in magnetized media and show general properties.
In Sec.~\ref{sec:Anom_Dyn_Res:Collective_modes}, we derive the dispersion relation of the transverse photon in both the weak and strong magnetic field limits, respectively.
In Sec.~\ref{sec:Anom_Dyn_Res:Imp_neutron_star}, we finally discuss implications for neutron star phenomenology.
This chapter is based on our original work~\cite{Hanai:2025pkw}.

\section{Electrodynamics in a background magnetic field}
\label{sec:Anom_Dyn_Res:EM_mag}
We here summarize the macroscopic electromagnetism in the presence of an external magnetic field (see, e.g., Ref.~\cite{Landau:8}).
The dynamical electromagnetic field is expressed as
\begin{align}
    \bs{E}=\delta\bs{E}\,,
    \qquad
    \bs{B}=\bs{B}_{\rm ex}+\delta\bs{B}\,,
\end{align}
where $\bs{B}_{\rm ex}$ is a uniform external magnetic field.
Without loss of generality, we set a uniform external magnetic field along the $z$-axis ($\bs{B}_{\rm ex}=B_{\rm ex}\bs{e}_z$).

The dynamics of electromagnetic fields in media are governed by the Maxwell equations.
To investigate the collective behavior, it is convenient to discuss in momentum space by the Fourier transformation $(t,\bs{x})\to(\omega,\bs{k})$.
Assuming that there is no external electric sources and the permeability is unity, the Maxwell equations lead to
\begin{align}
\label{eq:Anom_Dyn_Res:Maxwell_eqs_Fourier}
    \bs{k}\times\delta\bs{E}=\omega\delta\bs{B}\,,
    \qquad
    \bs{k}\times\delta\bs{B}=-\omega\delta\bs{E}-\im e\delta\bs{j}\,,
\end{align}
where $e\delta\bs{j}$ is the induced electric current.
Within the linear response theory, the electric current can be expressed using the electric field and the permittivity tensor $\varepsilon$ as
\begin{align}
\label{eq:Anom_Dyn_Res:Permittivity_current_general}
    (\varepsilon-1)\delta\bs{E}
    =
    \im\fr{1}{\omega}e\delta\bs{j}\,.
\end{align}
The details of the medium are encoded in the permittivity.
From the above, we arrive at the closed equation for the electric field 
\begin{align}
\label{eq:Anom_Dyn_Res:Maxwell_media_r}
    \bs{r}^2\delta\bs{E}-(\bs{r}\cdot\delta\bs{E})\bs{r}-\varepsilon\delta\bs{E}
    =
    0\,.
\end{align}
Here, $\bs{r}\equiv\bs{k}/\omega$ denotes the refractive vector.
Since the refractive index is, in general, complex, we write $\bs{r}^2$ rather than $|\bs{r}|^2$.
The eigenvalues of this equation encode the electromagnetic response of the medium.

The resulting behavior of the electromagnetic fields depends on the permittivity.
However, certain general properties of the dispersion relation can be inferred from symmetry considerations without any microscopic details.
In the presence of the background magnetic field, the permittivity tensor is invariant under the $\fr{\pi}{2}$-rotation about the $z$-axis, and we have $R\varepsilon R^{-1}=\varepsilon$ where $R$ is a rotation matrix,
\begin{align}
    R=
    \lb(
    \begin{array}{ccc}
        0 & -1 & 0  \\
        1 & 0 & 0 \\
        0 & 0 & 1
    \end{array}\rb)\,.
\end{align}
Hence, the permittivity tensor is characterized by three parameters as
\begin{align}
    \varepsilon
    =
    \lb(
    \begin{array}{ccc}
        \varepsilon_{xx} & \varepsilon_{xy} & 0  \\
        -\varepsilon_{xy} & \varepsilon_{xx} & 0 \\
        0 & 0 & \varepsilon_{zz}
    \end{array}\rb)\,.
\end{align}
The off-diagonal components are antisymmetric.
According to Onsager's theorem, the permittivity tensor satisfies $\varepsilon^{ij}(B_{\rm ex})=\varepsilon^{ji}(-B_{\rm ex})$, implying that the off-diagonal components are odd in the magnetic field, while the diagonal components are even.

We here focus on the electromagnetic waves propagating perpendicular to the background magnetic field and take the propagation direction along the $x$-axis $(\delta\bs{E}\propto\e^{-\im(\omega t-k_xx)})$.
In this case, Eq.~(\ref{eq:Anom_Dyn_Res:Maxwell_media_r}) becomes
\begin{align}
    \lb(
    \begin{array}{ccc}
        -\varepsilon_{xx} & -\varepsilon_{xy} & 0 \\
        \varepsilon_{xy} & r_x^2-\varepsilon_{xx} & 0 \\
        0 & 0 & r_x^2-\varepsilon_{zz}
    \end{array}\rb)
    \lb(
    \begin{array}{c}
        \delta E_x \\
        \delta E_y \\
        \delta E_z 
    \end{array}\rb)
    =0\,.
\end{align}
From this equation, we obtain the dispersion relations of the two modes as
\begin{align}
\label{eq:Anom_Dyn_Res:Dispersion_relation_general}
    r_{x}^2=\varepsilon_{zz}\,,
    \qquad
    r_{x}^2
    =
    \varepsilon_{xx}+\fr{\varepsilon_{xy}^2}{\varepsilon_{xx}}\,.
\end{align}
The first electromagnetic wave is referred to as the ordinary mode (O-mode), and the second one is called the extraordinary mode (X-mode).
The electric field of the O-mode, $\delta E_z$, is parallel to the external magnetic field, leading to $B_{\rm ex}\delta E_z\neq0$.
In quantum theory, the parallel component of the electric field gives rise to the chiral anomaly.
Therefore, the O-mode is modified by the quantum effect.
In the next section, we show that the dispersion relation of the O-mode is modified by the chiral anomaly.

\section{Collective modes of relativistic collisionless plasma}
\label{sec:Anom_Dyn_Res:Collective_modes}
In this section, we derive the O-mode dispersion relation in a relativistic plasma in the collisionless regime.
Including the fluctuation of the chiral charge density, the resulting transverse photon self-energy contains the effect of the back reaction induced by the chiral anomaly. 
We derive the electric current within the framework of kinetic theory in two regimes: the weak and strong magnetic field limits.
In the weak magnetic field limit, the chiral kinetic theory is applicable, while we employ the LLL approximation in the strong magnetic field limit.
Throughout this chapter, we assume that the typical energy scale of a plasma, such as temperature and density, is sufficiently larger than the fermion mass, and the fermions are treated as massless ones.
The fermion mass generally gives rise to corrections.
However, it can be neglected in our setup, as discussed later.

\subsection{Weak magnetic field limit}
We first discuss the O-mode in the weak field limit using the chiral kinetic theory.
Our derivation partially follows Refs.~\cite{Son:2012zy,Manuel:2013zaa}.
In this subsection, we assume the counting of the gauge field, coordinate derivative, coupling constant, and the chiral chemical potential as $A^\mu=\mcl{O}(\epsilon)$, $\del^{x}_{\mu}=\mcl{O}(\delta)$, $e=\mcl{O}(\delta)$, and $\mu_{5}=\mcl{O}(\delta^3\epsilon^2)$, respectively.
The assumption that the derivative and the coupling constant can be counted at the same order
is justified a posteriori from the resulting dispersion relation.
The power counting of the chiral chemical potential is taken to be consistent with the chiral anomaly discussed later.

In the collisionless regime, the Boltzmann equation becomes
\begin{align}
\label{eq:Anom_Dyn_Res:Boltzmann_collisionless}
    \fr{\del f_{\lambda}}{\del t}
    +\dot{\bs{x}}\cdot\fr{\del f_{\lambda}}{\del\bs{x}}
    +\dot{\bs{p}}\cdot\fr{\del f_{\lambda}}{\del\bs{p}}
    =0\,.
\end{align}
The equations of motion for chiral fermions are given in Eqs.~(\ref{eq:Chi_Ph:EoM_x}) and (\ref{eq:Chi_Ph:EoM_p}).
To solve the Boltzmann equation (\ref{eq:Anom_Dyn_Res:Boltzmann_collisionless}), we expand the distribution function as%
\footnote{There may exist a contribution $f_{\lambda}^{(\delta^3\epsilon^2)}$ arising from quantum effects.
However, for the O-mode, any current of even order in $\epsilon$ (or equivalently gauge fields) must vanish, since the permittivity depends only on even powers of the background magnetic field, as shown in Sec.~\ref{sec:Anom_Dyn_Res:EM_mag}.}
\begin{align}
    f_{\lambda}(t,\bs{x},\bs{p})
    &=
    \tilde{f}^{(0)}_{\lambda}
    +f^{(\delta\epsilon)}_{\lambda}
    +f^{(\delta^2\epsilon)}_{\lambda}
    +f_{\lambda}^{(\delta^2\epsilon^2)}
    +f_{\lambda}^{(\delta^3\epsilon^3)}\,,
\end{align}
where we defined
\begin{align}
    \tilde{f}^{(0)}_{\lambda}(\varepsilon_{\bs{p}},\bs{p})\equiv\fr{1}{\e^{\beta(\varepsilon_{\bs{p}}-\mu_{\lambda})}+1}\,,
\end{align}
with $\beta$ being the inverse temperature.
The magnetic moment in the dispersion relation~(\ref{eq:Chi_Ph:Dispersion_relation_chiral_magnetic_moment}) leads to an additional term of order $\mcl{O}(\delta\epsilon)$ as
\begin{align}
    \tilde{f}_{\lambda}^{(0)}(\varepsilon_{\bs{p}},\bs{p})
    &=
    f_{\lambda}^{(0)}(\bs{p})
    -e\bs{B}\cdot\bs{\Omega}_{\lambda}\fr{\del f_{\lambda}^{(0)}(\bs{p})}{\del|\bs{p}|}\,,
\end{align}
where
\begin{align}
    f^{(0)}_{\lambda}(\bs{p})\equiv\fr{1}{\e^{\beta(|\bs{p}|-\mu_{\lambda})}+1}\,.
\end{align}
By performing the Fourier transformation as $(t,\bs{x})\to(\omega,\bs{k})$, the kinetic equation leads to the distribution function of order $\mcl{O}(\delta\epsilon)$,
\begin{align}
    f^{(\delta\epsilon)}_{\lambda}(\omega,\bs{k},\bs{p})
    =
    -\im e\fr{\bs{v}\cdot\bs{E}}{v\cdot k+\im\eta}\fr{\del f^{(0)}_{\lambda}}{\del|\bs{p}|}\,,
\end{align}
where $v^{\mu}\equiv(1,\bs{v})$ and an infinitesimal positive constant $\eta$ is introduced to ensure causality.
The presence of the Berry curvature gives rise to the distribution function of order $\mcl{O}(\delta^2\epsilon)$ as
\begin{align}
    f_{\lambda}^{(\delta^2\epsilon)}(\omega,\bs{k},\bs{p})
    &=
    \lambda e\fr{\omega}{2|\bs{p}|}\fr{\bs{v}\cdot\bs{B}}{v\cdot k+\im\eta}\fr{\del f^{(0)}_{\lambda}}{\del|\bs{p}|}\,.
\end{align}
The distribution function $f^{(\delta^2\epsilon)}_{\lambda}$ finally gives a quantum correction to the permittivity as $\varepsilon_{zz} \propto B_{\rm ex}^2$, as shown later.
Meanwhile, even in the classical theory, the current $\delta\bs{j}\propto B_{\rm ex}^2\delta\bs{E}$ is induced by the distribution function $f_{\lambda}^{(\delta^3\epsilon^3)}$ and results in a contribution proportional to the square of the magnetic field.
Since the classical contribution from the background magnetic field comes only from the Lorentz force, we derive the corresponding distribution function by focusing only on the Lorentz force.
At order $\mcl{O}(\delta^2\epsilon^2)$, we obtain
\begin{align}
    f_{\lambda}^{(\delta^2\epsilon^2)}(\omega,\bs{k},\bs{p})
    &=
    -\im e\fr{(\bs{v}\times\bs{B}_{\rm ex})}{v\cdot k+\im\eta}\cdot\fr{\del f_{\lambda}^{(\delta\epsilon)}}{\del \bs{p}}
    \nom
    &=
    -e^2\fr{(\bs{v}\cdot\bs{E})[\bs{v}\cdot(\bs{B}_{\rm ex}\times\bs{k})]}{|\bs{p}|(v\cdot k+\im\eta)^3}\fr{\del f^{(0)}_{\lambda}}{\del|\bs{p}|}\,,
\end{align}
where we used $\bs{B}_{\rm ex}\parallel\bs{E}$ and $(\bs{v}\times\bs{B}_{\rm ex})\cdot\bs{k}=\bs{v}\cdot(\bs{B}_{\rm ex}\times\bs{k})$.
In the same way, we can derive the distribution function of order $\mcl{O}(\delta^3\epsilon^3)$ as
\begin{align}
    f_{\lambda}^{(\delta^3\epsilon^3)}(\omega,\bs{k},\bs{p})
    &=
    \im e^3\fr{\bs{v}\cdot\bs{E}}{|\bs{p}|^2}
    \lb[\fr{(\bs{v}\times\bs{B}_{\rm ex})\cdot(\bs{B}_{\rm ex}\times\bs{k})}{(v\cdot k+\im\eta)^4}+3\fr{[\bs{v}\cdot(\bs{B}_{\rm ex}\times\bs{k})]^2}{(v\cdot k+\im\eta)^5}\rb]
    \fr{\del f^{(0)}_{\lambda}}{\del|\bs{p}|}\,.
\end{align}

The desired currents are obtained by integrating the distribution functions in momentum space.
For our purpose to derive the O-mode, the permittivity tensor $\varepsilon_{zz}$ is needed, and we focus only on the $z$-component.
We first consider the current $j_{\lambda z}^{(\delta\epsilon)}$.
The integration over $|\bs{p}|$ can be performed using the result from the appendix of Ref.~\cite{Loganayagam:2012pz} as
\begin{align}
    \int\fr{\dif|\bs{p}|}{2\pi^2}|\bs{p}|^2\lb(-\fr{\del}{\del|\bs{p}|}\lb[f_{\lambda}^{(0)}+\bar{f}_{\lambda}^{(0)}\rb]\rb)
    =
    \fr{\mu_{\lambda}^2}{2\pi^2}+\fr{T^2}{6}\,.
\end{align}
The remaining angular integration (see App. \ref{app:Angular_Int}) leads to
\begin{align}
    j^{(\delta\epsilon)}_{\lambda z}(\omega,\bs{k})
    &=
    \int\fr{\dif^3\bs{p}}{(2\pi)^3}v_zf^{(\delta\epsilon)}_{\lambda}+(\text{antifermion})
    \nom
    &=
    \im\fr{m_{{\rm D}\lambda}^2}{e\omega}F(\xi)E_z\,,
\end{align}
where we defined the parameter $\xi\equiv\omega/|\bs{k}|$ and the dimensionless functions,
\begin{align}
    &F(\xi)
    \equiv
    \fr{1}{2}\lb[\xi^2+(1-\xi^2)L(\xi)\rb]\,,
    \\
\label{eq:Anom_Dyn_Res:Function_L}
    &L(\xi)
    \equiv\int\fr{\dif^2\bs{v}}{4\pi}\fr{\omega}{v\cdot k+\im\eta}
    =
    \fr{\xi}{2}\ln\lb|\fr{1+\xi}{1-\xi}\rb|-\im\fr{\pi\xi}{2}\theta(1-\xi)\,.
\end{align}
The Debye screening mass for the chirality $\lambda$ is given by
\begin{align}
    m_{{\rm D}\lambda}^2
    \equiv
    e^2\lb(\fr{\mu_{\lambda}^2}{2\pi^2}+\fr{T^2}{6}\rb)\,.
\end{align}
The current of order $\mcl{O}(\delta^2\epsilon)$ gives the CME~\cite{Son:2012zy,Manuel:2013zaa},
\begin{align}
    j_{\lambda z}^{(\delta^2\epsilon)}(\omega,\bs{k})
    &=
    \int\fr{\dif^3\bs{p}}{(2\pi)^3}
    \lb[v_zf^{(\delta^2\epsilon)}_{\lambda}+(\bs{v}\cdot\bs{\Omega}_{\lambda})eB_zf^{(0)}_{\lambda}-\im|\bs{p}|(\bs{\Omega}_{\lambda}\times\bs{k})_{z}f^{(\delta\epsilon)}_{\lambda}\rb]
    +(\text{antifermion})
    \nom
    &=
    \lambda\fr{e\mu_{\lambda}}{4\pi^2}G(\xi)B_{z}\,,
\end{align}
where the dimensionless function $G(\xi)$ is defined as
\begin{align}
    G(\xi)\equiv(1-\xi^2)[1-L(\xi)]\,.
\end{align}
From the symmetry argument in Sec.~\ref{sec:Anom_Dyn_Res:EM_mag}, the permittivity tensor $\varepsilon_{zz}$ is an even function of the magnetic field.
Therefore, the current of order $\mcl{O}(\delta^2\epsilon^2)$ must vanish.
Indeed, one can explicitly verify that $j_{\lambda z}^{(\delta^2\epsilon^2)}=0$ by the explicit calculation.
Finally, the current $j_{\lambda z}^{(\delta^3\epsilon^3)}$ is obtained as
\begin{align}
    j_{\lambda z}^{(\delta^3\epsilon^3)}(\omega,\bs{k})
    &=
    -\im e^3\int\fr{\dif^3\bs{p}}{(2\pi)^3}
    \fr{1}{|\bs{p}|^2}\fr{\del f^{(0)}_{\lambda}}{\del|\bs{p}|}
    \lb[\fr{v_xv_z^2}{(v\cdot k+\im\eta)^4}
    -3\fr{v_y^2v_z^2k_x}{(v\cdot k+\im\eta)^5}\rb]k_xB_{\rm ex}^2E_z
    \nom
        &\hspace{3em}+(\text{antifermion})
    \nom
    &=
    \im\fr{e^3B_{\rm ex}^2}{2\pi^2\omega|k_x|^2}I(\xi)E_{z}\,,
\end{align}
where the function $I(\xi)$ is introduced as
\begin{align}
    I(\xi)
    \equiv
    \xi\int\fr{\dif^2\bs{v}}{4\pi}\lb[\fr{v_xv_z^2}{(\xi-v_x+\im\eta)^4}-3\fr{v_y^2v_z^2}{(\xi-v_x+\im\eta)^5}\rb]\,.
\end{align}

We now derive the dispersion relation of the O-mode.
Substituting the dispersion relation (\ref{eq:Anom_Dyn_Res:Dispersion_relation_general}) into the $z$-component of the relation (\ref{eq:Anom_Dyn_Res:Permittivity_current_general}), we have
\begin{align}
\label{eq:Anom_Dyn_Res:Permittivity_current_z}
    (r_x^2-1)\delta E_z
    =
    \im\fr{1}{\omega}e\delta j_z\,.
\end{align}
In addition to the electromagnetic wave, we assume the fluctuation of the chiral charge density in chirally symmetric matter as $n_5=\delta n_5$, while the total number density $n$ is assumed not to fluctuate.
This is justified because the number density fluctuation is decoupled within the linear analysis, and does not affect the result.
In our counting, the chiral charge density is counted as $n_5\propto \mu_5=\mcl{O}(\delta^3\epsilon^2)$.

Combining the obtained currents together with the fluctuations of the electric field and the chiral charge density, we obtain the total induced electric current as
\begin{align}
\label{eq:Anom_Dyn_Res:Current_fluctuation_z}
    e\delta j_{z}(\omega,k_x)
    &=
    e\sum_{\lambda=\pm}
    \lb[\delta j_{\lambda z}
    ^{(\delta\epsilon)}
    +\delta j_{\lambda z}^{(\delta^2\epsilon)}
    +\delta j_{\lambda z}^{(\delta^3\epsilon^3)}\rb]
    \nom
    &=
    \im\fr{m_{\rm D}^2}{\omega}F(\xi)\delta E_z
    +\fr{e^2B_{\rm ex}}{2\pi^2\chi}G(\xi)\delta n_5
    +\im\fr{e^4B_{\rm ex}^2}{\pi^2\omega|k_x|^2}I(\xi)\delta E_{z}\,.
\end{align}
As already explained in Sec.~\ref{sec:Chi_Ph:CMW}, in a chirally symmetric system, the chiral charge susceptibility, $\chi$, is equal to the particle number susceptibility.

To close the equations, we also need the continuity equation describing the dynamics of the chiral charge.
The anomaly relation for the chiral fermions (\ref{eq:Chi_Ph:Chiral_anomaly}) yields
\begin{align}
\label{eq:Anom_Dyn_Res:Chiral_anomaly_fluctuation}
    \omega\delta n_5
    =
    \im\fr{e^2B_{\rm ex}}{2\pi^2}\delta E_z\,.
\end{align}
Note that the spatial derivative term in Eq.~(\ref{eq:Chi_Ph:Chiral_anomaly}) vanishes since there is no current parallel to $k_x$ in our setup.%
\footnote{In general, we also have the chiral electric separation effect (CESE) \cite{Huang:2013iia}, giving a nonlinear contribution as $\delta n_5\delta E_z$. However, there is no such term in the $x$-component of the axial current in our configuration.}
Combining Eqs.~(\ref{eq:Anom_Dyn_Res:Permittivity_current_z}), (\ref{eq:Anom_Dyn_Res:Current_fluctuation_z}), and (\ref{eq:Anom_Dyn_Res:Chiral_anomaly_fluctuation}), the wave equation is derived:
\renewcommand{\arraystretch}{2}
\begin{align}
    \lb(
    \begin{array}{cc}
        \displaystyle r_x^2-1+\fr{m_{\rm D}^2}{\omega^2}F(\xi)
        +\fr{e^4B_{\rm ex}^2}{\pi^2\omega^2|k_x|^2}I(\xi)& \displaystyle -\im\fr{e^2 B_{\rm ex}}{2\pi^2\chi\omega}G(\xi) \\
        \displaystyle -\im\fr{e^2 B_{\rm ex}}{2\pi^2} & \displaystyle \omega
    \end{array}
    \rb)
    \lb(
    \begin{array}{c}
        \delta E_z \\
        \delta n_5  
    \end{array}
    \rb)
    =0\,.
\end{align}
The eigenvalue of this equation leads to the O-mode dispersion relation including the photon self-energy $\Pi_{\rm O}$.
Therefore, our main result is given by
\begin{align}
\label{eq:Anom_Dyn_Res:Dispersion_relation_anomalous_general}
    \omega^2
    &=
    |k_x|^2
    +\Pi_{\rm O}(\xi)\,,
    \qquad
    \Pi_{\rm O}(\xi)
    \equiv
    m_{\rm D}^2F(\xi)
    +\fr{e^4B_{\rm ex}^2}{\pi^2|k_x|^2}I(\xi)
    +m_{\rm a}^2G(\xi)\,,
\end{align}
where we defined ``anomalous screening mass'' analogous to the Debye screening mass as
\begin{align}
\label{eq:Anom_Dyn_Res:Anomalous_screening_mass}
    m_{\rm a}^2
    \equiv
    \fr{e^4B_{\rm ex}^2}{4\pi^4\chi}\,.
\end{align}
The second and third terms in the photon self-energy are proportional to $B_{\rm ex}^2$ in agreement with the argument of symmetry in Sec.~\ref{sec:Anom_Dyn_Res:EM_mag}.
Eliminating $\delta n_{5}$ in Eq.~(\ref{eq:Anom_Dyn_Res:Current_fluctuation_z}) by using Eq.~(\ref{eq:Anom_Dyn_Res:Chiral_anomaly_fluctuation}), we can obtain the magnetoresistance.
This is an extension of the result in Ref.~\cite{Son:2012bg} for general frequency and wavenumber, including the classical magnetic contribution.

For analytical discussion, we consider appropriate limiting cases.
We first take the quasi-static limit $(\xi\ll1)$ and expand the dispersion relation in $\xi$, which is independent of the expansion based on the counting of $\delta$ and $\epsilon$ used previously.
Expanding the dimensionless functions up to order $\mcl{O}(\xi)$ (see also App.~\ref{app:Expansion_functions}), the photon self-energy becomes
\begin{align}
\label{eq:Anom_Dyn_Res:Photon_self-energy_O-mode_static}
    \Pi_{\rm O}(\xi)
    \simeq
    -\im\fr{\pi m_{\rm D}^2}{4}\xi
    -\im\fr{e^4B_{\rm ex}^2}{16\pi|k_x|^2}\xi
    +m_{\rm a}^2
    +\im\fr{\pi m_{\rm a}^2}{2}\xi\,.
\end{align}
The self-energy retains a real part even in the static limit $(\omega\to0)$.
This is in contrast to the classical result, in which the transverse photon
is gapless.
In the weak magnetic field limit, however, the strictly static limit
$\omega=0$ cannot be taken, because the anomalous mass term is of higher order
than the coordinate derivative.
As a result, the gap appears only at the nonzero frequency,
and we refer to this phenomenon as ``anomalous dynamical screening.''
Note that the X-mode is not
affected by the anomaly and remains gapless; consequently, dynamical magnetic
fields are \emph{partially} screened.

Next, we examine the quasi-long-wavelength limit $(\xi\gg1)$.
Using the expansion of the dimensionless functions in App.~\ref{app:Expansion_functions}, we have
\begin{align}
\label{eq:Anom_Dyn_Res:Photon_self-energy_O-mode_long}
    \Pi_{\rm O}(\xi)
    \simeq
    \omega_{\rm p}^2
    +\omega_{\rm a}^2
    +\fr{\omega_{\rm p}^2}{5}\fr{1}{\xi^2}
    -\fr{2\omega_{\rm a}^2}{5}\fr{1}{\xi^2}\,,
\end{align}
where $\omega_{\rm p}\equiv m_{\rm D}/\sqrt{3}$ is the plasma frequency and $\omega_{\rm a}\equiv m_{\rm a}/\sqrt{3}$ is the ``anomalous plasma frequency.'' 
Unlike the quasi-static limit, the classical magnetic correction is not included up to order $\mcl{O}(1/\xi^2)$.
This result is consistent with the the well-known behavior that the O-mode is not affected by the background magnetic field in the long-wavelength limit $(\bs{k}\to0)$.
Thus, the magnetic dependence comes only from the anomaly.
The dispersion relation of the plasma oscillation is then written as
\begin{align}
\label{eq:Anom_Dyn_Res:Dispersion_relation_anomalous_plasma_osillation}
    \omega^2
    &\simeq
    \omega_{\rm p}^2(1+\zeta^2)
    +\fr{6}{5}|k_x|^2\lb(1-\fr{4}{3}\zeta^2\rb)\,,
\end{align}
where $\zeta\equiv m_{\rm a}/m_{\rm D}=\omega_{\rm a}/\omega_{\rm D}$
.
From this expression, we can find that the chiral anomaly gives rise to the correction to both the gap and the velocity of the plasma oscillation.

Let us comment on the effect of fermion masses.
While we have used the chiral kinetic theory, real fermions such as electrons and quarks are massive, and this limits the applicability of our discussion.
Although the CME is unaffected by fermion masses \cite{Fukushima:2008xe}, the anomaly equation (\ref{eq:Anom_Dyn_Res:Chiral_anomaly_fluctuation}) is modified through the explicit breaking of the U(1)$_{\rm A}$ symmetry.%
\footnote{Unlike the CME, the CSE is affected by fermion masses \cite{Metlitski:2005pr,Gorbar:2013upa,Guo:2016dnm}. However, since the anomaly relation (\ref{eq:Anom_Dyn_Res:Chiral_anomaly_fluctuation}) does not include the axial current in our setup, our result is independent of the CSE.}
This mass effect induces the chirality flipping, which violates chiral charge conservation.
In the collisionless regime, however, scattering processes are negligible, and our results remain valid.
In Sec.~\ref{sec:Anom_Dyn_Res:Imp_neutron_star}, we estimate the chirality flipping time in neutron star matter, which turns out to be sufficiently longer than the relaxation time.

\subsection{Strong magnetic field limit}
So far, we have focused on the weak magnetic field regime using the chiral kinetic theory.
In this subsection, we turn to the strong magnetic field regime, in which the fermions occupy only the LLL.
In contrast to the chiral kinetic theory, the background magnetic field is taken to be of order unity $B_{\rm ex}=\mcl{O}(1)$, while the dynamical electromagnetic fields are counted as $\delta\bs{E},\delta\bs{B}=\mcl{O}(\delta\epsilon)$.
Accordingly, we adopt the power counting as $\delta A^{\mu}=\mcl{O}(\epsilon), \del^x_\mu=\mcl{O}(\delta)$, $e=\mcl{O}(\delta^{2/3})$, and $\mu_5=\mcl{O}(\delta^{4/3}\epsilon)$, where $\delta A^{\mu}$ is the gauge field fluctuation.
The coupling constant and the chiral chemical potential are counted so as to be compatible with the derived dispersion relation and the chiral anomaly, respectively.

Since the dispersion relation of the LLL is $\varepsilon_{\lambda}=\lambda p_z$, we can treat the motion of the chiral fermions in (1+1)-dimensional spacetime effectively.
The Boltzmann equation takes the form
\begin{align}
\label{eq:Anom_Dyn_Res:Boltzmann_eq_LLL}
    \fr{\del f_{\lambda}}{\del t}
    +\dot{z}\fr{\del f_{\lambda}}{\del z}
    +\dot{p}_z\fr{\del f_{\lambda}}{\del p_z}
    =
    0\,,
\end{align}
where the velocity and the equation of motion for the fermions are given by
\begin{align}
    \dot{z}
    =
    \lambda\,,
    \qquad
    \dot{p}_z
    =
    eE_z\,.
\end{align}
The transverse degrees of freedom are replaced by the Landau degeneracy $|eB_{\rm ex}|/(2\pi)$.

Similarly to the case of the chiral kinetic theory, we can solve the Boltzmann equation by expanding the distribution function as
\begin{align}
    f_{\lambda}(t,z,p_z)
    &=
    f^{(0)}_{\lambda}
    +f^{(\delta\epsilon)}_{\lambda}\,,
\end{align}
where we defined
\begin{align}
    f^{(0)}_{\lambda}(p_z)\equiv\fr{1}{\e^{\beta(|p_z|-\mu_{\lambda})}+1}\,.
\end{align}
The distribution function of order $\mcl{O}(\delta\epsilon)$ is derived as
\begin{align}
    f_{\lambda}^{(\delta\epsilon)}(\omega,k_z,p_z)
    &=
    -\im\fr{eE_z}{\omega-\lambda k_z+\im\eta}\fr{\del f_{\lambda}^{(0)}}{\del p_z}\,.
\end{align}

The currents of orders $\mcl{O}(\delta)$ and $\mcl{O}(\delta^2\epsilon)$ are given by, respectively,
\begin{align}
    j_{\lambda z}^{(\delta)}
    &=
    \fr{|eB_{\rm ex}|}{2\pi}\int\fr{\dif p_z}{2\pi}\dot{z}f_{\lambda}^{(0)}+(\text{antifermion})
    \nom
    &=
    \lambda\fr{\mu_{\lambda}}{4\pi^2}|eB_{\rm ex}|\,,
\end{align}
and
\begin{align}
    j_{\lambda z}^{(\delta^2\epsilon)}(\omega,k_z)
    &=
    \fr{|eB_{\rm ex}|}{2\pi}\int\fr{\dif p_z}{2\pi}\dot{z}f_{\lambda}^{(\delta\epsilon)}+(\text{antifermion})
    \nom
    &=
    \im\fr{|eB_{\rm ex}|}{2\pi^2}\fr{k_z}{\omega^2-k_z^2}eE_z\,.
\end{align}
In our setup, we set $k_z=0$, and only the current $j_{\lambda z}^{(\delta)}$ remains.
Thus, the fluctuation of the total electric current becomes
\begin{align}
\label{eq:Anom_Dyn_Res:Current_fluctuation_LLL}
    e\delta j_z
    &=
    e\sum_{\lambda=\pm}\delta j_{\lambda z}^{(\delta)}
    =
    \fr{e|eB_{\rm ex}|}{2\pi^2\chi}\delta n_5\,.
\end{align}
The counting of the chiral charge density is the same as that of the chiral chemical potential $\delta n_5=\mcl{O}(\delta^{4/3}\epsilon)$.

From Eqs.~(\ref{eq:Anom_Dyn_Res:Permittivity_current_z}), (\ref{eq:Anom_Dyn_Res:Chiral_anomaly_fluctuation}), and (\ref{eq:Anom_Dyn_Res:Current_fluctuation_LLL}), we obtain the O-mode dispersion relation,
\begin{align}
\label{eq:Anom_Dyn_Res:Dispersion_relation_LLL}
    \omega^2
    &=
    |k_x|^2+\Pi_{\rm O}\,,
    \qquad
    \Pi_{\rm O}=m_{\rm a}^2\,.
\end{align}
Since the susceptibility is $\chi\sim|eB_{\rm ex}|$, the anomalous screening mass becomes $m_{\rm a}\sim e^{3/2}\sqrt{B_{\rm ex}}$, giving rise to the soft scale.%
\footnote{In the LLL approximation, the system effectively becomes (1+1)-dimensional. Thus, the symmetry argument presented in Sec.~\ref{sec:Anom_Dyn_Res:EM_mag} is no longer valid.}
This is consistent with the derivative expansion.
We thus find that the anomalous screening emerges even in the strong magnetic field limit.
In contrast to the weak field case, however, the static limit can be taken in this regime, and the O-mode is genuinely screened.

\section{Implications for neutron star phenomenology}
\label{sec:Anom_Dyn_Res:Imp_neutron_star}
So far, we have discussed a general relativistic plasma.
In this section, as a phenomenological example, we consider degenerate electron matter in the core of neutron stars and give some implications.
We assume that the electron chemical potential is much larger than both the temperature and the electron mass $(\mu_{\rm e}\gg T, m_{\rm e})$.

\subsection{Relaxation time}
The screening characterizes the infrared (IR) scale of scattering processes.
We then discuss the effect of the anomalous dynamical screening on the relaxation time.
Here, we focus on the electron--electron scattering as a photon-mediated process.
Indeed, the neutron superfluid emerges at temperature $T\lesssim10^{9}~{\rm K}$~\cite{Page:2010aw}, where the electron--electron scattering becomes the dominant process.
The scattering amplitude $M$ includes the photon propagator as
\begin{align}
    M
    =
    \fr{C_1}{\omega^2-|\bs{k}|^2-\Pi_{\rm L}}
    +\fr{C_2}{\omega^2-|\bs{k}|^2-\Pi_{\rm T}}\,,
\end{align}
where $\Pi_{\rm L},\Pi_{\rm T}$ are the longitudinal and transverse components of the photon self-energy.
The details of the coefficients $C_1, C_2$ are unimportant in this discussion.
The scattering is dominated by the IR cutoff of momentum $(k_{\rm IR}^2\sim \Pi_{\rm L}, \Pi_{\rm T})$.
In general, due to the presence of the background magnetic field, these are further decomposed into more components.
However, we focus only on the O-mode and set $\Pi_{\rm T}=\Pi_{\rm O}, |\bs{k}|=|k_x|$ in the following discussion.
Due to the Pauli blocking, only fermions near the Fermi surface contribute to the scattering, and the typical energy transfer is of the order of the temperature, $\omega\sim T$.
Therefore, it is sufficient to use the photon self-energy in the quasi-static limit.
A simple dimensional analysis then leads to the following parameter dependence of the relaxation time: 
\begin{align}
    \fr{1}{\tau}\sim\alpha^2\fr{T^2}{k_{\rm IR}}\,,
\end{align}
where $\alpha=e^2/(4\pi)$ is the fine structure constant.
However, the Landau damping can modify the temperature behavior of the relaxation time, meaning that the non-Fermi liquid arises \cite{Heiselberg:1993cr}.
Instead of tracing the detailed discussion in Ref.~\cite{Heiselberg:1993cr}, we provide a schematic argument by focusing on the parameters.
When the background magnetic field is absent, the photon self-energy (\ref{eq:Anom_Dyn_Res:Photon_self-energy_O-mode_static}) becomes $\Pi_{\rm O}\simeq-\im\fr{\pi m_{\rm D}^2}{4}\xi$, resulting in the classical IR cutoff as $k_{\rm IR}\sim(m_{\rm D}^2\omega)^{1/3}\equiv k_{\rm c}$.
From this expression, we can reproduce the conventional parameter dependence of the relaxation time presented in Ref.~\cite{Heiselberg:1993cr} as 
\begin{align}
\label{eq:Anom_Dyn_Res:Relaxation_time_Landau}
    \fr{1}{\tau_{\rm c}}
    \sim
    \alpha^2\fr{T^2}{k_{\rm c}}
    \sim
    \alpha^2\fr{T^{5/3}}{m_{\rm D}^{2/3}}\,.
\end{align}

We now turn on a weak background magnetic field and examine the resulting correction.
Writing the IR cutoff as $k_{\rm IR}\sim k_{\rm c}+\delta k$, we solve $k_{\rm IR}^2\sim\Pi_{\rm O}(k_{\rm IR})$ using Eq.~(\ref{eq:Anom_Dyn_Res:Photon_self-energy_O-mode_static}) up 
to linear order in $\delta k$.
The resulting momentum is
\begin{align}
    k_{\rm IR}
    \sim
    k_{\rm c}\lb[1+\zeta^2+\lb(\fr{e^2B_{\rm ex}^2}{m_{\rm D}^4}+\zeta^2\rb)\fr{m_{\rm D}^2}{k_{\rm c}^2}\rb]\,.
\end{align}
Here, the numerical coefficients, which are not essential for the present discussion, are omitted.
The relaxation time in the weak magnetic field limit can be parametrically expressed as
\begin{align}
\label{eq:Anom_Dyn_Res:Relaxation_time_weak_field}
    \fr{1}{\tau_{\rm w}}
    \sim
    \alpha^2\fr{T^{5/3}}{m_{\rm D}^{2/3}}(1+\zeta^2)
    +\alpha^2 T\lb(\fr{e^2B_{\rm ex}^2}{m_{\rm D}^4}+\zeta^2\rb)\,.
\end{align}
The quantum correction is small $(\zeta\ll1)$ and does not substantially affect the relaxation time.
Although the magnetic field modifies the temperature dependence, the leading behavior remains proportional to $T^{5/3}$.

On the other hand, the result in the strong magnetic field (\ref{eq:Anom_Dyn_Res:Dispersion_relation_LLL}) leads to the IR cutoff $k_{\rm IR}\sim m_{\rm a}$.
Therefore, the relaxation time is characterized by the anomalous screening mass as
\begin{align}
\label{eq:Anom_Dyn_Res:Relaxation_time_strong_field}
    \fr{1}{\tau_{\rm s}}
    \sim
    \alpha^2\fr{T^2}{m_{\rm a}}\,.
\end{align}
From the above, we can expect that the temperature dependence of the relaxation time changes as the background magnetic field increases.

The relaxation time governs the dissipative transport.
Among them, we focus on the shear viscosity.
The shear viscosity of neutron stars was first derived in Refs.~\cite{Flowers:1979,Cutler:1987}, and the result is modified by the dynamical screening, which is more dominant~\cite{Shternin:2008ia,Shternin:2008es}.
The shear viscosity is given by $\eta=\fr{1}{5}n_{\rm e}\mu_{\rm e}\tau$.
To investigate the effect of the anomalous screening, we evaluate the ratio between Eqs.~(\ref{eq:Anom_Dyn_Res:Relaxation_time_Landau}) and (\ref{eq:Anom_Dyn_Res:Relaxation_time_strong_field}):
\begin{align}
\label{eq:Anom_Dyn_Res:Relaxation_time_ratio}
    \fr{\tau_{\rm s}}{\tau_{\rm c}}
    \sim
    \fr{m_{\rm a}}{m_{\rm D}^{2/3}T^{1/3}}
    \sim
    10~\lb(\fr{T}{10^{6}~{\rm K}}\rb)^{-1/3}
    \lb(\fr{\vphantom{T}\mu_{\rm e}}{10^{2}~{\rm MeV}}\rb)^{-2/3}
    \lb(\fr{B_{\rm ex}}{10^{18}~{\rm G}}\rb)^{1/2}\,,
\end{align}
where we substituted the susceptibility of the electron gas in the LLL approximation $\chi_{\rm e}=|eB_{\rm ex}|/(2\pi^2)$.
The magnetic fields of order $10^{18}~{\rm G}$ corresponds to the theoretical upper limit in neutron stars~\cite{Shapiro:1991,Cardall:2000bs,Ferrer:2010wz}.
This indicates that sufficiently strong magnetic fields tend to enhance the shear viscosity, whereas extrapolating our result to the intermediate magnetic field region ($B_{\rm ex} \lesssim 10^{16}~\rm{G}$) suggests that the shear viscosity is instead suppressed.

The shear viscosity plays an important role, e.g., in stellar oscillations.
In a rotating star, a collective excitation can arise due to the Coriolis force.
This is called the r-mode.
Due to the Chandrasekhar--Friedman--Schutz (CFS) mechanism~\cite{Chandrasekhar:1970pjp,Friedman:1978hf}, the r-mode is known to be unstable, emitting gravitational waves~\cite{Andersson:1997xt}.
The r-mode instability is suppressed by the viscosity, and the critical angular velocity is determined by the competition between the timescales of gravitational radiation and viscous damping.
Therefore, in the presence of intermediate magnetic fields, the r-mode instability may become more likely.

\subsection{Screening length}
The presence of the anomalous screening mass implies that a finite energy scale is required to excite dynamical magnetic fluctuations perpendicular to the background magnetic field.
We here focus on the result in the weak field limit~(\ref{eq:Anom_Dyn_Res:Photon_self-energy_O-mode_static}), because the magnetic field strengths accessible in this regime are more relevant for realistic descriptions of neutron stars.
In the framework of the chiral kinetic theory, the anomalous screening mass can appear at finite frequency $(\omega\neq 0)$.
However, we here extrapolate our result to the static limit $(\omega=0)$.
In this case, the $z$-component of the gauge field, which contributes to the O-mode, behaves as
\begin{align}
    A_{z}(x)\sim\e^{-|x|/l_{\rm a}}\,,
\end{align}
where the screening length is defined as $l_{\rm a}\equiv 1/m_{\rm a}$.

We now evaluate the screening length.
In the ideal gas approximation, the electron number susceptibility is given by $\chi_{\rm e}=\mu_{\rm e}^2/\pi^2$.
Substituting this into Eq.~(\ref{eq:Anom_Dyn_Res:Anomalous_screening_mass}), we obtain
\begin{align}
    l_{\rm a}
    =
    \fr{\mu_{\rm e}}{2\alpha B_{\rm ex}}
    \sim
    10^{-9}~{\rm m}~\lb(\fr{B_{\rm ex}}{10^{14}~{\rm G}}\rb)^{-1}\lb(\fr{\vphantom{T}\mu_{\rm e}}{10^{2}~{\rm MeV}}\rb)\,.
\end{align}
The typical size of a neutron star is much larger than the screening length.
Consequently, the magnetic degree of freedom associated with the O-mode becomes irrelevant for macroscopic electromagnetic dynamics.

Magnetohydrodynamics (MHD) assumes that magnetic fields are not screened.
The anomalous dynamical screening may therefore require a modification of the standard MHD description of magnetized relativistic plasmas.
Moreover, the above discussion is not limited to neutron stars and can be applied to other contexts, such as the early Universe.

\subsection{Chirality flipping time}
We finally comment on the chirality flipping induced by the electron mass.
Since the chirality flipping is caused by the longitudinal photon \cite{Schafer:2001za,Hanai:2022yfh}, the dynamical screening does not need to be taken into account.
In this case, the electron-proton scattering is the dominant process, and the chirality flipping time is \cite{Grabowska:2014efa, Dvornikov:2015iua}
\begin{align}
    &\tau_{\rm m}
    \simeq
    \fr{\pi\mu_{\rm e}^3}{\alpha^2 m_{\rm e}^2m_{\rm p}T}\lb(\ln\fr{4\mu_{\rm e}^2}{m_{\rm D}^2}-1\rb)^{-1}
    \sim
    10^{18}~({\rm MeV})^{-1}\lb(\fr{T}{10^6~{\rm K}}\rb)^{-1}\,,
\end{align}
where $m_{\rm e}$ denotes the electron mass.
However, in the parameter region of interest, the chirality flipping time is larger than the relaxation time $(\tau_{\rm m}\gg \tau_{\rm c},\tau_{\rm w},\tau_{\rm s})$.
Thus, in the collisionless regime, the chirality flipping can be ignored.

\chapter{Conclusion and Outlook}
\label{chap:Summary}

\section{Conclusion}

In this thesis, we have studied collective excitations induced by chiral transport phenomena in magnetized plasmas and their implications for neutron star phenomenology.

In Chap.~\ref{chap:Chi_Astero}, we discussed novel types of seismic oscillations and gravitational waves, named the CM- and CV-modes, in neutron stars and supernovae.
These modes are induced by the CME and CVE, respectively.
We have shown that the CM-mode can appear in two-flavor quark matter, while it is strongly damped in electron matter due to dynamical electromagnetic fields.
We further included the quark mass effects and derived the chirality-flipping rate.
If a sufficiently strong magnetic field exists inside a neutron star, the CM-mode can propagate.
We also estimated the frequency and amplitude of the gravitational waves generated by the CM-mode in the 2SC phase.
The typical frequency of the CM-mode is higher than that of conventional modes, which allows us to distinguish it observationally.
Although the CV-mode can exist in neutrino matter in the core of supernovae, it is diffusive because the typical angular velocity of supernovae is too small.

Since the frequency of the CM-mode depends on magnetic fields, such gravitational wave signals can provide clues about internal magnetic field of neutron stars, which has not yet been directly observed.
Detection of the CM-mode would also indicate the presence of quark matter in neutron stars.
To observe gravitational waves associated with the CM-mode, future detectors with high sensitivity at frequencies much higher than $10^4$~Hz are required.
Because seismic oscillations can also induce electromagnetic emission, the CM-mode might be observable as electromagnetic radiation as well.%
\footnote{A related example is the X-ray quasi-periodic oscillation, which is believed to originate from seismic oscillations during giant flares~\cite{Israel:2005av,Strohmayer:2006py}.}

In Chap.~\ref{chap:Anom_Dyn_Res}, we investigated the collective modes of relativistic magnetized plasmas in the collisionless regime.
We have shown that the O-mode dispersion relation receives the corrections from the back reaction induced by the chiral anomaly in two limiting cases: the weak and strong magnetic field limits.
In the weak magnetic field limit, using chiral kinetic theory, we found that the O-mode photon self-energy acquires a real part that persists in the static limit.
Similarly, in the strong magnetic field limit, the O-mode also gains a gap.
This anomalous dynamical screening is in contrast to the classical result, in which the transverse photon self-energy remains gapless.

Finally, we discussed the implications for neutron star phenomenology.
First, we investigated the electron-electron scattering mediated by photons.
In the weak magnetic field, the conventional Landau damping still dominates, while the anomalous screening can modify the relaxation time.
By extrapolating our result to the intermediate magnetic field region, the relaxation time may be suppressed relative to the conventional Landau damping contribution.
This implies that the r-mode instability could be enhanced in the intermediate magnetic field strengths.
Such an enhancement may have observable consequences for neutron star oscillations and gravitational-wave emission.
Second, we also evaluated the typical screening length associated with anomalous dynamical screening.
Dynamical magnetic fields are partially screened on length scales larger than this screening length, potentially affecting the evolution and observational signatures of magnetic fields in neutron stars.

\section{Outlook}
In Chap.~\ref{chap:Chi_Astero}, we focus only on the fluctuations of number density, chiral charge density, and energy density for simplicity.
To make the discussion more complete, we have to include all hydrodynamic variables.
We can then consider the coupling, for instance, between the CM-mode and p-mode.
Moreover, for future observation, numerical computation would be important to analyze the hydrodynamic equations with the chiral transport numerically without the Cowling approximation.
Although we do not include the effect of magnetic fields to derive the chirality flipping rate, one should examine this correction because the spin of chiral fermions subject to the magnetic field.
From an observational perspective, the CM-mode remains a target for future gravitational-wave detections.
The typical frequency of the CM-mode lies above the sensitivity band of current ground-based detectors such as LIGO, making its direct detection challenging at present.
Nevertheless, there has been growing interest in exploring gravitational waves at higher frequencies~\cite{Arvanitaki:2012cn,Aggarwal:2020olq,Ito:2022rxn,Aggarwal:2025noe}.
Many of these studies primarily focus on signals originating from physics beyond the Standard Model, including primordial black holes and QCD axions.
In contrast, the CM-mode is rooted in well-established neutron-star physics, and its detection would provide a unique probe of chiral transport phenomena and the microscopic properties of dense nuclear matter.

In Chap.~\ref{chap:Anom_Dyn_Res}, we employ the chiral kinetic theory and the LLL approximation.
On the other hand, it would be interesting to study collective modes in the intermediate magnetic field regime, which cannot be treated rigorously within our framework.
The anomalous effects may become comparable to or even larger than the classical contributions in this region.
In the context of QCD and Dirac semimetals, it is shown that a quadratic collective mode is induced by the chiral anomaly~\cite{Sogabe:2019gif}.
There, in a phase with broken chiral symmetry, the resulting NG mode (e.g., the pion) plays the role of the chiral chemical potential in our setup.
Therefore, we can expect that a similar collective mode arises.
Another important direction is the application to quark matter, such as color superconductivity in the cores of neutron stars and quark-gluon plasma (QGP) in the heavy-ion collision experiments.
In the case of the 2SC, one of the three color degrees of freedom does not participate in Cooper pairing.
The unpaired quark remains gapless and behaves as a relativistic fermion.
A strong magnetic field of $\sim 10^{18}~{\rm G}$ can also be created in the heavy-ion collision experiments \cite{Kharzeev:2007jp,Skokov:2009qp}.
The chiral anomaly can modify the transport properties in these systems.

\appendix

\chapter{Derivations of Chiral Anomaly in Quantum Field Theory}
\label{app:Chi_Anom_Der}
In this appendix, we derive the chiral anomaly in two ways: the triangle diagram~\cite{Adler:1969gk,Bell:1969ts} and Fujikawa's method~\cite{Fujikawa:1980eg}.
For notional convenience, we write integrals as
\begin{align}
    \int_{x}\equiv\int\dif^4 x\,,
    \qquad
    \int_{p}\equiv\int\fr{\dif^4 p}{(2\pi)^4}\,.
\end{align}

\section{Triangle diagram}
\label{sec:Chi_Anom_Der:Triangle}
In this section, we perform a perturbative calculation using the Feynman diagram known as the triangle diagram~\cite{Adler:1969gk,Bell:1969ts}.%
\footnote{While the chiral anomaly is also referred to as the ``ABJ anomaly,'' after the initials of Adler, Bell, and Jackiw, more early discussions are found in Refs.~\cite{Fukuda:1949,Steinberger:1949wx}.}

\subsection{Computation of axial current}
\begin{figure}[ht]
\centering
\begin{tikzpicture}
\begin{feynhand}
    \vertex[particle] (a) at (-0.86,0) {};
    \vertex[particle] (a) at (-0.86,0);
    \draw (-1,0) node[above] {$\gamma^{\mu}\gamma_5$};
    \vertex[particle] (b) at (0.86,1);
    \draw (0.86,1) node[above] {$\gamma^{\nu}$};
    \vertex[particle] (c) at (0.86,-1);
    \draw (0.86,-1) node[below] {$\gamma^{\rho}$};
    \vertex[crossdot] (d) at (2.3,1) {};
    \vertex[crossdot] (e) at (2.3,-1) {};
    \vertex[crossdot] (f) at (-2.3,0) {};
    \draw (-1.7,0) node [below] {$p\to$};
    \propag[fer] (a) to [edge label=$l+k$] (b);
    \propag[fer] (b) to [edge label=$l$] (c);
    \propag[fer] (c) to [edge label=$l-q$] (a);
    \propag[bos] (b) to [edge label=$k\to$] (d);
    \propag[bos] (c) to [edge label'=$q\to$] (e);
    \propag[bos] (f) to (a);
\end{feynhand}
\end{tikzpicture}
\hspace{3em}
\begin{tikzpicture}
\begin{feynhand}
    \vertex[particle] (a) at (-0.86,0) {};
    \vertex[particle] (a) at (-0.86,0);
    \draw (-1,0) node[above] {$\gamma^{\mu}\gamma_5$};
    \vertex[particle] (b) at (0.86,1);
    \draw (0.86,1) node[above] {$\gamma^{\rho}$};
    \vertex[particle] (c) at (0.86,-1);
    \draw (0.86,-1) node[below] {$\gamma^{\nu}$};
    \vertex[crossdot] (d) at (2.3,1) {};
    \vertex[crossdot] (e) at (2.3,-1) {};
    \vertex[crossdot] (f) at (-2.3,0) {};
    \draw (-1.7,0) node [below] {$p\to$};
    \propag[fer] (a) to [edge label=$l+q$] (b);
    \propag[fer] (b) to [edge label=$l$] (c);
    \propag[fer] (c) to [edge label=$l-k$] (a);
    \propag[bos] (b) to [edge label=$q\to$] (d);
    \propag[bos] (c) to [edge label'=$k\to$] (e);
    \propag[bos] (f) to (a);
\end{feynhand}
\end{tikzpicture}
    \caption{The Feynman diagram contributing to the axial current. When we fix the external momenta, there are two ways for the photon to go out}
\label{fig:Chi_Anom_Der:Triangle_daiagram}
\end{figure}
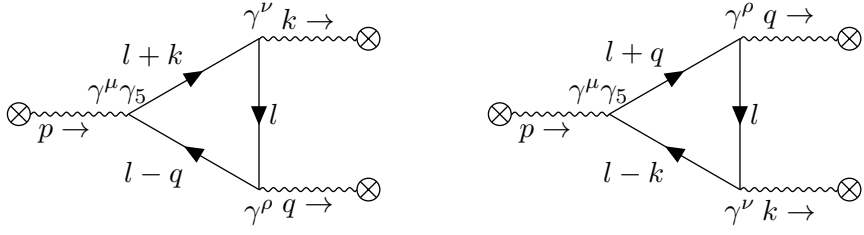

In the chiral anomaly, the conservation of the axial current is violated by the quantum effect.
Thus, we can examine the chiral anomaly by computing $\del_{\mu}j_5^{\mu}$ in the framework of QFT.
The vector current is induced even at first order in the gauge field as $j^{\mu}=-\Pi^{\mu\nu}A_{\nu}$.
On the other hand, in the case of the axial current, noting that the gauge field changes sign under the charge conjugation transformation while the axial current does not, one finds that it is not induced at first order in the gauge field in the vacuum.
Thus, we have to expand the axial current at second order.
The axial current can be written in the form of
\begin{align}
\label{eq:Chi_Anom_Der:Axial_current_momentum-space}
    j_{5}^{\mu}(x)
    &=
    -\int_{p,k,q}\Gamma^{\mu\nu\rho}(p,k,q)A_{\nu}(k)A_{\rho}(q)\delta(p-k-q)\e^{-\im p\cdot x}\,,
\end{align}
where we defined%
\footnote{Since the axial current here is not electric, the number of the coupling $e$ is square.}
\begin{align}
    \im\Gamma^{\mu\nu\rho}(p,k,q)
    &\equiv
    -(-\im )^3e^2\int_{l}\tr\lb[\gamma^{\mu}\gamma_{5}\fr{\im}{\slashed{l}+\slashed{k}}\gamma^{\nu}\fr{\im}{\slashed{l}}\gamma^{\rho}\fr{\im}{\slashed{l}-\slashed{q}}\rb]
    +(\nu\leftrightarrow\rho, k\leftrightarrow q)
    \nom
    &=
    -\im j_{5}^{\mu}(p)j^{\nu}(k)j^{\rho}(q)\,.
\end{align}
Note that in the second term, the indices and momenta are interchanged.
Since the gauge field $A^{\mu}$ is induced by the current $j^{\mu}$, $\Gamma^{\mu\nu\rho}$ can be expressed as a three-point correlation function of the current, as in the final equation.
The corresponding Feynman diagram is shown in Fig.~\ref{fig:Chi_Anom_Der:Triangle_daiagram}.

What we are interested in is not the axial current itself but its divergence (or continuity equation). 
Thus, in the following, we consider $-\im p_{\mu}j_{5}^{\mu}(p)$, which is the Fourier transformation of the continuity equation.
From the energy-momentum conservation, $p^{\mu}=k^{\mu}+q^{\mu}$, we can write
\begin{align}
    \im p_{\mu}\Gamma^{\mu\nu\rho}(p,k,q)
    &=
    -e^2\int_{l}\lb[(l_{\mu}+k_{\mu})-(l_{\mu}-q_{\mu})\rb]
    \tr\lb[\gamma^{\mu}\gamma_{5}\fr{1}{\slashed{l}+\slashed{k}}\gamma^{\nu}\fr{1}{\slashed{l}}\gamma^{\rho}\fr{1}{\slashed{l}-\slashed{q}}\rb]
    +(\nu\leftrightarrow\rho, k\leftrightarrow q)
    \nom
    &=
    e^2\int_{l}\tr\lb[\gamma_{5}\gamma^{\nu}\fr{1}{\slashed{l}}\gamma^{\rho}\fr{1}{\slashed{l}-\slashed{q}}
    -\gamma_{5}\gamma^{\nu}\fr{1}{\slashed{l}+\slashed{q}}\gamma^{\rho}\fr{1}{\slashed{l}}
    \rb]
    +e^2\int_{l}\tr\lb[\gamma_{5}\gamma^{\rho}\fr{1}{\slashed{l}}\gamma^{\nu}\fr{1}{\slashed{l}-\slashed{k}}
    -\gamma_{5}\gamma^{\rho}\fr{1}{\slashed{l}+\slashed{k}}\gamma^{\nu}\fr{1}{\slashed{l}}
    \rb]
    \nom
    &\equiv
    \Delta^{\nu\rho}_{1}(p,k,q)
    +\Delta^{\nu\rho}_{2}(p,k,q)\,.
\end{align}

We now naively try to shift the integral variable $l$.
In the second term of $\Delta^{\mu\nu}_{1}(p,k,q)$, we shift $l^{\mu}\to l^{\mu}-q^{\mu}$, and in the second term in $\Delta^{\mu\nu}_{2}(p,k,q)$, we shift $l^{\mu}\to l^{\mu}-k^{\mu}$.
It turns out that each of the integrals vanishes.
This means, apparently, that the Ward--Takahasi identity for the axial current is satisfied even in the quantum theory.
However, as we mentioned before, the axial current should not be conserved due to the quantum effect.
In the following discussion, we examine this problem in detail.

\subsection{Problem of the integration variable shift}
By examining $\Delta^{\mu\nu}_{1,2}$ computed above, we observe that the integral involves $\dif^4 l$ and that the denominator contains two momentum terms, which suggests a quadratic divergence.
A detailed calculation reveals that, due to the tensor structure of the numerator, the divergence is in fact linear.
It is important to note here that there is a subtraction of divergent terms.
Moreover, the final shift of the integration variable is a subtle operation that requires careful consideration.

In the case of divergent integrals, a shift of the integration variable can become problematic, as can be confirmed by a simple one-dimensional calculation.
We examine whether the value of an integral of a given function $f(x)$ remains unchanged under a shift of the variable, $x \to x + a$.

We then consider the difference between the integrals before and after the shift,
\begin{align}
    \Delta(a)
    &\equiv
    \int_{-\infty}^{\infty}\dif x\lb[f(x+a)-f(x)\rb]
    \nom
    &=
    \int_{-\infty}^{\infty}\dif x \lb[
    a\fr{\dif f(x)}{\dif x}
    +\fr{1}{2}a^2\fr{\dif^2f(x)}{\dif x^2}
    +\cdots\rb]
    \nom
    &=
     a[f(\infty)-f(-\infty)]
     +\fr{1}{2}a^2\lb[f'(\infty)-f'(-\infty)\rb]
     +\cdots\,.
\end{align}
It should be noted that higher derivatives lead to weaker divergences.
When dimensional analysis shows that the integral is either convergent or logarithmically divergent, the conditions $0 = f(\pm\infty) = f'(\pm\infty) = \cdots$ imply that $\Delta(a) = 0$, and hence a shift of the integration variable is allowed.
On the other hand, when the integral exhibits a linear divergence, $f(\pm\infty) \neq 0$; furthermore, in the case of stronger divergences, derivatives such as $f'(\pm\infty)$ also attain finite values, and care must be taken in such cases.
As an example, we suppose $f(x)=x$ leading to
\begin{align}
    \Delta(a)
    &=
    \int_{-\infty}^{\infty}\dif x[(x+a)-x]
    \nom
    &=
    \int_{-\infty}^{\infty}\dif xa
    \nom
    &=
    [a]_{x=-\infty}^{x=\infty}\,.
\end{align}
However, if we shift the variable as $x+a\to x$ before the integration, the integration becomes zero.
In the case that the integration diverges, we have to care about the shift of the integration variable.

Now, as we will see below, in the case of the chiral anomaly, the divergence is linear, and thus the first-order term in the variable shift becomes important.
The term $[f(\infty) - f(-\infty)]$ that appears here serves as a ``surface term'' in one dimension.
When this is generalized to higher dimensions, it becomes proportional to the surface area.

To prepare for the following calculations, we consider the four-dimensional momentum integral
\begin{align}
    \Delta(a)
    &=
    \int\fr{\dif^4 k}{(2\pi)^4} \lb[f(k+a)-f(k)\rb] \nom
    &\simeq
    a^{\mu}\int\fr{\dif^4 k}{(2\pi)^4} \fr{\del f(k)}{\del k^{\mu}}\,.
\end{align}
Let us now perform a Wick rotation ($k^0 \to \im k^4, a^0 \to \im a^4$), which turns the expression into a total derivative in the four-dimensional Euclidean space.
Then, by Gauss's divergence theorem, we obtain \begin{align}
    \Delta(a)
    &=
    \im a^{\mu}\int\fr{\dif^4 k_{\rm E}}{(2\pi)^4} \fr{\del f(k_{\rm E})}{\del k_{\rm E}^{\mu}} \nom &=
    \im a^{\mu}\lim_{k_{\rm E}\to\infty}\int\fr{\dif^3 S^{\mu}}{(2\pi)^4}f(k_{\rm E})\,,
\end{align} where the index $\mu$ runs from 1 to 4, and $\dif^3 S^{\mu}=k_{\rm E}^2k_{\rm E}^{\mu}\dif\Omega_{4}$ denotes the vector of surface element in four-dimensional space%
\footnote{Since we are working in Euclidean space, we do not distinguish between upper and lower indices.}.

In what follows, we consider the case in which the function $f(k_{\rm E})$ has the form
\begin{align}
    f^{\alpha}(k_{\rm E})
    =
    A\fr{k_{\rm E}^{\alpha}}{k_{\rm E}^4}\,,
\end{align}
with $A$ being a constant independent of $k_{\rm E}$, and the integral exhibits a linear divergence.

In this case, the integral yields
\begin{align} \label{eq:Chi_Anom_Der:Difference_linear_divergence}
    \Delta^{\alpha}(a)
    &=
    \im a^{\mu}A\lim_{k_{\rm E}\to\infty}\int\fr{\dif\Omega_{4}}{(2\pi)^4}\fr{k_{\rm E}^{\mu}k_{\rm E}^{\alpha}}{k_{\rm E}^2}
    \nom
    &=
    \im a^{\alpha}A\fr{1}{32\pi^2}\,.
\end{align}
Here, we have used the fact that the surface area of the unit 3-sphere is given by $\Omega_4=2\pi^2$, and the angular average yields $\int\fr{\dif\Omega_{4}}{2\pi^2}\hat{k}_{\rm E}^{\mu}\hat{k}_{\rm E}^{\alpha}=\fr{1}{4}\delta^{\mu\alpha}$ with $\hat{k}_{\rm E}^{\mu}$ being the unit momentum vector.

\subsection{Ambiguity of the integration}
We go back to the derivation of $\Gamma^{\mu\nu\rho}$.
For later discussion, we here introduce a parameter $\beta^\mu$ in the right-hand Feynman diagram in Fig.~\ref{fig:Chi_Anom_Der:Triangle_daiagram}, by shifting $l \to l + \beta$.
We then have
\begin{align}
    \Delta^{\nu\rho}_{1}
    &=
    e^2\int_{l}\tr\lb[\gamma_{5}\gamma^{\nu}\fr{1}{\slashed{l}}\gamma^{\rho}\fr{1}{\slashed{l}-\slashed{q}}
    -\gamma_{5}\gamma^{\nu}\fr{1}{\slashed{l}+\slashed{\beta}+\slashed{q}}\gamma^{\rho}\fr{1}{\slashed{l}+\slashed{\beta}}
    \rb]
    \nom
    &=
    4\im e^2\int_{l}\epsilon^{\nu\alpha\rho\beta}\lb[
    \fr{(l+a)_{\alpha}(l+a-q)_{\beta}}{(l+a)^2(l+a-q)^2}
    -\fr{l_{\alpha}(l-q)_{\beta}}{l^2(l-q)^2}
    \rb]\,.
\end{align}
In the last equality, we defined $a\equiv\beta+q$ representing the difference between the first and second terms.
Here, from the relation,
\begin{align}
    \epsilon^{\nu\alpha\rho\beta}\fr{l_{\alpha}(l-q)_{\beta}}{l^2(l-q)^2}
    =
    -\epsilon^{\nu\alpha\rho\beta}\fr{l_{\alpha}q_{\beta}}{l^2(l-q)^2}\,,
\end{align}
we find that the order of $l$ in the numerator decreases by one, and the integral becomes linearly divergent.
Thus, we can perform the integration using Eq.~(\ref{eq:Chi_Anom_Der:Difference_linear_divergence})\,.
Since we take the limit as $l\to\infty$, using $l-q\to l$, we have
\begin{align}
    \Delta^{\nu\rho}_{1}
    &=
    -4e^2(\beta+q)_{\alpha}\fr{-\epsilon^{\nu\alpha\rho\beta}q_{\beta}}{32\pi^2}
    \nom
    &=
    \fr{e^2}{8\pi^2}\epsilon^{\nu\alpha\rho\beta}q_{\beta}\beta_{\alpha}\,.
\end{align}
$\Delta^{\nu\rho}_{2}$ can be obtained by the replacement $\nu\leftrightarrow\rho$, $k\leftrightarrow q$ in the above result:
\begin{align}
    \Delta^{\nu\rho}_{2}
    =
    \fr{e^2}{8\pi^2}\epsilon^{\rho\alpha\nu\beta}k_{\beta}\beta_{\alpha}\,.
\end{align}
Therefore, we obtain 
\begin{align}
\label{eq:Chi_Anom_Der:Three_point_axial}
    \im p_{\mu}\Gamma^{\mu\nu\rho}(p,k,q)
    &=
    \Delta^{\nu\rho}_{1}+\Delta^{\nu\rho}_{2}
    \nom
    &=
    \fr{e^2}{8\pi^2}\epsilon^{\nu\rho\alpha\beta}\beta_{\alpha}(k-q)_{\beta}\,.
\end{align}
Due to the ambiguity of the parameter $\beta$, the integration also has ambiguity.

To remove this ambiguity, we have to consider the conservation law not only of the axial current but also of the vector current,
$\del_{\mu}j^{\mu}$.
In the triangle diagram, there are two vertices coupling to the vector current, and each of them corresponds to $\im k_{\nu}\Gamma^{\mu\nu\rho}(p,k,q)$ and $\im q_{\rho}\Gamma^{\mu\nu\rho}(p,k,q)$.
The former one is
\begin{align}
    \im k_{\nu}\Gamma^{\mu\nu\rho}(p,k,q)
    &=
    -e^2\int_{l}\lb[(l_{\nu}+k_{\nu})-l_{\nu}\rb]
    \tr\lb[\gamma^{\mu}\gamma_{5}\fr{1}{\slashed{l}+\slashed{k}}\gamma^{\nu}\fr{1}{\slashed{l}}\gamma^{\rho}\fr{1}{\slashed{l}-\slashed{q}}\rb]
    \nom
        &\quad
        -e^2\int_{l}\lb[(l_{\nu}+\beta_{\nu})-(l_{\nu}+\beta_{\nu}-k_{\nu})\rb]
        \tr\lb[\gamma^{\mu}\gamma_{5}\fr{1}{\slashed{l}+\slashed{\beta}+\slashed{q}}\gamma^{\rho}\fr{1}{\slashed{l}+\slashed{\beta}}\gamma^{\nu}\fr{1}{\slashed{l}+\slashed{\beta}-\slashed{k}}\rb]
    \nom
    &=
    -\fr{e^2}{8\pi^2}\epsilon^{\mu\nu\alpha\beta}k_{\alpha}\beta_{\beta}\,,
\end{align}
and the latter one is
\begin{align}
    \im q_{\rho}\Gamma^{\mu\nu\rho}(p,k,q)
    &=
    -e^2\int_{l}\lb[l_{\rho}-(l_{\rho}-q_{\rho})\rb]
    \tr\lb[\gamma^{\mu}\gamma_{5}\fr{1}{\slashed{l}+\slashed{k}}\gamma^{\nu}\fr{1}{\slashed{l}}\gamma^{\rho}\fr{1}{\slashed{l}-\slashed{q}}\rb]
    \nom
        &\quad
        -e^2\int_{l}\lb[(l_{\rho}+\beta_{\rho}+q_{\rho})-(l_{\rho}+\beta_{\rho})\rb]
        \tr\lb[\gamma^{\mu}\gamma_{5}\fr{1}{\slashed{l}+\slashed{\beta}+\slashed{q}}\gamma^{\rho}\fr{1}{\slashed{l}+\slashed{\beta}}\gamma^{\nu}\fr{1}{\slashed{l}+\slashed{\beta}-\slashed{k}}\rb]
    \nom
    &=
    -\fr{e^2}{8\pi^2}\epsilon^{\mu\nu\alpha\beta}(2k_{\alpha}-\beta_{\alpha})q_{\beta}\,.
\end{align}
If we here impose that the vector current must be conserved so as not to violate the gauge invariance, both of them become zero.
The choice satisfying this condition is
\begin{align}
    \beta^{\mu}=2k^{\mu}\,.
\end{align}
Substituting this into the equation for axial current (\ref{eq:Chi_Anom_Der:Three_point_axial}), we have
\begin{align}
    \im p_{\mu}\Gamma^{\mu\nu\rho}(p,k,q)
    &=
    -\fr{e^2}{4\pi^2}\epsilon^{\nu\rho\alpha\beta}k_{\alpha}q_{\beta}\,.
\end{align}
Therefore, the continuity equation of the axial current is
\begin{align}
    \del_{\mu}j_5^{\mu}(x)
    &=
    -\fr{e^2}{8\pi^2}F_{\mu\nu}\tilde{F}^{\mu\nu}\,,
\end{align}
and the chiral anomaly is derived.
What we can learn from the derivation here is that it is impossible to satisfy the conservation laws of the vector and axial currents simultaneously in quantum theory.
Since the conservation of the vector current must hold due to the gauge invariance, the axial current has to violate the conservation law.
In the discussion in Sec.~\ref{sec:Chi_Ph:Chi_Anom}, the conservation of the vector current was assumed.
In the present perturbative analysis, without this assumption, the parameter~$\beta$ cannot be determined, and the relation for the chiral anomaly remains undefined.

\section{Fujikawa's method}
\label{sec:Chi_Anom_Der:Fujikawa}
We derive the chiral anomaly using the path integral formalism.  
This is a non-perturbative derivation known as the so-called Fujikawa's method~\cite{Fujikawa:1980eg}.  
According to the symmetry of the classical action, a conservation law was derived via Noether's theorem.  
In quantum theory, a conservation law associated with a symmetry also exists, and it is expressed as the Ward--Takahashi identity.  
In the path integral formulation, quantum effects are encoded in the integration measure.  
Fujikawa's method shows that, under an axial transformation, the transformation of the integration measure (or Jacobian) becomes non-trivial, leading to a modification of the Ward--Takahashi identity.  

We consider the path integral of the massless Dirac fermion,
\begin{align}
    Z
    =
    \int\mcl{D}\psi\mcl{D}\bar{\psi}\exp\lb[\im\int_x\bar{\psi}\im\slashed{D}\psi\rb]\,.
\end{align}
Under the infinitesimal $\rm U(1)_{A}$ transformation,%
\footnote{For convenience in the later discussion of the Ward--Takahashi identity, we assume that the infinitesimal parameter is a local quantity.}
\begin{align}
\label{eq:Chi_Anom_Der:small_axial_trans}
    \psi\to\psi+\im \epsilon(x) \gamma_{5}\psi\,,
    \qquad
    \bar{\psi}\to\bar{\psi}+\im \epsilon(x) \bar{\psi}\gamma_{5}\,,
\end{align}
the action transforms as
\begin{align}
    \int_x\bar{\psi}\im\slashed{D}\psi
    &\to
    \int_x\lb[\bar{\psi}\im\slashed{D}\psi+\epsilon(x)\del_{\mu}\bar{\psi}\gamma^{\mu}\gamma_5\psi\rb]
    \nom
    &=
    \int_x\lb[\bar{\psi}\im\slashed{D}\psi+\epsilon(x)\del_{\mu}j_5^{\mu}\rb]\,.
\end{align}
Thus, to make the action invariant only, the axial current is conserved.
As we will see below, the transformation of the path integral measure of the fermion fields under the $\rm U(1)_{A}$ transformation is not invariant. 

We first consider eigenfunctions $\varphi_{n}$ which are c-numbers satisfying
\begin{align}
\label{eq:Chi_Anom_Der:Dirac_op_eigen}
    \im\slashed{D}\varphi_{n}
    =
    \lambda_{n}\varphi_{n}\,.
\end{align} 
Using these eigenfunctions, we can expand the fermion fields as
\begin{align}
    \psi
    =
    \sum_{n}a_{n}\varphi_{n},
    \qquad
    \bar{\psi}
    =
    \sum_{n}\bar{b}_{n}\varphi_{n}^{\dagger}\,,
\end{align}
where the coefficients, $a_{n},~\bar{b}_{n}$, are Grassmann numbers.
In addition, the eigenspinors satisfy the orthogonality,
\begin{align}
    \int_x\varphi_{m}^{\dagger}\varphi_{n}
    =
    \delta_{m,n}\,.
\end{align}
Thus, the integration measure is expressed as
\begin{align}
    \mcl{D}\psi\mcl{D}\bar{\psi}
    =
    \prod_{n}\dif a_{n}\dif \bar{b}_{n}\,.
\end{align}

Since the information of the fermion field is included in the coefficients, the change in the field is expressed as
\begin{align}
    \delta \psi
    =
    \delta\sum_{n}a_{n}\varphi_{n}
    =
    \sum_{n}(\delta a_{n})\varphi_{n}\,.
\end{align}
The change in the coefficients can be obtained from the inner product of $\delta\psi$ and $\varphi_{n}$.
Under the infinitesimal $\rm U(1)_{A}$ transformation (\ref{eq:Chi_Anom_Der:small_axial_trans}), we have
\begin{align}
    \delta \psi
    =
    \im \epsilon(x) \gamma_{5}\psi
    =
    \im \epsilon(x) \gamma_{5}\sum_{n}a_{n}\varphi_{n}\,.
\end{align}
It follows that 
\begin{align}
    \delta a_{n}
    &=
    \int_x\varphi_{n}^{\dagger}\delta \psi
    \nom
    &=
    \sum_{m}\im \int_x\epsilon(x)\varphi_{n}^{\dagger} \gamma_{5}\varphi_{m}a_{m}
    \nom
    &\equiv
    \sum_{m}X_{nm}a_{m}\,.
\end{align}
Therefore, the coefficient $a_{n}$ transforms as
\begin{align}
    a_{n}
    \to a_{n}'
    =
    (\delta_{nm}+X_{nm})a_{m}\,.
\end{align}
Since $a_{n}$ is a Grassmann number, and the Jacobian for a transformation is inverse compared with the ordinary number case, 
\begin{align}
    \mcl{D}\psi
    &=
    \prod_{n}\dif a_{n}
    \nom
    &\to
    \det{(\delta_{nm}+X_{nm})}\prod_{n}\dif a_{n}'
    \nom
    &\equiv
    J\mcl{D}\psi'\,,
\end{align}
where $J$ is the Jacobian.
The same applies to $\bar{\psi}$:
\begin{align}
    \mcl{D}\bar{\psi}\to J\mcl{D}\bar{\psi}'\,.
\end{align}
Thus, under the $\rm U(1)_{A}$ transformation, the integration measure changes as
\begin{align}
    \mcl{D}\psi\mcl{D}\bar{\psi}
    \to
    J^2\mcl{D}\psi'\mcl{D}\bar{\psi}'\,.
\end{align}
As this shows, we have the non-trivial factor with respect to the $\rm U(1)_{A}$ transformation.%
\footnote{In the case of $\rm U(1)_{V}$ transformation, since $a_{n}=(\delta_{nm}+Y_{nm})a_{m}',~\bar{b}_{n}=(\delta_{nm}-Y_{nm})\bar{b}_{m}'$, $Y_{nm}$ is canceled, and the change is trivial as $\mcl{D}\bar{\psi}\mcl{D}\psi\to \mcl{D}\bar{\psi}'\mcl{D}\psi'$.}

Next, we derive the explicit expression of the Jacobian.
Making use of the formula $\det(M)=\e^{\tr\ln{M}}$, we get
\begin{align}
    J
    &\simeq
    \exp\lb[\im\int_x\epsilon(x)\sum_{n}\varphi_{n}^{\dagger}\gamma_{5}\varphi_{n}\rb]\,,
\end{align}
where we applied the approximation $\ln(1+x)\simeq x~(x\ll 1)$.
There are two naive arguments for the Jacobian.
One is that $J=1$ since $\tr(\gamma_5)=0$, and the other is that $J$ is infinite since we have to sum over an infinite number of the eigenspinors.
However, neither of them is true.
We then regularize the sum in a gauge-invariant way, introducing a cutoff scale $\Lambda$ as
\begin{align}
    \sum_{n}\varphi_{n}^{\dagger}\gamma_{5}\varphi_{n}
    =
    \lim_{\Lambda\to\infty}\sum_{n}\varphi_{n}^{\dagger}\gamma_{5}\varphi_{n}\e^{\lambda_{n}^2/\Lambda^2}\,.
\end{align}
We note that the sign of $\lambda_{n}^2$ is negative after the Wick rotation, and the exponential factor makes the sum convergent.
Using $\varphi_{n}(x)=\braket{x|n}$ and $\sum\ket{n}\bra{n}=1$, we can rewrite
\begin{align}
    \lim_{\Lambda\to\infty}\sum_{n}\varphi_{n}^{\dagger}(x)\gamma_{5}\varphi_{n}(x)\e^{\lambda_{n}^2/\Lambda^2}
    &=
    \lim_{\Lambda\to\infty}\sum_{n}\braket{n|x}\gamma_{5}\bra{x}\e^{(\im\slashed{D})^2/\Lambda^2}\ket{n}
    \nom
    &=
    \lim_{\Lambda\to\infty}\tr\lb[\gamma_{5}\bra{x}\e^{(\im\slashed{D})^2/\Lambda^2}\ket{x}\rb]\,.
\end{align}
The square of the Dirac operator is
\begin{align}
    (\im\slashed{D})^2
    &=
    -D^2
    +\fr{e}{2}\sigma^{\mu\nu}F_{\mu\nu}\,,
\end{align}
where we used
\begin{align}
    \sigma^{\mu\nu}=\fr{\im}{2}[\gamma^{\mu},\gamma^{\nu}]\,,
    \qquad
    F_{\mu\nu}=\fr{\im}{e}[D_{\mu},D_{\nu}]\,.
\end{align}
Thus, the sum can be expressed as
\begin{align}
    \lim_{\Lambda\to\infty}\tr\lb[\gamma_{5}\bra{x}\e^{(\im\slashed{D})^2/\Lambda^2}\ket{x}\rb]
    &=
    \lim_{\Lambda\to\infty}\int_{p}\tr\lb[\gamma_{5}\bra{x}\exp\lb(-\fr{D^2}{\Lambda^2}+\fr{e}{2\Lambda^2}\sigma^{\mu\nu}F_{\mu\nu}\rb)\ket{p}\braket{p|x}\rb]
    \nom
    &\simeq
    \lim_{\Lambda\to\infty}\lb(\int_{p}\braket{x|p}\e^{-\del^2/\Lambda^2}\braket{p|x}\rb)\tr\lb[\gamma_{5}\fr{1}{2}\lb(\fr{e}{2\Lambda^2}\sigma^{\mu\nu}F_{\mu\nu}\rb)^2\rb]\,.
\end{align}
The integration over the momentum is
\begin{align}
    \int_{p}\braket{x|p}\e^{-\del^2/\Lambda^2}\braket{p|x}
    &=
    \im\fr{\Lambda^4}{16\pi^2}\,.
\end{align}
The trace part is
\begin{align}
    \tr\lb[\gamma_{5}\fr{1}{2}\lb(\fr{e}{2\Lambda^2}\sigma^{\mu\nu}F_{\mu\nu}\rb)^2\rb]
    &=
    \im\fr{e^2}{\Lambda^4}F_{\mu\nu}\tilde{F}^{\mu\nu}\,.
\end{align}
Therefore, the Jacobian reduces to
\begin{align}
    J
    =
    \exp\lb[-\im\int_x\epsilon(x)\fr{e^2}{32\pi^2}F_{\mu\nu}\tilde{F}^{\mu\nu}\rb]\,.
\end{align}

We now consider the U(1)$_{\rm A}$ transformation of the partition function.
The transformed partition function is
\begin{align}
    Z'
    &\simeq
    \int\mcl{D}\psi\mcl{D}\bar{\psi}\exp\lb[\im\int_x\bar{\psi}\im\slashed{D}\psi\rb]\lb[1+\im\int_x\epsilon(x)\lb(\del_{\mu}j_{5}^{\mu}+\fr{e^2}{16\pi^2}F_{\mu\nu}\tilde{F}^{\mu\nu}\rb)\rb]\,.
\end{align}
To satisfy $Z=Z'$ for arbitrary $\epsilon(x)$, we find the chiral anomaly
\begin{align}
    \del_{\mu}j_{5}^{\mu}
    =
    -\fr{e^2}{16\pi^2}F_{\mu\nu}\tilde{F}^{\mu\nu}\,.
\end{align}
This result is the same as that derived in the perturbative approach in Sec.~\ref{sec:Chi_Anom_Der:Triangle}.

\chapter{High Density Effective Theory and Mass Correction}
\label{app:HDET_Mass}

\section{High density effective theory}
\label{sec:Dense_QFT:HDET}
In this section, we construct an effective theory near the Fermi surface so-called the high-density effective theory (HDET)~\cite{Hong:1998tn,Hong:1999ru,Schafer:2003jn} (see also Refs.~\cite{Kogut:2004su,Casalbuoni:2018haw}).
When we discuss the chirality flipping later, we will utilize the HDET.

\subsection{Hierarchy of energy scales}
The field $\psi_{+}$ has the energy $p_0=-\mu+|\bs{p}|$ and $\psi_{-}$ has the energy $p_0=-\mu-|\bs{p}|$.
Near the Fermi surface $(|\bs{p}|\sim\mu)$, the $\psi_{+}$ can be excited easily while $\psi_{-}$ needs the energy $\sim2\mu$ to be excited. 
Thus, we can integrate out $\psi_{-}$ as long as we focus on the energy scale sufficiently smaller than the chemical potential $(E<\mu)$.
Although we consider QED for simplicity, the HDET itself is applicable when the density of the system is larger enough than any other scales.
This means that we can apply the HDET to QCD.

\subsection{Effective action}
We derive the action of the HDET by integrating out the high energy degree of freedom from the action of massless Dirac fermions with background electromagnetic fields.
We begin with the action
\begin{align}
\label{eqDenseQT:Action_massless_Dirac_density_gauge}
    S
    =
    \int_{x}\bar{\psi}(x)(\im\slashed{D}+\mu\gamma^{0})\psi(x)\,.
\end{align}
In the HDET, we focus on the momentum scale, it is convenient to discuss in momentum space.
Decomposing the momentum as
\begin{align}
    p_0=l_0\,,
    \qquad
    \bs{p}=\mu\bs{v}+\bs{l}\,,
\end{align}
where $l_0,|\bs{l}|\ll\mu$, and $\bs{v}$ is the Fermi velocity, we can write the integration over the momentum as
\begin{align}
    \int\fr{\dif^4 p}{(2\pi)^4}
    =
    \fr{\mu^2}{\pi}\int\fr{\dif\Omega_{\bs{v}}}{4\pi}\int\fr{\dif l^{0}}{2\pi}\int\fr{\dif |\bs{l}|}{2\pi}
    \equiv
    \fr{\mu^2}{\pi}\int_{l^2,\bs{v}}\,.
\end{align}
Thus, the Fourier transformation of the fermion field is
\begin{align}
    \psi(x)
    =
    \int_{p}\psi(p)\e^{-\im p\cdot x}
    =
    \fr{\mu^2}{\pi}\int_{\bs{v}}\e^{\im\mu \bs{v}\cdot\bs{x}}\int_{l^2}\psi(l,\bs{v})\e^{-\im l\cdot x}\,.
\end{align}
This expression shows that the fermion field is the sum of patches on the Fermi surface characterized by the orientation of the Fermi velocity.
At low energy, only the component perpendicular to the patch is relevant, which is a (1+1)-dimensional system practically. 

We introduce the projection operator as
\begin{align}
    \mcl{P}^{\alpha}_{\pm}(\bs{v})\equiv\fr{1\pm \alpha(\bs{v})}{2}\,,
\end{align}
and rewrite the fermion field as
\begin{align}
    \psi(l,\bs{v})
    =\mcl{P}^{\alpha}_{+}(\bs{v})\psi(l,\bs{v})+\mcl{P}^{\alpha}_{-}(\bs{v})\psi(l,\bs{v})
    \equiv\psi_{+}(l,\bs{v})+\psi_{-}(l,\bs{v})\,.
\end{align}
We have the following relations:
\begin{align}
\label{eq:Alpha_decompose_useful_formula_++}
    &\bar{\psi}_{+}(l,\bs{v})\gamma^{\mu}\psi_{+}(l,\bs{v})
    =
    v^{\mu}\bar{\psi}_{+}(l,\bs{v})\gamma^{0}\psi_{+}(l,\bs{v})\,,
    \\
\label{eq:Alpha_decompose_useful_formula_--}
    &\bar{\psi}_{-}(l,\bs{v})\gamma^{\mu}\psi_{-}(l,\bs{v})
    =
    \bar{v}^{\mu}\bar{\psi}_{-}(l,\bs{v})\gamma^{0}\psi_{-}(l,\bs{v})\,,
    \\
\label{eq:Alpha_decompose_useful_formula_+-}
    &\bar{\psi}_{+}(l,\bs{v})\gamma^{\mu}\psi_{-}(l,\bs{v})
    =
    \bar{\psi}_{+}(l,\bs{v})\gamma_{\perp}^{\mu}\psi_{-}(l,\bs{v})\,,
    \\
\label{eq:Alpha_decompose_useful_formula_-+}
    &\bar{\psi}_{-}(l,\bs{v})\gamma^{\mu}\psi_{+}(l,\bs{v})
    =
    \bar{\psi}_{-}(l,\bs{v})\gamma_{\perp}^{\mu}\psi_{+}(l,\bs{v})\,,
\end{align}
where $\gamma_{\parallel}^{\mu}=\lb(\gamma^{0},(\bs{v}\cdot\bs{\gamma})\bs{v}\rb)\,,~\gamma_{\perp}^{\mu}=\gamma^{\mu}-\gamma_{\parallel}^{\mu}$.

Now, we express the action (\ref{eqDenseQT:Action_massless_Dirac_density_gauge}) in momentum space.
The kinetic term is
\begin{align}
    S_{\rm kin}
    &=
    \int_{x}\bar{\psi}(x)(\im\slashed{\del}+\mu\gamma^{0})\psi(x)
    \nom
    &=
    \fr{\mu^2}{\pi}
    \int_{l^2,\bs{v}}
    \lb[
    \psi_{+}^{\dagger}(v\cdot l)\psi_{+}
    +\psi_{-}^{\dagger}(\bar{v}\cdot l+2\mu)\psi_{-}
    +\psi_{+}^{\dagger}\gamma^{0}\slashed{l}_{\perp}\psi_{-}
    +\psi_{-}^{\dagger}\gamma^{0}\slashed{l}_{\perp}\psi_{+}
    \rb]\,.
\end{align}
Here, we defined $\slashed{l}_{\perp}\equiv\gamma_{\perp}^{\mu}l_{\mu}$ and used
\begin{align}
    \int_{x}\e^{\im(\mu v'+l'-\mu v-l)\cdot x}
    =
    \fr{(2\pi)^4}{\mu^2}\delta^{(2)}(\bs{v}'-\bs{v})\delta^{(2)}(l'-l)\,.
\end{align}
The reason why this relation holds is that there is a separation between the Fermi energy and the excitation from it.
The interaction term is
\begin{align}
    S_{\rm g}
    &=
    -e\int_{x}\bar{\psi}(x)\slashed{A}(x)\psi(x)
    \nom
    &=
    -e\lb(\fr{\mu^2}{\pi}\rb)^{2}
    \int_{l^2,l'^2,\bs{v},\bs{v}'}
    \lb[\psi_{+}^{\dagger}(l',\bs{v}')\lb(A^{0}-\bs{\alpha}\cdot\bs{A}\rb)\psi_{+}(l,\bs{v})
    +\psi_{-}^{\dagger}(l',\bs{v}')\lb(A^{0}-\bs{\alpha}\cdot\bs{A}\rb)\psi_{-}(l,\bs{v})\rb.
    \nom
    &\hspace{3cm}\lb.+\psi_{+}^{\dagger}(l',\bs{v}')\lb(A^{0}-\bs{\alpha}\cdot\bs{A}\rb)\psi_{-}(l,\bs{v})
    +\psi_{-}^{\dagger}(l',\bs{v}')\lb(A^{0}-\bs{\alpha}\cdot\bs{A}\rb)\psi_{+}(l,\bs{v})\rb]\,.
\end{align}

To integrate out the negative energy fermions, we derive the equation of motion for $\psi_{-}$.
By performing the functional derivative of the full action $S = S_{\rm kin} + S_{\rm g}$, we obtain
\begin{align}
\fr{\delta S}{\delta\psi_{-}^{\dagger}(l,\bs{v})}
&=
(\bar{v}\cdot l+2\mu)\psi_{-}(l,\bs{v})
-e\fr{\mu^2}{\pi}
\int_{l'^2,\bs{v}'}\lb(A^{0}-\bs{\alpha}\cdot\bs{A}\rb)[\psi_{+}(l',\bs{v}')+\psi_{-}(l',\bs{v}')]\,.
\end{align}
Noting that the chemical potential $\mu$ is much larger than any other relevant scale, the equation of motion can be approximated as
\begin{align}
\psi_{-}(l,\bs{v})
&\simeq
\fr{e\mu}{2\pi}\int_{l'^2,\bs{v}'}\lb(A^{0}-\bs{\alpha}\cdot\bs{A}\rb)\psi_{+}(l',\bs{v}')\,.
\end{align}
By substituting this expression back into the original action, one can obtain the effective action.

In the free case, the equation of motion for $\psi_{+}$ reads
\begin{align}
    (v\cdot l)\psi_{+}(l,\bs{v})=0\,.
\end{align}
The general solution to this equation can be expressed as
\def\arraystretch{1}
\begin{align}
    \psi_{\rm R+\uparrow}(l,\bs{v})=\sqrt{2l^0}
    \lb(
    \begin{array}{c}
    0 \\
    0 \\
    1 \\
    0 \\
    \end{array}
    \rb)\,,
    \qquad
    \psi_{\rm L-\downarrow}(l,\bs{v})=\sqrt{2l^0}
    \lb(
    \begin{array}{c}
    0 \\
    1 \\
    0 \\
    0 \\
    \end{array}
    \rb)\,,
\end{align}
where the normalization is chosen such that $\psi_{\chi}^{\dagger}\psi_{\chi'}=2l^0\delta_{\chi\chi'}$. Here, $\chi$ and $\chi'$ denote chiralities.
Using these spinors, the completeness relation is given by
\begin{align}
    \sum_{\chi}\psi_{\chi+}\psi_{\chi+}^{\dagger}
    =
    2l^0\mcl{P}^{\alpha}_{+}\,.
\end{align}

\subsection{Effective mass interaction term}
In the preceding discussion, we have neglected the mass term. Here, we take into account the correction due to the mass.
This correction becomes important in the later discussion on chirality flipping.
The mass term is given by
\begin{align}
    S_{\rm m}
    &=
    -m\int_{x}\bar{\psi}(x)\psi(x)
    \nom
    &=
    -m\fr{\mu^2}{\pi}\int_{l^2,\bs{v}}
    \lb(\psi_{+}^{\dagger}\gamma^{0}\psi_{-}+\psi_{-}^{\dagger}\gamma^{0}\psi_{+}\rb)\,.
\end{align}
Including this term and integrating the negative-energy modes out, we obtain
\begin{align}
    S_{\rm mg}
    &=
    -2m\fr{\mu^2}{\pi}\fr{e\mu}{2\pi}
    \int_{l^2,l'^2,\bs{v},\bs{v}'}
    \lb[\psi_{+}^{\dagger}(l',\bs{v}')\gamma^{0}\lb(A^{0}-\bs{\alpha}\cdot\bs{A}\rb)\psi_{+}(l,\bs{v})\rb.
    \nom
    &\hspace{4cm}
    \lb.+\psi_{+}^{\dagger}(l',\bs{v}')\lb(A^{0}-\bs{\alpha}\cdot\bs{A}\rb)\gamma^{0}\psi_{+}(l,\bs{v})\rb]
    \nom
    &=-\lb(\fr{\mu^2}{\pi}\rb)^{2}
    \int_{l^2,l'^2,\bs{v},\bs{v}'}
    \fr{2em}{\mu}\lb[\psi_{+}^{\dagger}(l',\bs{v}')\gamma^{0}A^{0}\psi_{+}(l,\bs{v})\rb]\,,
\end{align}
where we have used ${\alpha^{i},\gamma^{0}}=0$.
The factor of 2 in the first expression originates from two contributions: one from replacing the mass term in the equation of motion of the negative-energy particles with the original gauge interaction term, and the other from replacing the gauge interaction term with the original mass term.

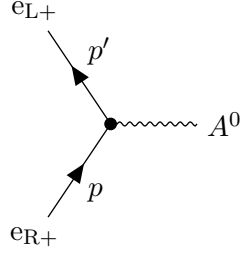
\begin{figure}[tb]
\centering
\begin{tikzpicture}
\begin{feynhand}
\vertex[dot] (a) at (0,0) {};
\vertex[particle] (b) at (-1,-1.5){$\rm{e_{R+}}$};
\vertex[particle] (c) at (-1,1.5){$\rm{e_{L+}}$};
\propag[fer] (b) to [edge label'=$p$] (a);
\propag[fer] (a) to [edge label'=$p'$] (c);
\vertex (d) at (1.5,0) {$A^{0}$};
\propag[photon] (a) to (d);
\end{feynhand}
\end{tikzpicture}
\caption{Feynman diagram for the vertex corresponding to the effective mass correction.}
\label{fig:Dense_QFT:Chi_flip_vertex}
\end{figure}

As can be seen from the above effective action, only the time component of the gauge field contributes to the interaction.
This implies that only the longitudinal mode of the photon contributes to chirality flipping.%
\footnote{This feature also holds at finite temperature. In that case, it can be demonstrated using the on-shell effective theory~\cite{Manuel:2014dza,Manuel:2016wqs}, whose logical structure is entirely analogous to that of the high-density effective theory.}
The corresponding Feynman diagram for this effective mass vertex is shown in Fig.~\ref{fig:Dense_QFT:Chi_flip_vertex}.
The Feynman rule for this effective mass correction vertex is given by
\begin{align}
    -\im e\fr{m}{\mu}\mcl{P}^{\alpha}{+}(\bs{v}')\gamma^{0}\delta{0}^{\mu}\mcl{P}^{\alpha}{+}(\bs{v})
    =
    -\im e\fr{m}{\mu}\delta{0}^{\mu}\gamma^{0}\mcl{P}^{\alpha}{-}(\bs{v}')\mcl{P}^{\alpha}{+}(\bs{v})\,.
\end{align}

\section{Rutherford scattering}
\begin{figure}[tb]
\centering
\begin{tikzpicture}
\begin{feynhand}
    \vertex[dot] (a) at (0,0) {}; 
    \vertex[particle] (b) at (-1,-1.5) {$\rm e_R$};
    \vertex[particle] (c) at (-1,1.5){$\rm e_L$};
    \propag[fer] (b) to (a);
    \propag[fer] (a) to (c);
    \vertex (d) at (2.5,0);
    \propag[photon] (a) to (d);
    \vertex[particle] (e) at (3.5,1.5) {p};
    \vertex[particle] (f) at (3.5,-1.5) {p};
    \propag[fer] (d) to (e);
    \propag[fer] (f) to (d);
\end{feynhand}

\node at (-0.9,-0.5) {$k$};
\node at (-0.9,0.5) {$k'$};
\node at (3.4,-0.5) {$p$};
\node at (3.4,0.5) {$p'$};
\node at (1.25,-0.7) {$q$};
\draw[->,>=stealth,thick] (1.6,-0.3)--(0.8,-0.3);
\end{tikzpicture}

\caption{Feynman diagram for Rutherford scattering.
The incoming and outgoing electrons (protons) carry momenta $k$ and $k'$ ($p$ and $p'$), respectively.}

\label{fig:e-q_diagram}
\end{figure}
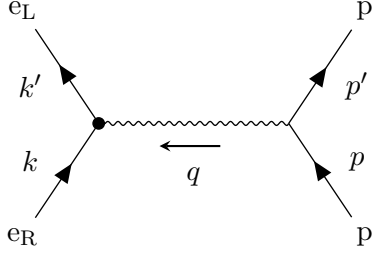

We consider chirality flipping via Rutherford scattering, namely electron-proton scattering (Fig.~\ref{fig:e-q_diagram}).
We neglect the proton recoil and employ the isoenergetic approximation $(q_0 = 0)$.
The notation for the momenta is chosen to be consistent with that in Chap.~\ref{chap:Chi_Ph}.

From the Feynman diagram, the amplitude can be written as
\begin{align}
    \im\mcl{M}_{\rm Rut}
    &=(-\im e)^2\fr{m_{\e}}{\mu_{\e}}\psi_{+}^{\dagger}(k')\gamma^{0}\mcl{P}^{\alpha}_{-}(\bs{v}')\mcl{P}^{\alpha}_{+}(\bs{v})\psi_{+}(k)
    D_{0\nu}(\bs{q})\bar{u}_{\rm p}(p')\gamma^{\nu}u_{\rm p}(p)\,,
\end{align}
where $u_{\rm p}$ denotes the proton spinor and $q^\mu = k'^{\mu}-k^{\mu}$.
Under the isoenergetic approximation, the projection operator takes a simple form:
\begin{align}
    &P_{{\rm l},00}=-1\,,
    \qquad
    P_{{\rm l},0i}=P_{{\rm l},ij}=0\,.
\end{align}
Then, we have
\begin{align}
    j^{\mu}\fr{P_{\rm{l},\mu\nu}}{q^2+\Pi_{\rm l}}j^{\nu}
    \simeq
    \fr{(j^0)^2}{|\bs{q}|^2-\Pi_{\rm l}}\,.
\end{align}
The squared amplitude is given by
\begin{align}
    \fr{1}{2}\sum_{\chi,s}|\mcl{M}_{\rm Rut}|^2
    &=
    \sum_{\chi,s}\fr{e^4m_{\e}^2}{2\mu_{\e}^2}\fr{1}{(|\bs{q}|^2-\Pi_{\rm l})^2}
    \tr\!\lb[
    \bar{u}_{\rm p}(p')\gamma^{0}u_{\rm p}(p)u_{\rm p}(p)\gamma^{0}\bar{u}_{\rm p}(p')\rb]
    \nom
    &\hspace{1cm}
    \times\tr\!\lb[\psi_{+}^{\dagger}(k')\gamma^{0}\mcl{P}^{\alpha}_{-}(\bs{v}')\mcl{P}^{\alpha}_{+}(\bs{v})\psi_{+}(k)\psi_{+}(k)\mcl{P}^{\alpha}_{+}(\bs{v})\mcl{P}^{\alpha}_{-}(\bs{v}')\gamma^{0}\psi_{+}^{\dagger}(k')\rb]
    \nom
    &=
    \fr{2e^4m_{\e}^2}{\mu_{\e}^2}\fr{l^0 l'^0}{(|\bs{q}|^2-\Pi_{\rm l})^2}
    (1-\cos\theta_{\bs{vv}'})
    \tr\!\lb(\gamma^{0}\slashed{p}\gamma^{0}\slashed{p}'\rb)
    \nom
    &=
    \fr{8e^4m_{\e}^2}{\mu_{\e}^2}\fr{l^0 l'^0}{(|\bs{q}|^2-\Pi_{\rm l})^2}
    (1-\cos\theta_{\bs{vv}'})
    \lb(E_{p}E_{p'}+\bs{p}\cdot\bs{p}'\rb)
    \nom
    &\simeq
    \fr{128\pi^2\alpha^2m_{\e}^2}{\mu_{\e}^2}\fr{l^0 l'^0E_{p}E_{p'}}{(|\bs{q}|^2-\Pi_{\rm l})^2}(1-\cos\theta_{\bs{vv}'})\,.
\end{align}
The factor $1/2$ accounts for the $\mathrm{R}\!\to\!\mathrm{L}$ process only.
We have also summed over the arbitrary proton spin $s$.
Since the proton is nonrelativistic, we used $E_{\rm p}\simeq m_{\rm p}\gg|\bs{p}|, |\bs{p}'|$ in the last line.
By setting $l^0\simeq l'^0\simeq\mu_{\e}$ and $E_{k}\simeq E_{k'}$, we reproduce the same result as in Ref.~\cite{Grabowska:2014efa}, leading to the chirality-flipping rate shown in App.~\ref{app:CM-mode_Electron}.

\section{Coulomb scattering}
\begin{figure}[htpb]
\centering
\begin{tikzpicture}
\begin{feynhand}
    \vertex[dot] (a) at (0,0) {}; 
    \vertex[particle] (b) at (-1,-1.5) {$\rm e_R$};
    \vertex[particle] (c) at (-1,1.5){$\rm e_L$};
    \propag[fer] (b) to (a);
    \propag[fer] (a) to (c);
    \vertex (d) at (2.5,0);
    \propag[photon] (a) to (d);
    \vertex[particle] (e) at (3.5,1.5) {e};
    \vertex[particle] (f) at (3.5,-1.5) {e};
    \propag[fer] (d) to (e);
    \propag[fer] (f) to (d);
\end{feynhand}

\node at (-0.9,-0.5) {$k$};
\node at (-0.9,0.5) {$k'$};
\node at (3.4,-0.5) {$p$};
\node at (3.4,0.5) {$p'$};
\node at (1.25,-0.7) {$q$};
\draw[->,>=stealth,thick] (1.6,-0.3)--(0.8,-0.3);
\end{tikzpicture}

\caption{Feynman diagram for Coulomb scattering.}

\label{fig:e-e_diagram}
\end{figure}
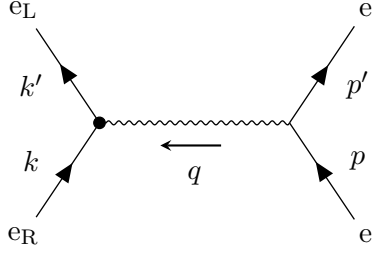

We next consider chirality flipping via Coulomb scattering.
The corresponding diagram (Fig.~\ref{fig:e-e_diagram}) is almost identical to that of the Rutherford scattering.

From the Feynman diagram, the amplitude is written as
\begin{align}
    \im\mcl{M}_{\rm Cou}
    &=
    (-\im e)^2\fr{m_{\e}}{\mu_{\e}}\psi_{+}^{\dagger}(k')\gamma^{0}\mcl{P}^{\alpha}_{-}(\bs{v}')\mcl{P}^{\alpha}_{+}(\bs{v})\psi_{+}(k)
    D_{0\nu}(q_0,\bs{q})\bar{\psi}_{+}(p')\gamma^{\nu}\psi_{+}(p)\,.
\end{align}
Following the same procedure as for the Rutherford scattering, the squared amplitude becomes
\begin{align}
    \fr{1}{2}\sum_{\chi,s}|\mcl{M}_{\rm Cou}|^2
    &=
    \fr{8e^4m_{\e}^2}{\mu_{\e}^2}\fr{l^0 l'^0|\bs{p}||\bs{p'}|}{(|\bs{q}|^2-\Pi_{\rm l})^2}
    (1-\cos\theta_{\bs{vv}'})
    (1+\cos\theta_{\bs{pp}'})\,.
\end{align}
By replacing the coupling $e\to g$ and the fermion species $\e\to\mathrm{q}$, and multiplying by the group-theoretical factor
$F_{\rm c}\equiv (N_{\rm c}^2-1)/(4N_{\rm c}^2)$,
we obtain the scattering amplitude for chirality flipping via one-gluon exchange:
\begin{align}
    |\mcl{M}_{\rm m,L}|^2
    =F_{\rm c}\fr{128\pi^2\alpha_{\rm s}^2m_{\rm q}^2}{\mu_{\rm q}^2}
    \fr{l^0 l'^0|\bs{p}||\bs{p'}|}{(|\bs{q}|^2-\Pi_{\rm l})^2}
    (1-\cos\theta_{\bs{vv}'})
    (1+\cos\theta_{\bs{pp}'})\,.
\end{align}
This expression coincides with the amplitude obtained in Subsec.~\ref{subsec:Chi_Astero:chirality_flipping}, under the assumption that the typical momentum scale is the Fermi momentum.

\chapter{CM-mode in Electron Matter}
\label{app:CM-mode_Electron}

In this appendix, we show that the CMW in electron matter is strongly damped by the dynamical electromagnetic fields~\cite{Rybalka:2018uzh,Shovkovy:2018tks}.
In the following, we put the subscript ``e'' for quantities related to electrons, such as the number density $n_{\rm e}$.
Unlike the quark number current discussed in Sec.~\ref{sec:Chi_Astero:CM-mode_q}, we have to include the Ohmic current induced by electric fields as
\begin{align}
    &\bs{j}_{\rm e}
    =-\fr{e\mu_{5,\rm {e}}}{2\pi^2}\bm{B}
    -D_{\rm e}\bm{\nabla}n_{\rm e}
    -\fr{\sigma}{e}\bm{E}\,,
\end{align}
where $\sigma$ is the conductivity, while the axial current is not affected by the electric field:
\begin{align}
    &\bm{j}_{\rm 5,e}
    =
    -\fr{e\mu_{\rm e}}{2\pi^2}\bm{B}
    -D_{\rm e}\bm{\nabla}n_{\rm 5,e}\,.
\end{align}
Due to the presence of dynamical electric fields, we have to take the Gauss law and the chiral anomaly into account.
Thus, the fundamental equations are
\begin{align}
    &\nabla_{\mu}F^{t\mu}=-ej^{t}\,,
    \\
    &\nabla_{\mu}j_{\rm{e}}^{\mu}=0\,,
    \\
    &\nabla_{\mu}j_{5,\rm{e}}^{\mu}=-\Gamma_{\rm m}j_{5,{\rm e}}^{t}+\fr{e^2}{2\pi^2}\bs{E}\cdot\bs{B}\,.
\end{align}
Focusing only on the propagation along $z$-axis, we consider the solutions in the form of
\begin{align}
    &\delta n_{\rm e}(t,z)=\delta n_{\rm e}(\omega,\bs{k})\e^{-\im(\omega t-k_zz)}\,,
    \\
    &\delta\bs{E}(t,z)=\delta\bs{E}(\omega,\bs{k})\e^{-\im(\omega t-k_zz)}\,.
\end{align}
eliminating $\delta\bs{E}$, the linearized equations can be expressed as
\begin{align}
    \lb(
    \begin{array}{cc}
       \omega+\im\e^{\lambda-\rho}D_{\rm e} k_{z}^2+\im\e^{\lambda}\sigma  & \displaystyle\e^{\lambda-\rho}\fr{eB}{2\pi^2 \chi_{\rm e}}k_{z} \\
       \displaystyle\e^{\lambda-\rho}\fr{eB}{2\pi^2 \chi_{\rm e}}k_{z}+\e^{\rho}\fr{e^3B}{2\pi^2 k_{z}}  & \omega+\im\e^{\lambda-\rho}D_{\rm e} k_{z}^2+\im\Gamma_{\rm m}
    \end{array}\rb)
    \lb(
    \begin{array}{c}
        \delta n_{\rm e} \\
        \delta n_{\rm 5,e}
    \end{array}\rb)
    =0\,,
\end{align}
where the susceptibility is defined as $\chi_{\rm e}\equiv\fr{\del n_{\rm e}}{\del\mu_{\rm e}}$.
Thus, the dispersion relation of the CM-mode of electron matter yields
\begin{align}
    \omega_{\rm CM}
    &=
    \pm\sqrt{V_{\rm CM}^2\lb(k_{z}^2+\e^{\rho}e^2 \chi_{\rm e}\rb)-\lb(\e^{\lambda}\fr{\sigma}{2}-\fr{\Gamma_{\rm m}}{2}\rb)^2}
    \nom
    &\quad
    -\im\e^{\lambda}\fr{\sigma}{2}
    -\im\fr{\Gamma_{\rm m}}{2}
    -\im\e^{\lambda-\rho}D_{\rm e}k_{z}^2\,,
\end{align}
where we defined $V_{\rm CM}\equiv eB/(2\pi^2 \chi_{\rm e})$.
Due to the chiral anomaly, the CM-mode in electron matter acquires  the gap.
For the propagating CM-mode, the following inequality needs to be satisfied:
\begin{align}
\label{eq:Chi_Astero:Overdamped_Condition}
    V_{\rm CM}^2\lb(k_{z}^2+\e^{\rho}e^2 \chi_{\rm e}\rb)
    >\lb(\e^{\lambda}\fr{\sigma}{2}-\fr{\Gamma_{\rm m}}{2}\rb)^2\,.
\end{align}

To estimate $\sigma$ and $\Gamma_{\rm m}$, we have to identify the dominant process to the chirality flipping.
There are three candidates: the Rutherford scattering (electron-proton scattering), the Compton scattering, and the electron-electron scattering.
In dense matter of our interest, the photon density $(\sim T^3)$ is much smaller than the electron density $(\sim\mu_{\rm e}^3)$, and the electrons are more degenerate than the protons.
Therefore, we focus on the Rutherford scattering.
In this process, the proton recoil can be ignored since the typical electron energy is much smaller than the proton mass $(\mu_{\rm e}\ll m_{\rm p})$.
Then, we can employ the isoenegetic approximation for the protons.
We also two cases.
In the case of neutron stars $(T\sim 10^{6}\text{--}10^{9}~{\rm K})$, both electrons and protons are degenerate, and in the case of supernovae ($T\sim 10^{10}\text{--}10^{11}~{\rm K}$), electrons are degenerate while protons are non-degenerate and obey the Boltzmann distribution.
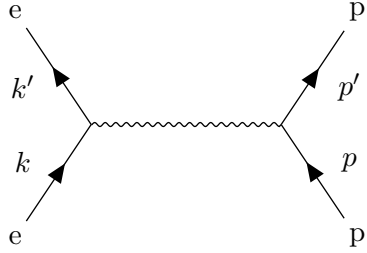
\begin{figure}[htpb]
\centering
\begin{tikzpicture}
\begin{feynhand}
    \vertex (a) at (0,0); 
    \vertex[particle] (b) at (-1,-1.5) {e};
    \vertex[particle] (c) at (-1,1.5){e};
    \propag[fer] (b) to (a);
    \propag[fer] (a) to (c);
    \vertex (d) at (2.5,0);
    \propag[photon] (a) to (d);
    \vertex[particle] (e) at (3.5,1.5) {p};
    \vertex[particle] (f) at (3.5,-1.5) {p};
    \propag[fer] (d) to (e);
    \propag[fer] (f) to (d);
\end{feynhand}

\node at (-0.9,-0.5) {$k$};
\node at (-0.9,0.5) {$k'$};
\node at (3.4,-0.5) {$p$};
\node at (3.4,0.5) {$p'$};
\end{tikzpicture}

\caption{The Feynman diagram of the Rutherford scattering.
In this figure, the incoming and outgoing electrons (protons) have the momenta $k$ and $k'$ ($p$ and $p'$), respectively.}

\label{fig:Chi_Astero:e-q_diagram}
\end{figure} 

We first consider the electric conductivity is given by
\begin{align}
\label{eq:Chi_Astero:Sigma}
    \sigma
    =
    \fr{e^2\bar{n}_{\rm e}\tau}{\bar{\mu}_{\rm e}}\,.
\end{align}
Under the assumption that the relaxation time is the same order of the mean free path $(\tau\sim l_{\rm e})$, the relaxation time of the Rutherford scattering becomes~\cite{Reddy:1997yr}
\begin{align}
\label{eq:Chi_Astero:mfp_e}
    \fr{1}{\tau}
    &\sim
    2\int\fr{\dif^3 \bm{k}'}{(2\pi)^3}\fr{\dif^3 \bm{p}}{(2\pi)^3}\fr{\dif^3 \bm{p}'}{(2\pi)^3}
    (2\pi)^4 \delta^{(4)}(k^{\mu}+p^{\mu}-k'^{\mu}-p'^{\mu})|M|^2
    \nom
    &\qquad\times
    f_{\rm p}(p)[1-f_{\rm e}(k')][1-f_{\rm p}(p')]
    \nom
    &\simeq
    \lb\{
    \begin{array}{ll}
    \displaystyle
    \fr{\alpha_{\rm e} m_{\rm p} T}{2\bar{\mu}_{\rm e}}
    \sim 
    10^{-5}~{\rm MeV}\lb(\fr{T}{10^{6}~{\rm K}}\rb)
    & \quad 
    (\text{neutron stars})
    \vspace{1em}
    \\
    \displaystyle
    \fr{\alpha_{\rm e} \bar{\mu}_{\rm e}}{6}
    \sim 
    10^{-1}~{\rm MeV}
    & \quad 
    (\text{supernovae})~~~~,
    \end{array}
    \rb.
\end{align}
where we set $\bar{\mu}_{\rm e}\sim 10^{2}~{\rm MeV}$.
When protons are non-degenerate in supernovae, the Pauli blocking of protons does not work, and the temperature dependence changes from $\sim T^1$ to $\sim T^0$.
The conductivity $\sigma$ reduces to
\begin{align}
    \sigma
    &\sim
    \lb\{
    \begin{array}{ll}
    \displaystyle
    10^{9}~{\rm MeV}\lb(\fr{T}{10^{6}~{\rm K}}\rb)^{-1}
    & \quad 
    (\text{neutron stars})
    \vspace{1em}
    \\
    \displaystyle
    10^{5}~{\rm MeV}
    & \quad 
    (\text{supernovae})~~~~.
    \end{array}
    \rb.
\end{align}
The chirality flipping rate in electron matter~\cite{Dvornikov:2015iua,Grabowska:2014efa}%
\footnote{The expression for neutron stars is different from the result in Ref.~\cite{Dvornikov:2015iua}. There, instead of the Debye mass, the plasma frequency is used for the IR cutoff. However, the Rutherford scattering is mediated by the the longitudinal photon in the static region, Thus, we use the Debye mass for the IR cutoff.} is given by
\begin{align}
    \Gamma_{\rm m}\simeq
    \lb\{
    \begin{array}{ll}
    \displaystyle
    \fr{\alpha^2 m_{\rm e}^2 m_{\rm p} T}{\pi \bar{\mu}_{\rm e}^3}\lb(\ln{\fr{4\bar{\mu}_{\rm e}^2}{m_{\rm De}^2}}-1\rb)
    \sim 
    10^{-12}~{\rm MeV}\lb(\fr{T}{10^{6}~{\rm K}}\rb)
    & \quad 
    (\text{neutron stars})
    \vspace{1em}
    \\
    \displaystyle
    \fr{\alpha^2 m_{\rm e}^2}{3\pi \bar{\mu}_{\rm e}}\lb(\ln{\fr{4\bar{\mu}_{\rm e}^2}{m_{\rm De}^2}}-1\rb)
    \sim 
    10^{-8}~{\rm MeV}
    & \quad 
    (\text{supernovae})~~~~\,,
    \end{array}
    \rb.
\end{align}
where $m_{\rm De}$ is the Debye mass of the photon.

From the above, the condition (\ref{eq:Chi_Astero:Overdamped_Condition}) is rewritten as
\begin{align}
    1
    \gtrsim
    \fr{\bar{\mu}_{\rm e}\sigma}{\alpha_{\rm e} B}
    \sim
    \lb\{
    \begin{array}{ll}
    \displaystyle
    10^{9}~\lb(\fr{B}{10^{18}~{\rm G}}\rb)^{-1}\lb(\fr{T}{10^{6}~{\rm K}}\rb)^{-1}
    & \quad 
    (\text{neutron stars})
    \vspace{1em}
    \\
    \displaystyle
    10^{5}~\lb(\fr{B}{10^{18}~{\rm G}}\rb)^{-1}
    & \quad 
    (\text{supernovae})~~~~\,,
    \end{array}
    \rb.
\end{align}
where we used $k_{z}\ll 2\pi/l_{\rm mfp}\ll \bar{\mu}_{\rm e}$ and $\sigma\gg \Gamma_{\rm m}$.
To satisfy the inequality, the sufficiently strong magnetic field of order $10^{23}~{\rm G}$ is needed.
Such a situation is unrealistic, and the CM-mode in electron matter is strongly damped.

\chapter{Helicity Eigenspinor}
\label{app:Helicity_Spinor}

In this appendix, we derive the helicity eigenspinors $u_{h}(k)$ of fermions which are used to obtain the currents in Sec.~\ref{sec:Chi_Astero:CM-mode_q}.
The eigenspinors satisfy the Dirac equation and the helicity equation as
\begin{align}
\label{eq:Chi_Astero:App:scat-amp:Dirac-eq}
    &(\slashed{k}-m)u_{h}(k)
    =
    0,
    \\
\label{eq:Chi_Astero:App:scat-amp:helicity-eq}
    &\lb(\fr{\bm{\Sigma}\cdot\bm{k}}{|\bm{k}|}-h\rb)u_{h}(k)
    =
    0\,,
\end{align}
where $m$ is the fermion mass, $h=\pm1$ is the helicity, and $\Sigma^i$ is defined as
\begin{align}
    \Sigma^i
    =
    \lb(
    \begin{array}{cc}
        \sigma^{i} & 0 \\
        0 & \sigma^{i}
    \end{array}
    \rb),
\end{align}
with $\sigma^{i}$ ($i=1,2,3$) being the Pauli matrices.
Here, we impose the normalization as $\bar u_{h} u_{h'} = 2m \delta_{h h'}$.

Using the Dirac representation of the gamma matrices,
\begin{align}
    \gamma^{0}
    =
    \lb(
    \begin{array}{cc}
        1 & 0 \\
        0 & -1
    \end{array}
    \rb),
    \qquad
    \gamma^{i}
    =
    \lb(
    \begin{array}{cc}
        0 & \sigma^{i} \\
        -\sigma^{i} & 0
    \end{array}
    \rb)\,,
\end{align}
Eq.~(\ref{eq:Chi_Astero:App:scat-amp:Dirac-eq}) can be written as
\begin{align}
    \lb(
    \begin{array}{cc}
        |\bs{k}|-m & -\bm{\sigma}\cdot\bm{k} \\
        \bm{\sigma}\cdot\bm{k} & -|\bs{k}|-m
    \end{array}
    \rb)
    u_{h}(k)
    =
    0\,.
\end{align}
The solution of this equation is given by
\begin{align}
    u_{h}(k)
    =
    N(k)
    \lb(
    \begin{array}{c}
        \xi_{h} \\
        hF(k)\xi_{h}
    \end{array}
    \rb)\,,
\end{align}
where $\xi_{h}$ is a two-component spinor satisfying 
\begin{align}
    \lb(\bm{\sigma}\cdot\bm{k}-h|\bs{k}|\rb)\xi_{h}
    =
    0\,.
\end{align}
The coefficients are defined as
\begin{align}
    N(k)
    \equiv
    \sqrt{\varepsilon_{\bs{k}}+m}\,,
    \qquad
    F(k)
    \equiv
    \sqrt{\fr{\varepsilon_{\bs{k}}-m}{\varepsilon_{\bs{k}}+m}}\,,
\end{align}
where $\varepsilon_{\bs{k}} = \sqrt{|\bs{k}|^2 + m^2}$.
In the spherical coordinates, the solution of this equation is given by
\begin{align}
    &\xi_{+}
    =
    \lb(
    \begin{array}{c}
        \e^{-\im\phi/2}\cos{\fr{\theta}{2}}
        \vspace{0.5em}
        \\
        \e^{\im\phi/2}\sin{\fr{\theta}{2}} 
    \end{array}
    \rb)\,,
    \\
    &\xi_{-}
    =
    \lb(
    \begin{array}{c}
        -\e^{-\im\phi/2}\sin{\fr{\theta}{2}}
        \vspace{0.5em}
        \\
        \e^{\im\phi/2}\cos{\fr{\theta}{2}} 
    \end{array}
    \rb)\,.
\end{align}

\chapter{Angular Integrals}
\label{app:Angular_Int}
In this appendix, we derive the angular integral used in Sec.~\ref{sec:Anom_Dyn_Res:Collective_modes} and summarize the formulae.
We first consider the integral
\begin{align}
    I^{i}(k)
    \equiv
    \int\fr{\dif^2\bs{v}}{4\pi}\fr{v^i}{v\cdot k}\,.
\end{align}
Since the only available vector is $k^i$, this integral can be expressed as
\begin{align}
    I^i(k)
    =
    k^iA(k)\,,
\end{align}
where $A(k)$ is a function to be determined.
Multiplying both sides by $k^i$, we have
\begin{align}
    A(k)
    &=
    \fr{1}{|\bs{k}|^2}\int\fr{\dif^2\bs{v}}{4\pi}\fr{\bs{v}\cdot\bs{k}}{v\cdot k}
    \nom
    &=
    \fr{1}{|\bs{k}|^2}\int\fr{\dif^2\bs{v}}{4\pi}\lb(-1+\fr{\omega}{v\cdot k}\rb)
    \nom
    &=
    \fr{1}{|\bs{k}|^2}\lb[-1+L(k)\rb]\,.
\end{align}

We next compute
\begin{align}
    I^{ij}(k)
    \equiv
    \int\fr{\dif^2\bs{v}}{4\pi}\fr{v^iv^j}{v\cdot k}\,.
\end{align}
The available tensors are $\delta^{ij}$ and $k^{i}k^{j}$.
Thus, we can write
\begin{align}
    \int\fr{\dif^2\bs{v}}{4\pi}\fr{v^iv^j}{v\cdot k}
    =
    \delta^{ij}B(k)+k^ik^jC(k)\,.
\end{align}
The products with $\delta^{ij}$ and $k^ik^j$ lead to the following coefficients respectively:
\begin{align}
    &B(k)
    =
    \fr{\omega}{2|\bs{k}|^2}\lb[1-\fr{k^2}{\omega^2}L(k)\rb]\,,
    \\
    &C(k)
    =
    -\fr{\omega}{2|\bs{k}|^4}\lb[1-\fr{k^2}{\omega^2}L(k)\rb]
    -\fr{\omega}{|\bs{k}|^4}\lb[1-L(k)\rb]\,.
\end{align}
Collecting the results, the angular integrals are summarized as
\begin{align}
    &\int\fr{\dif^2\bs{v}}{4\pi}=1\,,
    \\
    &\int\fr{\dif^2\bs{v}}{4\pi}v^i=0\,,
    \\
    &\int\fr{\dif^2\bs{v}}{4\pi}v^iv^j
    =\fr{1}{3}\delta^{ij}\,,
    \\
    &\int\fr{\dif^2\bs{v}}{4\pi}\fr{\omega}{v\cdot k}
    =
    L\,,
    \\
    &\int\fr{\dif^2\bs{v}}{4\pi}\fr{v^i}{v\cdot k}
    =
    -\fr{k^i}{|\bs{k}|^2}(1-L)\,,
    \\
    &\int\fr{\dif^2\bs{v}}{4\pi}\fr{v^iv^j}{v\cdot k}
    =
    \lb(\delta^{ij}-\fr{k^ik^j}{|\bs{k}|^2}\rb)\fr{\omega}{2|\bs{k}|^2}\lb(1-\fr{k^2}{\omega}L\rb)
    -\fr{k^ik^j}{|\bs{k}|^2}\fr{\omega}{|\bs{k}|^2}(1-L)\,.
\end{align}

\chapter{Expansion of the dimensionless functions}
\label{app:Expansion_functions}
In this appendix, we derive the limiting expressions of the dimensionless functions, $F(\xi)$, $G(\xi)$, and $I(\xi)$, which are introduced in Sec.~\ref{sec:Anom_Dyn_Res:Collective_modes}.
The function $L(\xi)$ can be expanded as
\begin{align}
    L(\xi)
    \simeq
    \lb\{
    \begin{array}{lc}
        \displaystyle -\im\fr{\pi}{2}\xi+\xi^2+\mcl{O}(\xi^3) & (\xi\ll1) \\
        \displaystyle 1+\fr{1}{3\xi^2}+\fr{1}{5\xi^4}+\mcl{O}(\xi^{-6})& (\xi\gg1)
    \end{array}\rb.\,.
\end{align}
From these expressions, one can easily expand the functions, $F(\xi)$ and $G(\xi)$, as
\begin{align}
    &F(\xi)
    \simeq
    \lb\{
    \begin{array}{lc}
        \displaystyle -\im\fr{\pi}{4}\xi+\mcl{O}(\xi^2) & (\xi\ll1) \\
        \displaystyle \fr{1}{3}+\fr{1}{15\xi^2}+\mcl{O}(\xi^{-4})& (\xi\gg1)
    \end{array}\rb.\,,
    \\
    &G(\xi)
    \simeq
    \lb\{
    \begin{array}{lc}
        \displaystyle 1+\im\fr{\pi}{2}\xi+\mcl{O}(\xi^2) & (\xi\ll1) \\
        \displaystyle \fr{1}{3}-\fr{2}{15\xi^2}+\mcl{O}(\xi^{-4})& (\xi\gg1)
    \end{array}\rb.\,.
\end{align}

We now consider the function $I(\xi)$.
In our setup, it is convenient to express the velocity as $\bs{v}=(\cos\theta, \sin\theta\cos\phi,\sin\theta\sin\phi)$, and we can easily perform the integration over $\phi$ as
\begin{align}
    I(\xi)
    &=
    \fr{1}{4\pi}\xi
    \int_{-1}^{1}\dif v_x\int_{0}^{2\pi}\dif\phi\lb[\fr{v_x\sin^2\theta\sin^2\phi}{(\xi-v_x+\im\eta)^4}
    -3\fr{\sin^4\theta\cos^2\phi\sin^2\phi}{(\xi-v_x+\im\eta)^5}\rb]
    \nom
    &=
    \fr{1}{4}\xi
    \int_{-1}^{1}\dif v_x\lb[\fr{v_x(1-v_x^2)}{(\xi-v_x+\im\eta)^4}-\fr{3}{4}\fr{(1-v_x^2)^2}{(\xi-v_x+\im\eta)^5}\rb]\,.
\end{align}
Rather than evaluating the remaining integral exactly, it is sufficient for our purposes to examine the behavior of $I(\xi)$ in two limiting cases.
In the quasi-static limit, the pole gives rise to an imaginary part, since for $|x|<1$, we have
\begin{align}
    \fr{1}{(x+\im\eta)^{n}}
    =\mcl{P}\fr{1}{x^n}-\im\pi\fr{(-1)^{n-1}}{(n-1)!}\delta^{(n-1)}(x)\,,
\end{align}
where $n$ is a positive integer, $\delta^{(n)}(x)$ denotes the $n$-th derivative of the delta function, and $\mcl{P}$ stands for the principal value.
Up to order $\mcl{O}(\xi)$, the real part vanishes as
\begin{align}
    {\rm Re}[I(\xi)]
    &\simeq
    \fr{1}{4}\xi
    \mcl{P}\int_{-1}^{1}\dif v_x\lb[\fr{v_x(1-v_x^2)}{v_x^4}-\fr{3}{4}\fr{(1-v_x^2)^2}{v_x^5}\rb]
    +\mcl{O}(\xi^2)
    \nom
    &=
    \mcl{O}(\xi^2)\,.
\end{align}
The imaginary part becomes
\begin{align}
    {\rm Im}[I(\xi)]
    &\simeq
    -\fr{\pi}{16}\xi\,.
\end{align}
In the quasi-long-wavelength limit, the integrand does not have any poles, and the function gives just higher order terms as $I(\xi)\simeq\mcl{O}(\xi^{-3})$.
From the above, up to the order of interest, the function $I(\xi)$ can be expanded as
\begin{align}
    I(\xi)\simeq
    \lb\{
    \begin{array}{lc}
        \displaystyle -\im\fr{\pi}{16}\xi+\mcl{O}(\xi^2) & (\xi\ll1) \\
        \displaystyle \mcl{O}(\xi^{-3})& (\xi\gg1)
    \end{array}\rb.\,.
\end{align}

\chapter*{Acknowledgment}
\addcontentsline{toc}{chapter}{Acknowledgment}
First of all, I would like to express my deepest gratitude to my supervisor, Prof. Naoki Yamamoto.
In my third year as an undergraduate, he provided me with the opportunity to engage in studies that led to my current research theme, and from my fourth-year group assignment through to the present, I have been under his guidance for a total of eight years.
I do not believe I will ever have the privilege of being guided by a single person for such a long period, either in the past or in the future.
Through discussions with him, I have gained invaluable insights that will serve me throughout my life: curiosity toward my current research topic, a physical perspective on natural phenomena, and a dedicated approach to research.
I am sincerely grateful for his mentorship.

I would also like to express my sincere appreciation to the referees of my Ph.D. defense, Profs. Tomoharu Oka, Kazuo Hoshino, and Tetsutaro Higaki.
I am grateful to Hajime Sotani and Yoshimasa Hidaka for useful conversations and valuable comments with regard to our work.
I also would like to thank Koichi Hattori, with whom I am still conducting collaborative research, for sharing a great deal of his knowledge.

I am also deeply thankful to Di-Lun Yang and Jin Matsumoto, who were assistant professors in our group at that time.
In particular, I would like to thank Matsumoto for his extensive support, including advice on presentations, application documents, and many aspects of my daily research life.

I am grateful to Noriyuki Sogabe and Kentaro Nishimura, who are my senior colleagues in our group, for their kind advice.
I also shared many meaningful times, not only at our group events but also in everyday life, with both former and current members of the group, Akihiro Yamada, Yuka Miyauchi, Issei Seki,  Keidai Akiba, and Rento Takasaki.
I would also like to thank all other members of the theoretical physics group for enjoyable interactions.
In particular, Shumpei Iwasaki, who has walked the same path with me since our first undergraduate year, has been someone with whom I could easily talk about both research and personal matters, and who has always been an inspiring companion.
I would like to express my sincere gratitude to him here.

Finally, this Ph.D. thesis could not have been completed without my family, who have generously provided both mental and financial support and have always allowed me to follow the path I wished to pursue.
I would like to express my heartfelt thanks to them once again.

\bibliography{Ref.bib}
\bibliographystyle{unsrt} 
\addcontentsline{toc}{chapter}{Reference}

\end{document}